\documentclass[12pt]{article}
\usepackage[style=numeric, backend=bibtex]{biblatex}
\usepackage[parfill]{parskip}
\usepackage{multicol}
\usepackage{amsthm}
\usepackage{array}

\usepackage[margin=1in]{geometry}
\usepackage{amsfonts, amsmath, amssymb, enumerate, float, graphicx, lastpage, booktabs, mathtools, xcolor, caption, verbatim,enumitem}
\usepackage{hyperref}
\usepackage[margin=20pt]{subfig}

\begin{document}
\begin{center}
    {\large
    Prevalence and Propagation of Fake News\footnote[1]{Preprint submitted for publication review on 7 May 2021}}
    
    \vspace{5mm}
    
    \begin{multicols}{5}
    {\small Banafsheh Behzad$^{a,}$\footnote[2]{Corresponding author}}\\
    {\tiny Banafsheh.Behzad@csulb.edu}
    
    {\small Bhavana Bheem$^b$}\\
    {\tiny bbheem@g.hmc.edu}

    {\small Daniela Elizondo$^b$}\\
    {\tiny delizondo@g.hmc.edu}

    {\small Deyana Marsh$^b$}\\
    {\tiny dmarsh@g.hmc.edu}
    
    {\small Susan Martonosi$^b$}\\
    {\tiny martonosi@g.hmc.edu}

        \end{multicols}
    
    {\it \small   $^a$Department of Information Systems, College of Business Administration at California State University, Long Beach\\
    $^b$Department of Mathematics, Harvey Mudd College}
\end{center}

\vspace{10mm}

\hrule 

{\bf Abstract}

In recent years, scholars have raised concerns on the effects that unreliable news, or ``fake news,'' has on our political sphere, and our democracy as a whole. For example, the propagation of fake news on social media is widely believed to have influenced the outcome of national elections, including the 2016 U.S. Presidential Election, and the 2020 COVID-19 pandemic. What drives the propagation of fake news on an individual level, and which interventions could effectively reduce the propagation rate? Our model disentangles bias from truthfulness of an article and examines the relationship between these two parameters and a reader's own beliefs. Using the model, we create policy recommendations for both social media platforms and individual social media users to reduce the spread of untruthful or highly biased news. We recommend that platforms sponsor unbiased truthful news, focus fact-checking efforts on mild to moderately biased news, recommend friend suggestions across the political spectrum, and provide users with reports about the political alignment of their feed. We recommend that individual social media users fact check news that strongly aligns with their political bias and read articles of opposing political bias. 

\section{Introduction}
In recent years, social media networks such as Facebook and Twitter have become major news sources; an estimated $62\%$ of Americans claim to get news from social media (\cite{gottfried_shearer_gottfried_shearer_2017}). Concurrently, social media has seen a surge of \textit{fake news} which, for the purpose of this paper, is news containing a high frequency of falsehoods. In March and April 2021, the following examples were some of the top fake news claims circulating through social media.
\begin{quotation} 
\textit{``Per the CDC There Are Nearly Twice As Many Vaccine Related Deaths SO FAR in 2021 (1,755) Than All the Vaccine Deaths this Past Decade (994)"-March 2021, (\cite{Hoft_2021(1)}),} and \textit{``Stanford Study Results: Face masks are Ineffective to Block Transmission of COVID-19 and Actually Can Cause Health Deterioration and Premature Death."-April 2021 (\cite{Hoft_2021(2)})}
\end{quotation}

\noindent Both of these claims were proven to be false, and their only purpose is to misinform people. Considering the high ramifications of just one of these claims propagating, mitigating the spread of untruthful news is of paramount importance (\cite{grinberg_2017, lazer_baum_benkler_2018, vicario_bessi_zollo_petroni_scala_caldarelli_stanley_quattrociocchi_2016}).

Social media has the unique characteristic of its users playing a decisive role in which content gets propagated. Alarmingly, research shows that users are less likely to fact check the news they receive on social media than the news they receive through other sources (\cite{allcott_gentzkow_2017, Jun5976}). Furthermore, fake news takes advantage of the political echo chambers within social media to further its propagation (\cite{vicario_bessi_zollo_petroni_scala_caldarelli_stanley_quattrociocchi_2016}). With active users continuously making decisions on whether to propagate news (e.g. share, re-tweet), unverified and unreliable sources can potentially reach as large an audience on social media as national news sources (\cite{allcott_gentzkow_2017}).

Current techniques for mitigating the spread of fake news include increasing the spread of truthful news and flagging untruthful news articles or accounts (\cite{pmlr-v70-farajtabar17a, lazer_baum_benkler_2018}). The former technique increases the amount of truthful news in a social media user's feed, and the latter diminishes the credibility of fake news sources. One major source of fake news propagation are social bots, computer generated and controlled social media accounts. Since social bots target particular social media users whose usage characteristics make them more likely to further propagate fake news (\cite{Sharma_2019, shu_bernard_liu_2018}), one tactic for fake news mitigation is having more methods of detecting the difference between bot accounts and human accounts in order to deactivate bot accounts (\cite{chu_gia_wan_2012, DBLP:journals/corr/abs-1805-10244, inu_lip_kor_2018}). Although current techniques are beneficial, social media users are still being exposed to fake news during crucial times like the 2016 US presidential election and the 2020 COVID-19 pandemic (\cite{allcott_gentzkow_2017, mikkelson_2020}). 
 
The initial step in finding techniques to mitigate fake news is understanding how fake news spreads on both the macroscopic and microscopic levels. The macroscopic spread refers to the network level propagation of content. One study uses a network graph model to analyze how people believe and spread news; they conclude that social media cultivates echo chambers in which low quality news is likely shared (\cite{Brooks2019AMF}). Vicario \textit{et al.} find that cascade dynamics of fake news depend on polarized echo chambers (\cite{vicario_bessi_zollo_petroni_scala_caldarelli_stanley_quattrociocchi_2016}). Since early detection is vital to diminishing the effects of fake news, Louni and Subbalakshmi focus on detecting the source of news within a cluster (\cite{louni_subbalakshmi_2014}). Other studies effectively detect fake news on a network level by using propagation paths to classify different types of news (\cite{AAAI1816826, pierri2019topology}). Vosoughi \textit{et al.} graph data from Twitter and conclude through that fake news spreads faster and more broadly than true news when analyzing a social media network (\cite{vosoughi_roy_aral_2018}). Jang \textit{et al.} utilize evolutionary tree analysis to similarly conclude that fake news spreads deeper through social media networks and undergoes frequent modifications (\cite{jang_geng_li_xia_huang_kim_tang_2018}). Research on the macroscopic spread of content provides an understanding of which clusters facilitate fake news spread and also of news propagation patterns. Both findings inform techniques for early detection of fake news which is crucial for mitigation. 

Research on content spread at the microscopic level focuses on an individual user's decision to further propagate information from their social media feeds. Findings at the microscopic level on fake news propagation do not necessarily align with findings at a macroscopic level since an individual level decision to share news does not take into consideration the network effects. Some microscopic level studies utilize user surveys to understand user behavior, and other studies create models to pinpoint social media attributes that correlate to a higher likelihood of propagating fake news. Guess \textit{et al.} use Poisson regressions to link a representative online survey to the respondents' sharing history on Facebook. They find that fake news is more likely to propagate through conservatives and older users (\cite{gue_nag_tuc_2019}). Allcott and Gentzkow similarly conduct surveys on a portion of the voting population and find that people with higher education, who spend more time on social media, and have no partisan attachment are the most accurate in distinguishing fake from real news (\cite{allcott_gentzkow_2017}). Furthermore, Papanastasiou utilizes an agent-based sequential model to conclude that social media users tend to correlate popularity with credibility (\cite{papa_2017}). Lee \textit{et al.} successfully create a feature-based prediction model and a time estimation model to characterize the likelihood an account retweets a particular message based on certain traits of the Twitter account, such as its sociability and follower base (\cite{Lee:2014}). Characterizing the probability that a social media user will further spread untruthful or polarizing news based on individual factors, like political ideology, is vital in understanding how to stop the spread. 

This paper further expands the research within the microscopic level. We develop a model for the probability an individual social media user with a certain political belief chooses to share a news article having a certain political bias and truthfulness. Different from previous work, our model defines both truthfulness and bias on a continuous numerical spectrum and also does not conflate them for a given article. Our model estimates the probability that a particular population will share a certain article. Numerical probabilities provide a more accurate understanding of how different social media populations share content relative to each other which helps inform recommendations for platforms and individual users to reduce the spread of untruthful or highly biased news. 

From the analysis of the data, we find that if social media users engage only with users of similar political beliefs, then the probability of sharing news, fake or true, increases. If users engage with users of opposing political beliefs, then the probability of sharing decreases, and the decrease is more pronounced for fake news. This paper recommends social media platforms: sponsor unbiased truthful news; focus existing fact-checking and flagging algorithms on moderately liberal- and conservative-biased news, rather than flagging extremely biased articles; suggest content and connections from across the political spectrum; and supply social media users with a score reflecting the political alignment of their feed. For individual users, this paper recommends fact-checking news that aligns strongly with their belief, as people are more susceptible to propagating this type of news, and to engage with users with opposing political beliefs.

The methodology of this paper follows a three step approach: developing a probabilistic model for the propagation of political news; analyzing the model to determine which attributes of content and user population contribute to the propagation of untruthful or biased content; finally, using the findings to recommend interventions against fake news. In the next section, we explain our model and outline the assumptions that are built into it.  Section \ref{s:optbehav} analyzes the model from an optimization perspective to determine characteristics of content that a malicious agent seeking to propagate untruthful or highly biased news might target. Section \ref{s:data} details the data set used to validate our model. Section \ref{s:empirical} presents an empirical analysis of the propagation rate of political news within populations of readers that have varying political beliefs. The analyses of Sections \ref{s:optbehav} and \ref{s:empirical} inform the recommended techniques for mitigating the spread of fake news discussed in Section \ref{s:discussion}. Lastly, we provide our ideas for future work and conclusions in Sections \ref{s:future} and \ref{s:conclude}, respectively.

\section{Probability Model of Propagation}
\label{S:probmodel}
This section describes how we model the probability $p$ that a reader propagates an encountered article (e.g. ``share," ``retweet," ``like," or ``upvote," depending on the social media platform of interest) as a function of the article's bias and truthfulness, and the reader's belief.

An article $i$ is characterized by two attributes: bias $b_i$ and truthfulness $t_i$. Bias is a continuous value on $[-1,1]$ which reflects the political ideology represented by the topics, facts, and claims included in the article. A bias equal to $-1$ corresponds to a very liberal (``left-biased'') article, a value of $+1$ corresponds to a very conservative (``right-biased'') article, and 0 corresponds to a politically unbiased article. The truthfulness $t_i$ representing both the frequency and severity of falsehoods in the article, ranges from $0$ (completely false news) to $1$ (perfectly true). The article's reader $j$ is characterized by their political belief $B_j$ which also ranges from $-1$ (very liberal, or ``left-political'') to $+1$ (very conservative, or ``right-political'').

The shape of the function is based on  two assumptions, which are informed by previous works, as noted below: 

\begin{enumerate}
    \item A person will be more likely to propagate an article when its bias is close to their beliefs, even when the article exhibits low truthfulness (\cite{bak_mes_ada_2015, grinberg_2017, Iyengar_Hahn_2009, manickam_lan_dasarathy_baraniuk_2019}). That is, $p(b_i, t_i, B_j)$ should increase as $\lvert b_i - B_j \rvert$ decreases.  Users often share content as a means of political persuasion, and we assume that a user would not wish to disseminate content that could persuade others to adopt a dramatically different political belief.

    \item The probability of propagation increases with article truthfulness, $t_i$. Articles that are too untruthful are unlikely to be propagated because they are too outlandish (\cite{grinberg_2017}). In other words, we assume that given two articles exhibiting the same bias, a reader will be more likely to share the more truthful article\footnote{Readers may not always be able to distinguish truth from untruth with high precision.  However, we are modeling the \textit{likelihood} of sharing an article.  External factors such as the reputability of the source can serve as signals to readers as to the truthfulness of content.}. However, we also assume diminishing marginal increases in sharing rate as $t_i$ increases.  Because probabilities are bounded by 1, it is natural to assume this rate of increase will decline.
    
\end{enumerate}

\begin{figure}[H]
\centering
    \subfloat[Probability of sharing as a function of truthfulness; $B_j$ = 0.45, $f_r$ = 1, $k_r$ = 10]{
    \includegraphics[width = 0.32\textwidth]{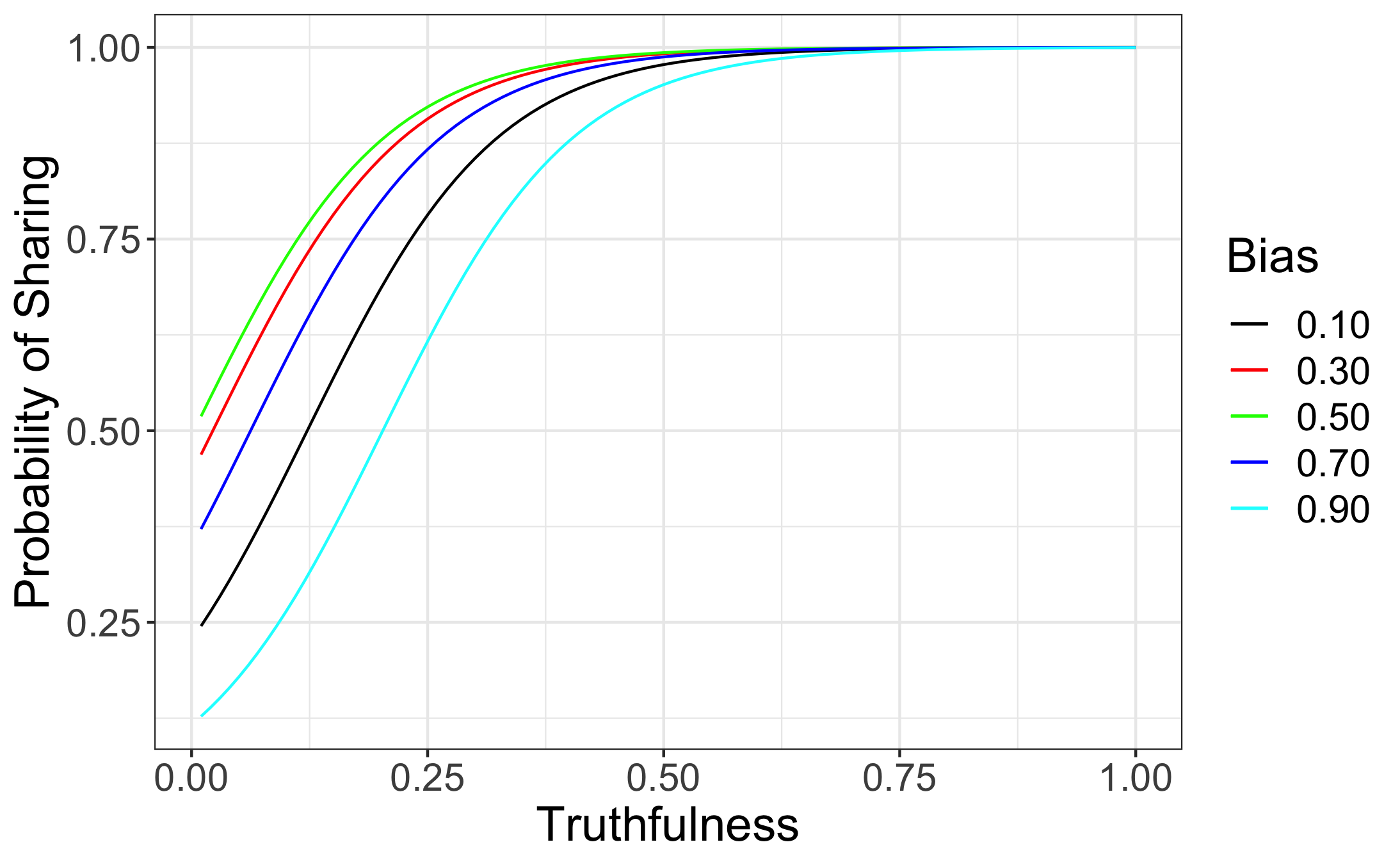}
    }
    \subfloat[Probability of sharing as a function of bias; $B_j$  = 0.45, $f_r$ = 1, $k_r$ = 10]{
    \includegraphics[width = 0.32\textwidth]{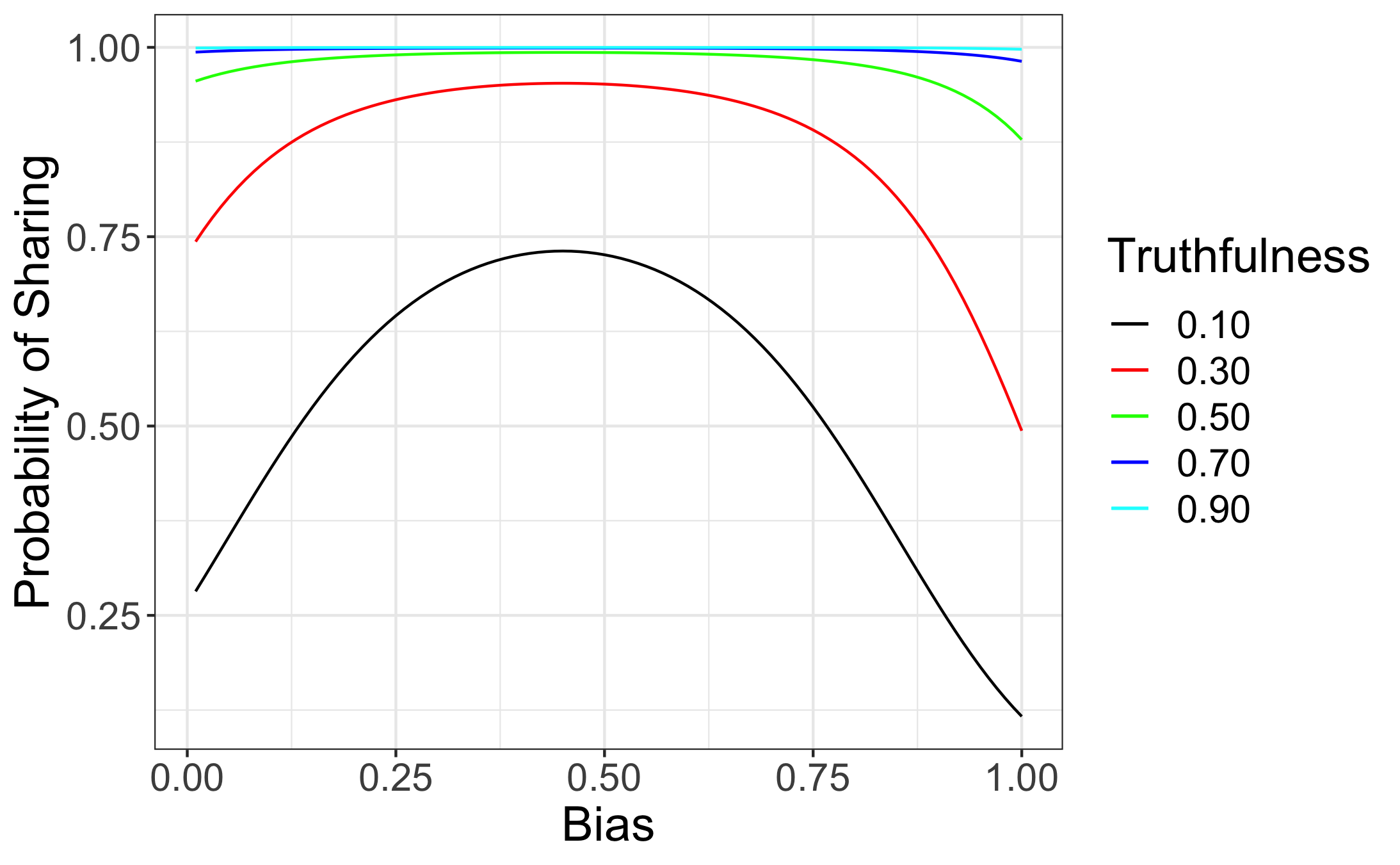}
    }
    \subfloat[Probability of sharing as a function of reader belief; $t_i$ = 0.45, $f_r$ = 1, $k_r$ = 10]{
    \includegraphics[width = 0.32\textwidth]{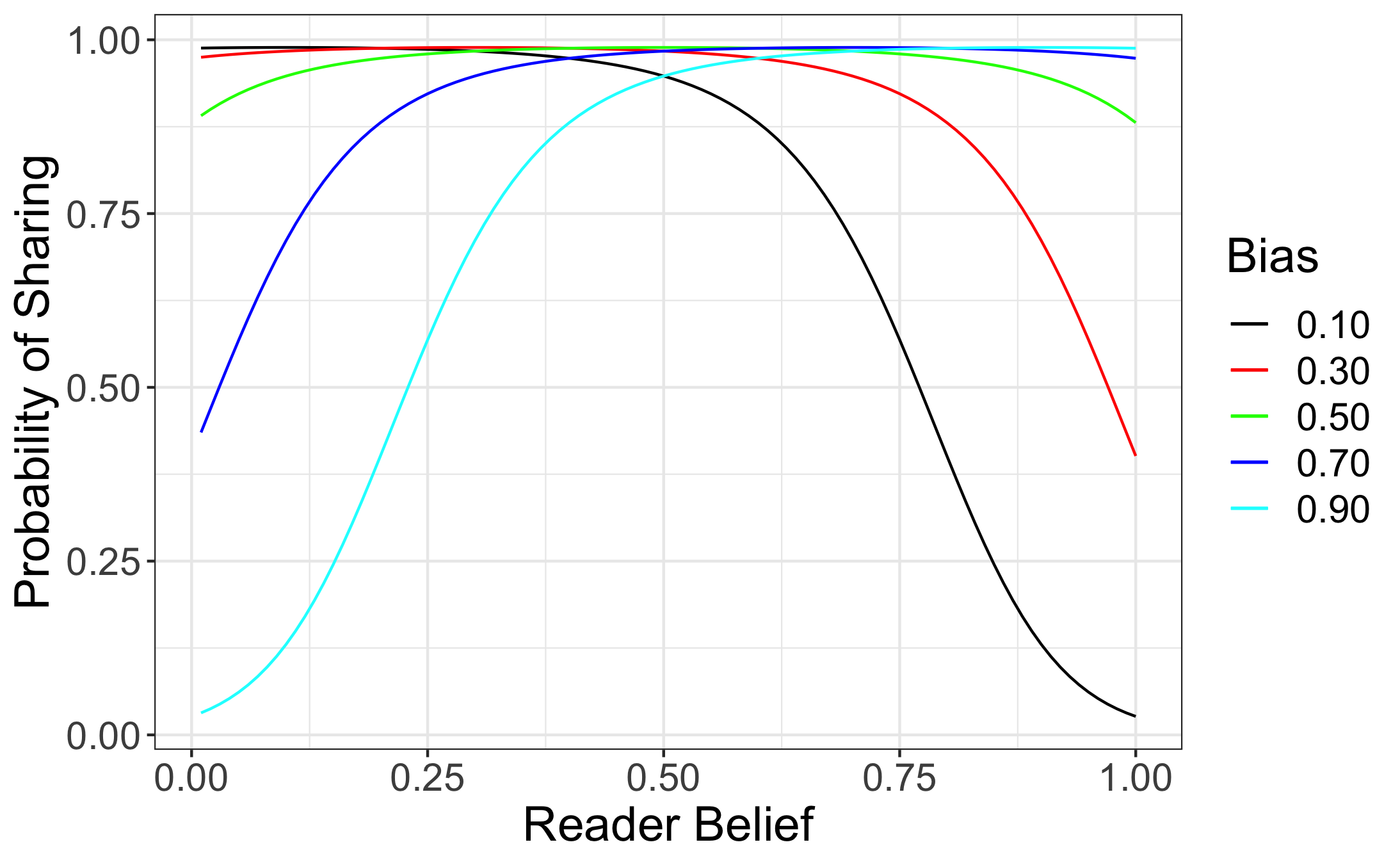}
    }
    \caption{Level curves of the probability model of equation (\ref{eq:probfcn})}
    \label{fig:levelCurves}
\end{figure}

\noindent Additionally, we require a valid probability function, bounded between $0$ and $1$ for $(b_i, t_i, B_j) \in [-1,1]\times[0,1]\times[-1,1]$.  A probability function that satisfies these assumptions and is reasonably tractable is the logistic curve:
\begin{equation} \label{eq:probfcn}
p(b_t, t_i, B_j) = \begin{cases} 
        \frac{f_l}{1+e^{-k_l(t_i-(b_i-B_j)^2)}} & \quad \mbox{ if } B_j < 0 \\
        \frac{f_r}{1+e^{-k_r(t_i-(b_i-B_j)^2)}} & \quad \mbox{ if } B_j \geq 0 
        \end{cases}
\end{equation} where $f_l, f_r, k_l,$ and $k_r$ are scaling parameters.  $f_l$ and $f_r$ scale the overall rate of sharing for left-political and right-political readers, respectively.  $k_l$ and $k_r$ scale the rate of decay in sharing probability associated with the squared deviation between article bias and reader belief and the truthfulness of the article. Several plots of level curves of the probability function are shown in Figure \ref{fig:levelCurves}.

Because of its tractability and suitability in modeling a probability function satisfying our assumptions, we will use the model given by equation (\ref{eq:probfcn}) throughout the remainder of the paper.

\section{Characterizing the Optimal Behavior of a Malicious Agent}
\label{s:optbehav}

We first examine the model from a mathematical perspective to characterize the optimal behavior of a malicious agent wishing to spread ``fake'' news. We assume that a malicious agent seeks to propagate untruthful (low $t_i$) or highly biased (high $|b_i|$) news at the highest possible rate, assuming a single reader having belief $B_j$ (in Section \ref{ss:singlereader}), or a population of readers having a belief distribution (in Section \ref{ss:distnreaders}).  Specifically, we answer three questions:
\begin{enumerate}
    \item What general characteristics of articles (as reflected by $(b_i, t_i)$ pair) achieve the highest propagation rate?
    \item For untruthful articles, what degree of bias achieves the highest propagation rate?
    \item For highly biased articles, what degree of truthfulness achieves the highest propagation rate?
\end{enumerate}  By answering these questions, we can understand how a malicious agent might select or create political content to maximize propagation by readers of a particular belief.  Content that bears these optimal characteristics could then be prioritized for fact-checking or other interventions.

Although not an assumption integrated into the model, for the purposes of optimization, this section assumes that more biased news is inherently less truthful (\cite{xiang_sarvary_2007}). Thus, we constrain the feasible region of valid $(b_i, t_i)$ pairs to the region $\lvert b_i \rvert + t_i \leq 1$. This assumption is not only reasonable but also necessary to make the optimization nontrivial; otherwise, the optimal strategy for maximizing propagation would always be to produce perfectly truthful articles, which contradicts what we observe on social media.  

\subsection{Single Reader} \label{ss:singlereader}

We first address each of the three questions above in the context of a malicious agent seeking to propagate content to a single reader having belief $B_j$.

\begin{itemize}
    \item \textbf{What general characteristics of articles achieve the highest propagation rate? (Single reader case)}

The probability function of equation (\ref{eq:probfcn}) has no local maximum, indicating that the global maximum will lie along the boundaries of the feasible region defined by $\lvert b_i \rvert + t_i \leq 1$, $-1 \leq b_i \leq 1$, and $0 \leq t_i \leq 1$. To find the bias and truth combination that maximizes propagation for a single reader having a given belief, we optimize the function on all three boundaries separately. 

For the constraint $t_i = 1-\lvert b_i \rvert $, the function's maximum is a piecewise function split at $ \lvert B_j \rvert = 0.5$. For extreme conservative beliefs ($B_j \geq 0.5) $, the function's maximum is at $b_i = - \frac{1}{2} + B_j$ and $t_i = \frac{3}{2} - B_j$. As the function is symmetric, for extreme liberal beliefs ($B_j \leq - 0.5$), the function's maximum is at $b_i = \frac{1}{2} + B_j$ and $t_i = \frac{3}{2} + B_j$. When reader belief is more moderate, i.e., $\lvert B_j \rvert < 0.5$, the maximum value of the function occurs at maximum truthfulness: $b_i=0$ and $t_i = 1$. Although the probability of sharing increases as $\lvert b_i - B_j \rvert$ decreases, the constraint $ \lvert b_i \rvert + t_i \leq 1$ causes the optimal article bias to be substantially lower in magnitude than the reader's belief. 

On the constraint boundary $b_i = 0$, the maximum propagation rate is attained when $t_i=1$. Unbiased news achieves maximum propagation when it is perfectly truthful. On the other hand, along the constraint $t_i=0$, the maximum is at $b_i = B_j$. Untruthful news achieves maximum propagation when its bias matches the belief of the reader. 

From this analysis, we provide the first of our recommendations for social media platforms (P) and users (U)\footnote{A summary of all recommendations made in this paper can be found in Table \ref{t:recommend} in Section \ref{s:discussion}.}.

\textit{Recommendation P1: Populate users' feeds with unbiased truthful news as sponsored articles.}  Such articles achieve the highest propagation rate among users with moderately biased political views.

\item \textbf{For untruthful articles that have low $t_i$, what degree of bias achieves the highest propagation rate? (Single reader case)}

We saw above that when $t_i=0$, maximum propagation is achieved when $b_i = B_j$.  This holds more generally for fixed $t_i$, provided that $t_i + |B_j| <= 1$.  When $t_i + |B_j| > 1$, there is no local maximum along the cross sectional probability curve.  Thus, maximum propagation is attained at the endpoint $|b_i| = 1-t_i$ (with the sign of $b_i$ chosen to match the sign of $B_j$).  To propagate a low-truth article to a single user, the malicious agent will try to match the article's bias to the user's belief, unless the user's belief is so extreme that doing so would violate the truth-bias constraint. In this latter case, the optimal bias of the article will be chosen along the constraint boundary (i.e., as extreme as possible).  From this, we infer our next recommendation.

\textit{Recommendation U1: Fact check news that aligns strongly with a user's beliefs by using a third party fact-checking website for individual facts within the article.}  Because a malicious agent acting optimally would align false content with a user's political beliefs, users are more susceptible to sharing untruthful news that aligns with their beliefs.

\item \textbf{For polarizing articles that have high $|b_i|$, what degree of truthfulness achieves the highest propagation rate? (Single reader case)}

For any fixed reader belief and article bias, the probability of sharing an article increases with article truthfulness, by assumption.  Thus in our constrained optimization framework, the malicious agent wishing to propagate a highly biased article would select article truthfulness along the constraint boundary: $t_i = 1-|b_i|$.
\end{itemize}

\subsection{Distribution of Readers}
\label{ss:distnreaders}
Expanding on the case of a single reader, we examine the maximal propagation rate over a distribution of readers. To start, given a discrete distribution of reader belief $f(B_j)$, the probability of a reader sharing an article having bias $b_i$ and truthfulness $t_i$ is given by: 
\begin{equation} 
p(b_i, t_i) = \sum_{B_j} {p(b_i, t_i, B_j) * f(B_j)}.
\label{eq:probofbiandti}
\end{equation}
With the probability function described in equation (\ref{eq:probofbiandti}), we  solve numerically for the values of $b_i$ and $t_i$ that attain maximum propagation over systematically varied distributions of user belief having a variety of expectations and variances. For each distribution, we find the combination of article bias and truthfulness that maximizes the probability of propagation to answer each of the three questions posed earlier.     We use the BARON solver (\cite{sahinidis:baron:17.8.9, ts:05}) on the NEOS server (\cite{czyzyk_et_al_1998, dolan_2001, gropp_more_1997}) to obtain the results. 

\begin{itemize}
\item \textbf{What general characteristics of articles achieve the highest propagation rate? (Distribution of readers case)}

Plots of numerical solutions, shown in Figure \ref{graph:ampldist}, display the effects of the reader belief distribution's expected value and variance on both optimal bias and the optimal probability of sharing.  As the expected value of $B_j$ increases, the optimal bias increases and the probability of sharing decreases.  Additionally, in all cases examined, the optimal truthfulness of the article was found to lie on the constraint boundary $|b_i| + t_i = 1$. Interestingly, the relationship between the expected value and optimal bias is approximately the same as the closed form solution for the single reader case, but variance also affects optimal bias. As the variance of the distribution of $B_j$ increases, the optimal bias increases modestly for expected values greater than $0.5$, and the probability of sharing decreases for all expected values.

Thus, the propagation-maximizing article parameter values for the population of readers resemble those of the case of a single reader. However, increasing the variance of reader beliefs decreases the overall sharing probability and modestly increases the bias required to attain maximum propagation. This allows us to make our second recommendation for social media platforms:

\textit{Recommendation P2: Recommend friends and connections across the political spectrum.}  Social media platforms generally recommend connections, content and advertising based on similarities.  The opposite tendency, however, may help reduce the sharing rate of biased, untruthful news by increasing the variation in political belief of a user population.

\begin{figure}
    \centering
    \subfloat[Optimal article bias $b_i$ as a function of the expectation ($x$-axis) and variance (gradient) of the distribution of reader beliefs]{\includegraphics[width = 0.45\textwidth]{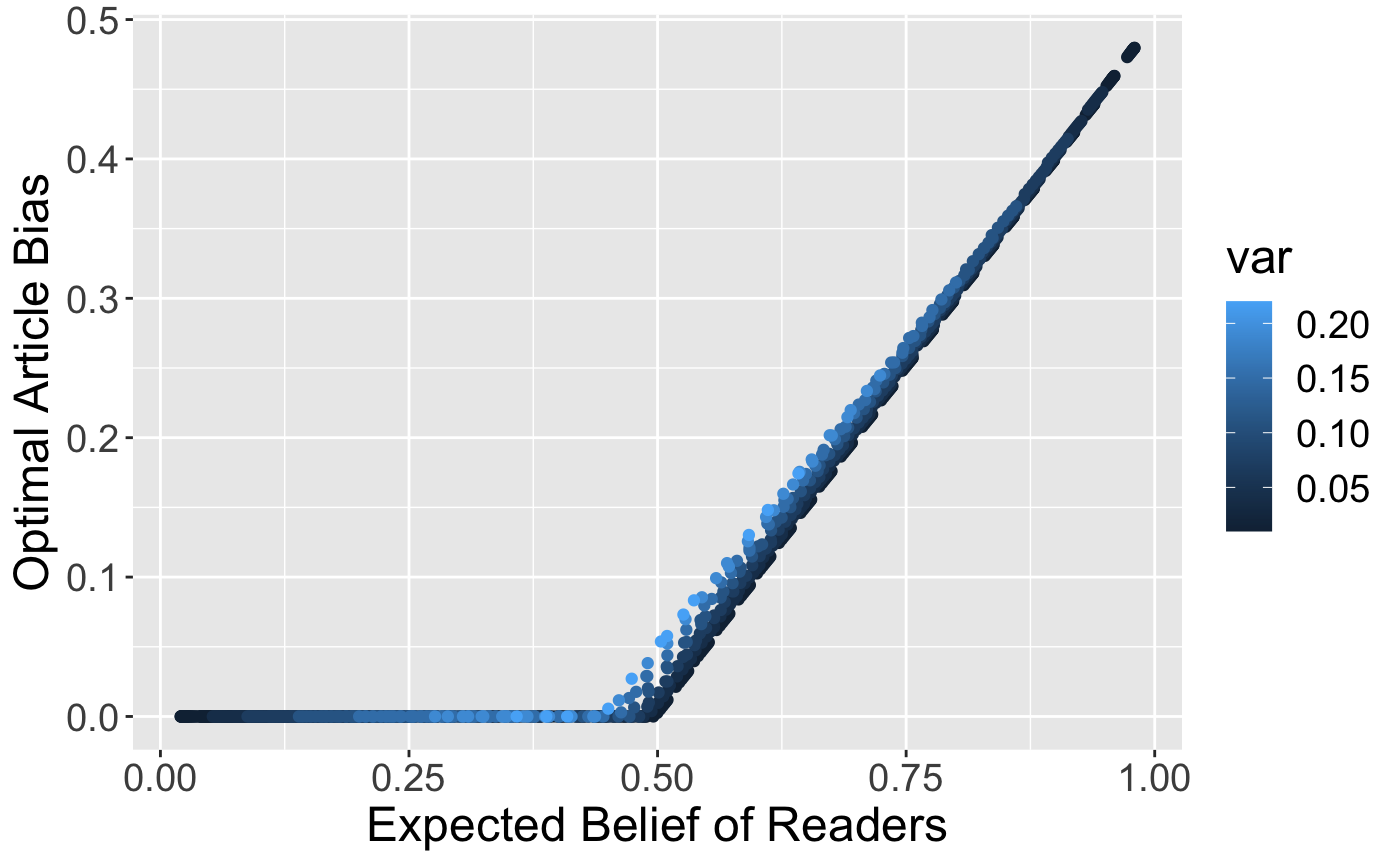}}
    \subfloat[Probability of sharing an article as a function of expectation ($x$-axis) and variance (gradient) of the distribution of reader beliefs]{
  \includegraphics[width = 0.45\textwidth]{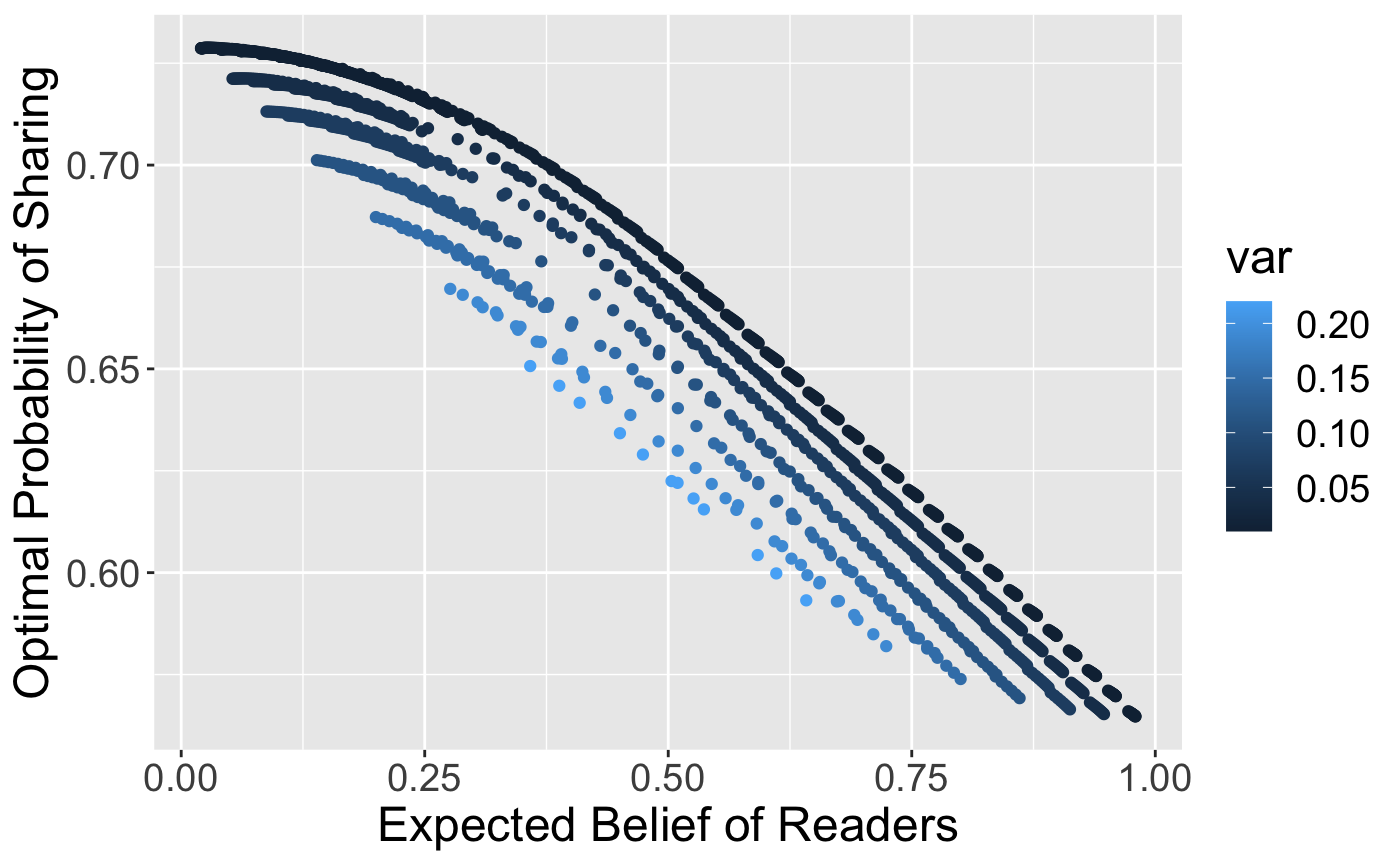}}
    \caption{Optimal article bias and probability of sharing over all $(b_i, t_i)$ pairs, for numerous discrete probability distributions of reader belief having given expected value and variance.}
  \label{graph:ampldist}
\end{figure}

\item  \textbf{For untruthful articles that have low $t_i$, what degree of bias achieves the highest propagation rate? (Distribution of readers case)}

In the single-user case, we saw that for content having a low $t_i$ the choice of $b_i$ that maximizes propagation is equal to the user's belief $B_j$ unless constrained by the boundary $|B_j|+t_i \leq 1$.  For the case of a distribution of readers, Figure \ref{graph:ampldistlowtruth}(a) displays the optimal article bias, when $t_i = 0.1$ as a function of the expected reader belief, shaded by the distribution variance; Figure \ref{graph:ampldistlowtruth}(b) does the same for optimal probability of sharing.  We see that when truth is fixed, the optimal bias equals $E[B_j]$ until the constraint boundary is reached, at which point the optimal bias equals $1 - t_i$.  The optimal bias is unaffected by variance.  As in Figure \ref{graph:ampldist}(b), we see that the optimal sharing rate decreases as reader belief variance increases.  Once again, we conclude that recommending contacts across the political spectrum could reduce the rate at which untruthful content is shared (Recommendation P2).

\begin{figure}
    \centering
    \subfloat[Optimal article bias $b_i$ with low $t_i$ as a function of the expectation ($x$-axis) and variance (gradient) of the distribution of reader beliefs]{\includegraphics[width = 0.45\textwidth, scale = 0.25]{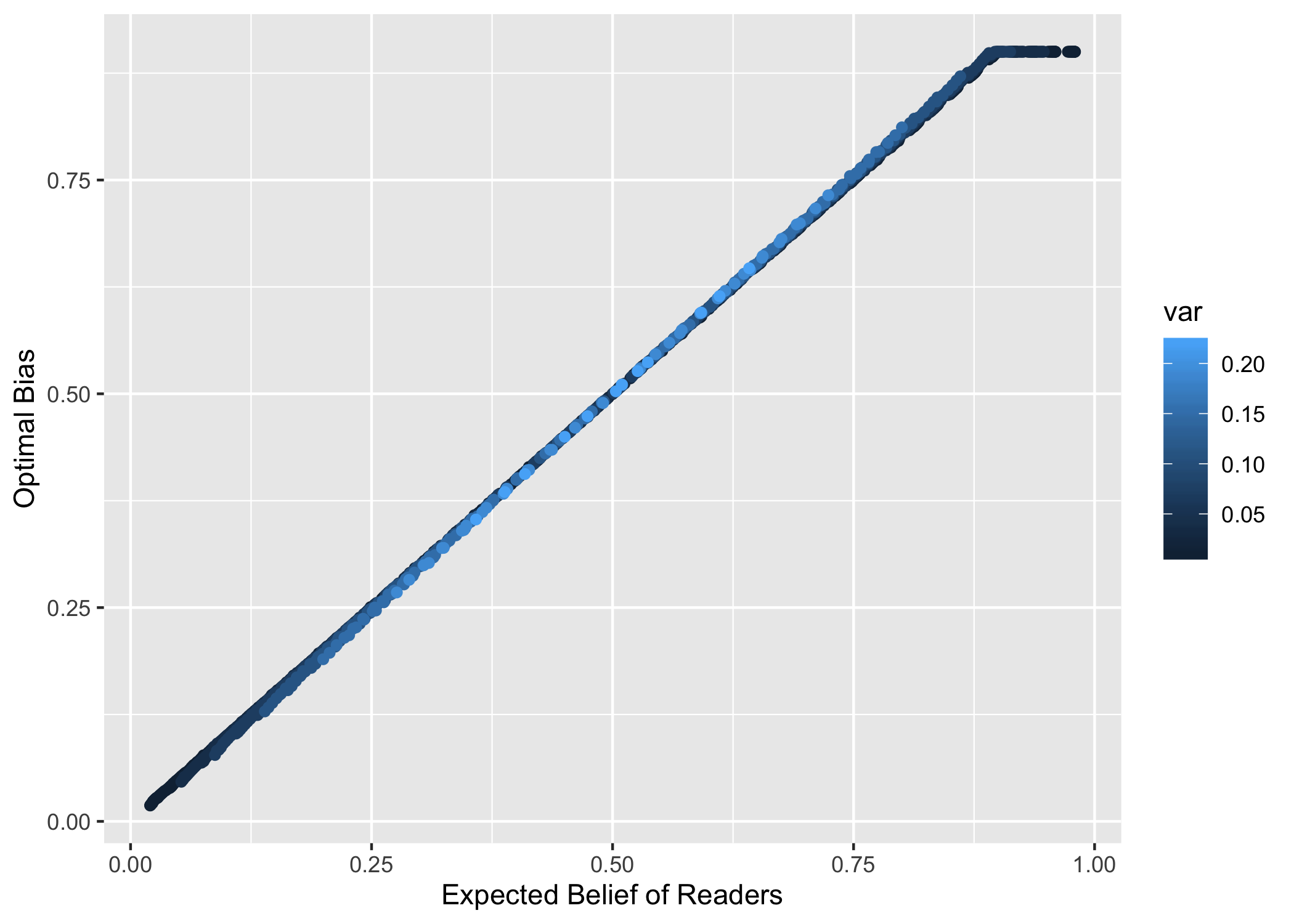}}
    \subfloat[Probability of sharing an article as a function of expectation ($x$-axis) and variance (gradient) of the distribution of reader beliefs]{
  \includegraphics[width = 0.45\textwidth, scale = 0.25]{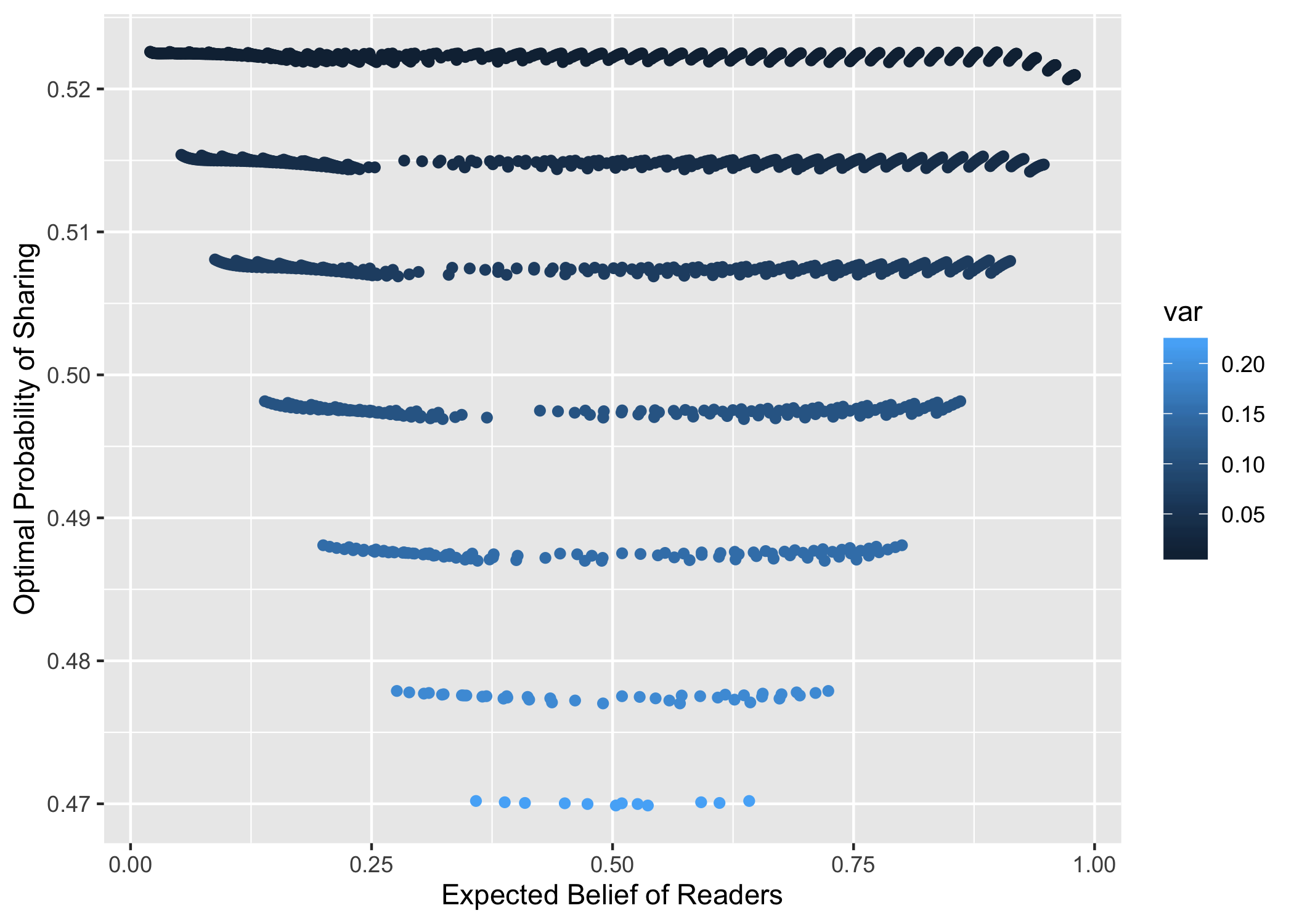}}
    \caption{Optimal article bias and probability of sharing over all $b_i$ values when $t_i = 0.1$, for numerous discrete probability distributions of reader belief having given expected value and variance}
    \label{graph:ampldistlowtruth}
\end{figure}

\item  \textbf{For polarizing articles that have high $|b_i|$, what degree of truthfulness achieves the highest propagation rate? (Distribution of readers case)}

As in the single reader case, sharing probability, for any value of $|b_i|$, increases with $t_i$, independent of the beliefs of the readers.  Thus, for highly biased content, the degree of truthfulness achieving the highest propagation rate lies along the constraint boundary, $t_i = 1 - |b_i|$.

\end{itemize}

Having explored the mathematical properties of our model in the context of a malicious agent seeking to propagate untruthful or polarizing content, we now introduce the data set we use to conduct an empirical analysis of the model.

\section{Model Validation}
\label{s:data}

We validate our model using social media data from \cite{grinberg_2017} and Media Bias Monitor (MBM) (\cite{rib_hen_ben_2018}) to estimate sharing probabilities and the three model parameters: bias, truthfulness and reader belief.  We then compare the proposed probability function to the propagation rates of articles found on social media and verify that the model assumptions and general shape of the functional curve are satisfied.

\subsection{Data}
\label{ss:data}
To obtain a set of readers along with their political beliefs, we utilize the data from Grinberg \textit{et al.} which contains files with information about \textit{panel members}, a set of Twitter users whose Twitter profiles can be uniquely linked to public voting records. There are 572 panel members, who are categorized by political affiliation: extreme left, left, lean left, center, lean right, right, or extreme right. In the original data set, the categories lean left and left were combined to form the category left, and the same was done for the right side; the groups are not merged together in this paper.  We assign a belief value $B_j$ to each panelist equal to the average political affiliation of reader $j$'s assigned belief group, which can be found in the last column of Table \ref{tab:dist}.  

In the raw data set, the panel members are heterogeneous in their social media behaviors, with some panelists categorized as \textit{supersharers} (sharing more than 922 political articles), \textit{superconsumers} (having more than 45,128 exposures to political articles), \textit{bots} (nonhuman automated users responsible for mass distribution of content), and \textit{apolitical} (having fewer than 100 exposures to political articles).  As in Grinberg \textit{et al.}, to mitigate against outliers, we filter out these users, leaving a more homogeneous population of panel members for our analysis.  As shown in Appendix \ref{ss:apptypical}, the overall sharing rate of the excluded panelists (governed by parameter $f$ in Equation (\ref{eq:probfcn})) is statistically different from that of the remaining panelists.  However, the model parameter $k$ governing the sharing probability curve's overall shape is not statistically different between the excluded and remaining panelists.  Therefore, to simplify our analysis we restrict our data to panel members of an average activity rate.

The data from Grinberg \textit{et al.} also holds a collection of tweets containing URLs and the political news sources of the URLs. The data has 3,141,106 tweets that the panel members were potentially exposed to, meaning the tweet was shared by an account the user follows. The URLs from these tweets are given unique identifiers making the actual link anonymous. Thus, it is impossible to attribute $b_i$ and $t_i$ values to individual URLs shared by the panel member. Instead, we use the recorded domain names of the news website from which the URL originates (\cite{grinberg_2017}). We make the assumption that every article on a particular news website has the same bias and truthfulness value as the entire domain. We can make this assumption without losing much detail, as many news sources have an editorial bias and targeted audiences which influence their published articles (\cite{budak_goel_rao_2016, grinberg_2017}).  As with user political beliefs, we categorize the article bias as extreme left (i.e., very liberal, having values of $b_i$ close to $-1$), left, lean left, center, lean right, right, or extreme right (i.e., very conservative, having values of $b_i$ close to $+1$).

To our knowledge there is currently no established research that calculates a numeric value for the content-based semantic bias for a wide range of domains. Therefore, we have to use other ways of estimating bias of an article. The Grinberg \textit{et al.} data includes \textit{alignment} scores, the weighted average of panelists exposed to a website that are registered with either the Democratic or Republican party, for only 245 distinct websites. Instead, we estimate article bias using MBM, a database that has over 20,000 news websites (\cite{rib_hen_ben_2018}). Ribeiro \textit{et al.} develop MBM as a novel scalable way of calculating domain bias that uses audience demographics (\cite{rib_hen_ben_2018}). Ribeiro \textit{et al.} find correlation coefficients between their bias estimates and those of Budak \textit{et al.} who use machine learning and crowdsourcing techniques to determine the bias of articles from fifteen domains (\cite{budak_goel_rao_2016}). The strong correlation makes MBM's bias estimates a sufficient proxy for article bias, $b_i$.

In order to estimate a numerical truthfulness score for each URL, we expand on color codes provided by Grinberg \textit{et al.} that assess the overall truthfulness of the domain (\cite{grinberg_2017}). Grinberg \textit{et al.} provide justifications for each domain's assigned coding which we use to develop a numerical conversion for each color. Domains that are coded as black are given a truthfulness value of zero. All other domains are assigned a value of $t_i$ from a range of values. These conversions are shown in Table \ref{table:convertcolor}. Note that a truthfulness rating above green refers to domains with rigorous editorial processes, such as academic journals. Since the data set did not have such domains, we assign no domain a $t_i$ in the interval $[0.8, 1]$. By assigning truthfulness ratings to each justification from Grinberg \textit{et al.}, we are able to average the domain's justification ratings to assign a $t_i$ value to each URL.  Lastly, for the purpose of the panel plot given in Figure \ref{graph:ppoverbelief}, we redistribute the $t_i$ values into the truthfulness categories of very low, low, mixed, high, and very high.

\begin{table}[H]
\centering
\begin{tabular}{ |p{2cm}|p{8cm}|p{2cm}|p{3cm}|}
 \hline
 Truthfulness Color & Grinberg \textit{et al.} Description (\cite{grinberg_2017}) & Numeric Range & Figure \ref{graph:ppoverbelief} Category \\
 \hline
 \hline
 Black   & ``Sites that published almost exclusively fabricated stories"   &   $0.0-0.1$ & Very Low\\
  \hline
 Red &   Sites that ``spread falsehoods that clearly reflected a flawed editorial process"   & $0.1-0.2$ & Low \\
  \hline
 Orange & Sites ``where annotators were less certain that the falsehoods stemmed from a systematically flawed process" &  $0.2-0.3$ & Low \\
  \hline
 Yellow & ``Mild or rare inaccuracies" & $0.3-0.6$ & Mixed ($0.3-0.5$) High ($0.5-0.6$)\\
  \hline
 Green  & ``Factual and sourced reporting"  & $0.6-0.8$ & High ($0.6-0.7$) Very High ($0.7-0.8$)\\
 \hline
\end{tabular}
\caption{Conversion of colors provided in Grinberg \textit{et al.} (\cite{grinberg_2017}) into a numerical range compatible with the truthfulness scale of the model.  To evenly distribute the data, the numerical scale was further categorized into truthfulness bins for the purposes of Figure \ref{graph:ppoverbelief}}
\label{table:convertcolor}
\end{table}

Lastly, we estimate the probability of propagation by dividing the number of people in each political affiliation group who shared URLs from a given website by the number of people in that group who were exposed to URLs from that website. The data from Grinberg \textit{et al.} categorizes tweets as being a retweet, quote, or original posting. On Twitter, retweets and quotes function as shares. When a URL is shared as an original post on Twitter, there is no record of the level of exposure the user had to the news source. So, in calculating the probability, we consider only retweets and quotes. The panel members shared 1,566 URL links from thirty-seven distinct websites.

\subsection{Validating Model Assumptions}
\label{ss:validate}

\begin{figure}[h]
\includegraphics[width =\textwidth]{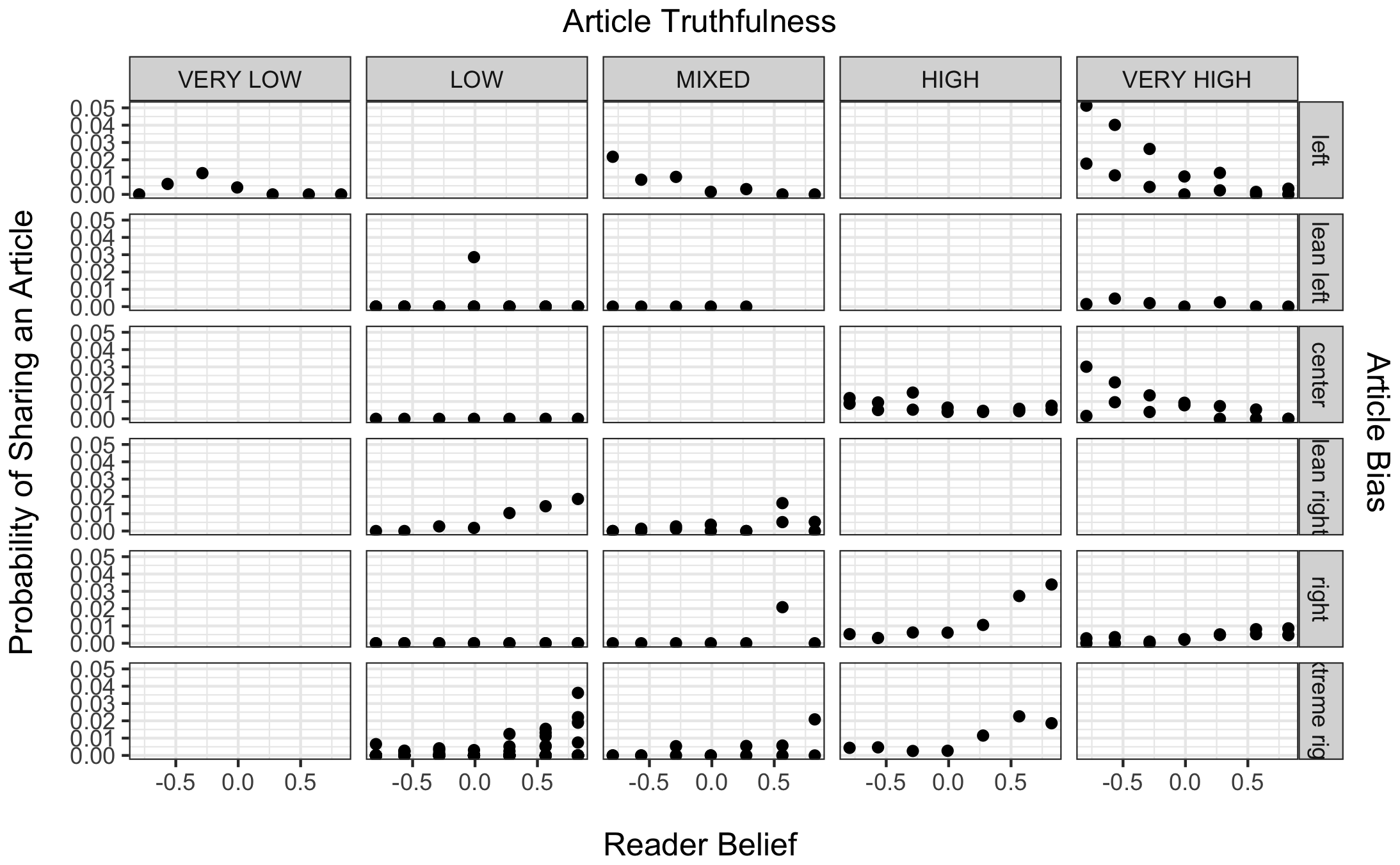}
\caption{Probability of sharing an article depending on the political belief of a reader ($x$-axis), truthfulness of an article (columns), and bias of an article (rows) (\cite{grinberg_2017, rib_hen_ben_2018}) }
\label{graph:ppoverbelief}
\end{figure}

The data set, as described above, is used to validate the assumptions built into the model. Figure \ref{graph:ppoverbelief} plots the propagation rate as a function of reader belief, across combinations of article truthfulness and bias. We use these plots as well as statistical comparisons to assert the validity of our assumptions outlined in Section \ref{S:probmodel}. 

The first assumption is that the probability of sharing increases as the bias of an article becomes more aligned with reader belief.  This can be seen in Figure \ref{graph:ppoverbelief} as panels in the top two rows (corresponding to liberal article bias) show a decreasing trend in sharing rate as reader belief increases (becomes more conservative), and panels in the bottom three rows (corresponding to conservative article bias) show an increasing trend as reader belief becomes more conservative.  A statistical analysis of the slope coefficient describing sharing rate as a function of $|b_i - B_j|$ reveals that this slope is statistically negative for both liberal ($p$-value: 0.00002) and conservative readers ($p$-value: 0.00001) further supporting this assumption. Moreover, readers are less likely to share an article having an opposing political bias. The sharing rate of liberals sharing liberal news is statistically higher than the sharing rate of liberals sharing conservative news ($p$-value: 0.00054), and likewise for the respective comparison applied to conservatives ($p$-value: 0.00856).

The model's second assumption has two components, the first being that higher truthfulness increases the probability of sharing, and the second being that the rate of increase with increasing truthfulness itself decreases (diminishing marginal increases in sharing rate). 
Figure \ref{graph:ppoverbelief} confirms that articles in the ``High" and ``Very High" truthfulness columns have higher probabilities of sharing than in the ``Low" and ``Very Low" truthfulness columns.  This is also supported by a statistical test of the slope coefficient describing sharing rate as a function of $t_i$.  This coefficient is statistically positive for both liberal and conservative readers ($p$-values 0.00000 and 0.02535, respectively).  However, the assumption of diminishing marginal increases in sharing rate on propagation rate as truthfulness increases cannot be validated here.   

Having validated the model on real social media data, we can now begin an empirical analysis of the model.

\section{Empirical Analysis of Sharing Rate Over Populations of Users}
\label{s:empirical}

In this section, we examine qualitative trends in model-estimated sharing probabilities over a population of users as a function of the bias and truthfulness of the article being propagated.  The purpose in this analysis is not to use the model for statistical prediction of propagation rates but to understand general characteristics of the population and the content being disseminated that result in larger or smaller propagation rates, as estimated by the model. We interrogate the types of populations over which the propagation of untruthful news will be highest, and for a given population, the types of news that the model suggests will be widely shared.  Understanding which content is more likely to be shared, and the characteristics of populations that are more susceptible to the sharing of untruthful or highly biased news, provides insights into how users and platforms could prioritize their interventions to mitigate the propagation of untruthful or highly biased news.  Using these qualitative results, we are able to make several recommendations for social media users and platforms.

\subsection{Estimating Model Parameters}
\label{sss:estimating}

To examine trends in model-estimated propagation rates of political news over a population of users, we must first estimate reasonable values for the scaling parameters $f_l$, $k_l$, $f_r$ and $k_r$.  The values used in the remainder of this section are given in the column labeled ``Base Scenario'' in Table \ref{tab:basevalues} and are obtained from a nonlinear least squares regression model fitted to the observed values of $(b_i, \ t_i, \ B_j)$ in our data set. (The fit summary for this regression is provided in Appendix \ref{ss:appnonlin}.)  To account for uncertainty in these parameter estimates, the column labeled ``Sensitivity Analysis'' provides low and high values of each parameter used to test the sensitivity of our recommendations to the parameters.  We find the recommendations resulting from our model analysis to be robust to changes in these parameter values; Appendix \ref{ss:appsensanal} contains the results of this sensitivity analysis.

\begin{table}[H]
\centering
\begin{tabular}{|c|c|c|}
\hline
& Base & Sensitivity \\
Parameter & Scenario & Analysis \\
\hline
\hline
$f_l$ & 0.010 & [0.005, 0.014] \\
$k_l$ & 4.465  & [2.232, 6.697]\\
\hline
$f_r$ & 0.007 & [0.004, 0.011]\\
$k_r$ & 5.581 & [2.791, 8.372]\\
\hline
\end{tabular}
\caption{Estimated parameter values used for model analysis.  The Base Case values are used in the analysis of Section \ref{ss:results}. The sensitivity analysis is found in Appendix \ref{ss:appsensanal}.}.
\label{tab:basevalues}
\end{table}

\subsection{Empirical and Synthetic Population Distributions}
\label{s:popdists}

Using these parameter estimates, we compute the model-estimated probability of sharing an article for all possible combinations of $b_i$ and $t_i$, and over a distribution of readers, using equation (\ref{eq:probofbiandti}). The probability distribution function $f(B_j)$ is the fraction of the population in the belief group having average political belief $B_j$. This process of computing probabilities is done separately for articles having left- and right-bias using their respective estimated parameters. 

We use the distribution found in the data from Grinberg \textit{et al.}, henceforth referred to as the \textit{empirical distribution}, and five more constructed distributions: two bimodal distributions (representing partisan and hyperpartisan populations) and three unimodal distributions (representing left-unimodal, centrist, and right-unimodal populations). Table \ref{tab:dist} gives the proportions of each distribution type having each political belief. 

The five constructed distributions of readers represent different types of social media populations. The two bimodal distributions reflect populations with a higher concentration of partisan and hyperpartisan beliefs. The three unimodal distributions represent populations concentrated on the left side of the political spectrum, in the center of the  spectrum, or on the right side of the spectrum. Comparing the trends in the model-estimated propagation rate over bimodal and unimodal distributions of readers' political beliefs will help us understand the effects of partisanship on the overall propagation rate of different types of political content.

\hyphenation{Distribution}
\newcolumntype{P}[1]{>{\raggedright\arraybackslash}p{#1}}
\begin{table}[t]
\centering
\resizebox{\columnwidth}{!}{\begin{tabular}{|l|P{2cm}|P{2cm}|P{2cm}|P{2cm}|P{2cm}|P{2cm}|P{2cm}|}
\hline
& Empirical Distribution & Partisan Distribution (Bimodal on the Left and Right) & Hyperparti-san Distribution (Bimodal on the Extreme Left and Extreme Right)  & Left-Unimodal Distribution & Centrist-Unimodal Distribution & Right-Unimodal Distribution & Belief Center Point \\
 \hline
 \hline
 Extreme Left & 0.092 & 0.080 & 0.400  & 0.200 & 0.020 & 0.020 & -0.857  \\
 \hline
 Left & 0.230 & 0.400 & 0.080  & 0.400 & 0.080 & 0.040 & -0.571 \\
 \hline 
 Lean Left & 0.225 & 0.020 & 0.020 & 0.200 & 0.200 & 0.060 & -0.286 \\
 \hline
 Center & 0.184 & 0 & 0 & 0.080 & 0.400 & 0.080 & 0 \\
 \hline
 Lean Right & 0.131 & 0.020 & 0.020 & 0.060 & 0.200 & 0.200  & 0.286 \\
 \hline 
 Right & 0.091 & 0.400 & 0.080 & 0.040 & 0.080 & 0.400 & 0.571 \\
 \hline
 Extreme Right & 0.046 & 0.080 & 0.400 & 0.020 & 0.020 & 0.200 & 0.857 \\
 \hline
 \hline
 Expectation & -0.146 & 0 & 0 & -0.400 & 0  & 0.400 & \\
 \cline{1-7}
 Variance & 0.214 & 0.382 & 0.643  & 0.167 & 0.114 & 0.167 & \\
 \hline
\end{tabular}}
\caption{The distribution of readers' political belief for the empirical and synthetic distributions}
\label{tab:dist}
\end{table}

\subsection{Results}
\label{ss:results}

\begin{figure}[t]
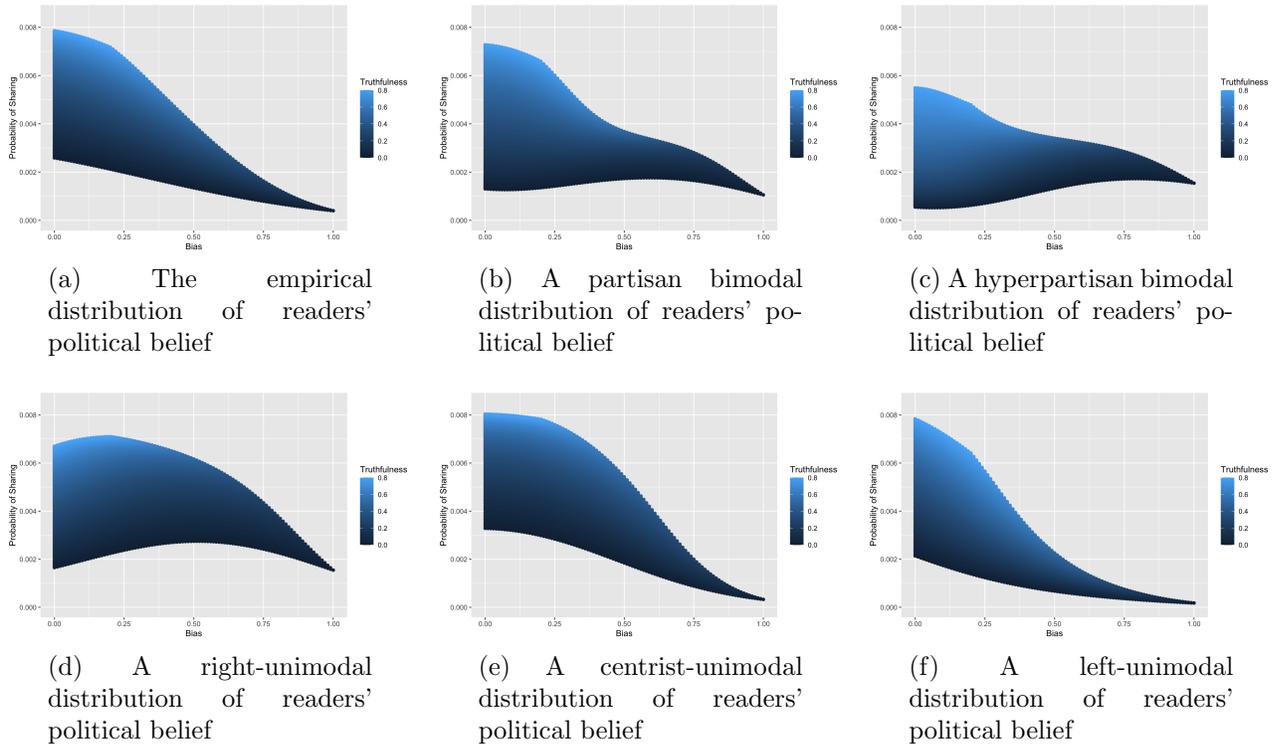

    \centering
    \subfloat[The empirical distribution of readers' political belief]{
    \includegraphics[width = 0.33\textwidth]{PublicationGraphs/mbm_right_real_bias.pdf}
    }
    \subfloat[A partisan bimodal distribution of readers' political belief 
    ]{
    \includegraphics[width = 0.33\textwidth]{PublicationGraphs/mbm_right_bim_bias.pdf}
    }
    \subfloat[A hyperpartisan bimodal distribution of readers' political belief 
    ]{
    \includegraphics[width = 0.33\textwidth]{PublicationGraphs/mbm_right_exbim_bias.pdf}
    }
    
    \subfloat[A right-unimodal distribution of readers' political belief 
    ]{
    \includegraphics[width = 0.33\textwidth]{PublicationGraphs/mbm_right_rightuni_bias.pdf}
    }
    \subfloat[A centrist-unimodal distribution of readers' political belief 
    ]{
    \includegraphics[width = 0.33\textwidth]{PublicationGraphs/mbm_right_uni_bias.pdf}
    }
    \subfloat[A left-unimodal distribution of readers' political belief 
    ]{
    \includegraphics[width = 0.33\textwidth]{PublicationGraphs/mbm_right_leftuni_bias.pdf}
    }
    \caption{The probability of sharing an article as a function of right political bias ($x$-axis) and truthfulness (gradient)}
    \label{fig:biasright}
\end{figure}

\begin{figure}[t]
    \centering
\subfloat[The empirical distribution of readers' political belief]{
    \includegraphics[width = 0.33\textwidth]{PublicationGraphs/mbm_right_real_truth.pdf}
    }
    \subfloat[A partisan bimodal distribution of readers' political belief 
    ]{
    \includegraphics[width = 0.33\textwidth]{PublicationGraphs/mbm_right_bim_truth.pdf}
    }
    \subfloat[A hyperpartisan bimodal distribution of readers' political belief 
    ]{
    \includegraphics[width = 0.33\textwidth]{PublicationGraphs/mbm_right_exbim_truth.pdf}
    }

    \subfloat[A right-unimodal distribution of readers' political belief 
    ]{
    \includegraphics[width = 0.33\textwidth]{PublicationGraphs/mbm_right_rightuni_truth.pdf}
    }
    \subfloat[A centrist-unimodal distribution of readers' political belief 
    ]{
    \includegraphics[width = 0.33\textwidth]{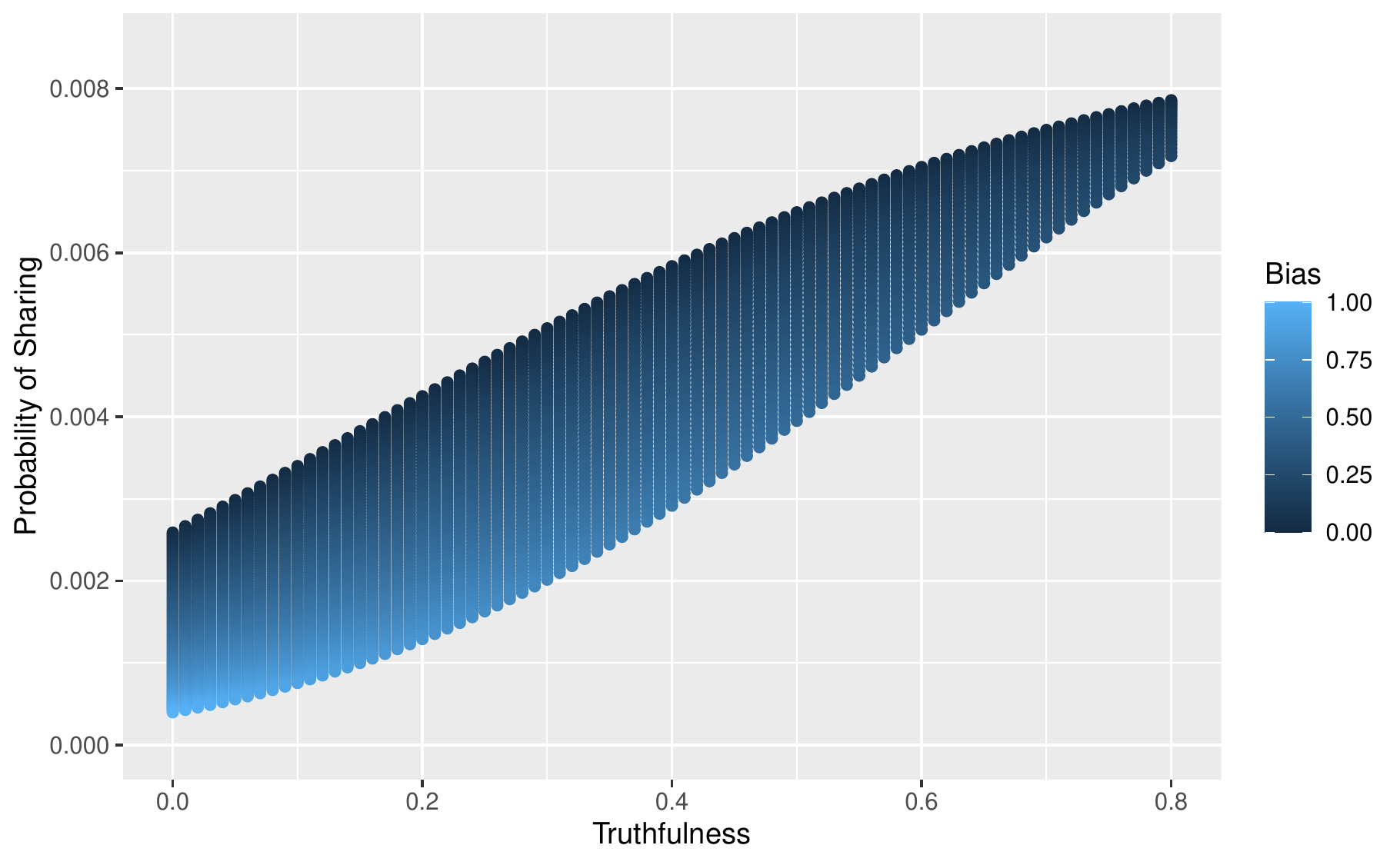}
    }
    \subfloat[A left-unimodal distribution of readers' political belief 
    ]{
    \includegraphics[width = 0.33\textwidth]{PublicationGraphs/mbm_right_leftuni_truth.pdf}
    }
    
    \caption{The probability of sharing an article as a function of truthfulness ($x$-axis) and right political bias (gradient)}
    \label{fig:truthright}
\end{figure}

In the sections that follow, we present population-wide model-estimated sharing rates of political content as a function of the content's bias and truthfulness, for each of the six population distributions described above.  We focus our analysis on the propagation of right-biased articles.  The analysis for left-biased articles is symmetric and therefore not presented in this paper.  Figures \ref{fig:biasright} and \ref{fig:truthright} present the probability of article propagation as a function of bias and truth, respectively, for right-biased articles over the six distributions.  

We examine the same three questions as were posed previously, now focused on both content and population characteristics: 1) What general characteristics of articles and populations achieve the highest propagation rate? 2) For untruthful articles, what characteristics yield the highest propagation rate? 3) For highly biased articles, what characteristics yield the highest propagation rate?

\subsubsection{What general characteristics of articles and populations achieve the highest propagation rate?}
\label{sss:overall}

To determine the conditions under which the highest overall propagation rate is achieved, we find the location of the maximum of each graph.  As seen most clearly in Figure \ref{fig:biasright}, in all distributions except the right-unimodal distribution (Fig. \ref{fig:biasright}(d)), 
overall propagation rate is maximized for unbiased, high-truth content. The centrist-unimodal population (Fig. \ref{fig:biasright}(e)) exhibits the highest sharing rate of this type of content.  In the right-unimodal population, the sharing rate is highest for \textit{mildly} biased, moderately truthful content.  Thus we conclude that unbiased, truthful content can achieve high propagation rates in most population types.  This is consistent with Recommendation P1.  However, bias in the population of readers promotes the propagation of biased and less truthful news.

We can compare the centrist population to the partisan and hyperpartisan populations to understand how the variability of beliefs in a population impacts propagation rate even as the average belief is held constant at $0$.  Contrasting Figures \ref{fig:biasright}(b), (c), and (e), we see that partisan and hyperpartisan populations share biased news at the highest rate of the distributions studied, while centrist populations share unbiased news at the highest rate.  Importantly, as seen in Figures \ref{fig:truthright}(b), (c), and (e), a centrist population shares highly truthful news at a higher rate than partisan and hyperpartisan populations.  This leads to the conclusion that partisanism incentivizes the propagation of biased, untruthful news, even when the mean belief of the population is centered at zero. 

Based on this analysis we can conclude that if users engage only with users of similar political beliefs, then the probability
of sharing news, fake or true, increases. If users engage with users of opposing political
beliefs, then the probability of sharing decreases, and the decrease is more pronounced
for fake news \footnote{This result can be supported mathematically; a proof is given in Appendix \ref{app:proof}.}. This suggests a trade-off between mitigating the spread of untruthful news and increasing the flow of information. On one hand, dissemination and flow of information is considered to be positive (\cite{Rosling2018}), and based on the aforementioned result, it increases when users engage only with users of similar political beliefs. On the other hand, spread of fake news is harmful (\cite{Giachanou2020}) and this as well increases when users engage only with users of similar political beliefs. Thus in the extreme,  in an echo chamber the level of information flow might be high but it might be predominated by misinformation.  This result lends additional support for Recommendation P2 (\textit{Recommend friends and connections across the political spectrum.})

\subsubsection{For untruthful articles, what characteristics yield the highest propagation rate?}
\label{sss:popuntruthful}

To determine the conditions under which the propagation of \textit{untruthful} (low $t_i$) content is highest, we can examine the dark regions of Figure \ref{fig:biasright} forming the bottom curve of each plot, and identify where the maximum occurs along this curve.  

We see that for the empirical (Fig. \ref{fig:biasright}(a)), centrist-unimodal (Fig. \ref{fig:biasright}(e)), and left-unimodal (Fig. \ref{fig:biasright}(f)) distributions, the maximum of the bottom curve occurs at the left side of the graph: propagation of untruthful right-leaning content is highest in these populations when the content is unbiased.  Thus, a unimodal population with centrist beliefs will prioritize  truthful unbiased news, but such a population is also susceptible to propagating untruthful unbiased news.   This suggests that fact-checking efforts should not focus only on extremely biased content; the greatest propagation of untruthful content in these populations will occur with unbiased content.

However, for the partisan bimodal (Fig. \ref{fig:biasright}(b)) and right-unimodal (Fig. \ref{fig:biasright}(d)) distributions, propagation of untruthful content is highest for moderately biased content ($|b_i| \approx 0.5$).  And for the hyperpartisan bimodal population (Fig. \ref{fig:biasright}(c)), propagation of untruthful content is highest for strongly biased content ($|b_i| \approx 0.75$).  The probability of sharing any partisan news, true or fake, increases when the expected belief is similarly partisan. Thus, a malicious agent trying to maximize the probability of sharing an untruthful article in a partisan population would choose to propagate biased content, with the optimal bias increasing with the partisanism of the population.

\textit{Recommendation P3: Prioritize the fact-checking of moderately biased news, based on the user population.}  In unimodal populations, untruthful content is most likely to be shared if it exhibits little to no bias.  However, in partisan bimodal populations, untruthful content is most likely to be shared if it exhibits moderate bias.  Strongly biased untruthful content is highly likely to be shared only in hyperpartisan populations.

\subsubsection{For highly biased articles, what characteristics yield the highest propagation rate?}
\label{sss:pophighbias}

The probability of sharing \textit{strongly-right-biased} content in populations having the empirical, centrist-unimodal and left-unimodal distributions is low overall.  Moreover, there is a general decreasing trend in propagation rate of strongly-right-biased content as we look across the tails of the graphs from right-unimodal (Fig. \ref{fig:biasright}(d)), to centrist-unimodal (Fig. \ref{fig:biasright}(e)), to left-unimodal populations (Fig. \ref{fig:biasright}(f)). This behavior is expected as it demonstrates an assumption built into the model: people are less likely to share articles that oppose their political beliefs.  The hyperpartisan population (Fig. \ref{fig:biasright}(c)) has a higher propagation rate of extremely biased, low truthfulness news, and a lower overall propagation rate of unbiased news of all truthfulness levels.  This further supports Recommendation P2 above (\textit{Recommend friends and connections across the political spectrum}) to reduce extreme bimodality in the population.  This also supports three additional recommendations:

\textit{Recommendation P4: Suggest related articles that present an issue from a diversity of viewpoints.}  In addition to permitting a user to assess the verity of the content when presented multiple ways, untruthful content will be less likely to be shared.

\textit{Recommendation P5: Give users a report about the alignment of their feed.}  This will help users understand if they are consuming a sufficiently varied diet of political content.

\textit{Recommendation U2: Read articles with opposing beliefs to mitigate the effects of confirmation bias.}  Once presented with content from across the political spectrum and given a metric on the political alignment of their own media use, users should consciously engage with content that runs counter to their beliefs.

From this analysis it is clear that the choice of content propagated by a malicious agent will be influenced by the population it is targeting.   
Even when the average population belief is centered at zero, bimodal partisanism incentivizes the propagation of biased, untruthful news, whereas a centrist population is more likely to share truthful news.  This suggests that interventions to reduce bimodality in the social media user population could reduce the propagation of biased and untruthful content. Moreover, fact-checking mildly or moderately biased news could have more impact than fact-checking strongly biased news, as users are more susceptible to sharing untruthful news when it exhibits mild-to-moderate bias.

\section{Discussion}
\label{s:discussion}
In this section, we further develop our recommendations to mitigate fake news. The spread of fake news is intricate and occurs at both the individual and network level. Our work is novel in that it characterizes the effect of a news article's bias and truthfulness on its likelihood of being propagated by individual users. While we do not consider how political content travels through a network of social media users, we have characterized how a population of individuals interacts with news once received.  Our policy recommendations come in two forms: recommendations for social media platforms and individual users. These recommendations, with references to the sections of this paper that support the recommendations, are summarized in Table \ref{t:recommend}.

\begin{table}[H]
\centering
\begin{tabular}{ |p{3cm}|p{9cm}|p{2cm}|  }
 \hline
 Platform  & & Paper \\
 Recommendation & Description & Section(s) \\
 \hline
 P1 & Populate users' feeds with unbiased truthful news as sponsored articles. & \ref{ss:singlereader}, \ref{ss:distnreaders}, \ref{sss:overall}\\
 P2 & Recommend friends and connections across the political spectrum. & \ref{ss:distnreaders}, \ref{sss:overall}, \ref{sss:pophighbias} \\
 P3 & Prioritize the fact-checking of moderately biased news, based on the user population. & \ref{sss:popuntruthful} \\
 P4 & Suggest related articles that present an issue from a diversity of viewpoints. & \ref{sss:pophighbias} \\
 P5 & Give users a report about the alignment of their feed & \ref{sss:pophighbias} \\
 \hline \hline
User  & & Paper\\
Recommendation & Description & Section(s) \\
\hline
U1 & Fact check news that aligns strongly with a user's beliefs. &  \ref{ss:singlereader}, \ref{ss:distnreaders} \\
U2 & Read articles with opposing beliefs to mitigate the effects of confirmation bias. & \ref{sss:pophighbias} \\
 \hline
\end{tabular}
\caption{Platform and user recommendations derived from the analysis of this paper, with references to sections of this paper's analysis that generate each.}
\label{t:recommend}
\end{table}

\noindent Recommendations for platforms:
We can further divide the recommendations for platforms into two groups. The first group seeks to change/decrease the supply of biased news. P1, P2, and P4 are in this group. With these interventions in place, the platform minimizes the access of the users to highly biased news. The second group operates through information. P3 and P5 are in this group. With these platform informational interventions, the users are being provided information intended to modify their behavior. In many domains informational interventions have been shown to result in substantial changes in behavior of individuals (\cite{Bettinger2019}). 

\begin{itemize}
\item \textit{[P1]: Populate users' feeds with unbiased truthful news as sponsored articles.}  \\
An unbiased truthful article maximizes the probability of sharing on social media for all distributions except the non-centrist-unimodal and hyperpartisan bimodal distributions. However, even in these distributions, the optimal article is only moderately biased with a relatively high truthfulness. An unbiased and highly truthful article will attain a high propagation rate among many populations of readers. By sponsoring unbiased, truthful content, users will have more opportunities to engage with, and potentially share,  legitimate content.

\item \textit{[P2]: Recommend friends and connections across the political spectrum.}  \\
From the analysis, bimodal and centrist-unimodal distributions decrease the overall probability of sharing content as compared to the non-centrist-unimodal distributions, particularly for untruthful news. This result suggests that a population that has readers with opposing beliefs or moderate beliefs will exhibit a lower probability of sharing untruthful news. When suggesting new contacts to a user (e.g. a friend of a friend), platforms should suggest contacts that are across the political spectrum, as this will give users a chance to expose themselves to different political ideologies. This recommendation is counter to the current practice of many social media platforms to rank recommended connections based on \textit{similarity} with the user.  

\item \textit{[P3]: Prioritize the fact-checking of moderately biased news, based on the user population.} \\ In populations having a centrist unimodal or partisan bimodal belief distribution, a malicious agent wishing to maximize the probability of sharing untruthful news will choose moderately biased, rather than extremely biased, content.  Algorithms should fact-check articles of mild or moderate bias, as these are the articles, when untruthful, readers are more susceptible to.  Only in hyperpartisan bimodal populations do untruthful articles exhibiting strong bias achieve maximum propagation.

It is worth noting that this suggestion aligns with research on persuasion.  According to \cite{johansen}, in environments with one-sided flow of information, significant persuasive effects would be expected. However, it is expected that the level of persuasion decreases in environments with users across the political spectrum and more specifically among users with opposing beliefs. In such environments, an extremely biased article is unlikely to influence the user’s views, because its contents are improbable under the user’s prior, whereas mildly biased news is significantly more likely to persuade the user. In other words, mildly biased news might update user's belief, frictionlessly (\cite{Bettinger2019}). With this recommendation, our model contributes to the studies which analyze the evolution of beliefs/opinions in social media (see \cite{Nordio2017} among all).

\item \textit{[P4]: Suggest related articles that present an issue from a diversity of viewpoints.} \\
When a user accesses an article on their social media feed, the platform can recommend related articles from multiple news sources, ideally from a variety of political alignments.  This allows the user to assess whether multiple news sources report similar statements as the initial article they have read, and directly assess the bias and truthfulness of the statements. It is noteworthy to mention that the current algorithm of the majority of the social media platforms including Facebook, Twitter, and Instagram is based on the  objective of showing users the content that is relevant to them (\cite{Forbes}). Our suggestion runs counter to this practice, and represents a tradeoff a social media platform might have to make between profitability and social good.

\item \textit{[P5]: Give users a report about the alignment of their feed.} \\
To reduce the polarity of a population, platforms can provide each user with a score measuring the alignment of their feed.  This will help users to be more cognizant of  whether they expose themselves to different types of news and allow them to broaden their networks and content beyond those that only work to confirm their beliefs.
\end{itemize}

\noindent Recommendations for users:
\begin{itemize}

\item \textit{[U1]: Fact check news that aligns strongly with a user's beliefs by using a third party fact-checking website for individual facts within the article.} \\
Users are more susceptible to untruthful news that aligns with their beliefs. In response, users should be more proactive with the news that they read, and fact-check articles of this nature. If the platform does not suggest related content, as suggested in Recommendation P4 above, users can also seek multiple news sources to corroborate the story.

\item \textit{[U2]: Read articles with opposing beliefs to mitigate the effects of confirmation bias.} \\
When users read articles that oppose their beliefs the probability of sharing an untruthful news article is relatively low, and they prioritize sharing truthful news. 
\end{itemize}

Our work supports the conclusion that both unimodal skew or bimodal partisanism in the reader population contributes to the spread of biased, untruthful news. Some platforms already have ways to mitigate the spread of fake news. Facebook has fact-checkers that review content and rate whether it is false or true (\cite{facebook}). Twitter has recently implemented a feature that flags tweets that could be misconstruing information or be misleading (\cite{roth_2020}).  These are important steps to mitigate spread of fake news, and our model provides a mechanism for prioritizing this fact-checking effort.  Additionally, while fact-checking content that has already been shared is a retroactive endeavor, several of our recommendations above serve to more proactively prevent the sharing of fake news.

\section{Limitations and Future Directions}
\label{s:future}

The model developed in this paper is a novel way of characterizing the spread of political content on social media. Unlike previous models, it disentangles the effects of article truthfulness and bias on the sharing rate. Whereas previous work treats bias and truthfulness as categorical or binary values, our model allows for more granularity as the parameters are continuous and are permitted to vary individually. This aspect of our model enables us to examine each parameter's respective relationship to the spread of fake news in more detail.
 
The model and similar work are limited by the lack of available data sources for measuring bias and truthfulness of news articles. Currently, work in the literature on calculating content-based semantic bias is not scalable, which limits the size of the data set used in this paper. Although alignment has a strong correlation with bias, future papers could continue the process of finding credible and scalable ways to calculate content-based semantic bias.  

Future work with the model provided in this paper could increase its dimensionality. Currently, the model describes the probability an individual user will propagate a certain article based on three parameters: bias of the article, truthfulness of the article and political belief of the reader. The model is based on accurate assumptions, as validated with data by Grinberg \textit{et al.} and MBM (\cite{rib_hen_ben_2018}); however, political news propagation is based on more than just these three variables. For example, the popularity of an article has a direct correlation to the probability that the article will be shared (\cite{papa_2017}). Additionally, the model makes an assumption that a reader's belief stays static even as they are consuming news. The model also does not differentiate between potentially different motivation behind sharing (i.e. retweeting, liking, or commenting). For example, a user can share a tweet in order to express their opinion, support a person/institution, or persuade others. The act of sharing, regardless of its motive, is treated the same in the model. 

Lastly, the current model does not address how fake news spreads in a network, nor how sharing behavior might change in response to changes in content. In reality, fake news is largely caused and spread by social bots among a network of users. Future research could use our model to characterize the propagation of fake news over a social network that includes social bots.  The network effects also govern the types of content an individual is likely to be exposed to, whereas our analysis has examined the conditional sharing rate assuming that an exposure to certain content has occurred. Moreover, the recommendations we make consider short-term impacts of interventions.  If the bias distribution of articles to which a reader is exposed changes as a result of interventions, it is possible that sharing behavior will change over time in response. Such evolution is not captured by our model; future research could explore this further.

\section{Conclusion}
\label{s:conclude}

This paper provides a probabilistic model that describes the likelihood of an individual social media user sharing an online news article based on the article's bias and truthfulness, as well as their own political belief.  The model is validated using data from \cite{grinberg_2017} and \cite{rib_hen_ben_2018}. 
We are able to examine the immediate potential impact of interventions to reduce the likelihood of untruthful news being shared, by modeling the decision of an individual user to share (or not) political content.  Future work would incorporate this model into a network framework.  Doing so would permit us to examine the time evolution of intervention effects. 

We characterize a malicious agent as someone who wants to spread untruthful and/or highly biased news, and we use the model to determine the agent's choice of bias and truthfulness that maximizes the probability of propagation. Understanding the content and population characteristics that a malicious agent would target permits us to prioritize among intervention strategies, such as which content to fact-check. Additionally, we examine model-estimated propagation rates in six different populations to characterize the conditions under which ``fake'' news is most likely to propagate.

Across all population types studied, our model suggests that readers generally prefer to share highly truthful news.  Thus social media platforms should promote unbiased, truthful news. Since users are more susceptible to sharing untruthful content when it is mildly or moderately biased, platforms should prioritize allocating fact-checking resources to this type of content, rather than extremely biased content. Social media platforms should suggest content and contacts that reflect differing political opinions, so that users can avoid politically homogeneous clusters and extricate themselves from populations that only confirm their beliefs.  Additionally, platforms should provide users with information about the political alignment of their feed to encourage users to seek out articles with opposing beliefs from their own.  Individuals should also attempt to fact check news that aligns strongly with their beliefs. 

Although these recommendations are relatively straightforward to implement, two in particular run counter to the business model of most social media platforms: P2 (\textit{Recommend friends and connections across the political spectrum}) and P4 (\textit{Suggest related articles that present an issue from a diversity of viewpoints}).  Platforms such as Facebook, Twitter, and Instagram make content and contact recommendations based on \textit{similarity} with the user; in doing so, they provide content that is perceived as being relevant to the user, which increases the user's enjoyment and use of the platform.  However, the recommendation that social media platforms should suggest content and contacts that oppose a user's belief presents a tradeoff that a platform must make in order to mitigate fake news spread.

In summary, our model characterizes an individual user's decision of whether or not to share political news on social media.  Using this model, we present policy recommendations for social media platforms and users.  Mitigating the spread of fake news has become increasingly urgent worldwide, as we've seen in recent elections and the global Covid-19 pandemic. Our model provides a useful tool for identifying realistic actions that platforms and users can take to mitigate the spread of fake news.

\newpage

\appendix
\section{Comparison of Model Parameters between Typical and Atypical Users}
\label{ss:apptypical}

To determine 
the influence of heterogeneous sharing rates, we conduct a statistical test on the fitting parameters of panel members with average sharing rates and extreme sharing rates (e.g. supersharers and superconsumers, as described in Section \ref{ss:data}). We use indicator variables to differentiate between panel members with average and extreme rates of sharing, as defined by \cite{grinberg_2017}. Then we fit a modified version of the model using non-linear least squares to understand if atypical users exhibit statistically different sharing rates. We fit equation (\ref{eq:e_sharing_rates}) in our analysis, where $f$ and $k$ are our fitting parameters for panel members with an average sharing rate and $f_e$ and $k_e$ are the effects on those fitting parameters of our panel members with extreme sharing rates. We use the indicator variable $I_e$ to flag when a panelist has an extreme sharing rate. 

\begin{equation}
    \label{eq:e_sharing_rates}
    p(b_i, t_i, B_j) = \frac{f + f_e * I_e}{1 + e^{(-(k + k_e * I_e)*(t_i - (b_i - B_j)^2))}}
\end{equation}

From our non-linear least squares regression, which can be found in Table \ref{tab:e_sharing_rate}, we see that $f_e$ for both left-political and right-political readers is statistically different from zero, meaning that the extreme users exhibit a different sharing rate from typical users, as expected. 
Parameter $k_e$ is not statistically different from zero. Since $k$  governs the shape of the probability curve we infer the general shape of the curve is unaffected by the type of user. Therefore, we restrict our analysis to panel members having an average activity rate.

\begin{table}[H]
    \centering
    \begin{tabular}{|l|c|c|c|c|}
    \hline
         & \multicolumn{2}{c}{Left Reader Belief} & \multicolumn{2}{|c|}{Right Reader Belief} \\
        \hline
         & $f_e$ & $k_e$ & $f_e$ & $k_e$ \\
         \hline
        Estimate & 0.152 & -1.129 & 0.253 & -3.395 \\
        \hline
        $p$ value & 1.32e-15 *** & 0.958 & 2e-16 *** & 0.945 \\
        \hline
    \end{tabular}
    \caption{The fitted parameters found using non-linear least squares regression for panel members with extreme sharing rates; Significance value: 0 `***'}
\label{tab:e_sharing_rate}
\end{table}

\section{Nonlinear Least Squares Fit}
\label{ss:appnonlin}
To find estimates of the fitting parameters for our model we use non-linear least squares regression. Table \ref{tab:fittingparams} includes these estimates, their $p$-values, standard errors, and the overall residual standard errors. These values are used in our analysis of population distributions as well as within our analysis of typical and atypical social media users.

\begin{table}[H]
    \centering
    \begin{tabular}{|l|c|c|c|c|}
    \hline
    & $f_l$ & $k_l$ & $f_r$ & $k_r$ \\
    \hline
    \hline
      Estimate      & 0.010 & 4.465 & 0.007 & 5.581 \\
         \hline
          \textit{p} value & 1.12e-12 *** & 0.036 * & 1.25e-9 *** & 0.11 \\
         \hline
         Standard Error & 0.001 & 2.108 & 0.001 & 3.475 \\
         \hline
         Residual Standard Error & \multicolumn{2}{c}{0.007} & \multicolumn{2}{|c|}{0.007}\\
         \hline
    \end{tabular}
    \caption{The fitted parameters found using a non-linear least squares regression; Significance value: 0 `***', 0.001 `**' }
\label{tab:fittingparams}
\end{table}

\section{Sensitivity of Population Results to Estimated Model Parameters}
\label{ss:appsensanal}

\begin{figure}[H]
    \centering
    \subfloat[The empirical distribution of readers' political belief]{
    \includegraphics[width = 0.33\textwidth]{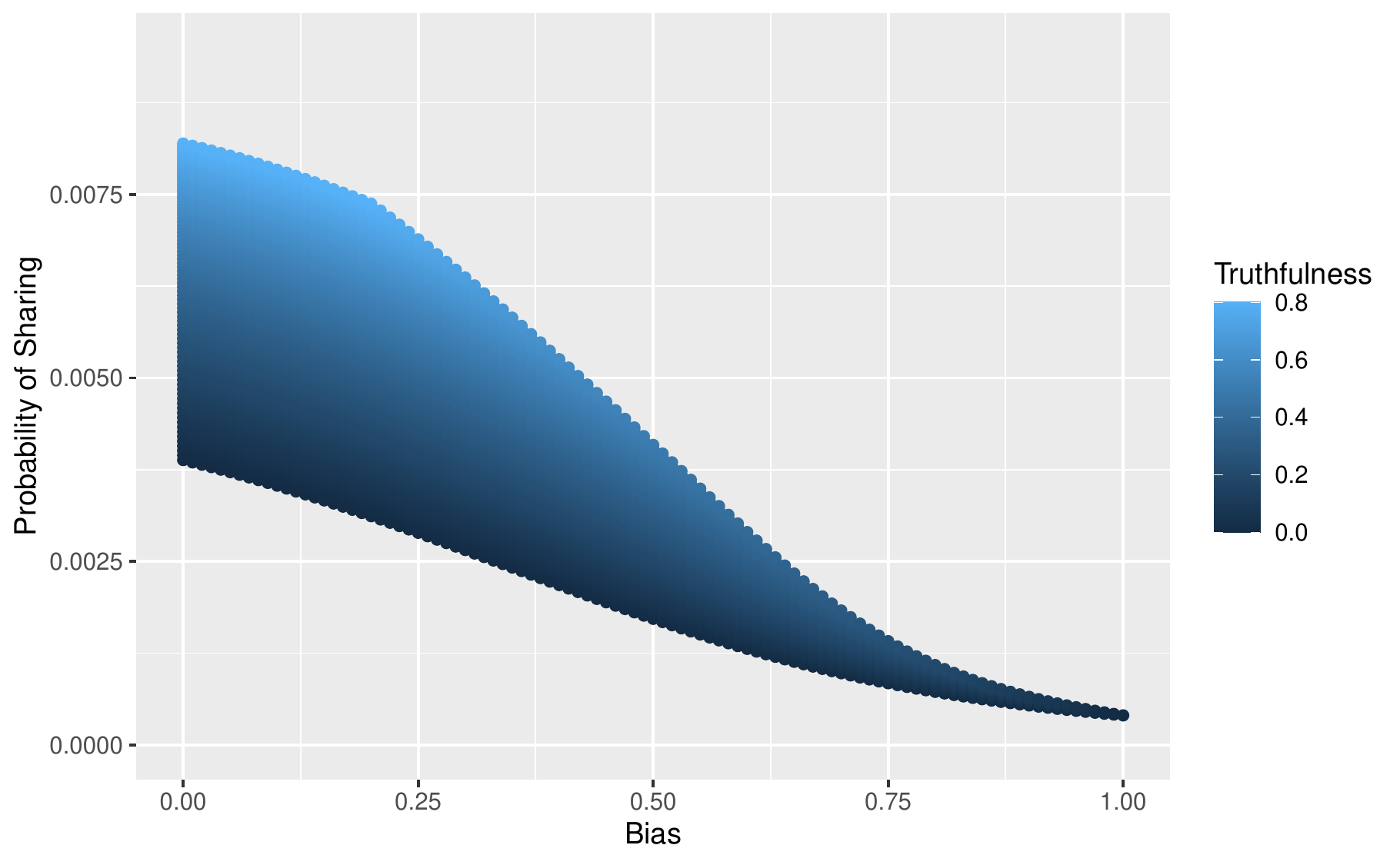}
    }
    \subfloat[A partisan bimodal distribution of readers' political belief 
    ]{
    \includegraphics[width = 0.33\textwidth]{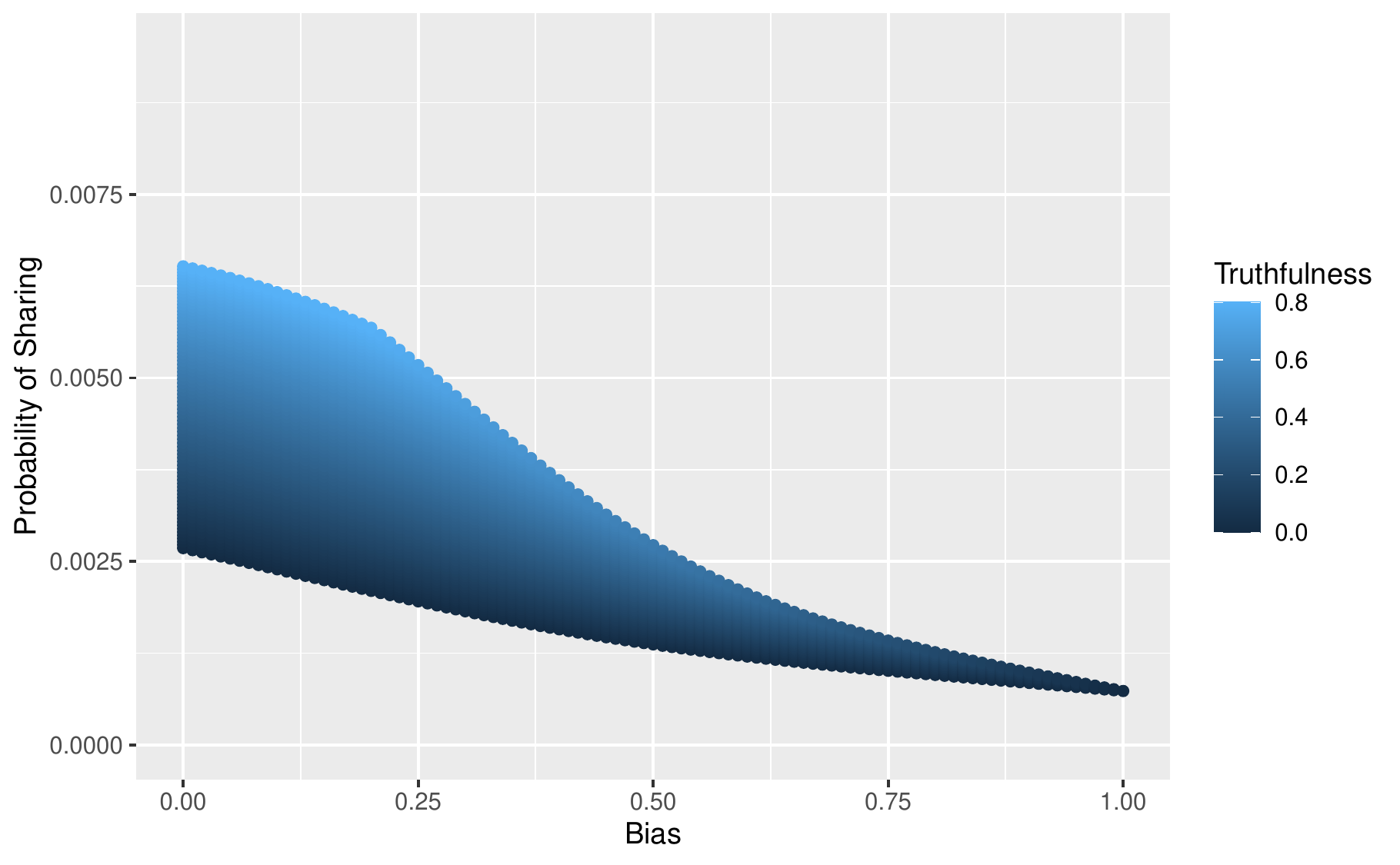}
    }
    \subfloat[A hyperpartisan bimodal distribution of readers' political belief 
    ]{
    \includegraphics[width = 0.33\textwidth]{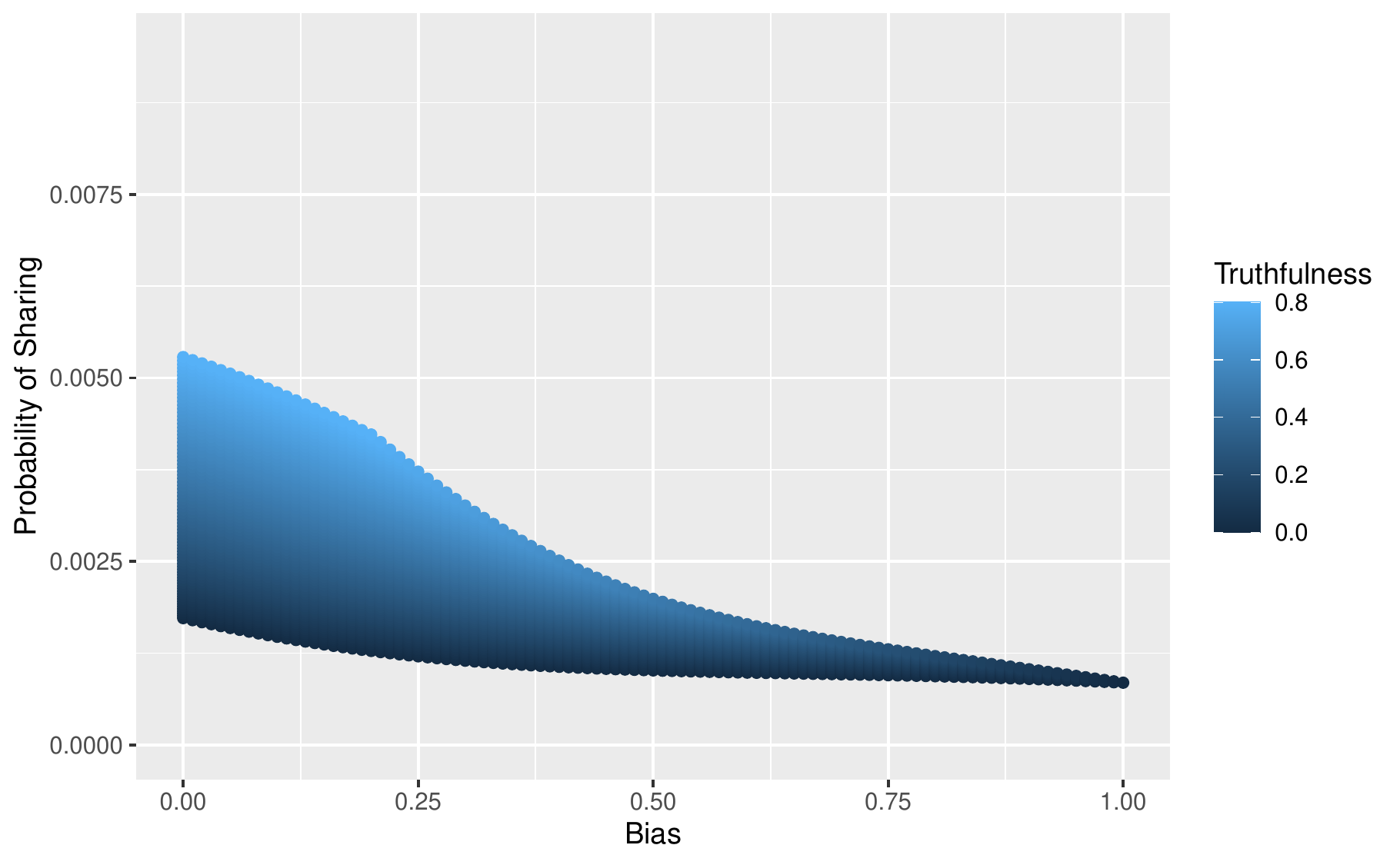}
    }
    
    \subfloat[A right-unimodal distribution of readers' political belief 
    ]{
    \includegraphics[width = 0.33\textwidth]{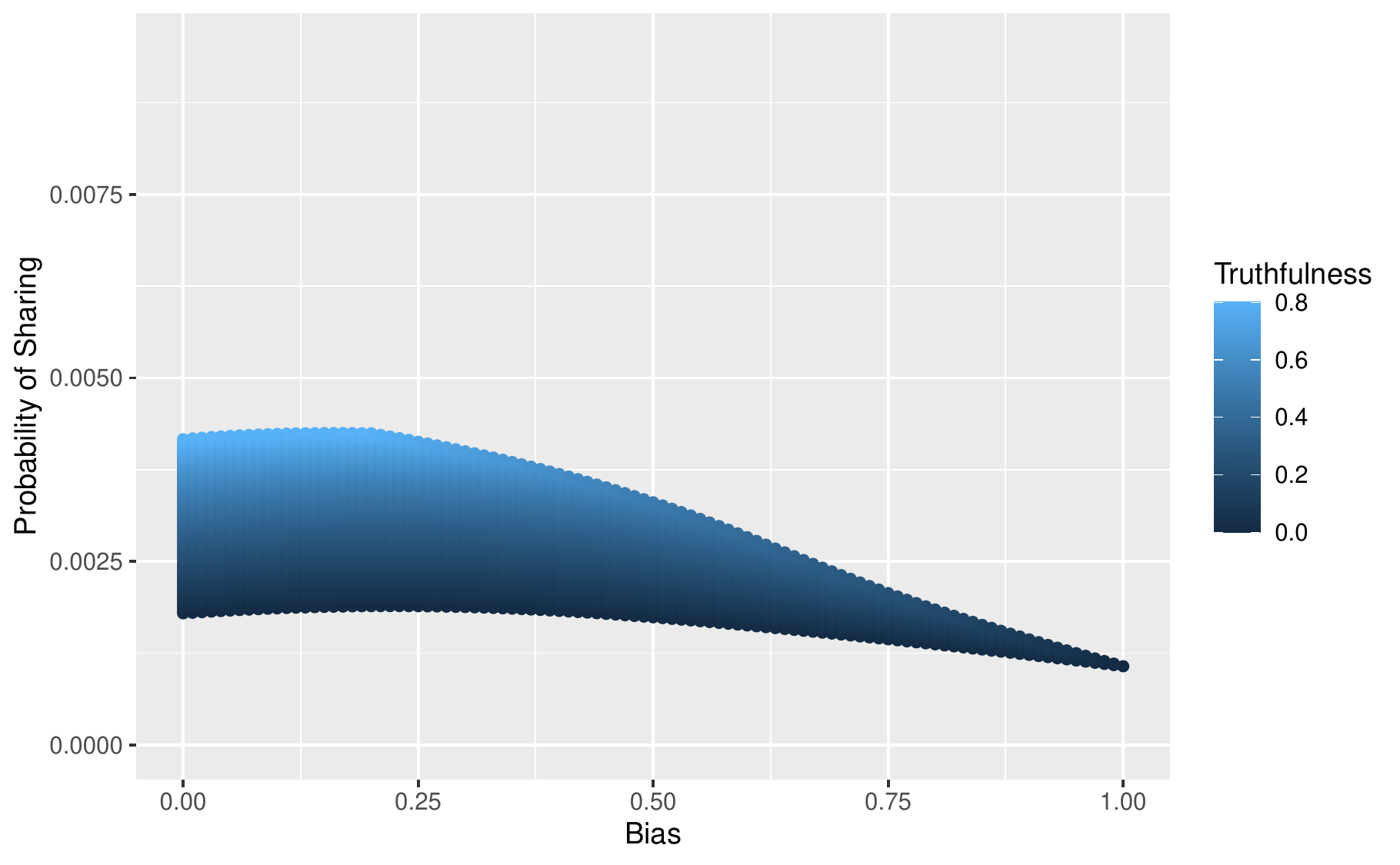}
    }
    \subfloat[A centrist-unimodal distribution of readers' political belief 
    ]{
    \includegraphics[width = 0.33\textwidth]{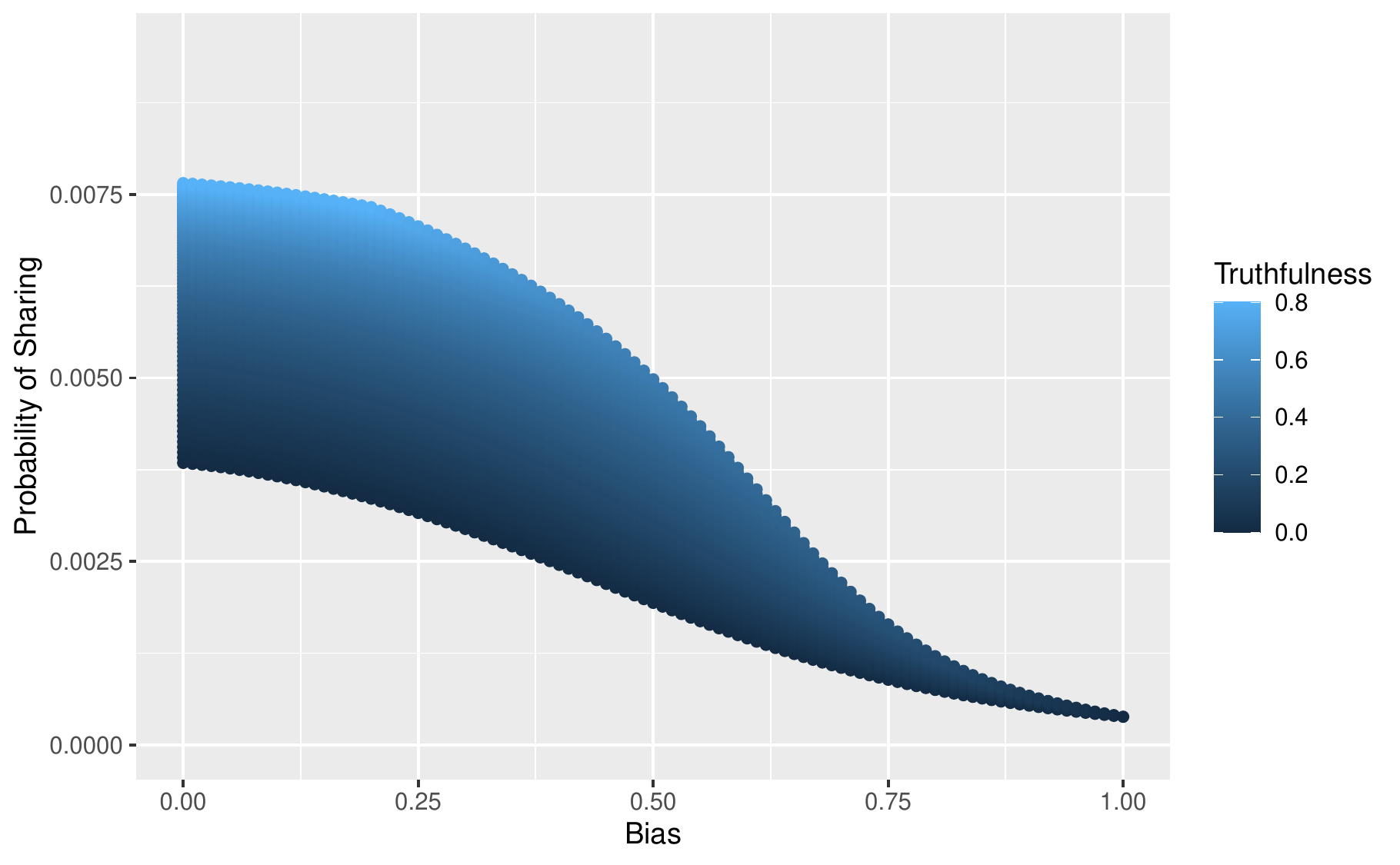}
    }
    \subfloat[A left-unimodal distribution of readers' political belief 
    ]{
    \includegraphics[width = 0.33\textwidth]{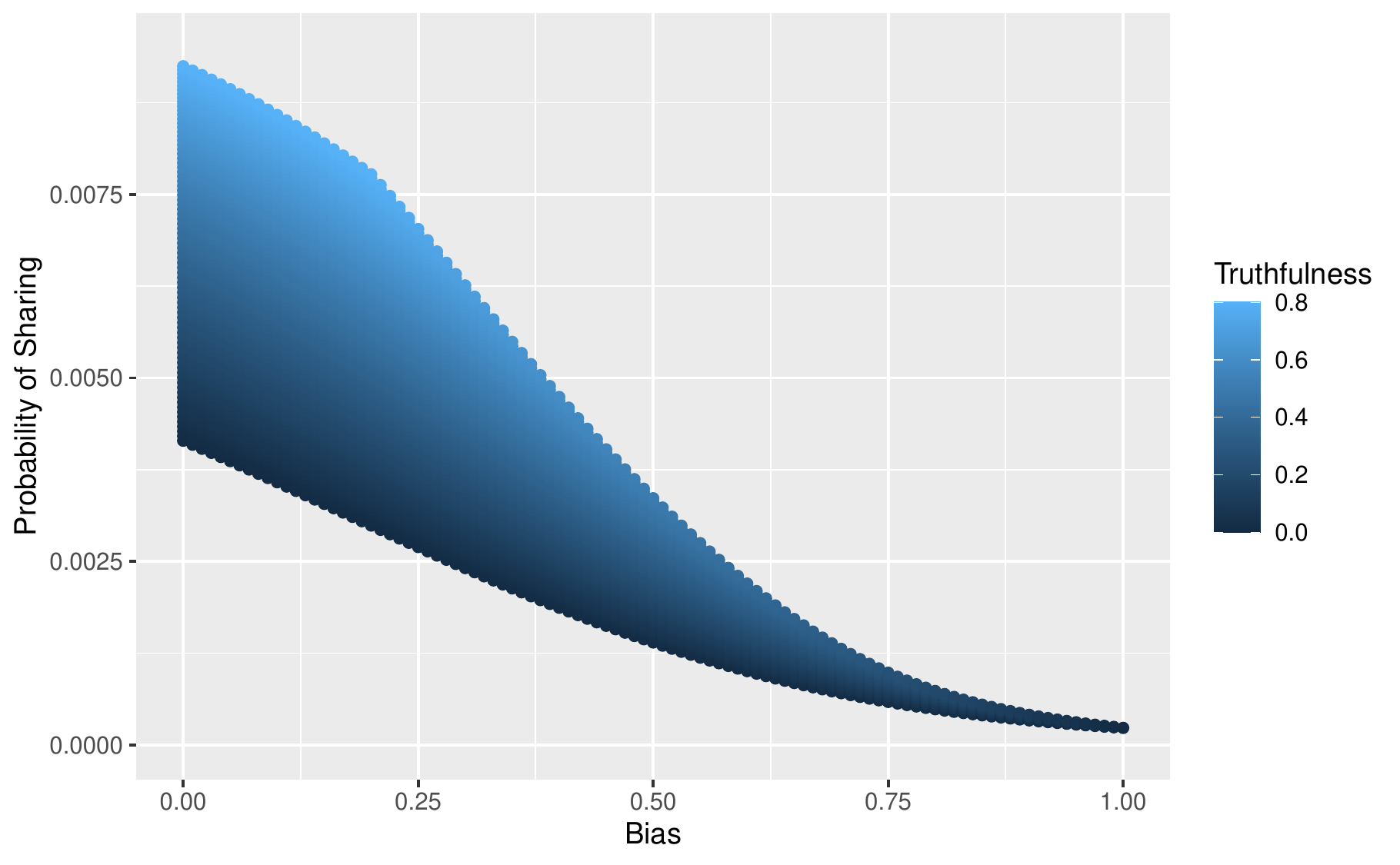}
    }
    \caption{The probability of sharing an article as a function of right political bias ($x$-axis) and truthfulness (gradient). Fitting parameters combination: $f_l$ - high, $k_l$ - low, $f_r$ - low, $k_r$ - low}
    \label{fig:biasbimrighthilololo}

\end{figure}

\begin{figure}[H]
    \centering
    \subfloat[The empirical distribution of readers' political belief]{
    \includegraphics[width = 0.33\textwidth]{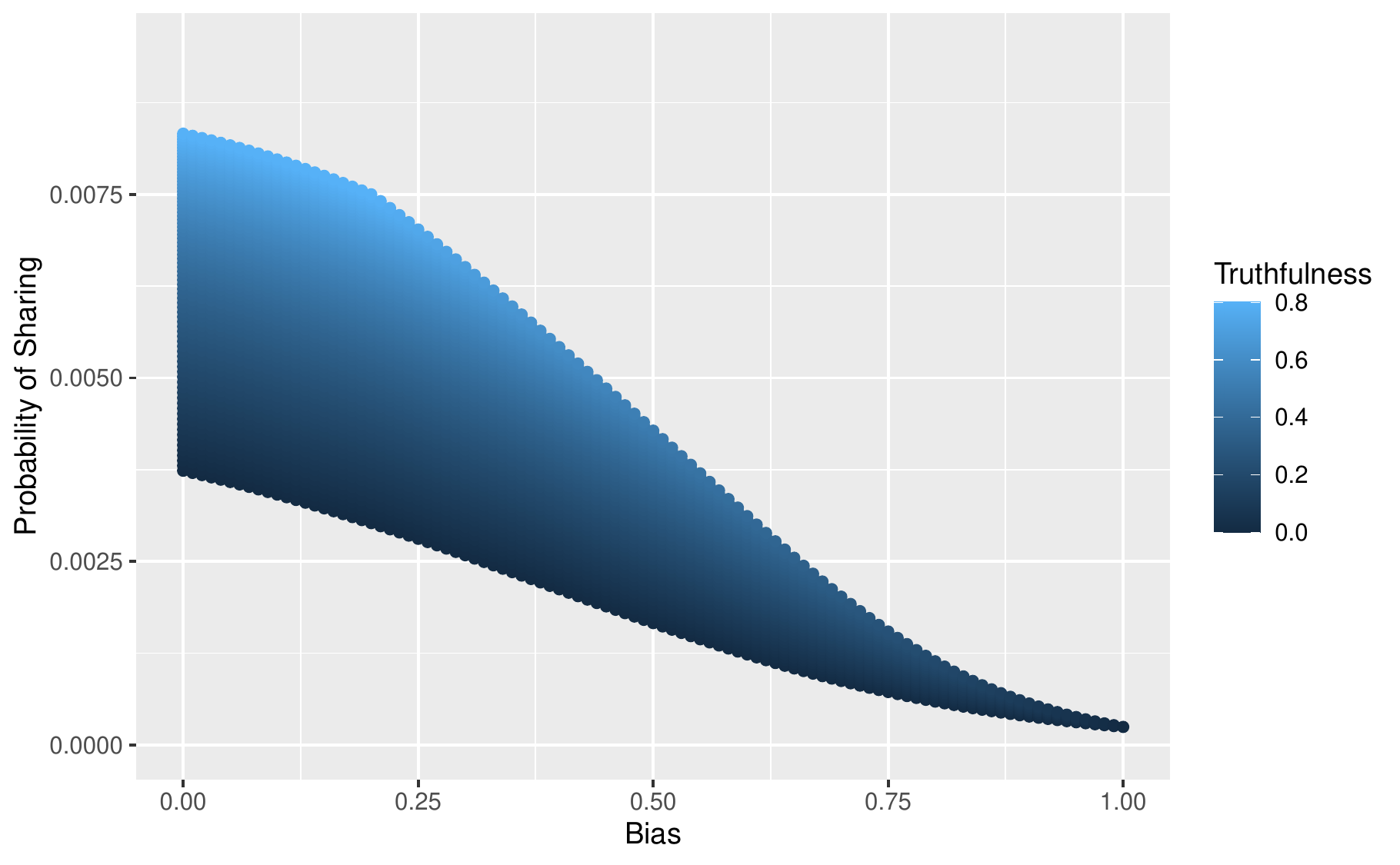}
    }
    \subfloat[A partisan bimodal distribution of readers' political belief 
    ]{
    \includegraphics[width = 0.33\textwidth]{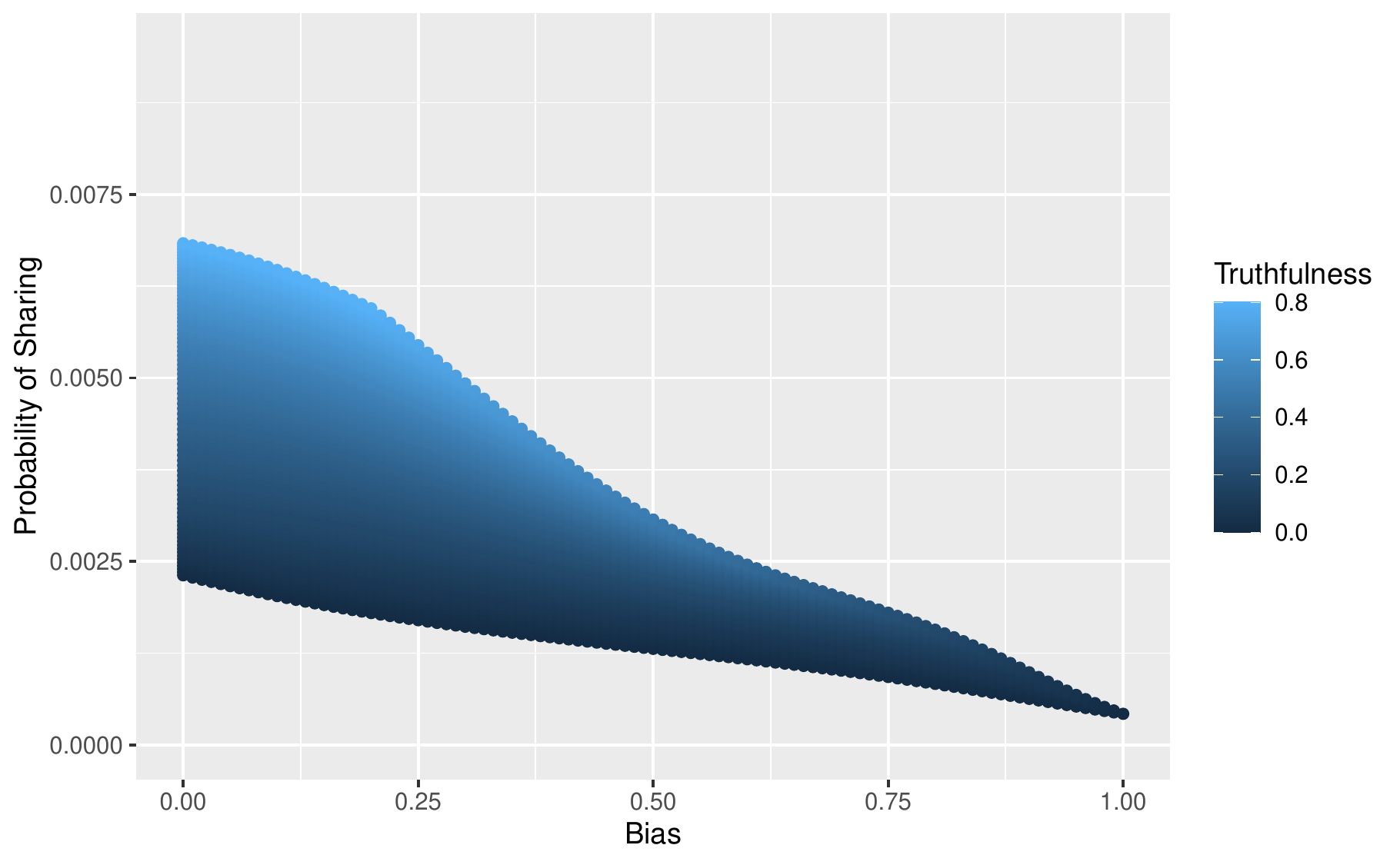}
    }
    \subfloat[A hyperpartisan bimodal distribution of readers' political belief 
    ]{
    \includegraphics[width = 0.33\textwidth]{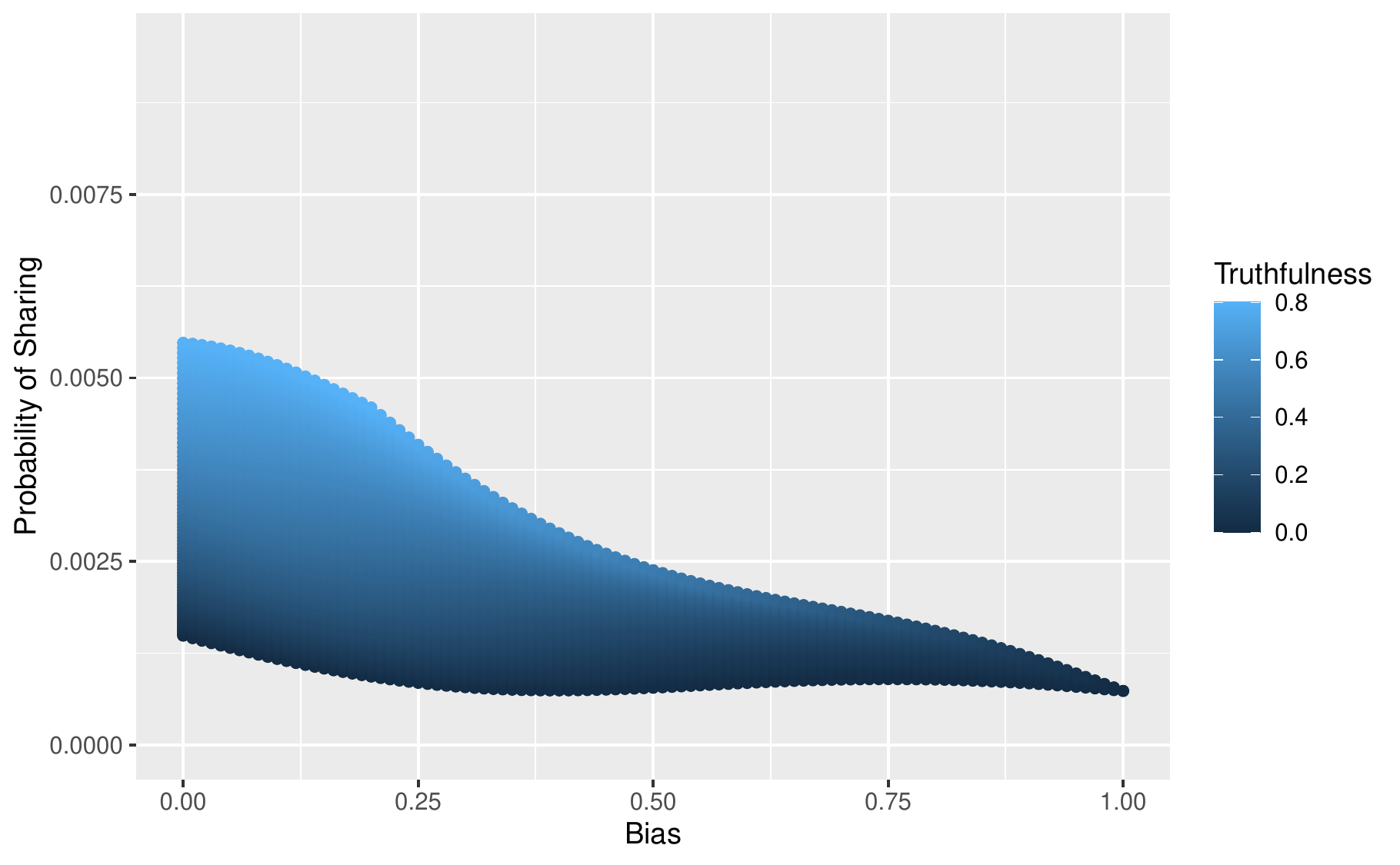}
    }
    
    \subfloat[A right-unimodal distribution of readers' political belief 
    ]{
    \includegraphics[width = 0.33\textwidth]{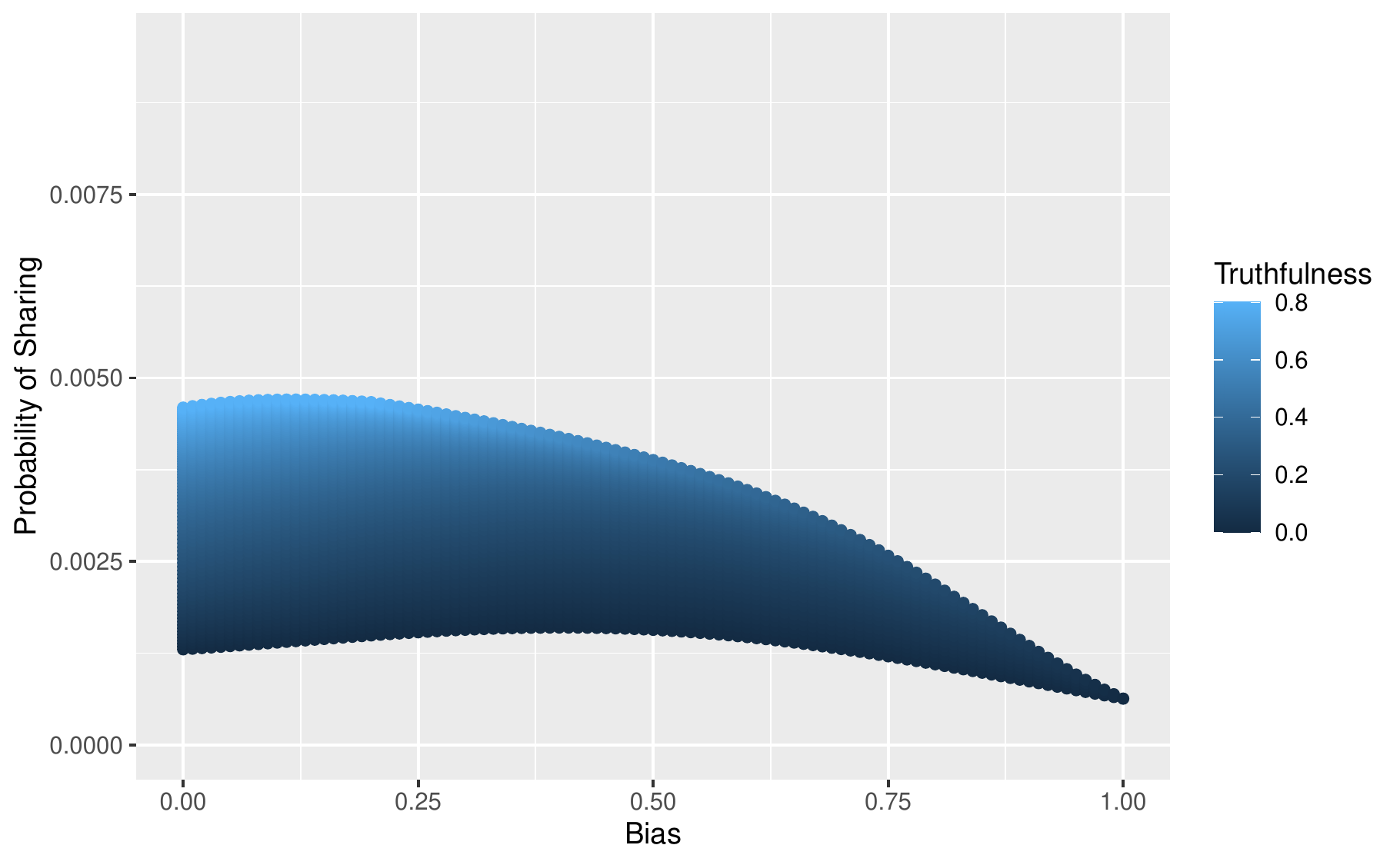}
    }
    \subfloat[A centrist-unimodal distribution of readers' political belief 
    ]{
    \includegraphics[width = 0.33\textwidth]{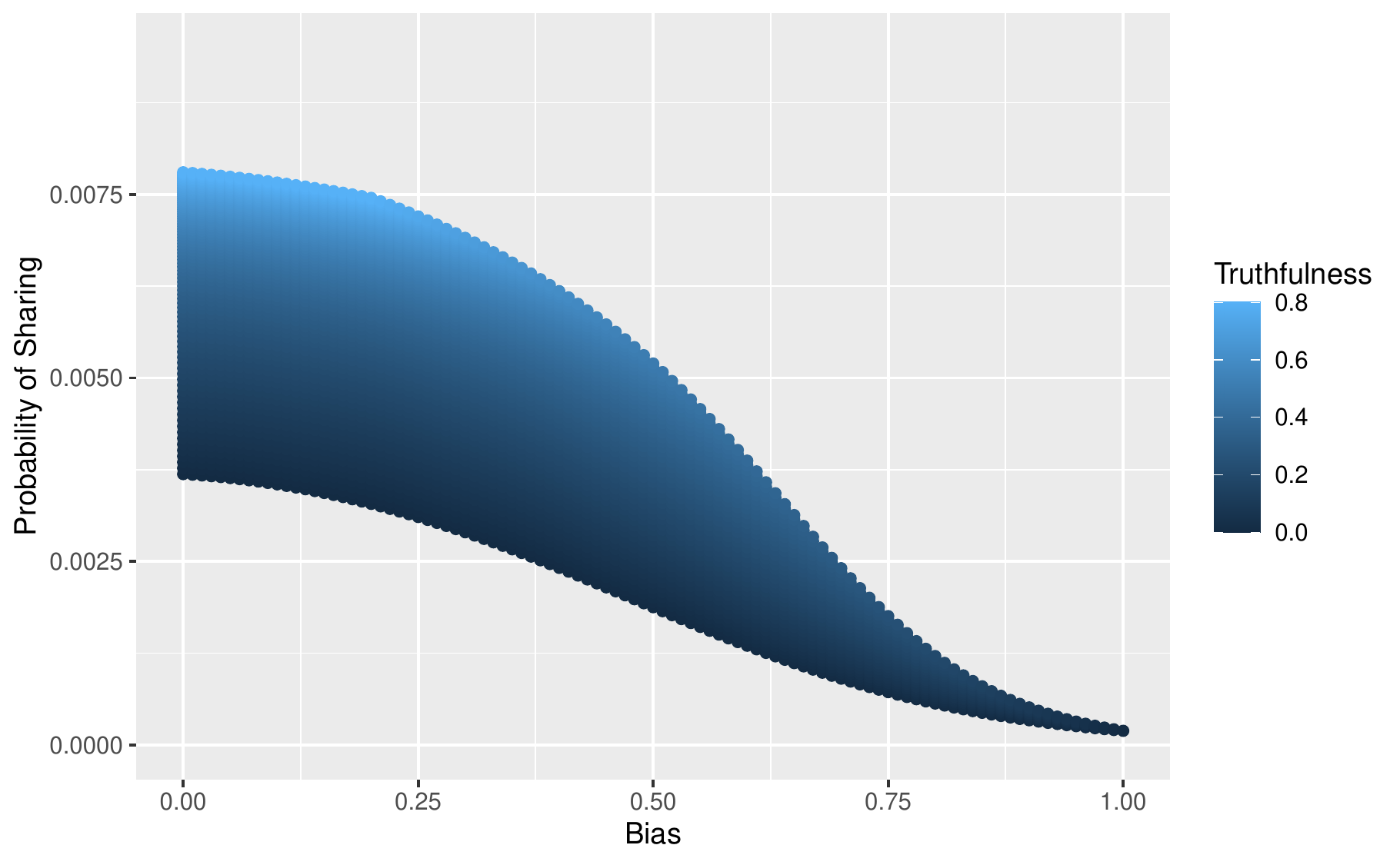}
    }
    \subfloat[A left-unimodal distribution of readers' political belief 
    ]{
    \includegraphics[width = 0.33\textwidth]{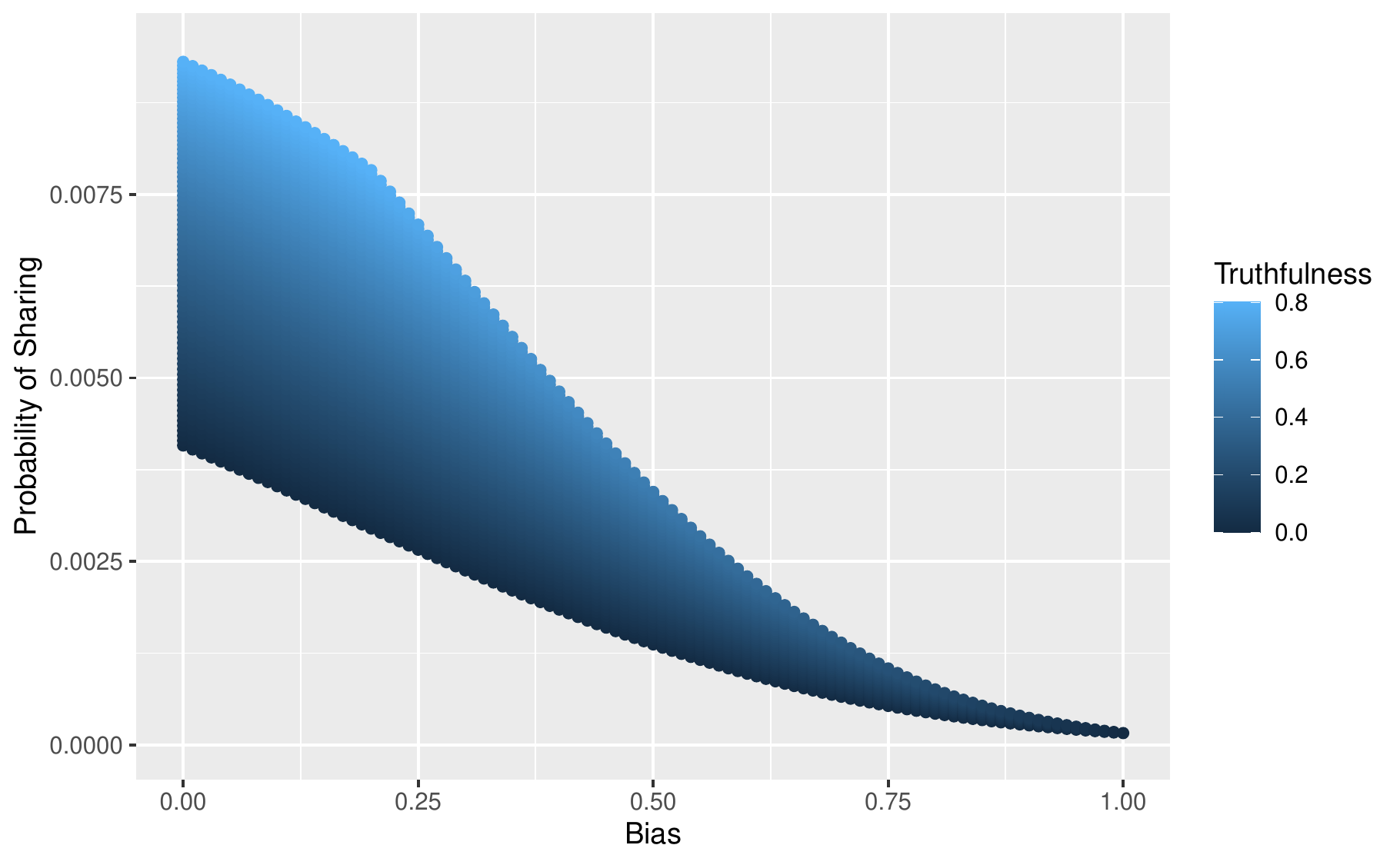}
    }
    \caption{The probability of sharing an article as a function of right political bias ($x$-axis) and truthfulness (gradient). Fitting parameters combination: $f_l$ - high, $k_l$ - low, $f_r$ - low, $k_r$ - high}
    \label{fig:biasbimrighthilolohi}

\end{figure}

We use our model to understand general trends of sharing political news on social media. The fitting parameters found using non-linear least squares may have some unknown bias. Therefore we conduct sensitivity analysis on the fitting parameters to understand how the trends identified in Section \ref{ss:results} change. The levels we use for $f_l$, $k_l$, $f_r$, and $k_r$ can be found in Table \ref{tab:basevalues}. For each synthetic distribution described in Table \ref{tab:dist} we find the probability of sharing using sixteen different combinations of $f_l$, $k_l$, $f_r$, and $k_r$. As seen in the graphs below, even though the actual probability of sharing does change, the conclusions we draw rarely change. The only combinations that result in a (modest) change in the conclusions are when we use the two combinations in which $f_l$ is at its high level, $k_l$ and $f_r$ are at their low levels, and $k_r$ is at either level (Figures \ref{fig:biasbimrighthilololo} and \ref{fig:biasbimrighthilolohi}). 
In these cases, the interpretation changes as follows: To propagate untruthful right-biased news, an adversary should propagate low-bias news to all populations except the right-centric population, to which it should propagate low-to-moderately biased news. This change is because for these parameter settings, the overall sharing rate of left-biased readers has increased while that of right-biased readers has decreased.  Thus, the sharing rate in the population is more heavily affected by the behavior of left-political readers when encountering right-biased news.

\begin{figure}[H]
    \centering
    \subfloat[The empirical distribution of readers' political belief]{
    \includegraphics[width = 0.33\textwidth]{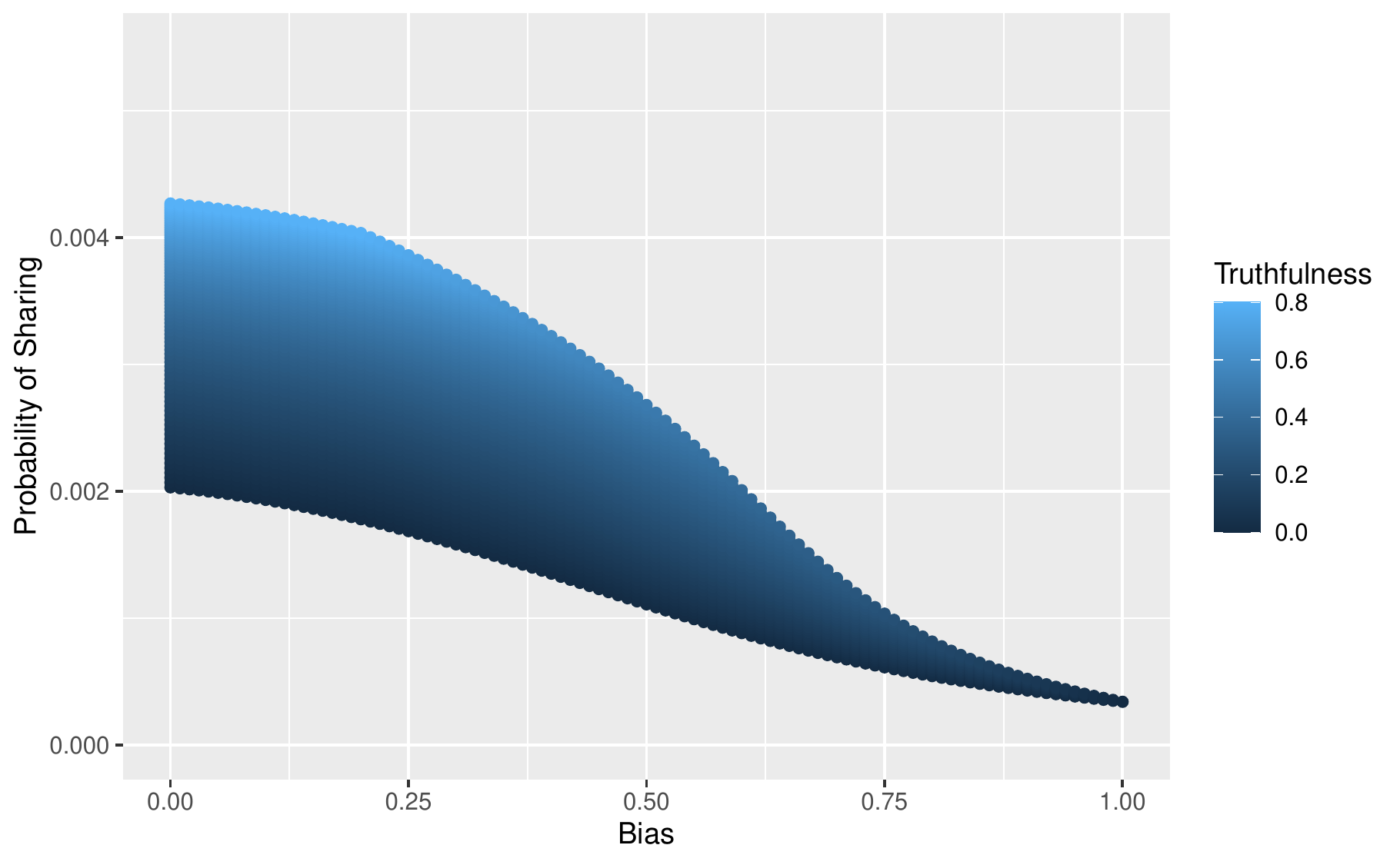}
    }
    \subfloat[A partisan bimodal distribution of readers' political belief 
    ]{
    \includegraphics[width = 0.33\textwidth]{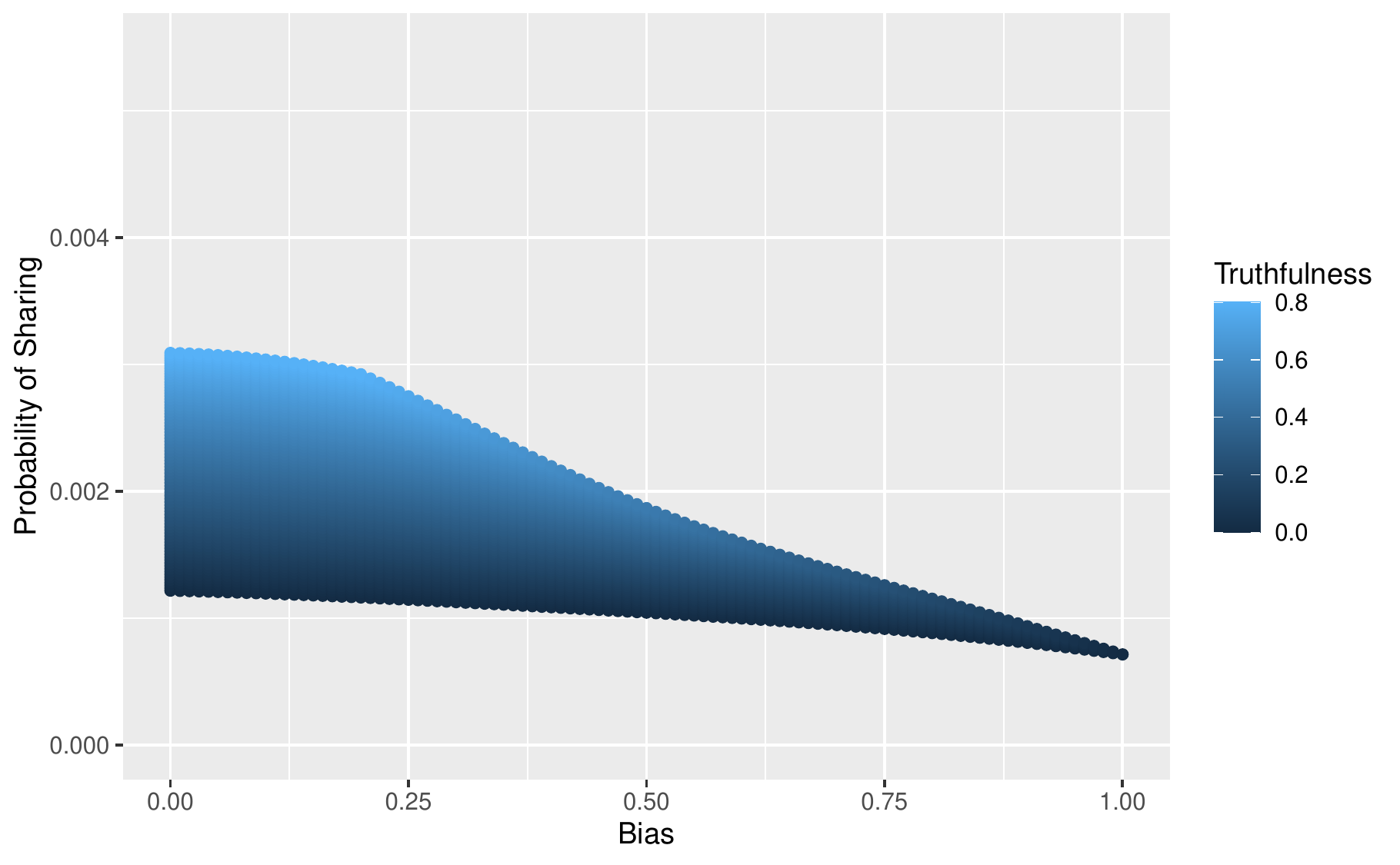}
    }
    \subfloat[A hyperpartisan bimodal distribution of readers' political belief 
    ]{
    \includegraphics[width = 0.33\textwidth]{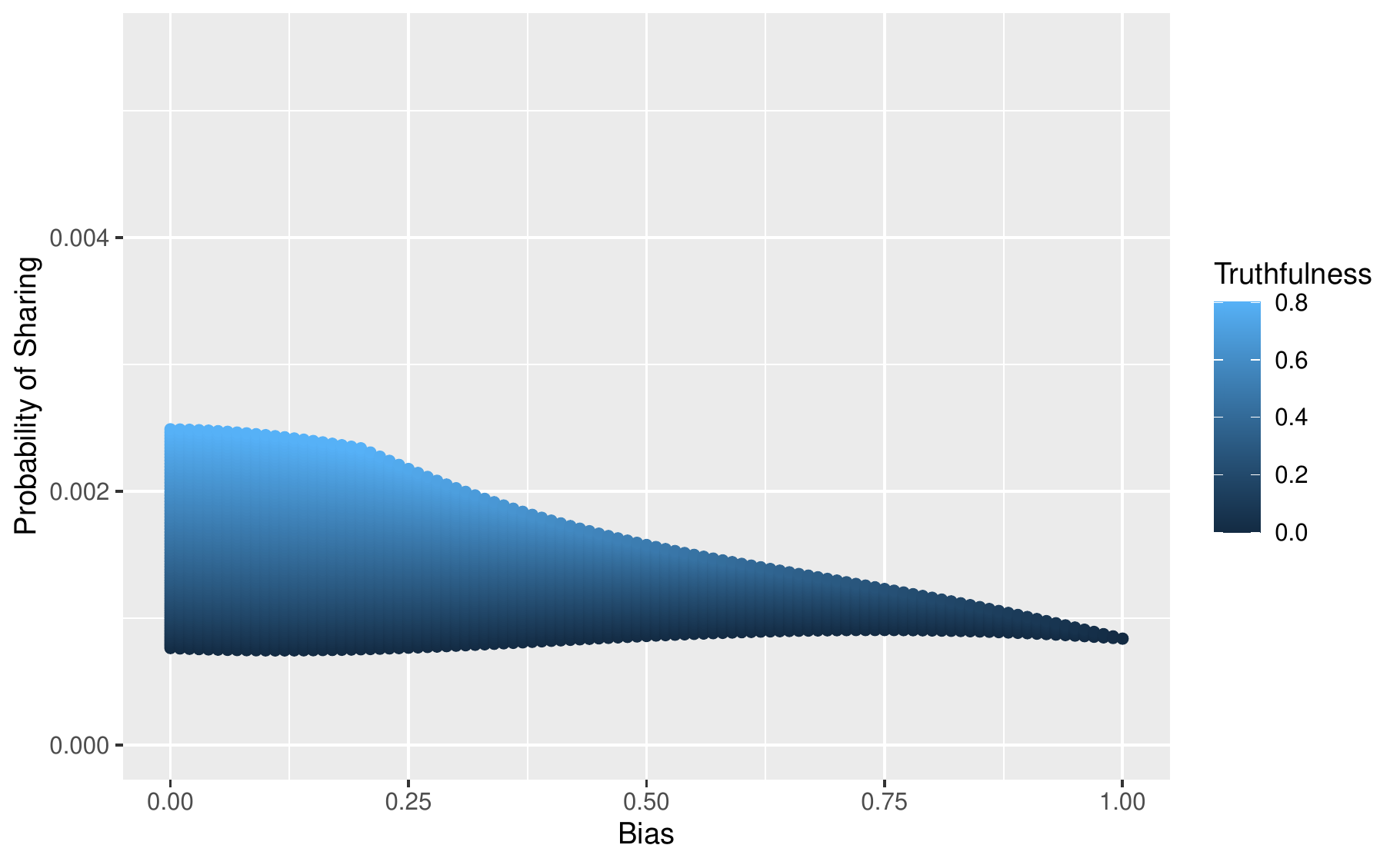}
    }
    
    \subfloat[A right-unimodal distribution of readers' political belief 
    ]{
    \includegraphics[width = 0.33\textwidth]{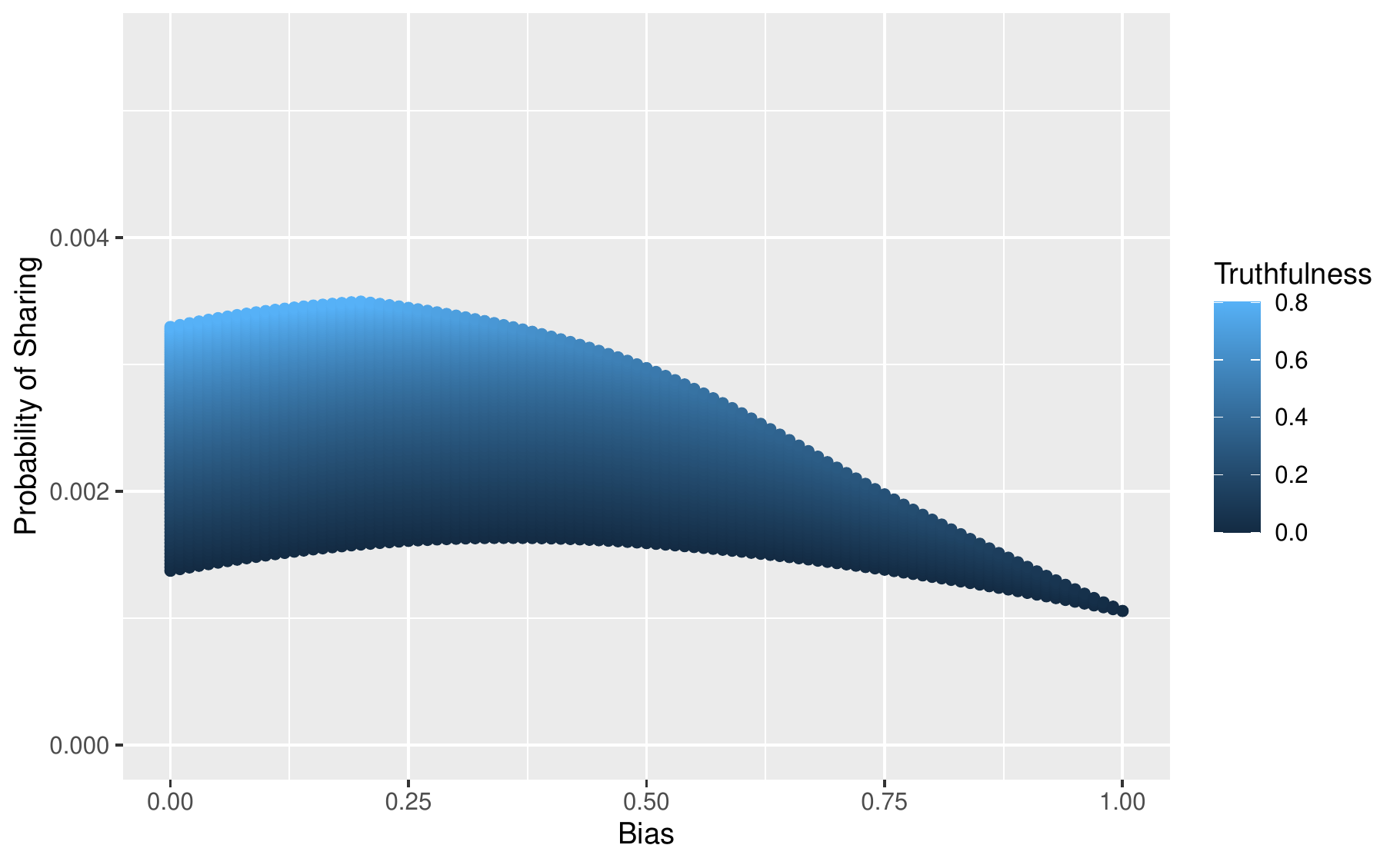}
    }
    \subfloat[A centrist-unimodal distribution of readers' political belief 
    ]{
    \includegraphics[width = 0.33\textwidth]{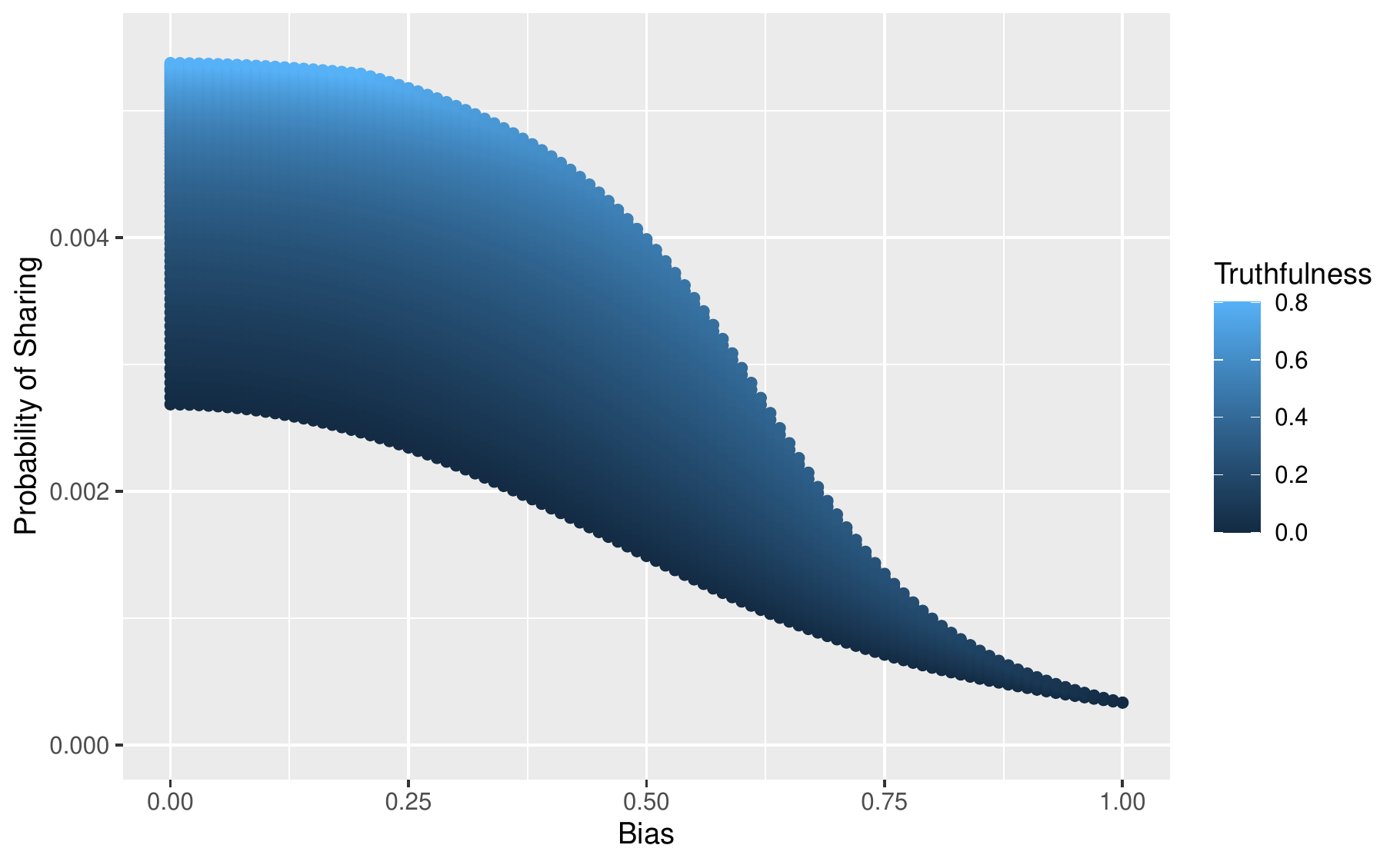}
    }
    \subfloat[A left-unimodal distribution of readers' political belief 
    ]{
    \includegraphics[width = 0.33\textwidth]{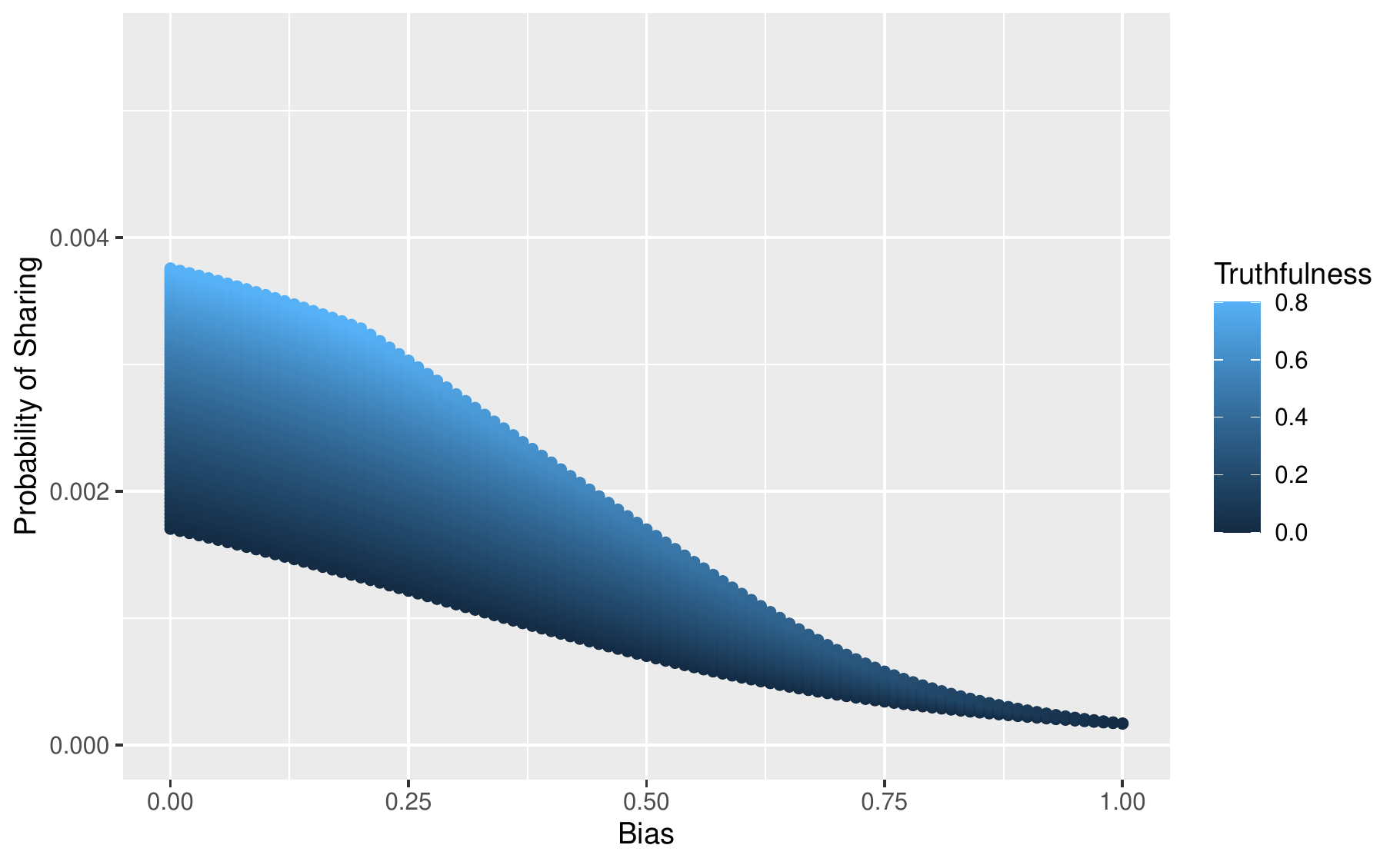}
    }
    \caption{The probability of sharing an article as a function of right political bias ($x$-axis) and truthfulness (gradient). Fitting parameters combination: $f_l$ - low, $k_l$ - low, $f_r$ - low, $k_r$ - low}
    \label{fig:biasbimrightlolololo}
    
\end{figure}

\begin{figure}[H]
    \centering
    \subfloat[The empirical distribution of readers' political belief]{
    \includegraphics[width = 0.33\textwidth]{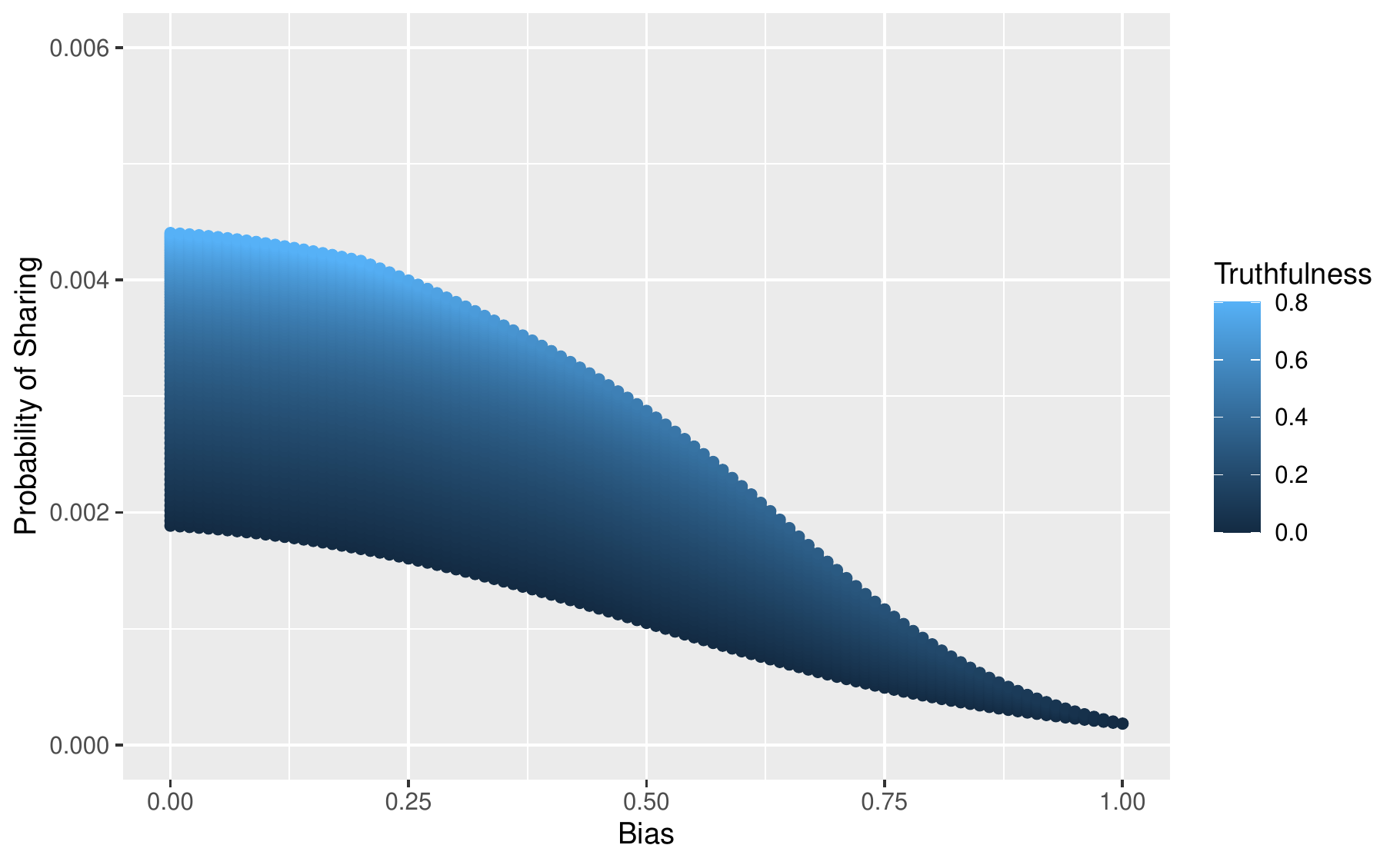}
    }
    \subfloat[A partisan bimodal distribution of readers' political belief 
    ]{
    \includegraphics[width = 0.33\textwidth]{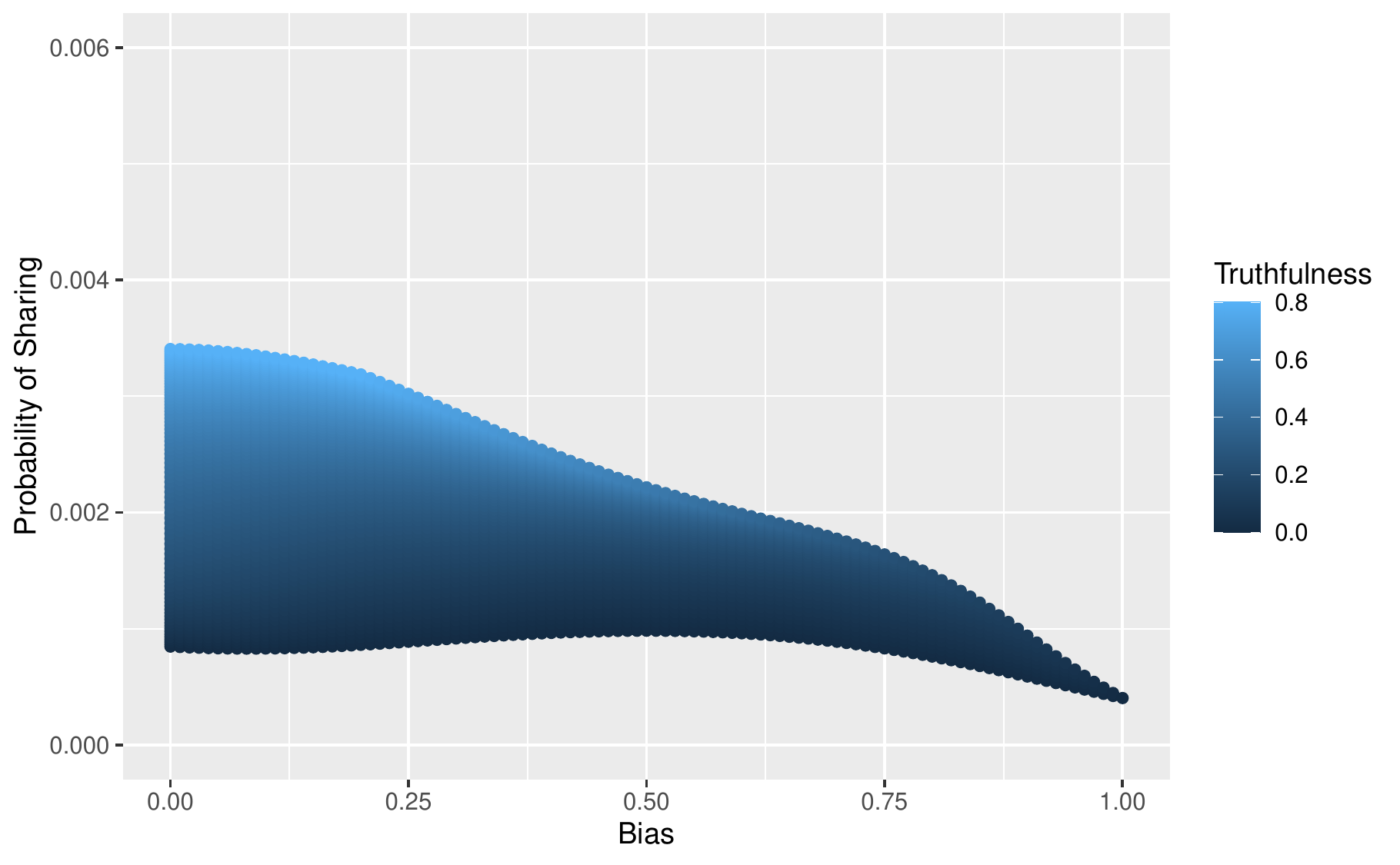}
    }
    \subfloat[A hyperpartisan bimodal distribution of readers' political belief 
    ]{
    \includegraphics[width = 0.33\textwidth]{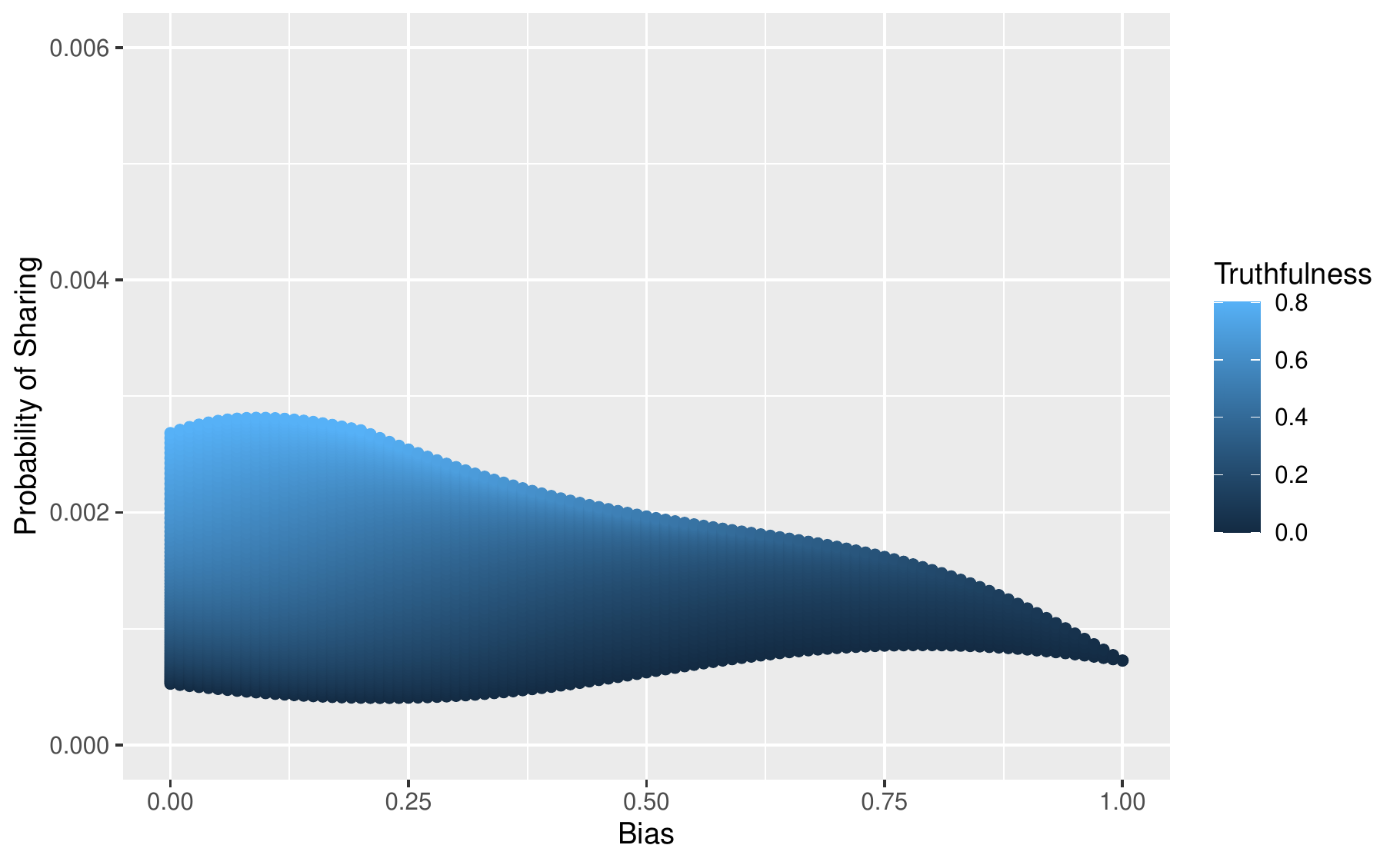}
    }
    
    \subfloat[A right-unimodal distribution of readers' political belief 
    ]{
    \includegraphics[width = 0.33\textwidth]{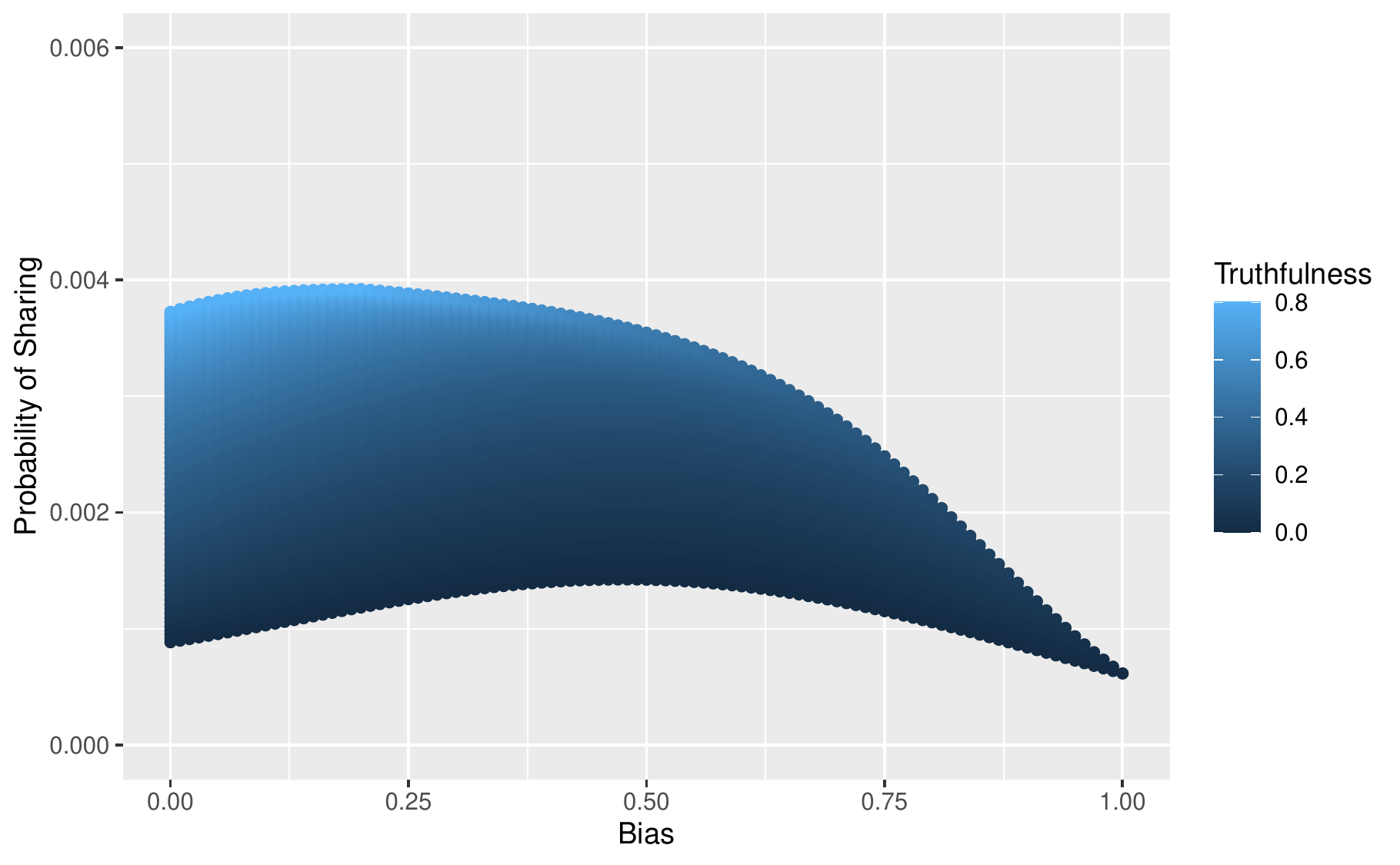}
    }
    \subfloat[A centrist-unimodal distribution of readers' political belief 
    ]{
    \includegraphics[width = 0.33\textwidth]{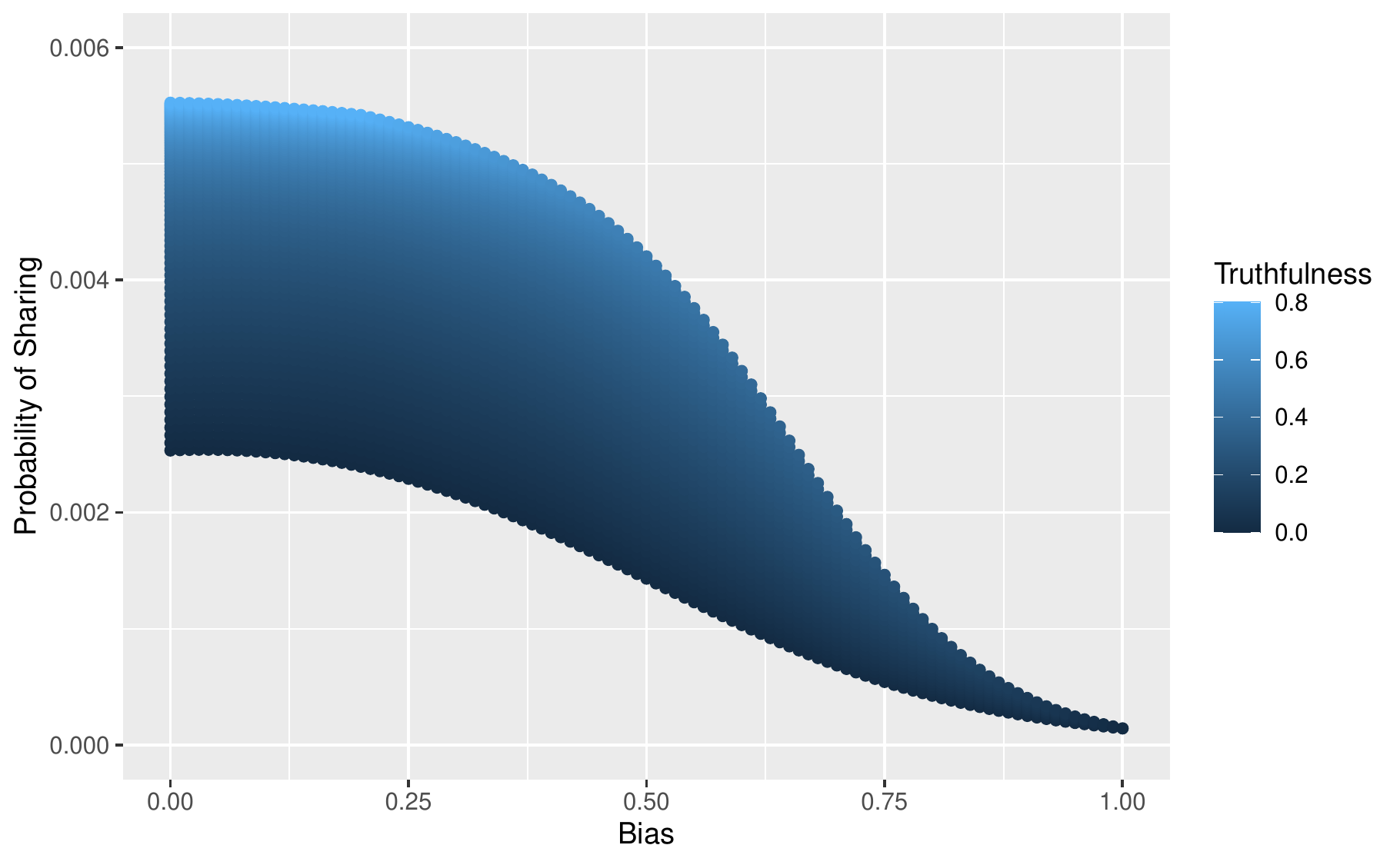}
    }
    \subfloat[A left-unimodal distribution of readers' political belief 
    ]{
    \includegraphics[width = 0.33\textwidth]{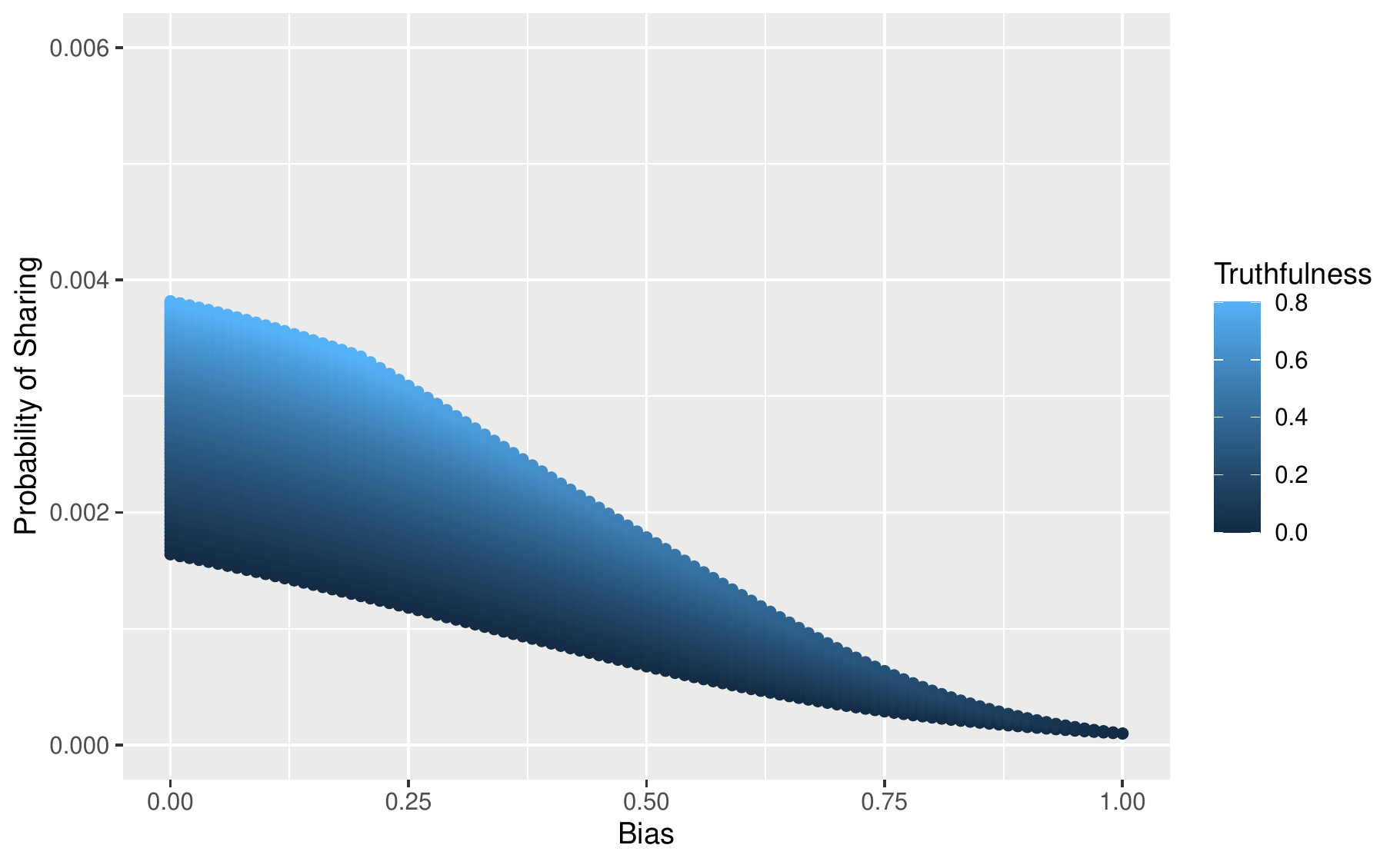}
    }
    \caption{The probability of sharing an article as a function of right political bias ($x$-axis) and truthfulness (gradient). Fitting parameters combination: $f_l$ - low, $k_l$ - low, $f_r$ - low, $k_r$ - high}
    \label{fig:biasbimrightlololohi}

\end{figure}

\begin{figure}[H]
    \centering
    \subfloat[The empirical distribution of readers' political belief]{
    \includegraphics[width = 0.33\textwidth]{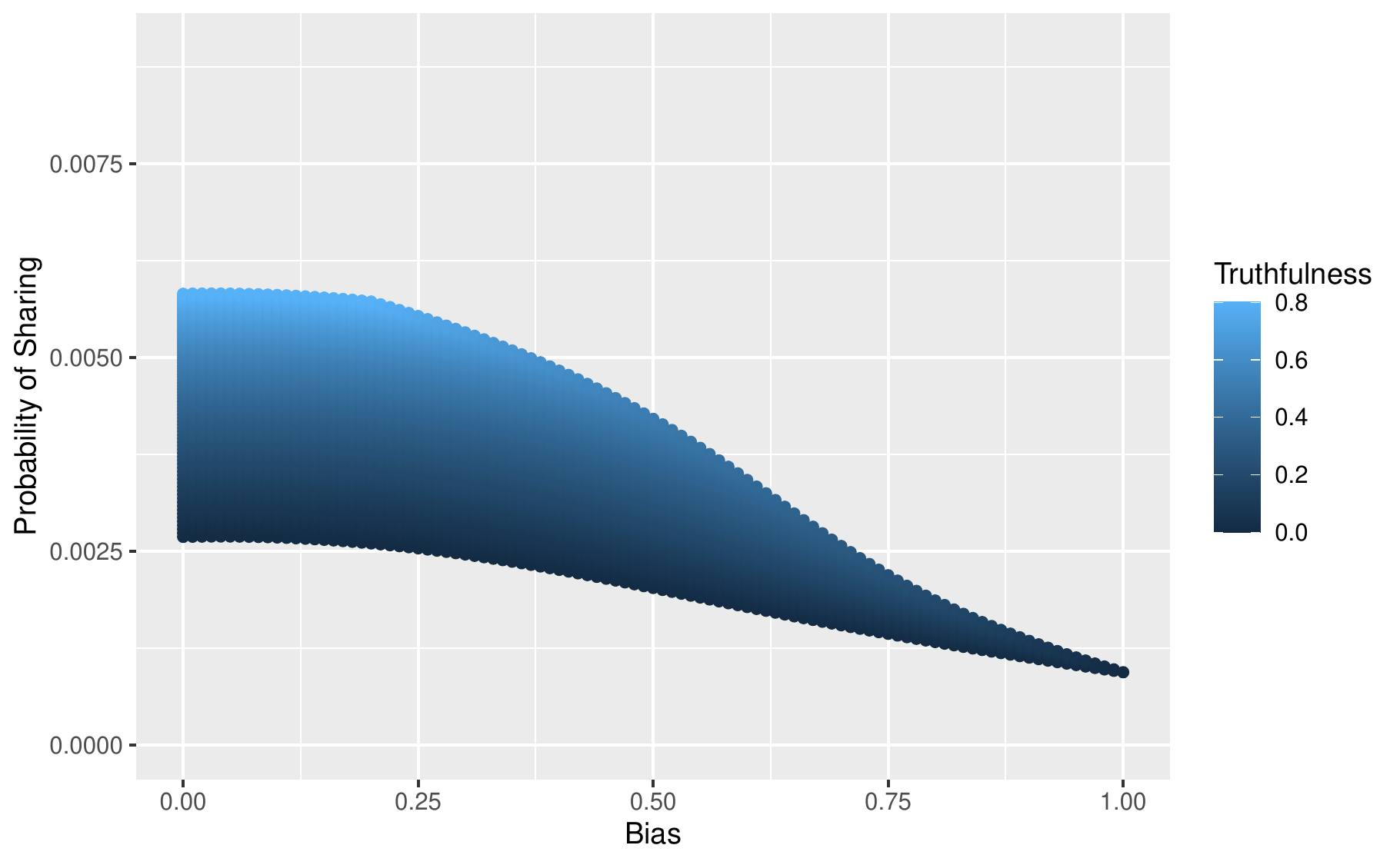}
    }
    \subfloat[A partisan bimodal distribution of readers' political belief 
    ]{
    \includegraphics[width = 0.33\textwidth]{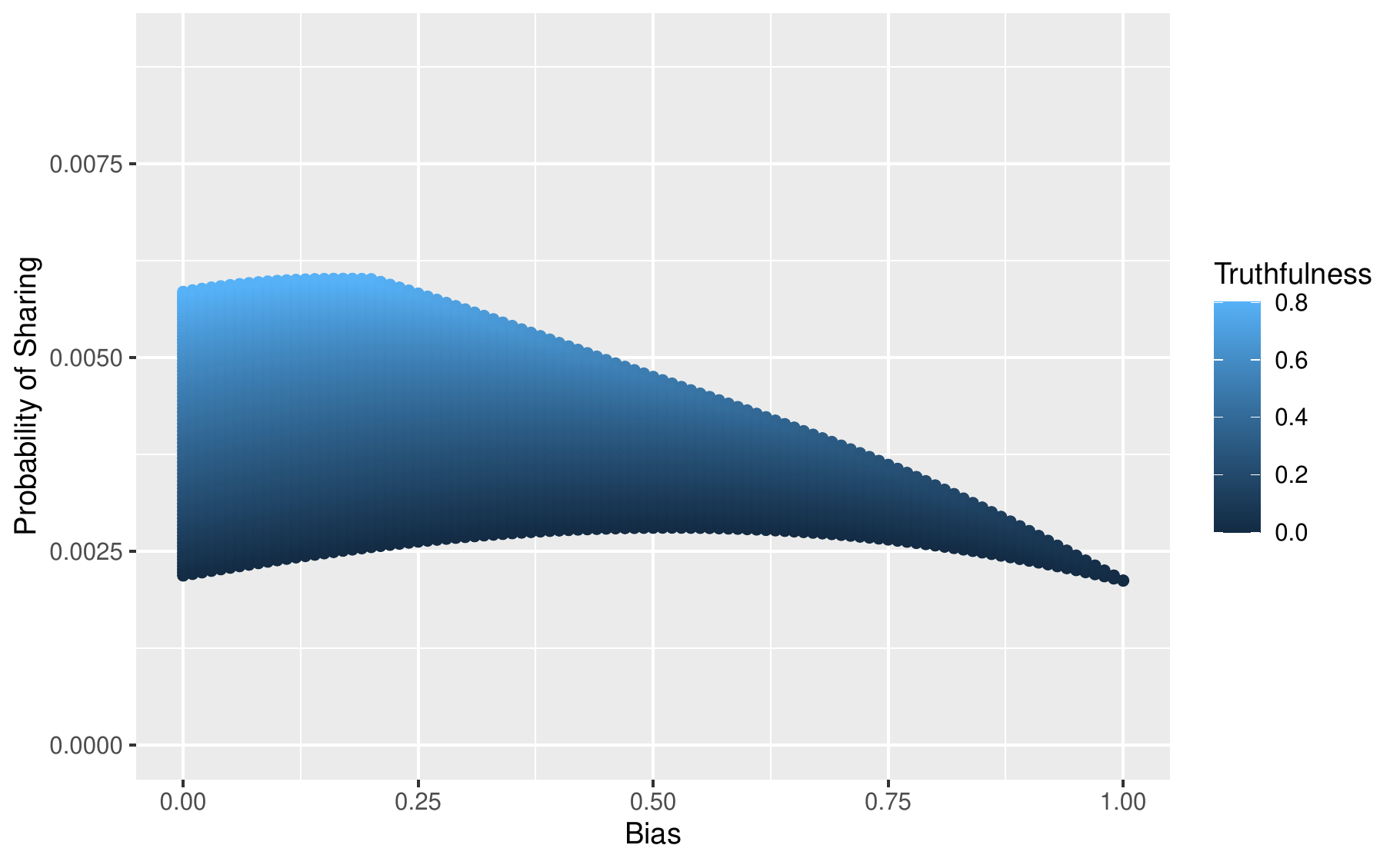}
    }
    \subfloat[A hyperpartisan bimodal distribution of readers' political belief 
    ]{
    \includegraphics[width = 0.33\textwidth]{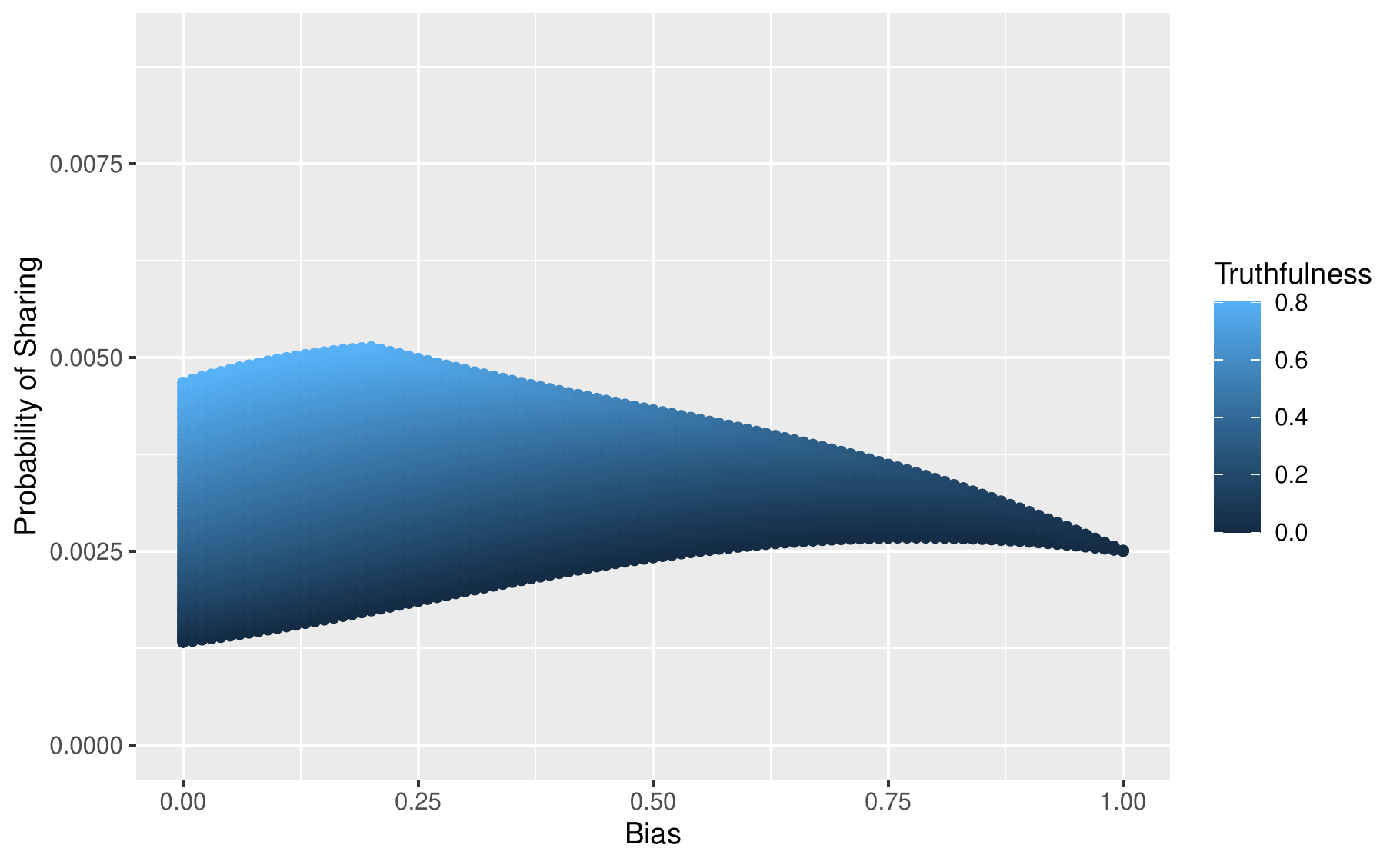}
    }
    
    \subfloat[A right-unimodal distribution of readers' political belief 
    ]{
    \includegraphics[width = 0.33\textwidth]{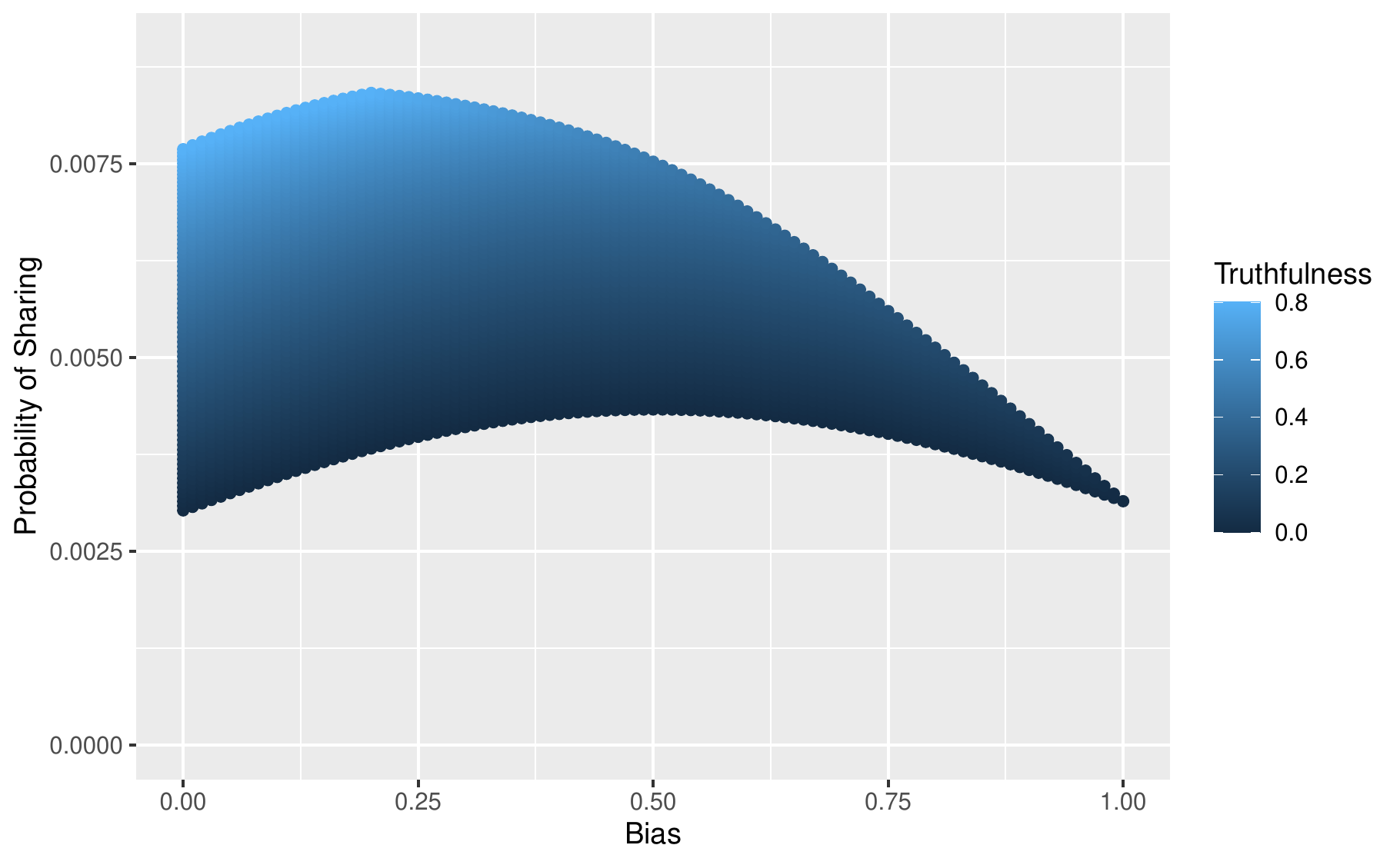}
    }
    \subfloat[A centrist-unimodal distribution of readers' political belief 
    ]{
    \includegraphics[width = 0.33\textwidth]{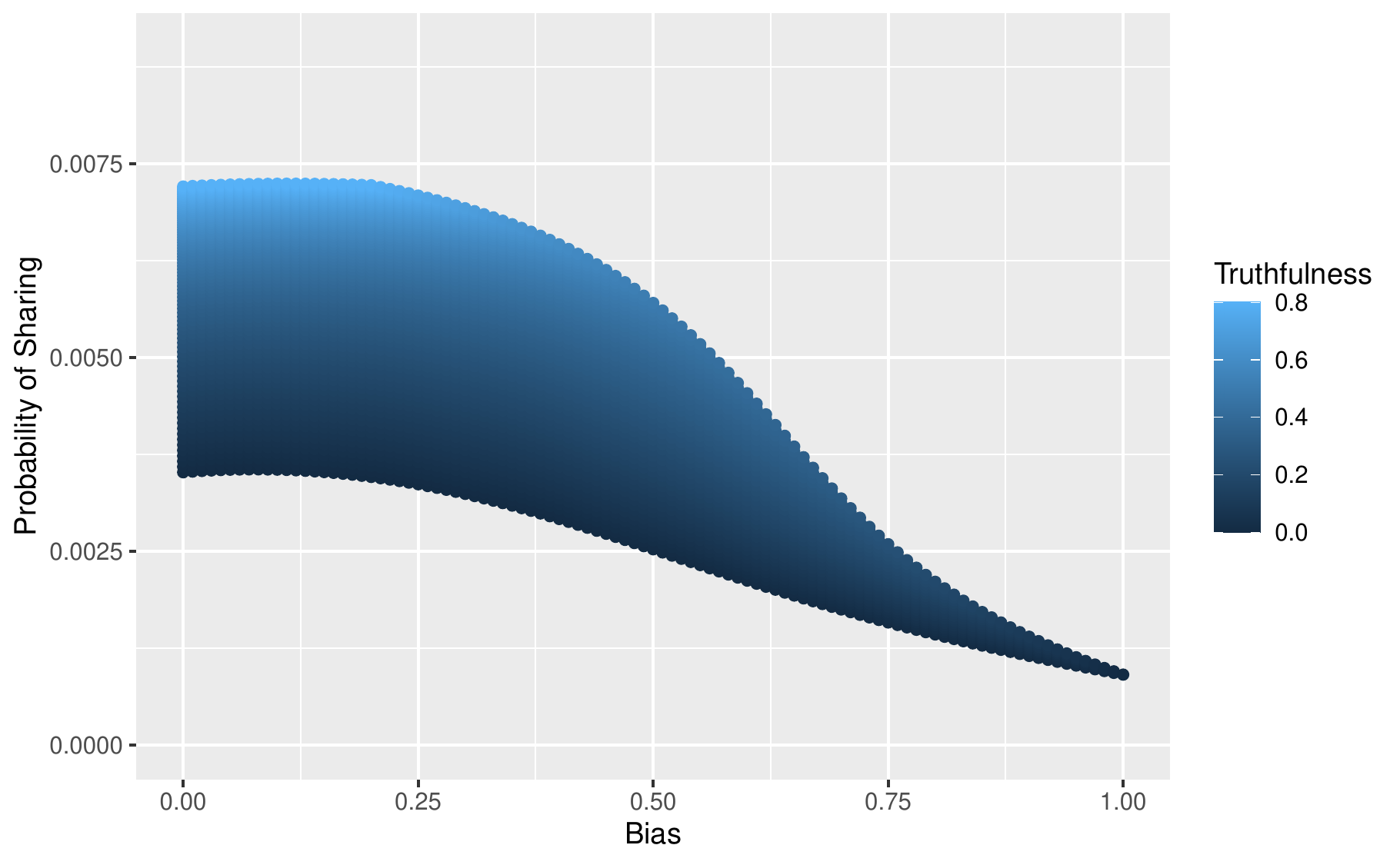}
    }
    \subfloat[A left-unimodal distribution of readers' political belief 
    ]{
    \includegraphics[width = 0.33\textwidth]{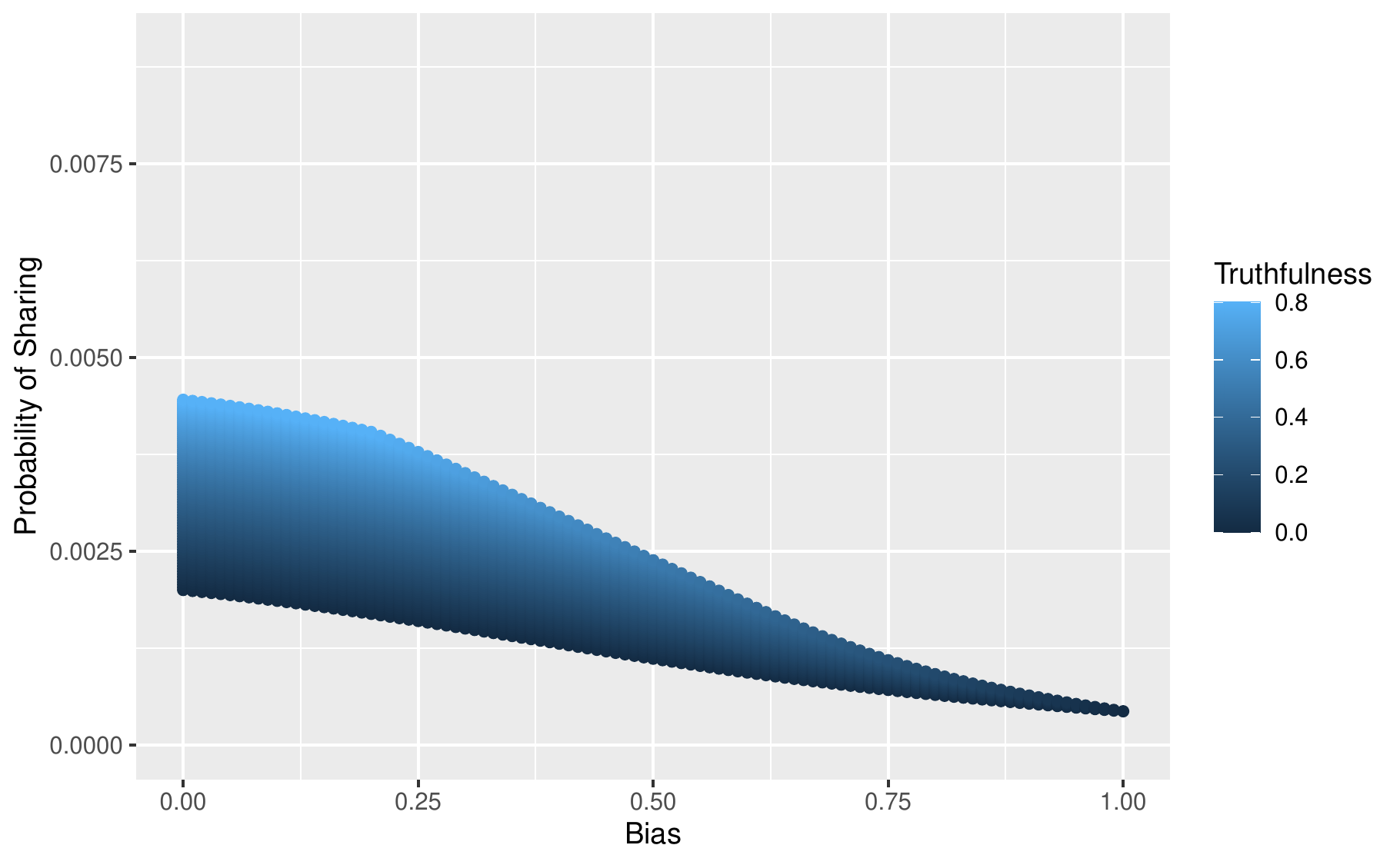}
    }
    \caption{The probability of sharing an article as a function of right political bias ($x$-axis) and truthfulness (gradient). Fitting parameters combination: $f_l$ - low, $k_l$ - low, $f_r$ - high, $k_r$ - low}
    \label{fig:biasbimrightlolohilo}

\end{figure}

\begin{figure}[H]
    \centering
    \subfloat[The empirical distribution of readers' political belief]{
    \includegraphics[width = 0.33\textwidth]{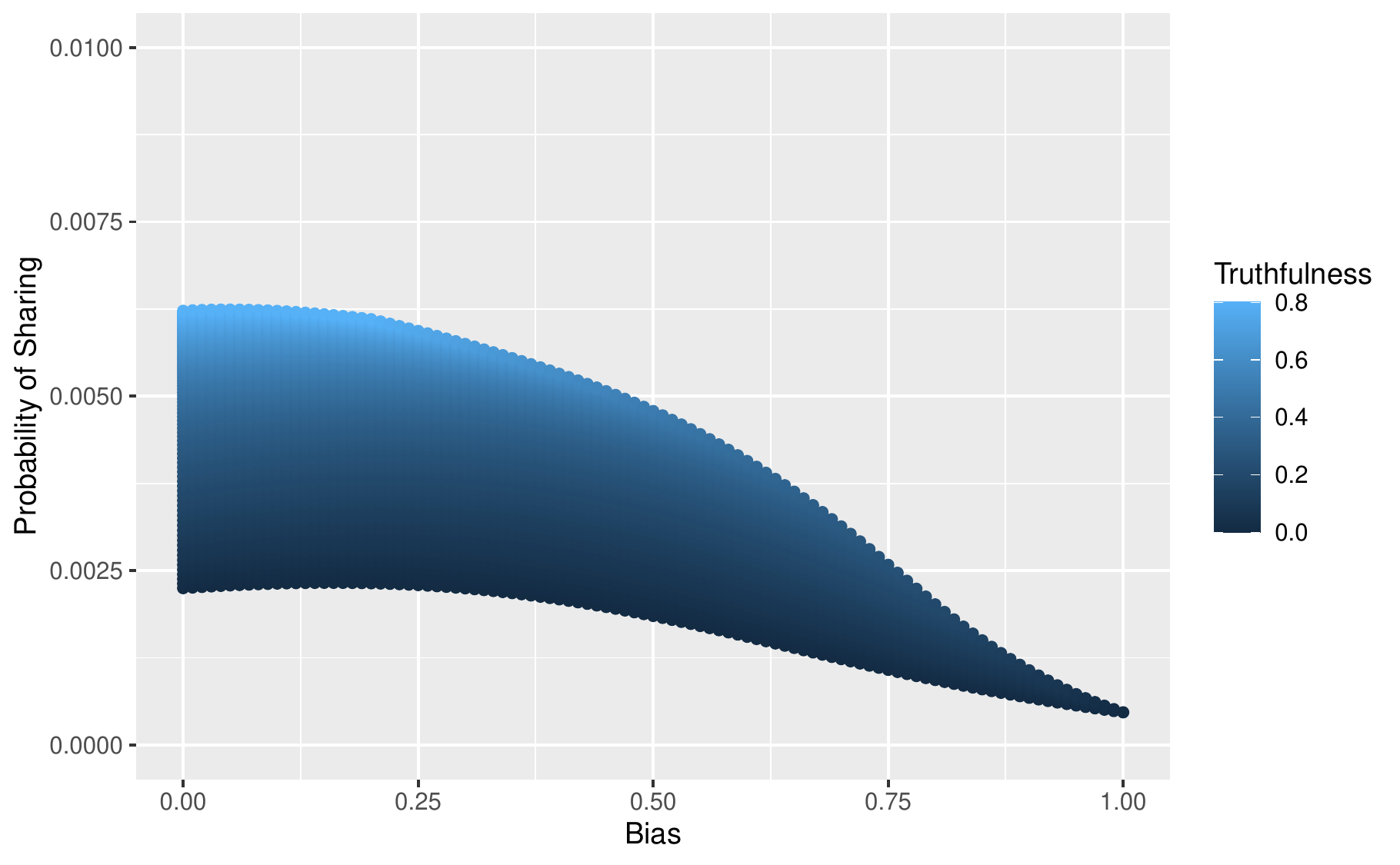}
    }
    \subfloat[A partisan bimodal distribution of readers' political belief 
    ]{
    \includegraphics[width = 0.33\textwidth]{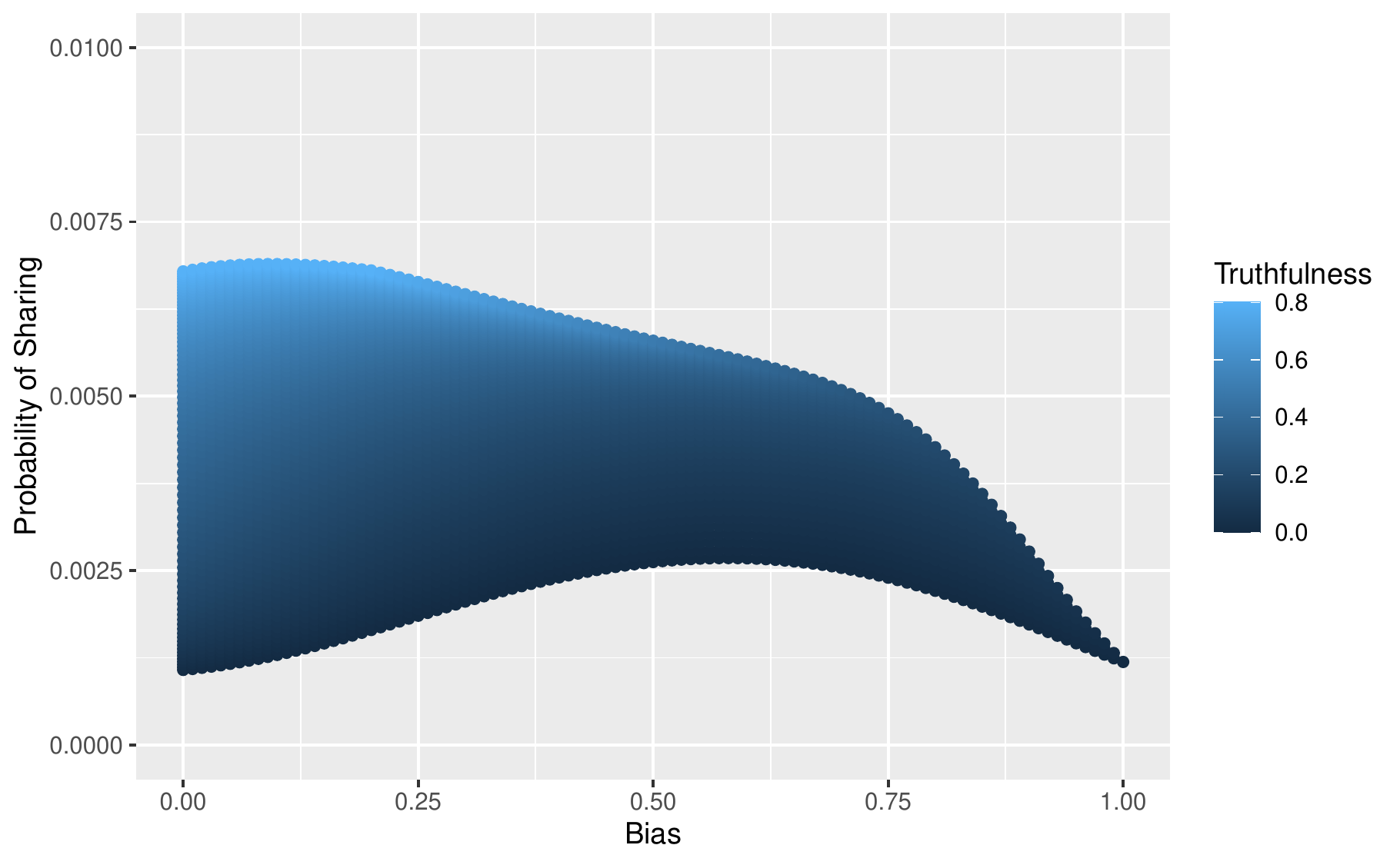}
    }
    \subfloat[A hyperpartisan bimodal distribution of readers' political belief 
    ]{
    \includegraphics[width = 0.33\textwidth]{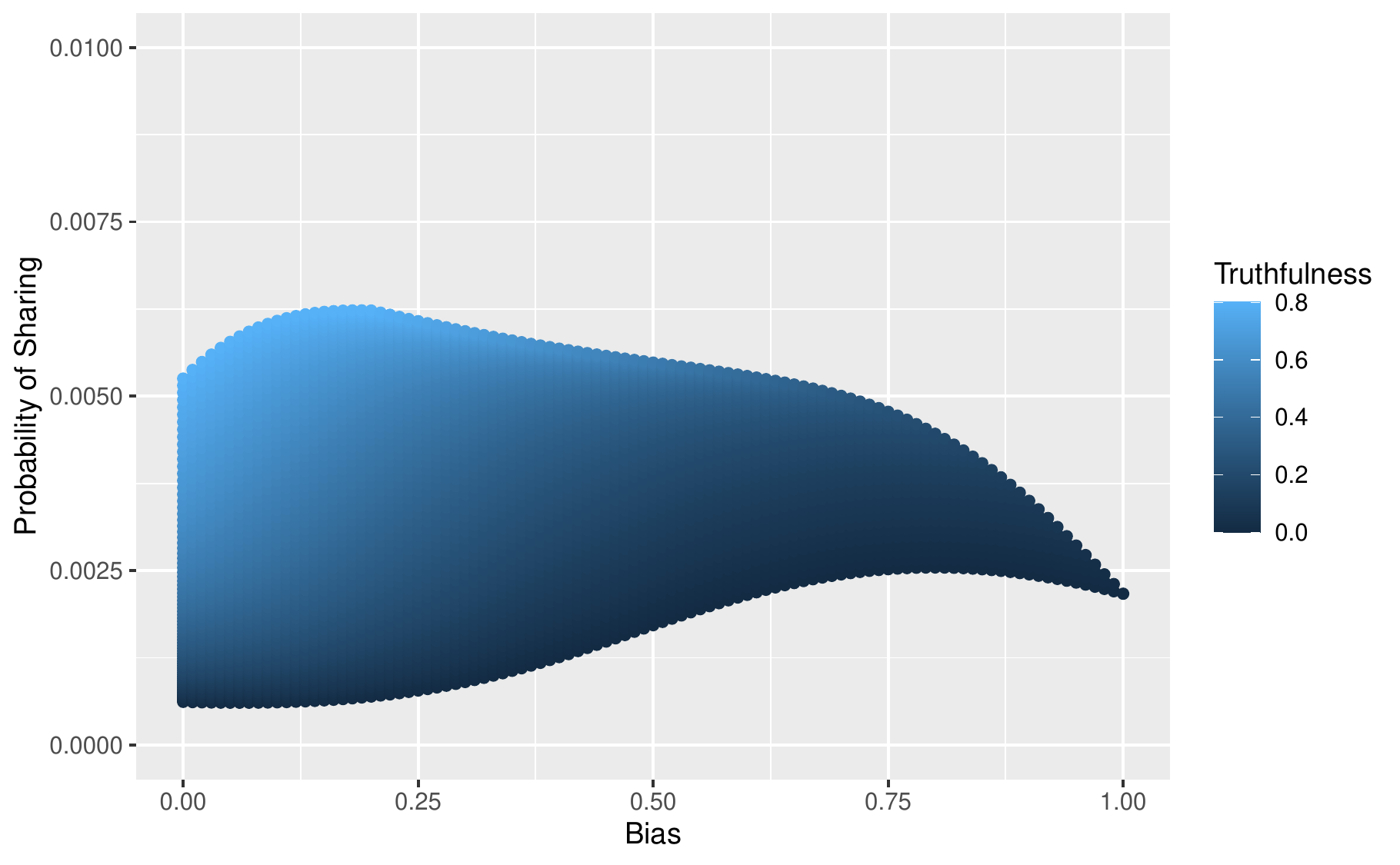}
    }
    
    \subfloat[A right-unimodal distribution of readers' political belief 
    ]{
    \includegraphics[width = 0.33\textwidth]{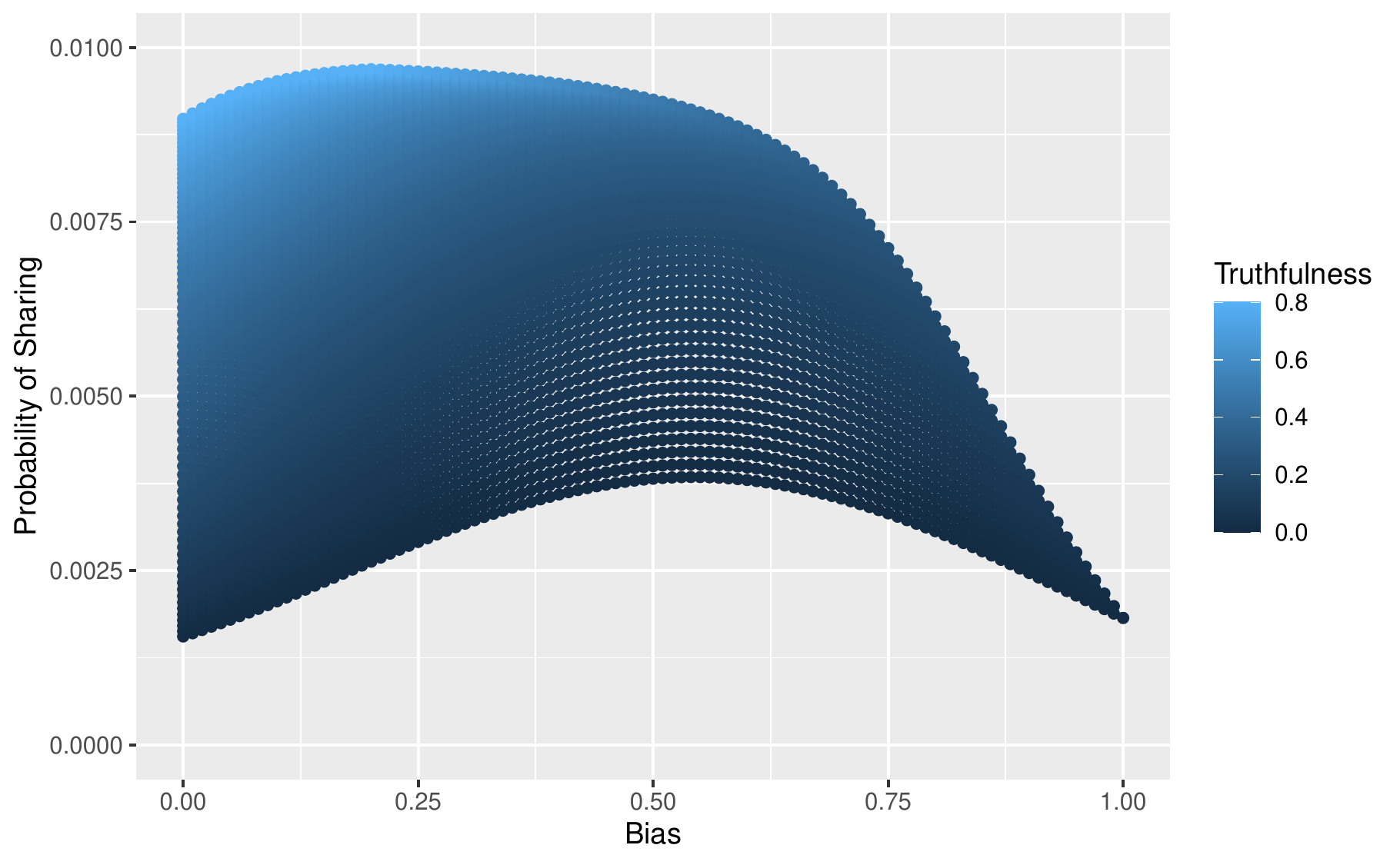}
    }
    \subfloat[A centrist-unimodal distribution of readers' political belief 
    ]{
    \includegraphics[width = 0.33\textwidth]{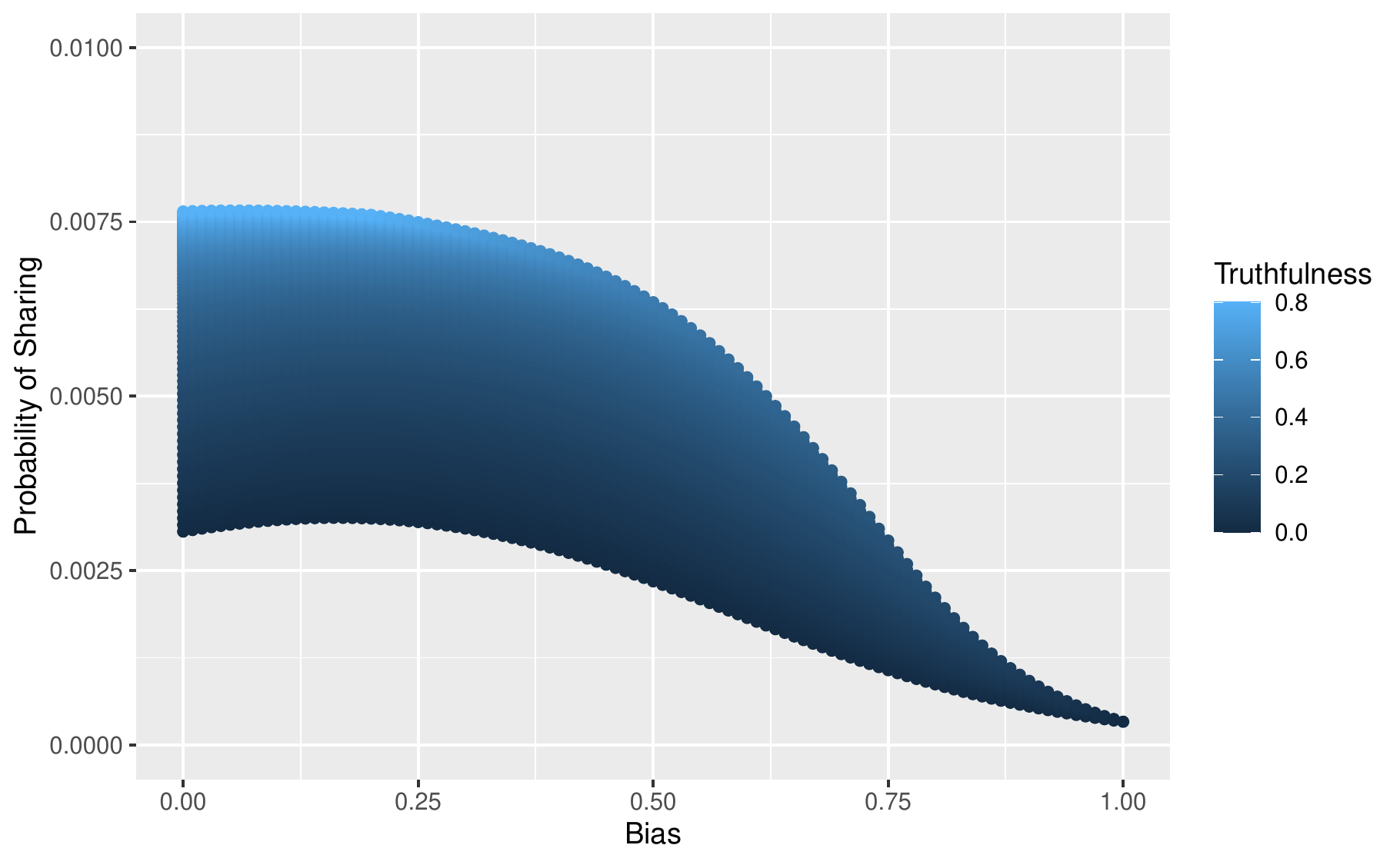}
    }
    \subfloat[A left-unimodal distribution of readers' political belief 
    ]{
    \includegraphics[width = 0.33\textwidth]{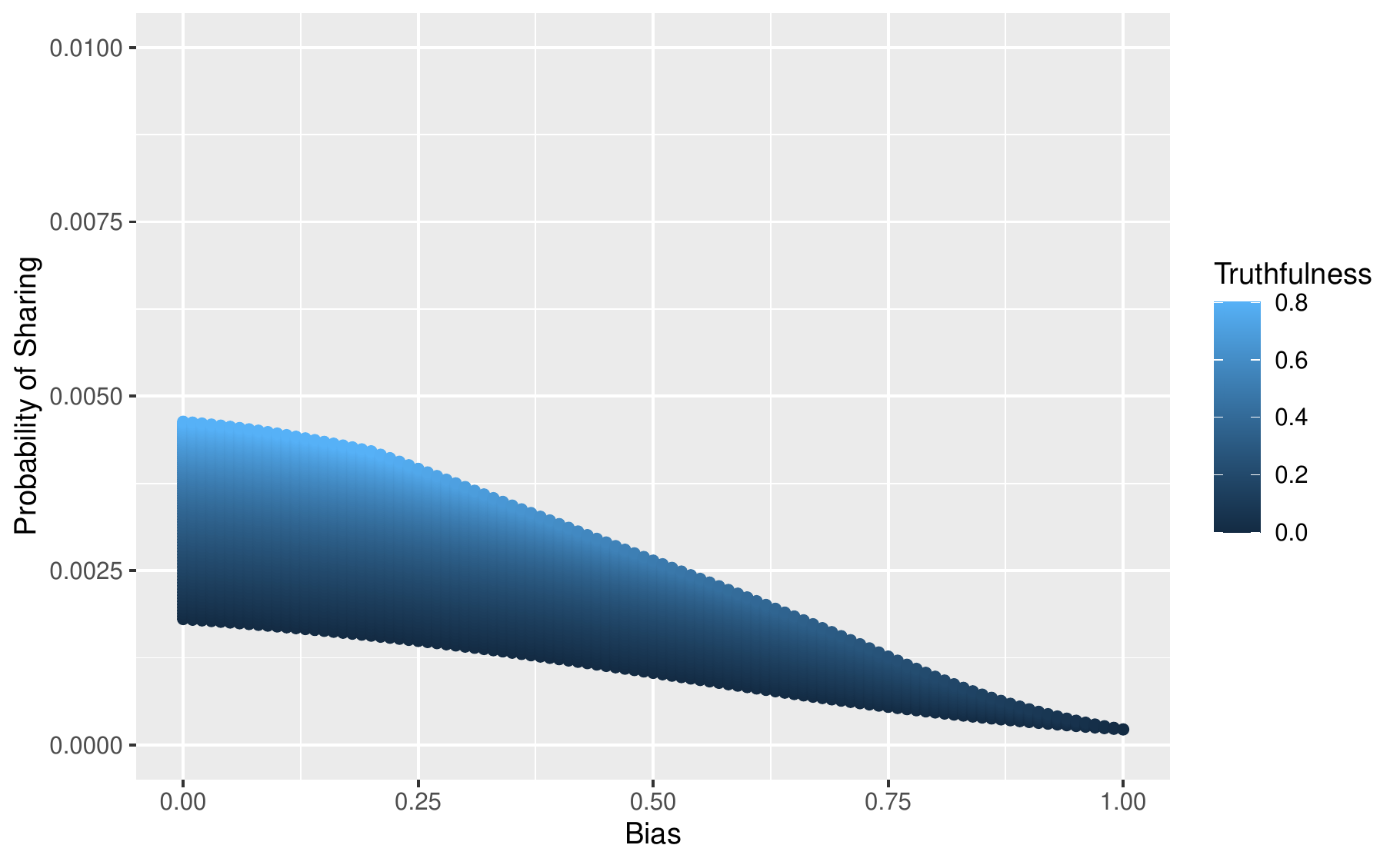}
    }
    \caption{The probability of sharing an article as a function of right political bias ($x$-axis) and truthfulness (gradient). Fitting parameters combination: $f_l$ - low, $k_l$ - low, $f_r$ - high, $k_r$ - high}
    \label{fig:biasbimrightlolohihi}
    
\end{figure}

\begin{figure}[H]
    \centering
    \subfloat[The empirical distribution of readers' political belief]{
    \includegraphics[width = 0.33\textwidth]{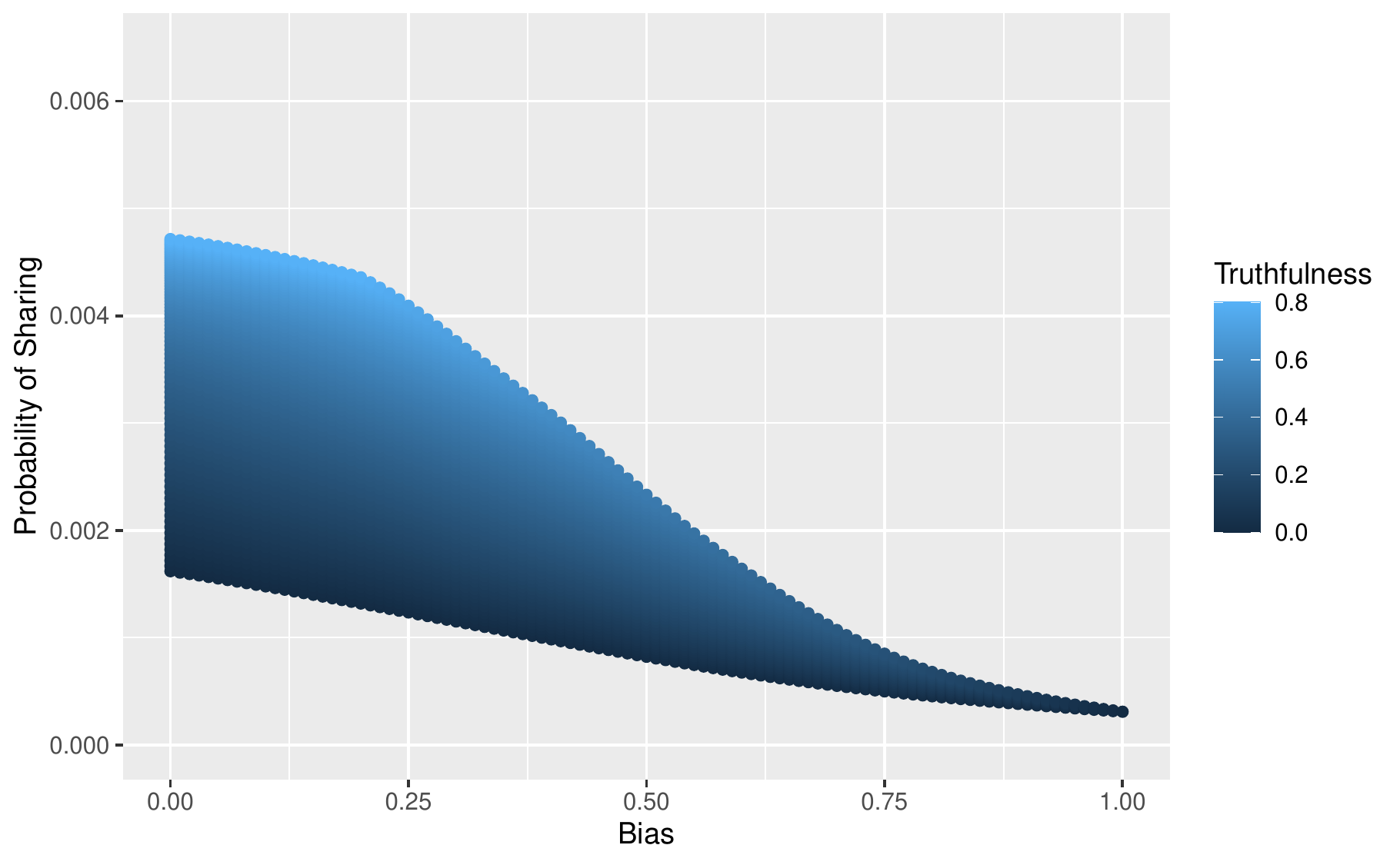}
    }
    \subfloat[A partisan bimodal distribution of readers' political belief 
    ]{
    \includegraphics[width = 0.33\textwidth]{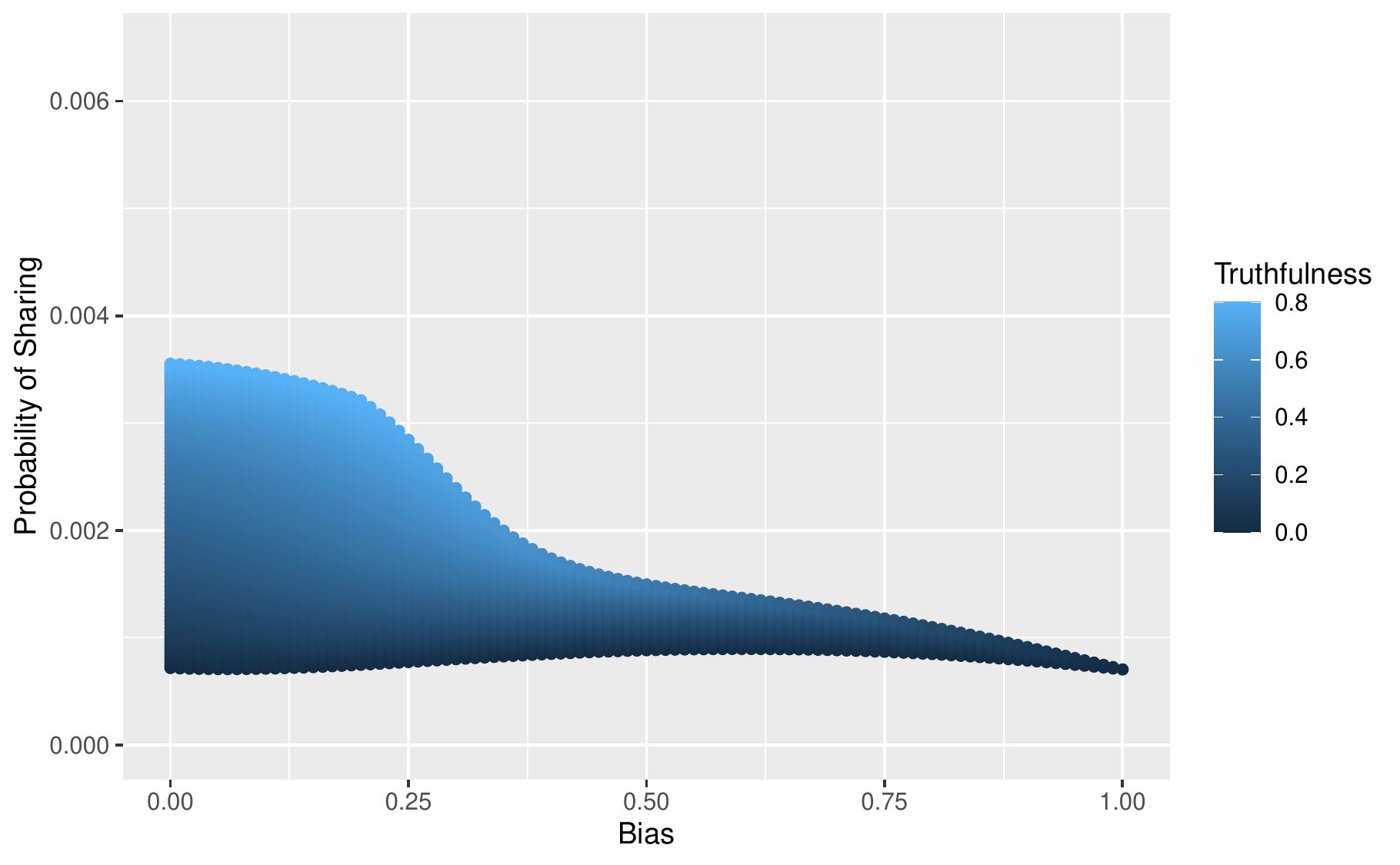}
    }
    \subfloat[A hyperpartisan bimodal distribution of readers' political belief 
    ]{
    \includegraphics[width = 0.33\textwidth]{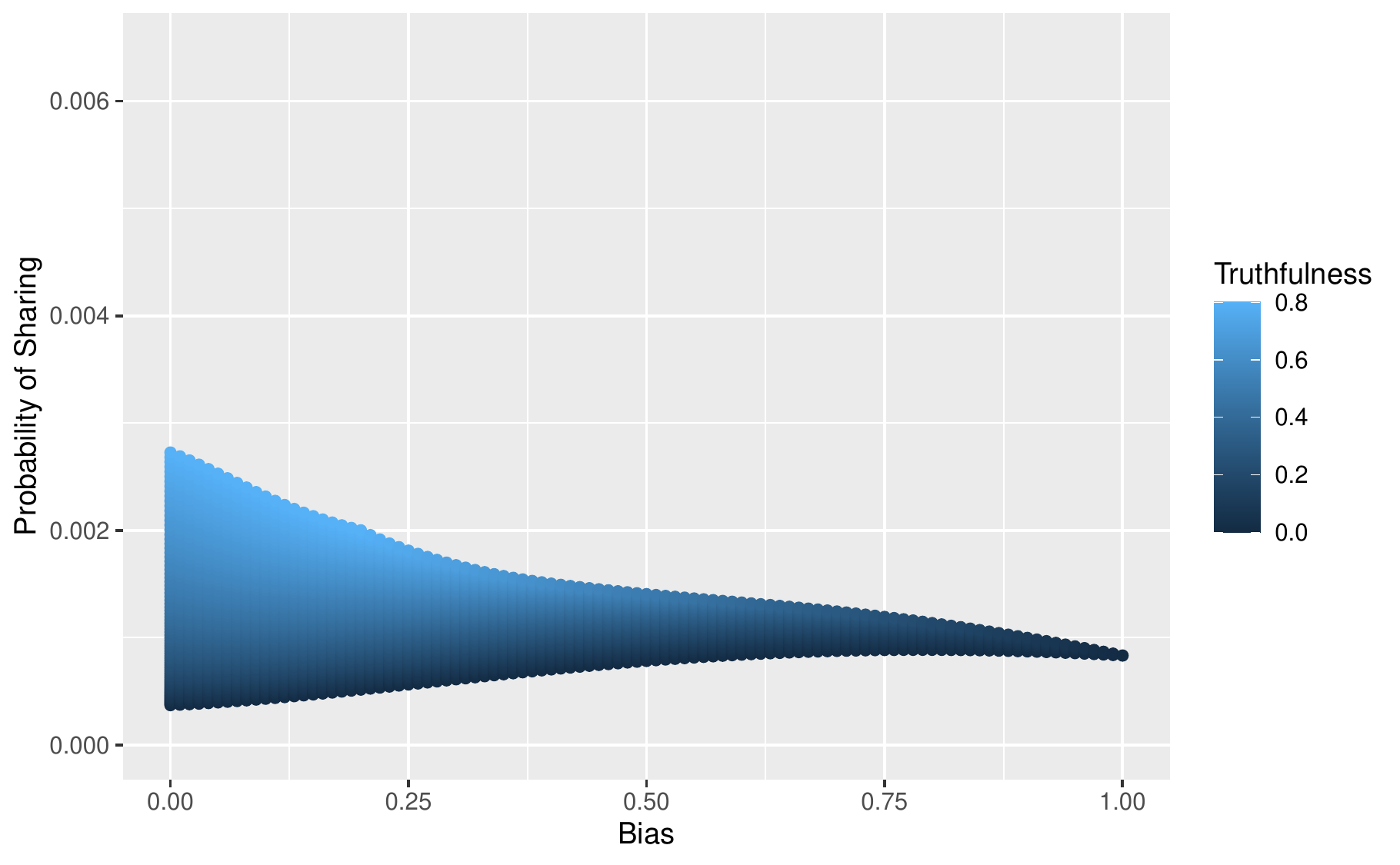}
    }
    
    \subfloat[A right-unimodal distribution of readers' political belief 
    ]{
    \includegraphics[width = 0.33\textwidth]{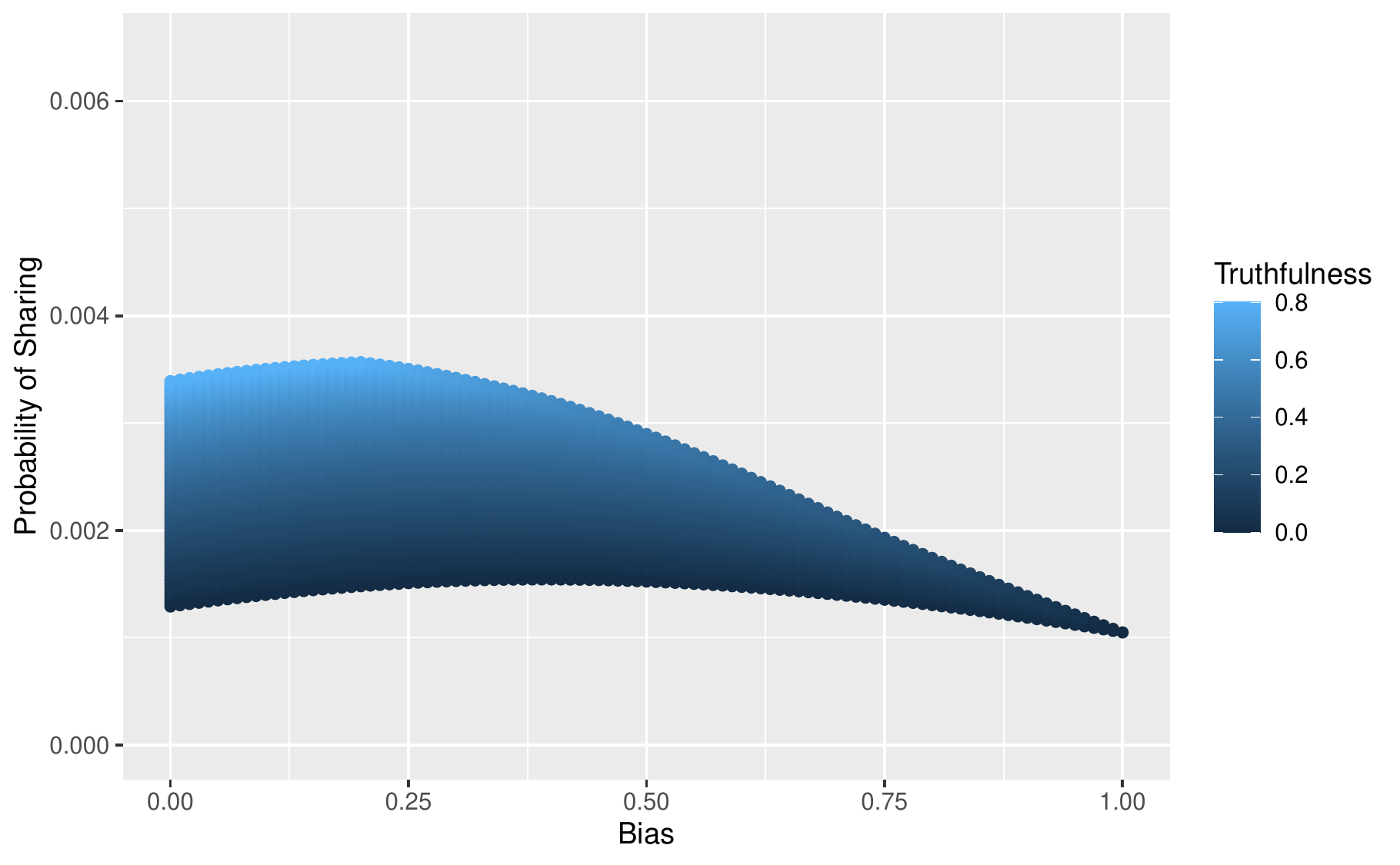}
    }
    \subfloat[A centrist-unimodal distribution of readers' political belief 
    ]{
    \includegraphics[width = 0.33\textwidth]{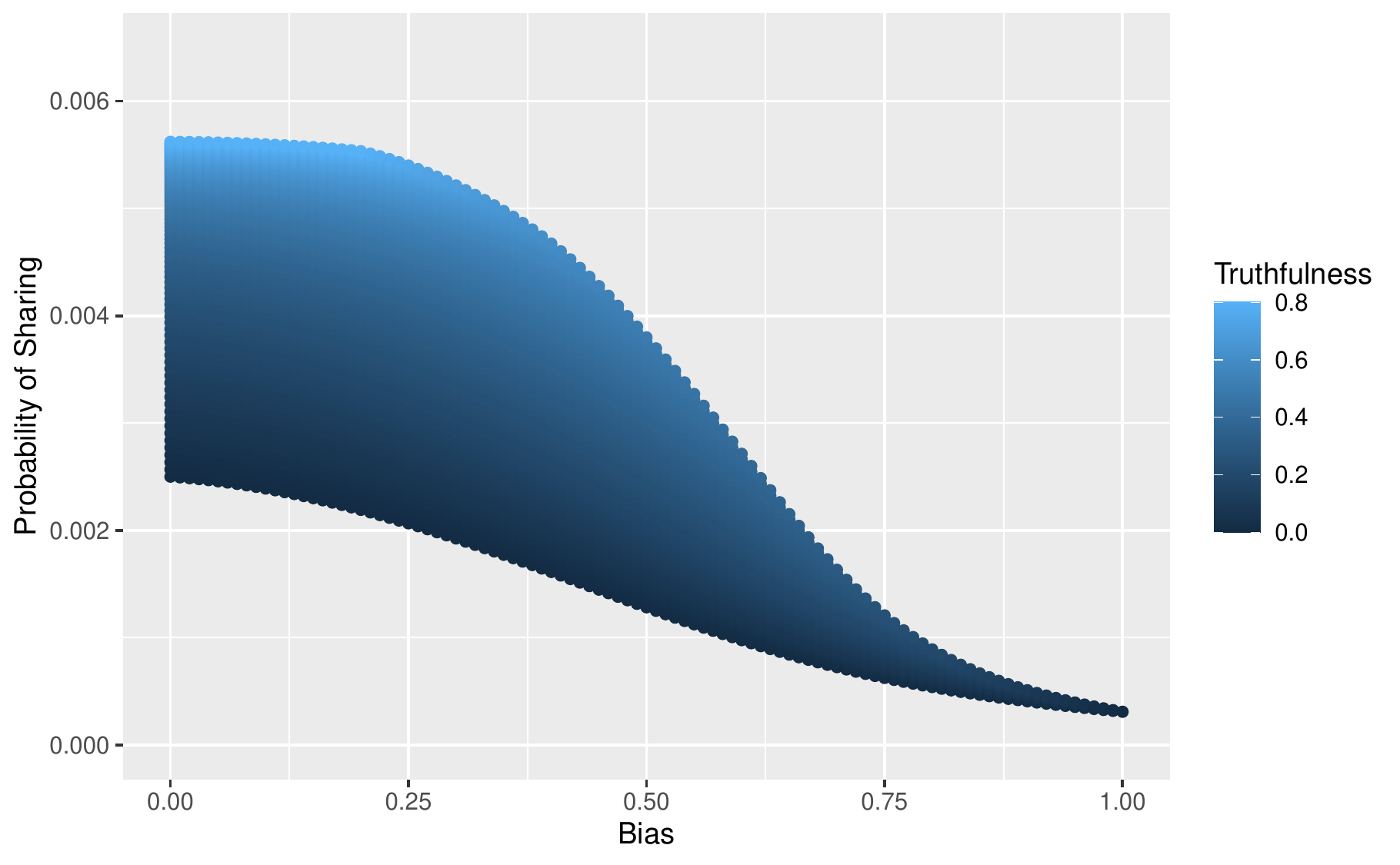}
    }
    \subfloat[A left-unimodal distribution of readers' political belief 
    ]{
    \includegraphics[width = 0.33\textwidth]{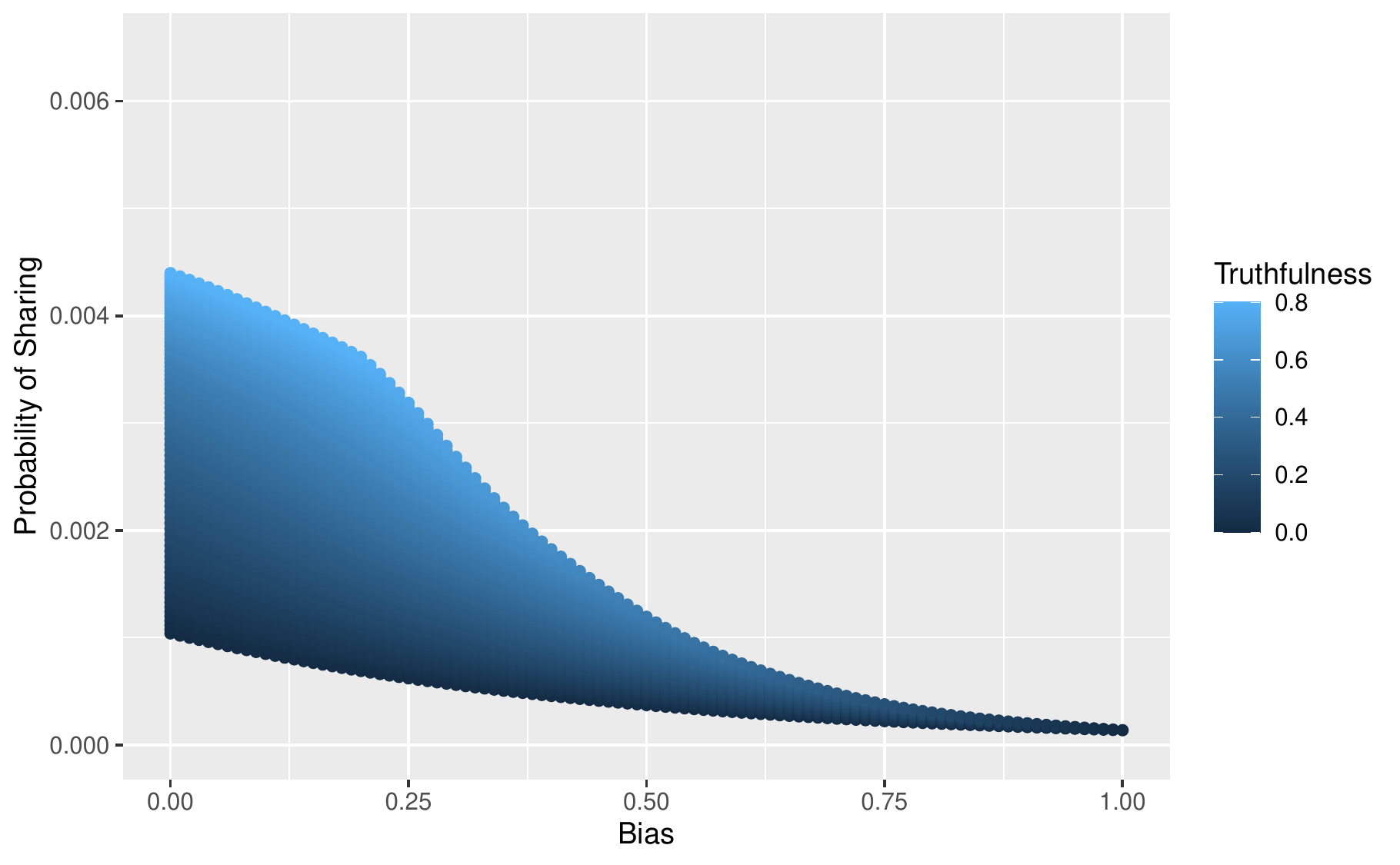}
    }
    \caption{The probability of sharing an article as a function of right political bias ($x$-axis) and truthfulness (gradient). Fitting parameters combination: $f_l$ - low, $k_l$ - high, $f_r$ - low, $k_r$ - low}
    \label{fig:biasbimrightlohilolo}

\end{figure}

\begin{figure}[H]
    \centering
    \subfloat[The empirical distribution of readers' political belief]{
    \includegraphics[width = 0.33\textwidth]{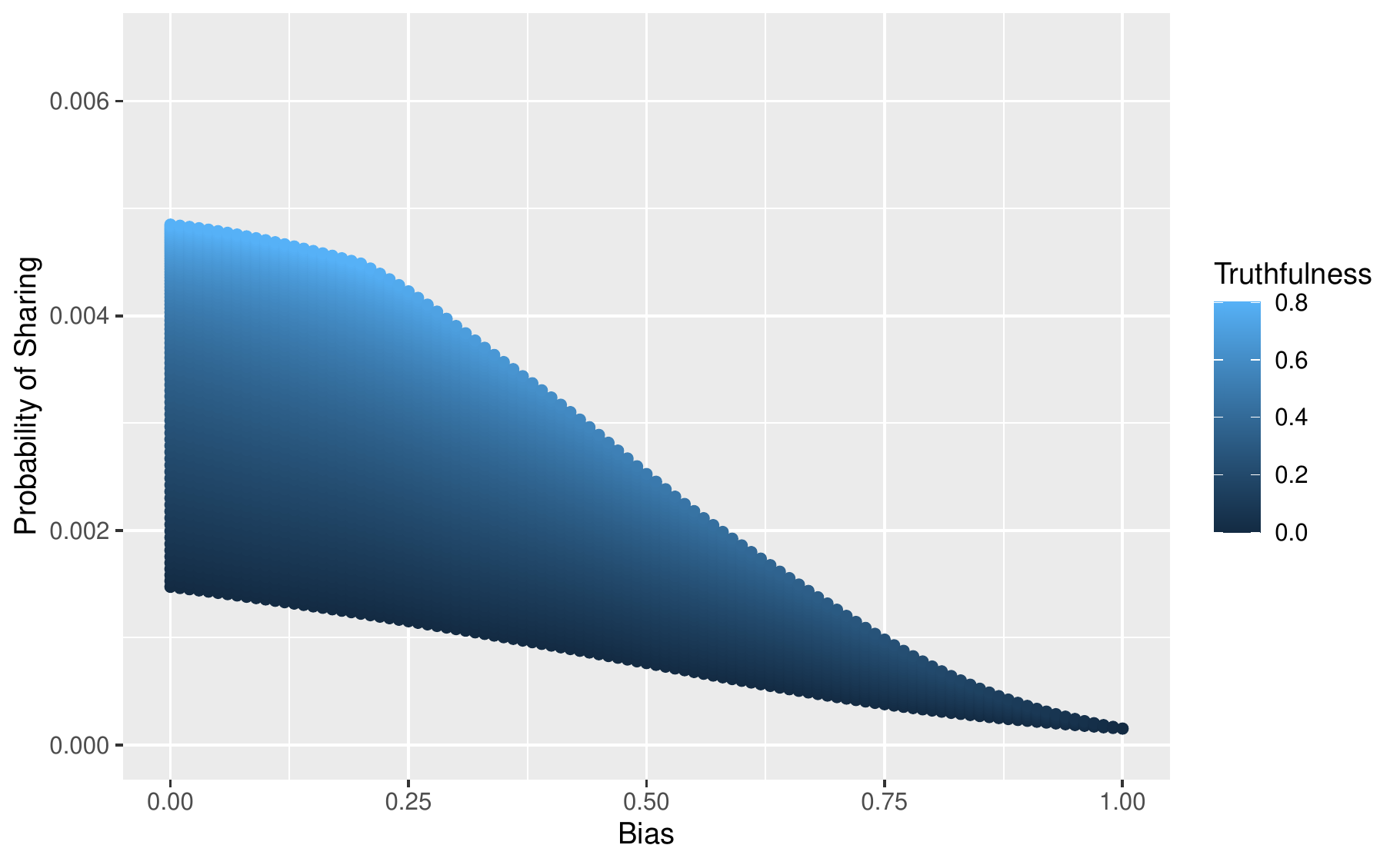}
    }
    \subfloat[A partisan bimodal distribution of readers' political belief 
    ]{
    \includegraphics[width = 0.33\textwidth]{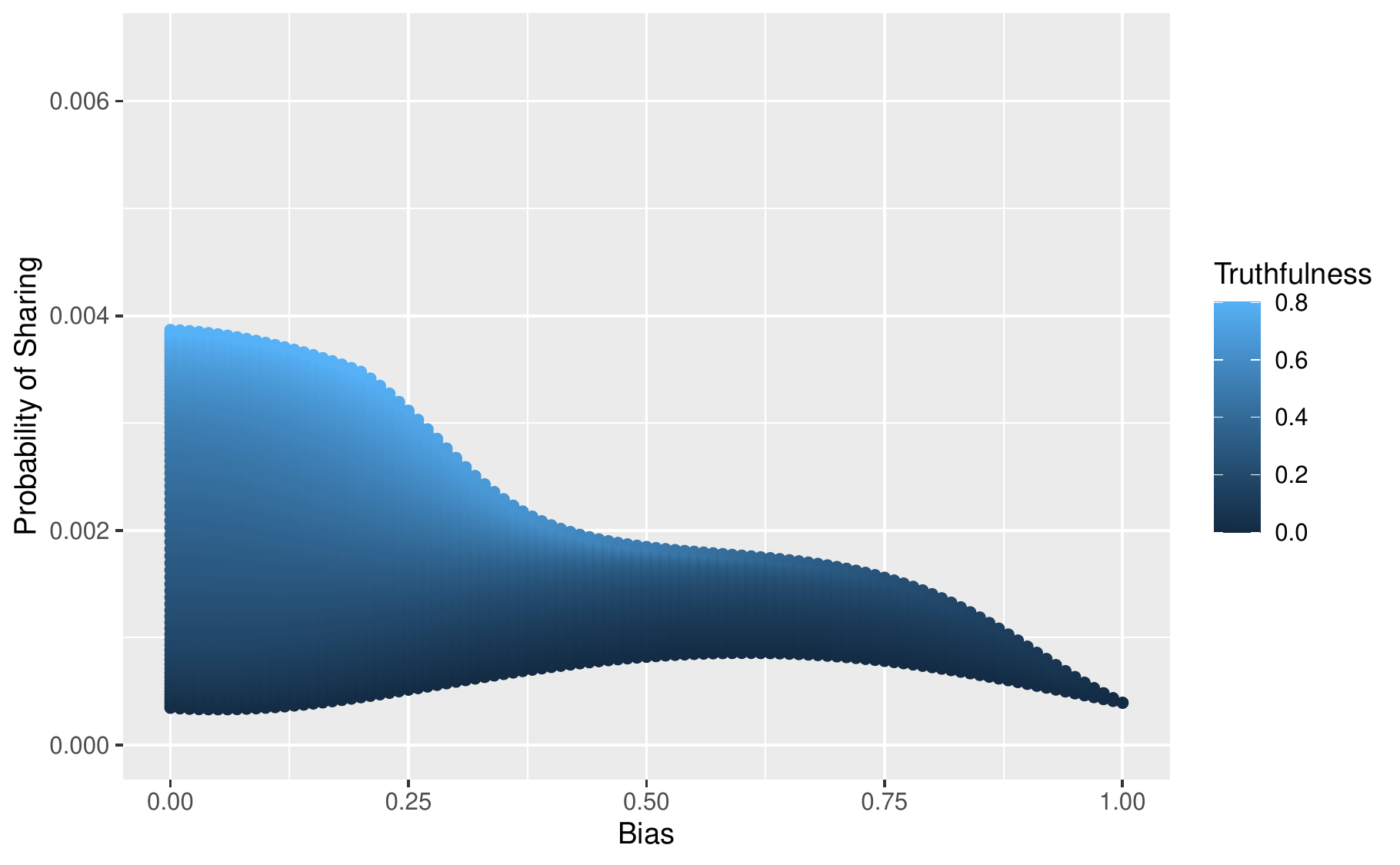}
    }
    \subfloat[A hyperpartisan bimodal distribution of readers' political belief 
    ]{
    \includegraphics[width = 0.33\textwidth]{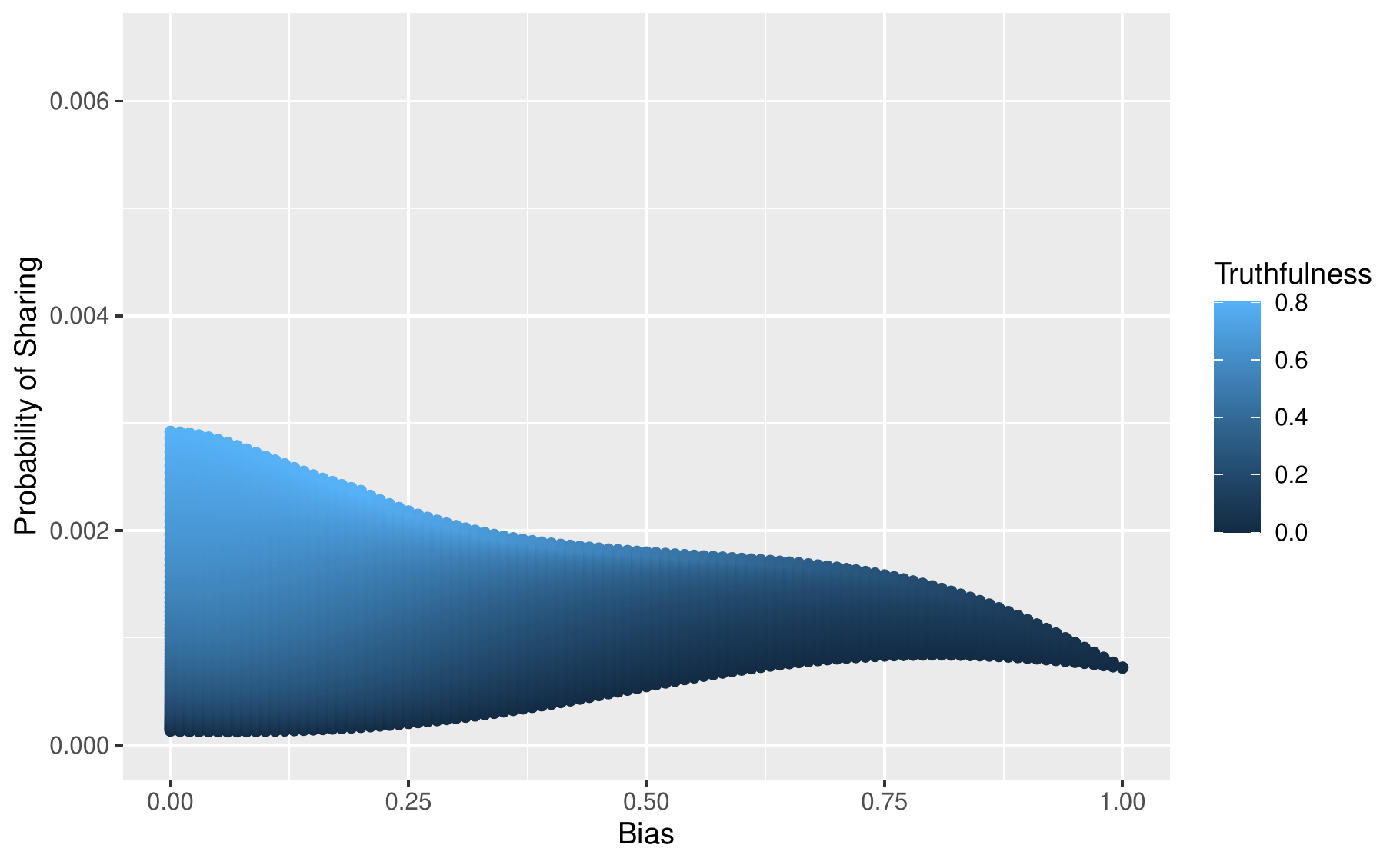}
    }
    
    \subfloat[A right-unimodal distribution of readers' political belief 
    ]{
    \includegraphics[width = 0.33\textwidth]{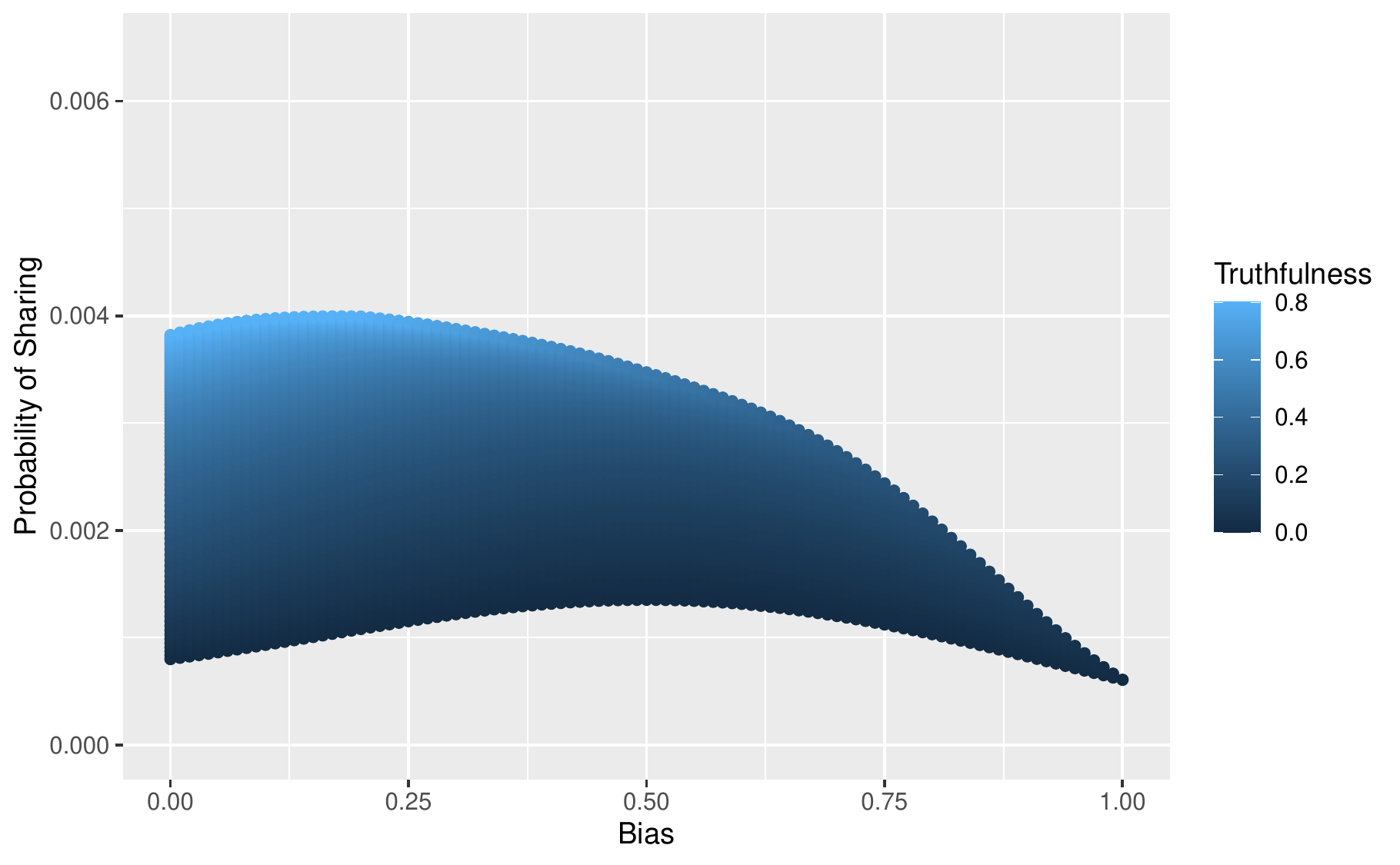}
    }
    \subfloat[A centrist-unimodal distribution of readers' political belief 
    ]{
    \includegraphics[width = 0.33\textwidth]{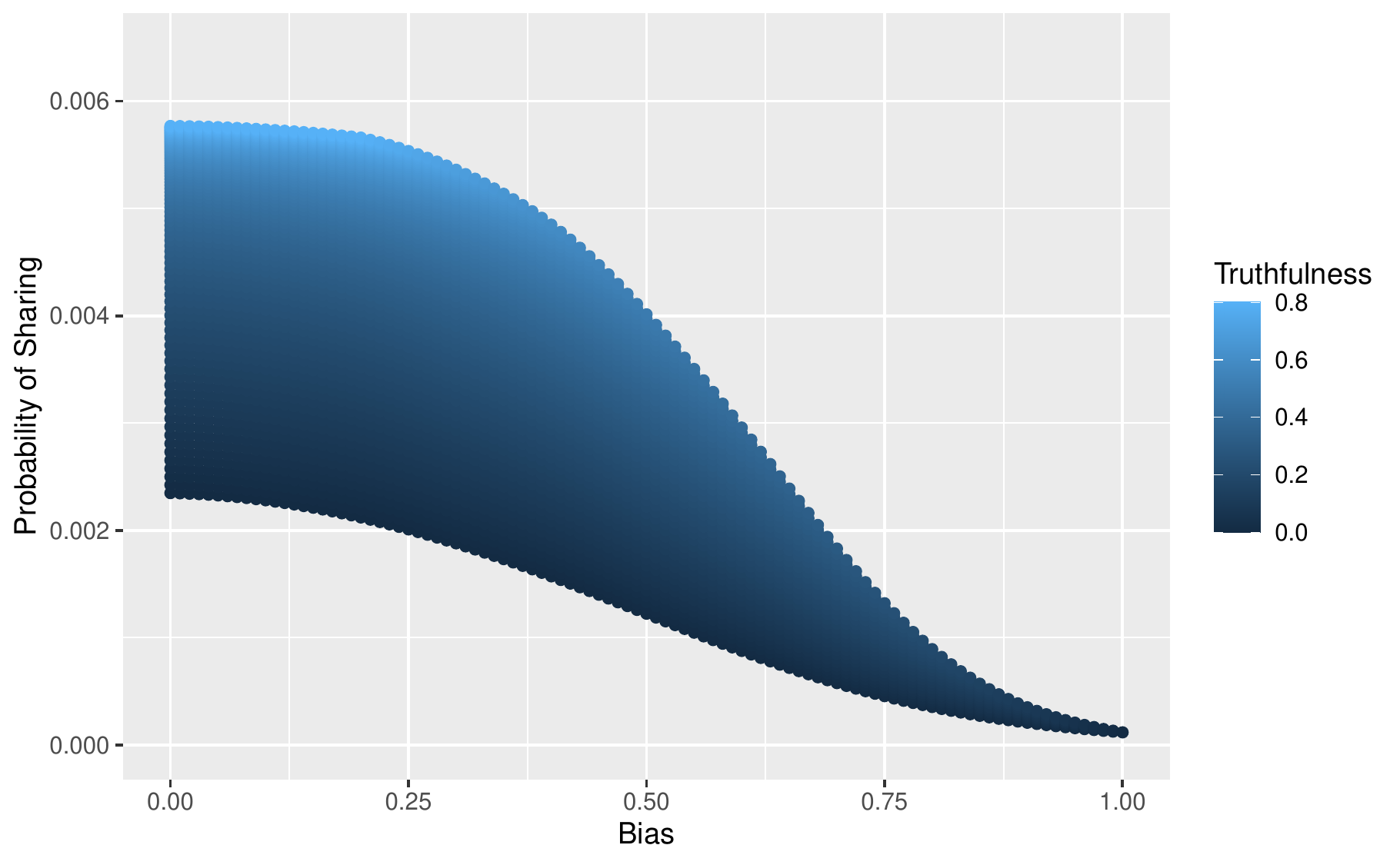}
    }
    \subfloat[A left-unimodal distribution of readers' political belief 
    ]{
    \includegraphics[width = 0.33\textwidth]{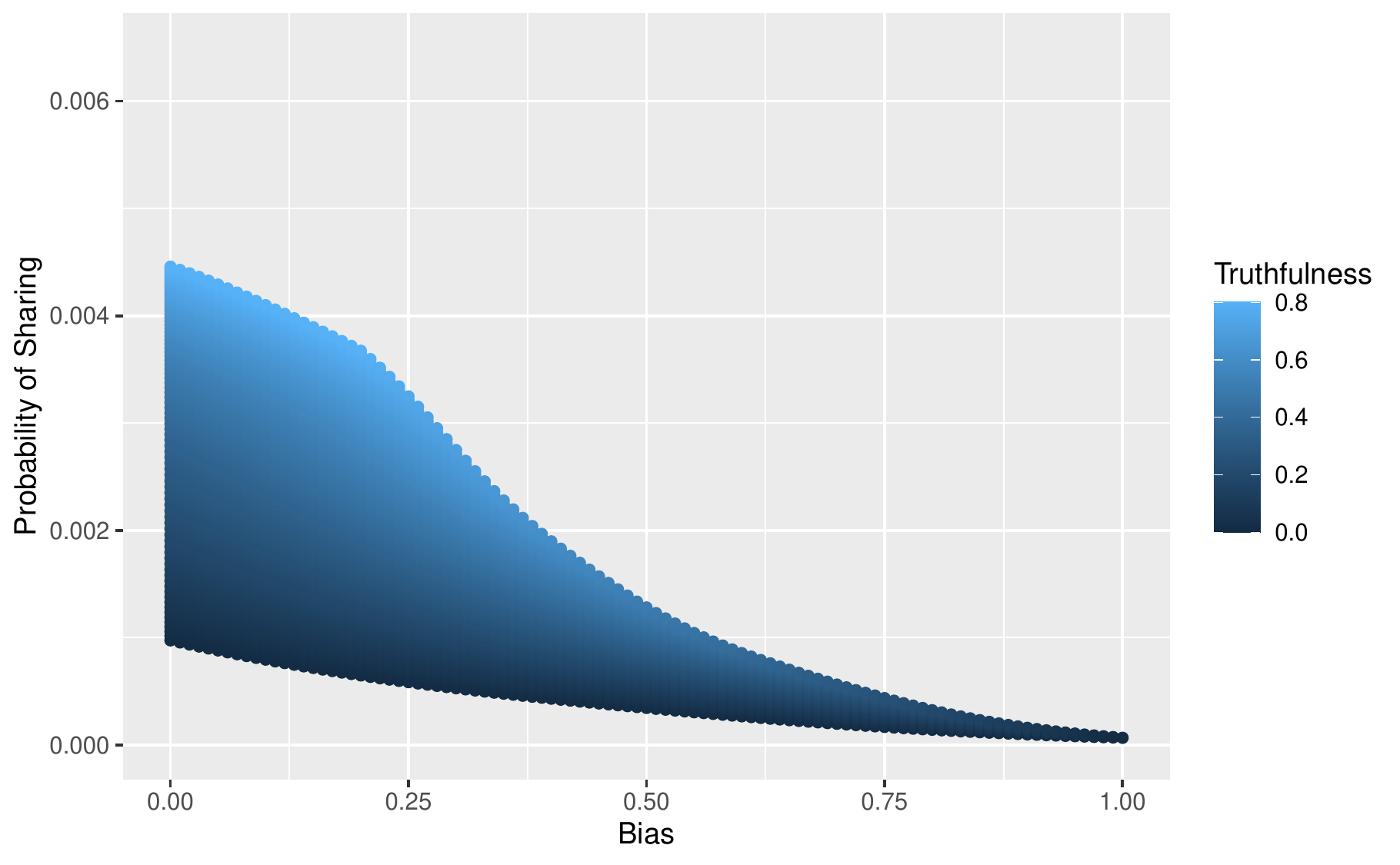}
    }
    \caption{The probability of sharing an article as a function of right political bias ($x$-axis) and truthfulness (gradient). Fitting parameters combination: $f_l$ - low, $k_l$ - high, $f_r$ - low, $k_r$ - high}
    \label{fig:biasbimrightlohilohi}

\end{figure}

\begin{figure}[H]
    \centering
    \subfloat[The empirical distribution of readers' political belief]{
    \includegraphics[width = 0.33\textwidth]{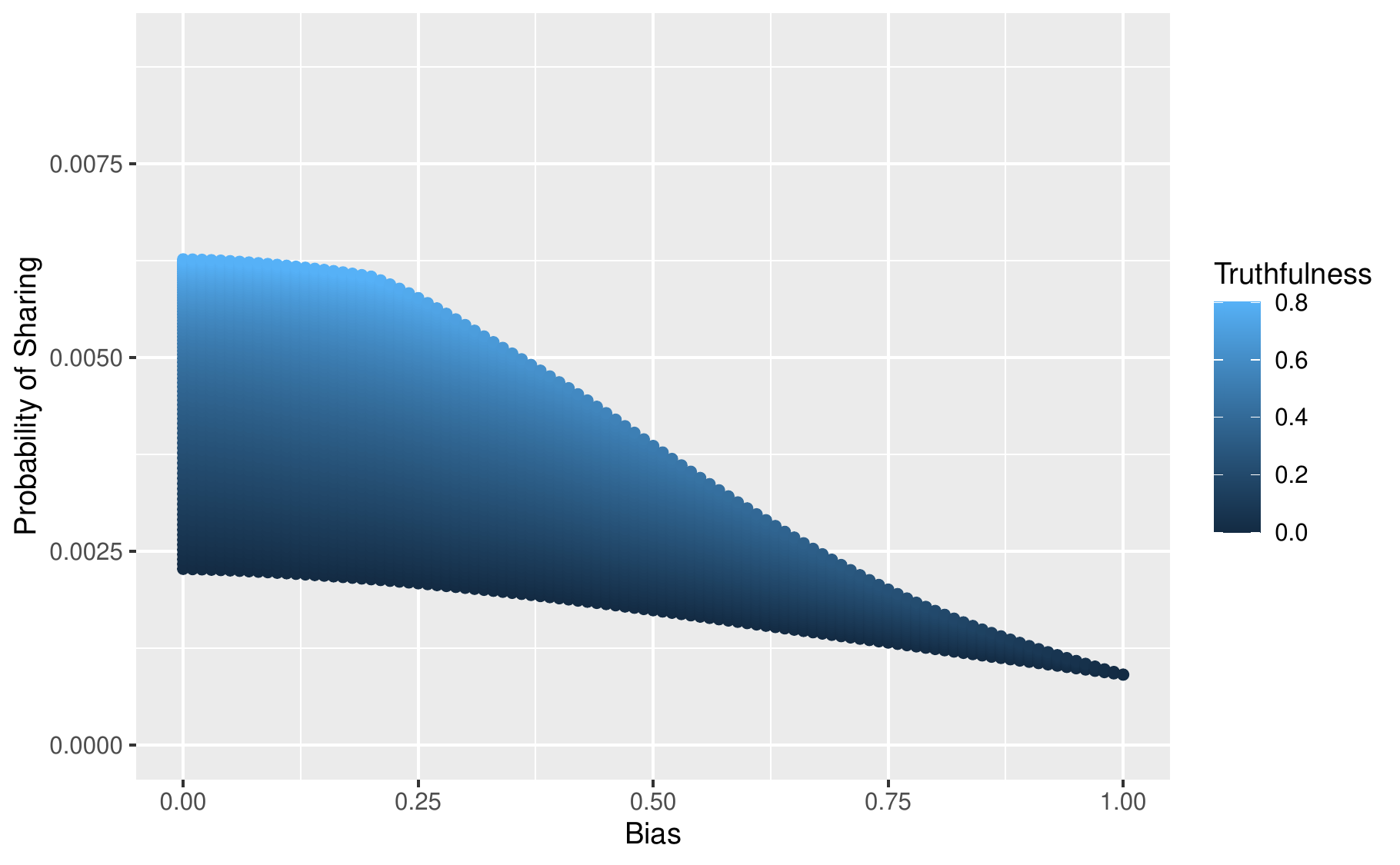}
    }
    \subfloat[A partisan bimodal distribution of readers' political belief 
    ]{
    \includegraphics[width = 0.33\textwidth]{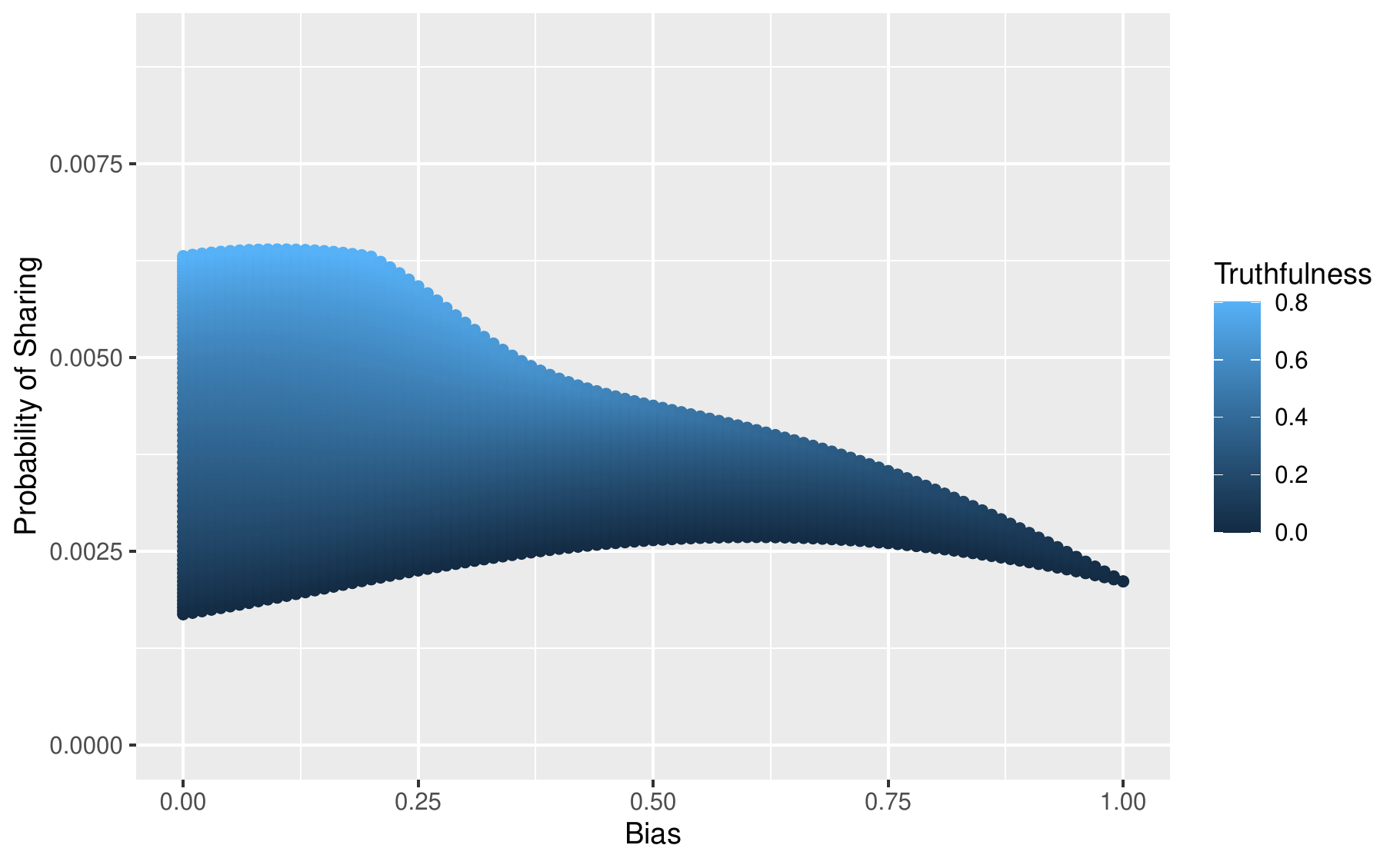}
    }
    \subfloat[A hyperpartisan bimodal distribution of readers' political belief 
    ]{
    \includegraphics[width = 0.33\textwidth]{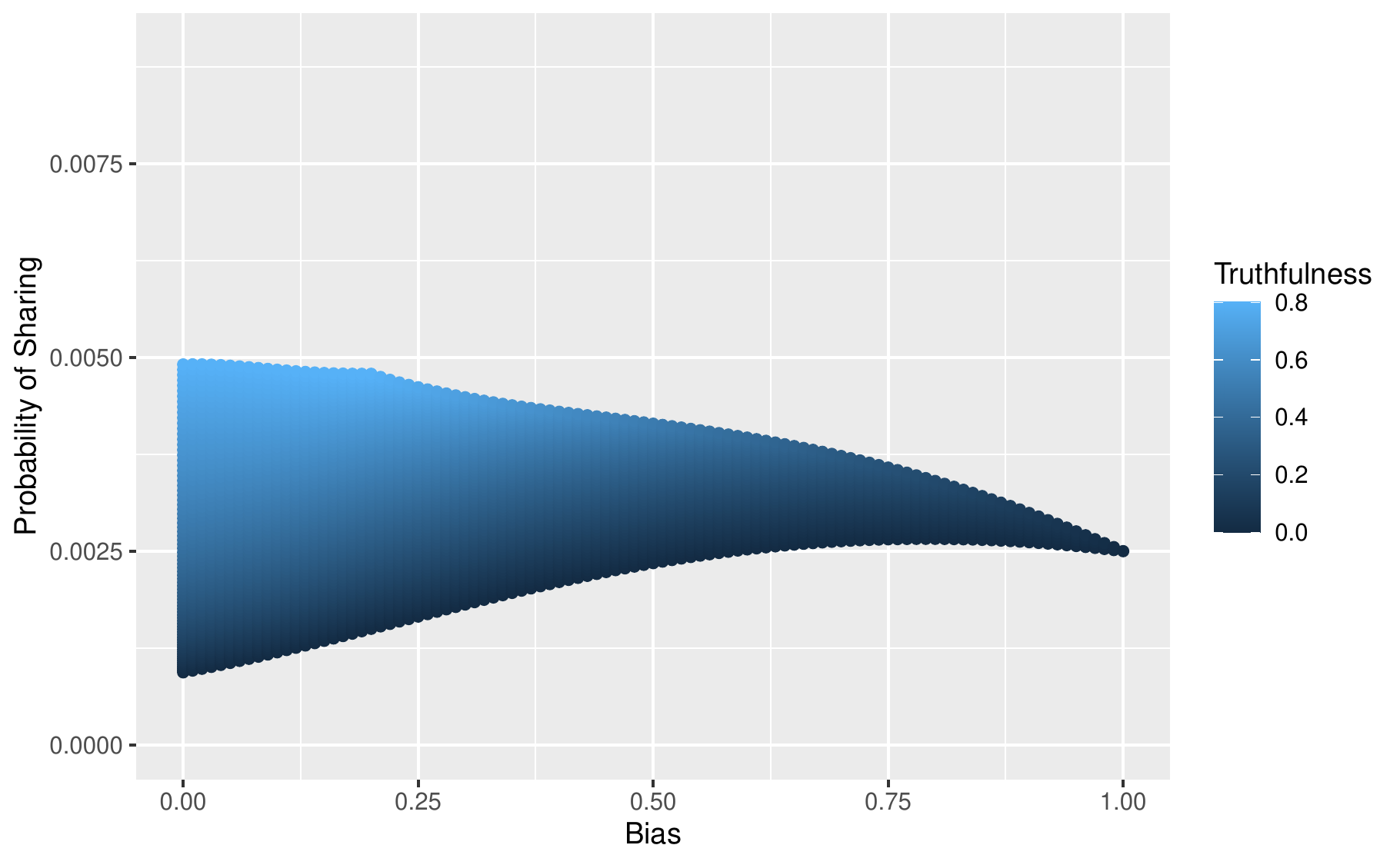}
    }
    
    \subfloat[A right-unimodal distribution of readers' political belief 
    ]{
    \includegraphics[width = 0.33\textwidth]{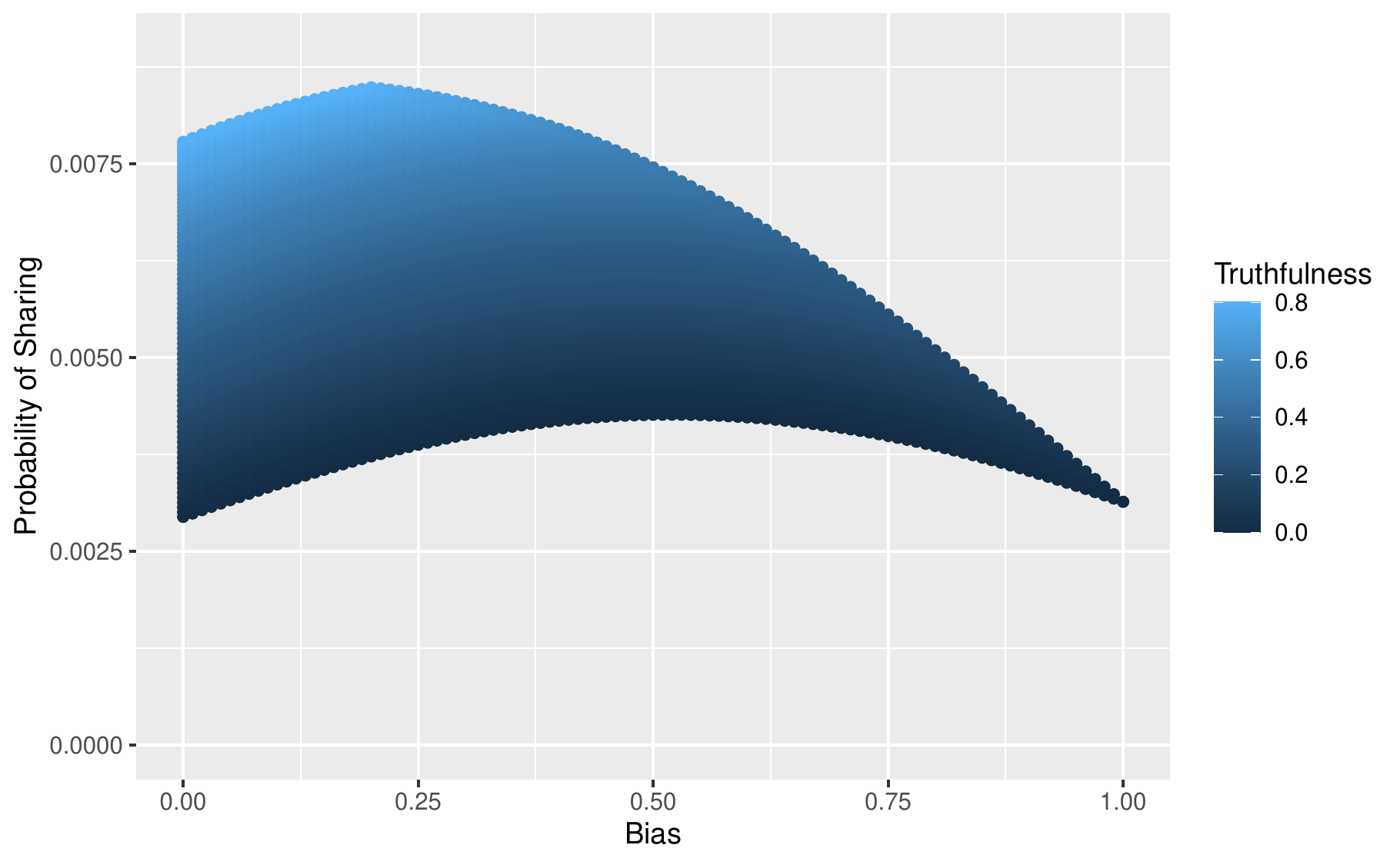}
    }
    \subfloat[A centrist-unimodal distribution of readers' political belief 
    ]{
    \includegraphics[width = 0.33\textwidth]{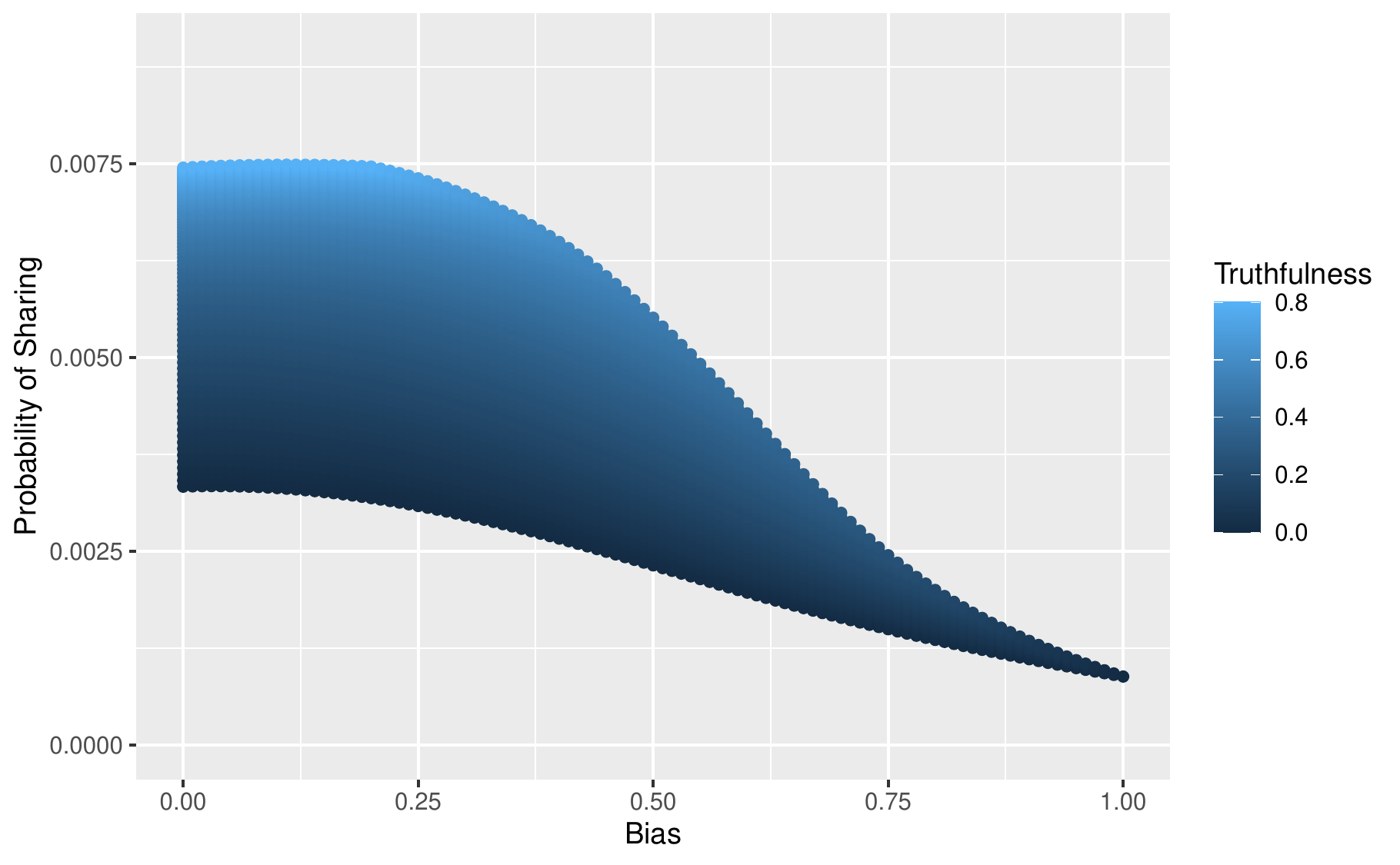}
    }
    \subfloat[A left-unimodal distribution of readers' political belief 
    ]{
    \includegraphics[width = 0.33\textwidth]{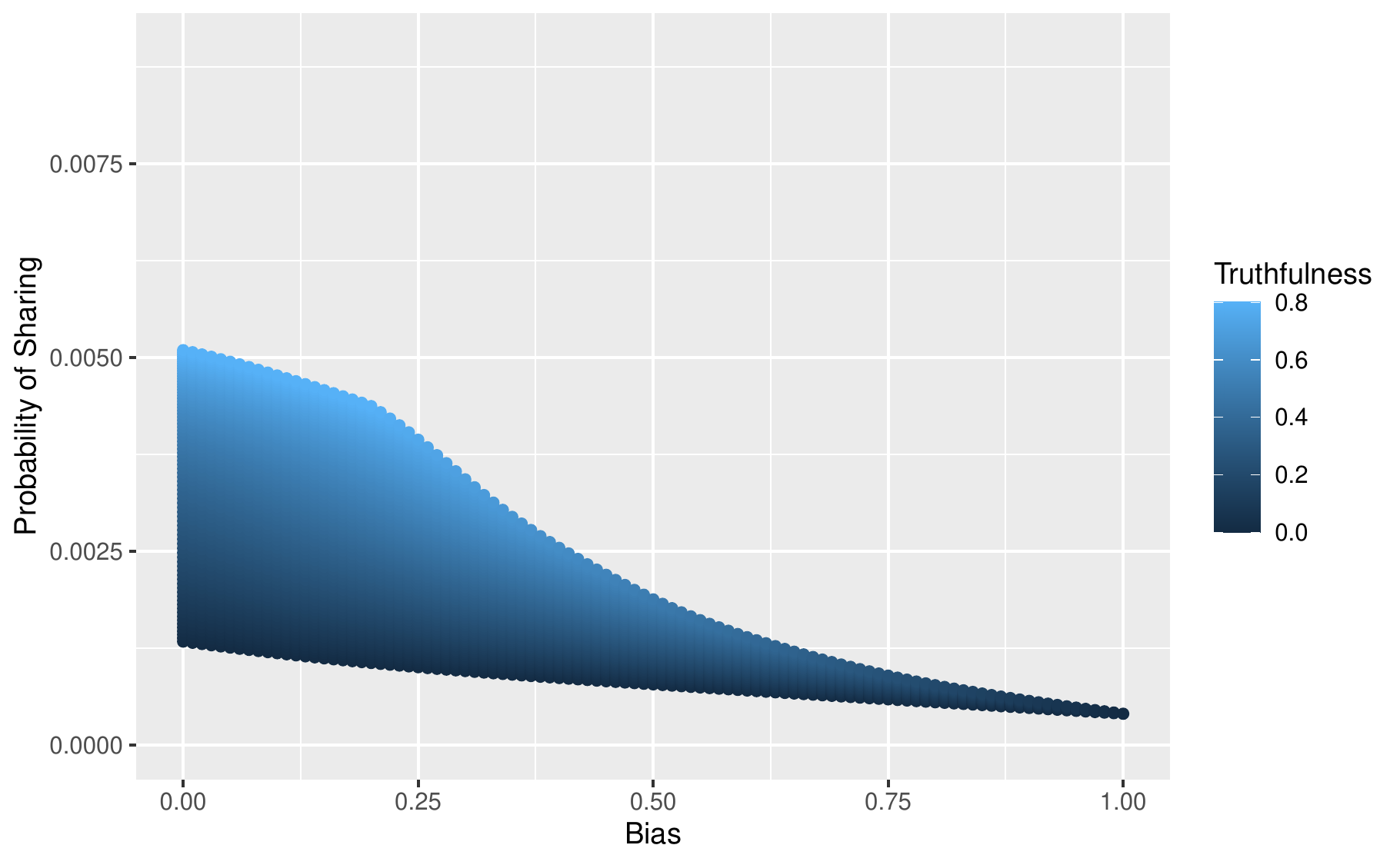}
    }
    \caption{The probability of sharing an article as a function of right political bias ($x$-axis) and truthfulness (gradient). Fitting parameters combination: $f_l$ - low, $k_l$ - high, $f_r$ - high, $k_r$ - low}
    \label{fig:biasbimrightlohihilo}

\end{figure}

\begin{figure}[H]
    \centering
    \subfloat[The empirical distribution of readers' political belief]{
    \includegraphics[width = 0.33\textwidth]{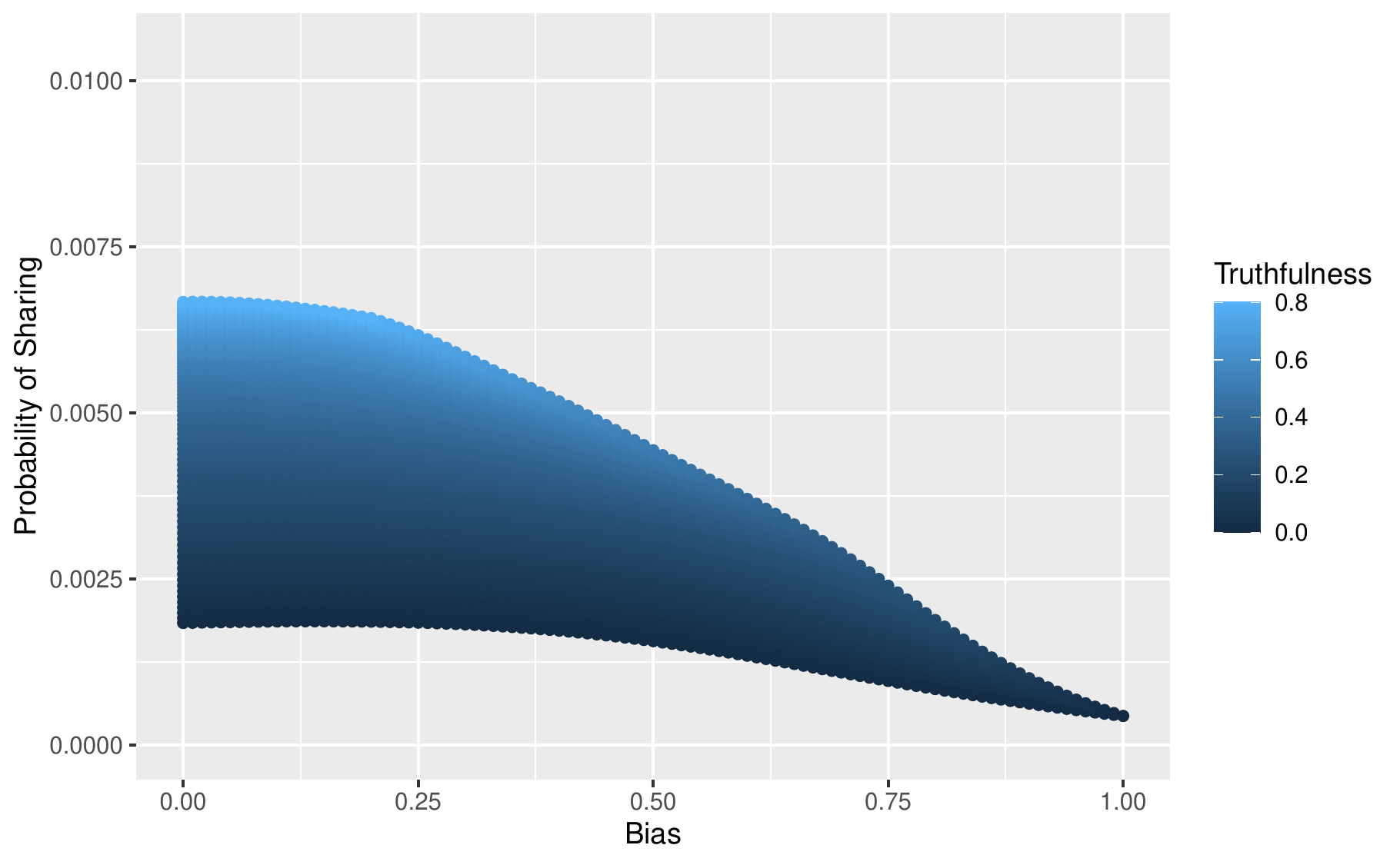}
    }
    \subfloat[A partisan bimodal distribution of readers' political belief 
    ]{
    \includegraphics[width = 0.33\textwidth]{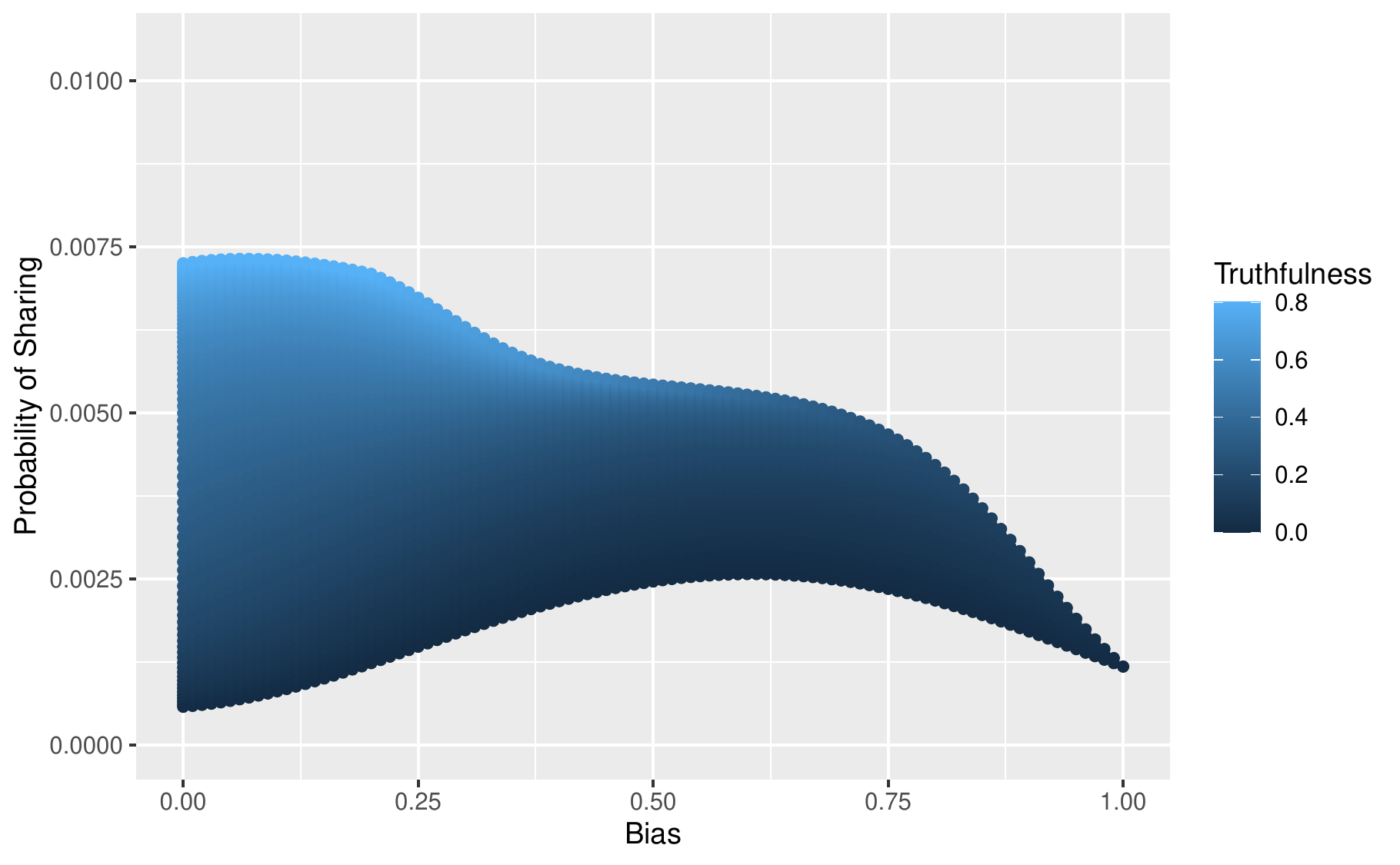}
    }
    \subfloat[A hyperpartisan bimodal distribution of readers' political belief 
    ]{
    \includegraphics[width = 0.33\textwidth]{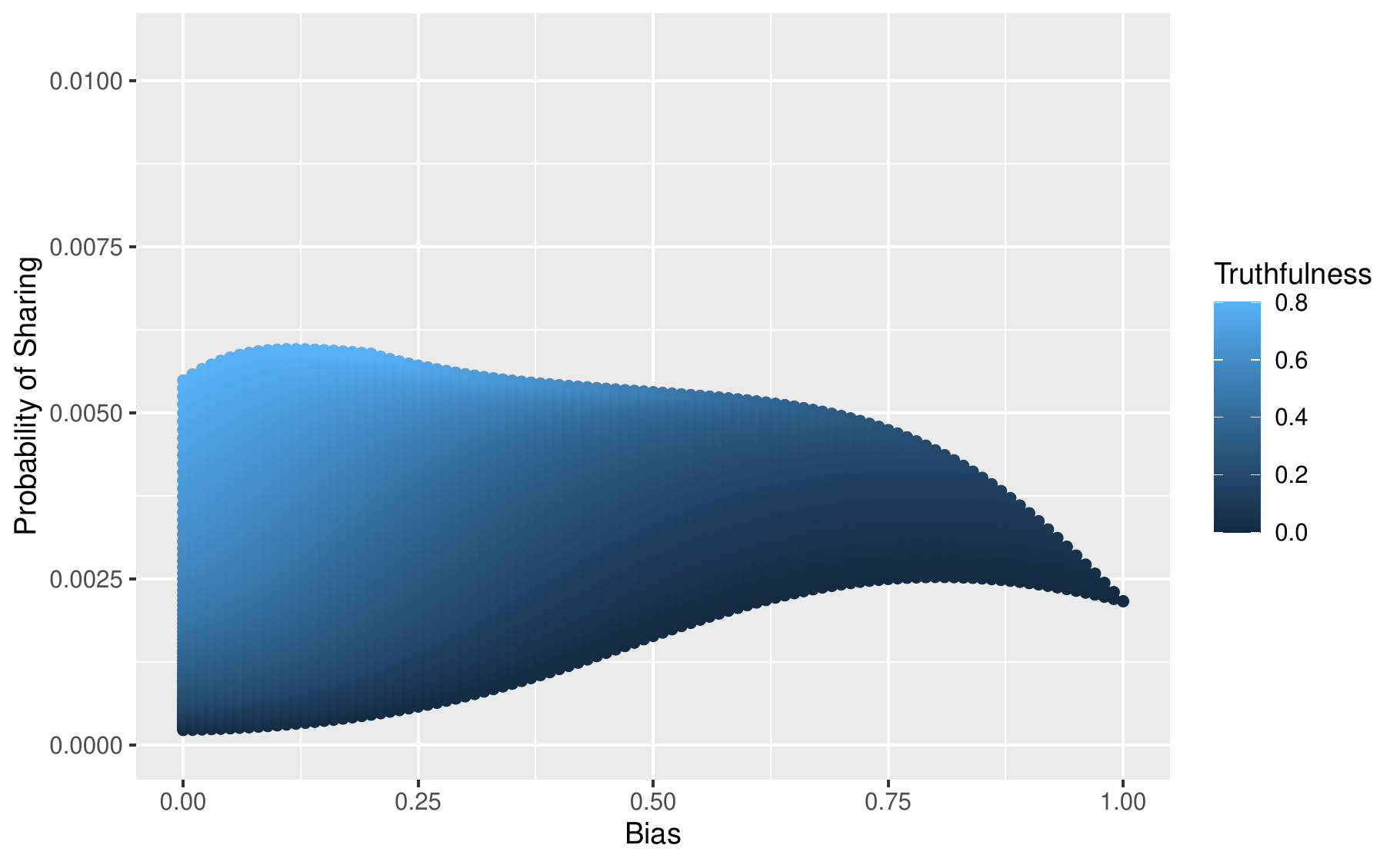}
    }
    
    \subfloat[A right-unimodal distribution of readers' political belief 
    ]{
    \includegraphics[width = 0.33\textwidth]{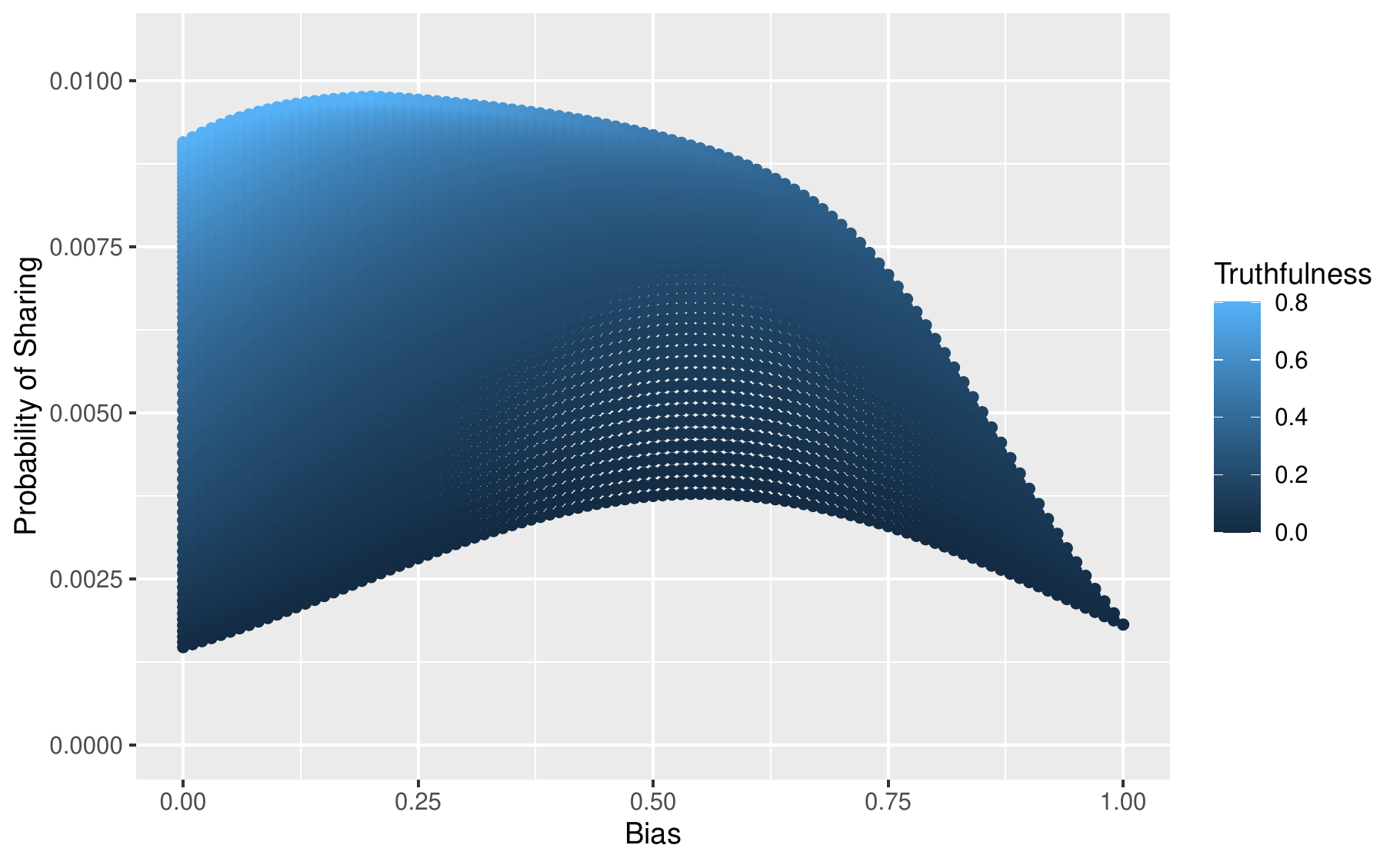}
    }
    \subfloat[A centrist-unimodal distribution of readers' political belief 
    ]{
    \includegraphics[width = 0.33\textwidth]{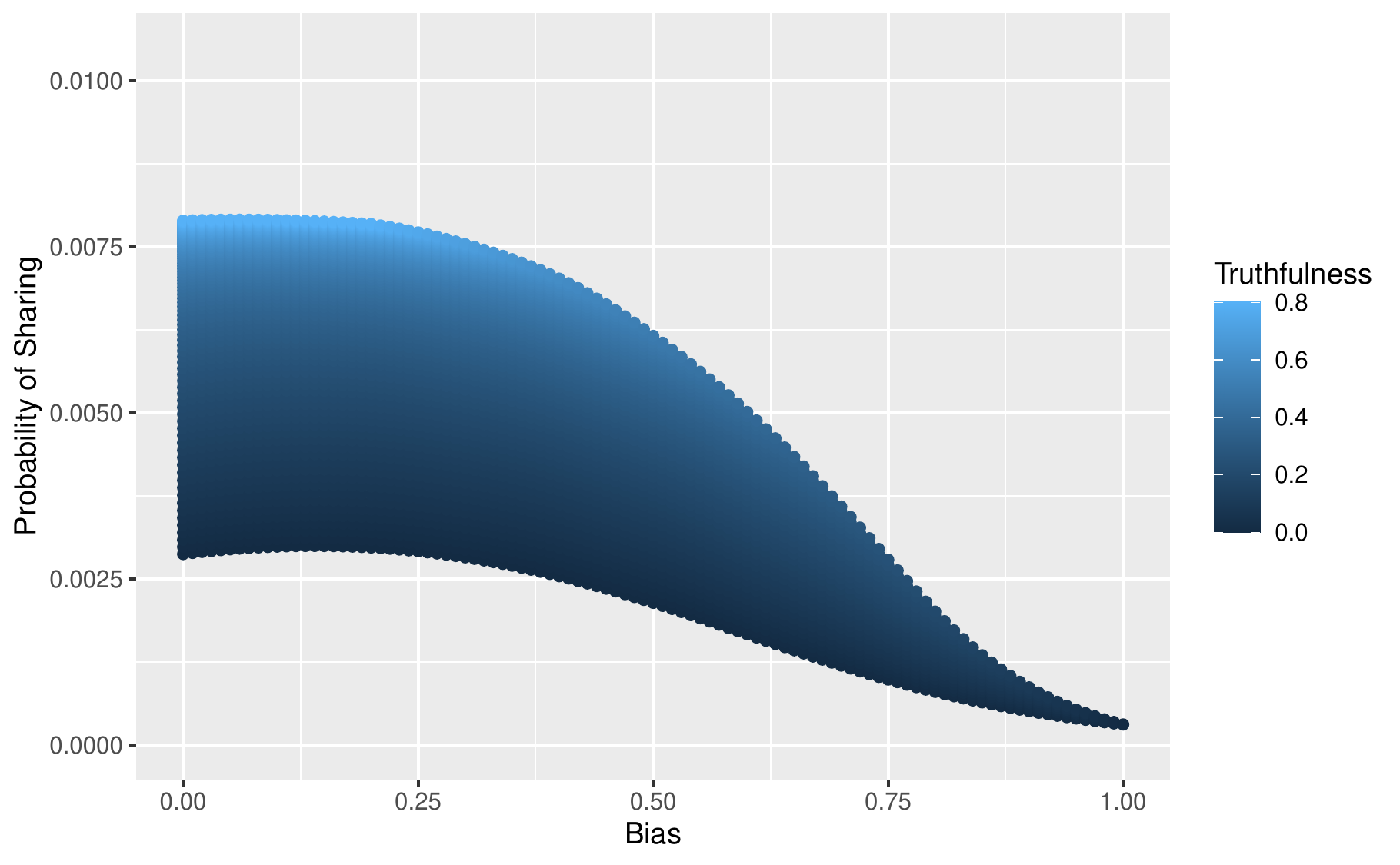}
    }
    \subfloat[A left-unimodal distribution of readers' political belief 
    ]{
    \includegraphics[width = 0.33\textwidth]{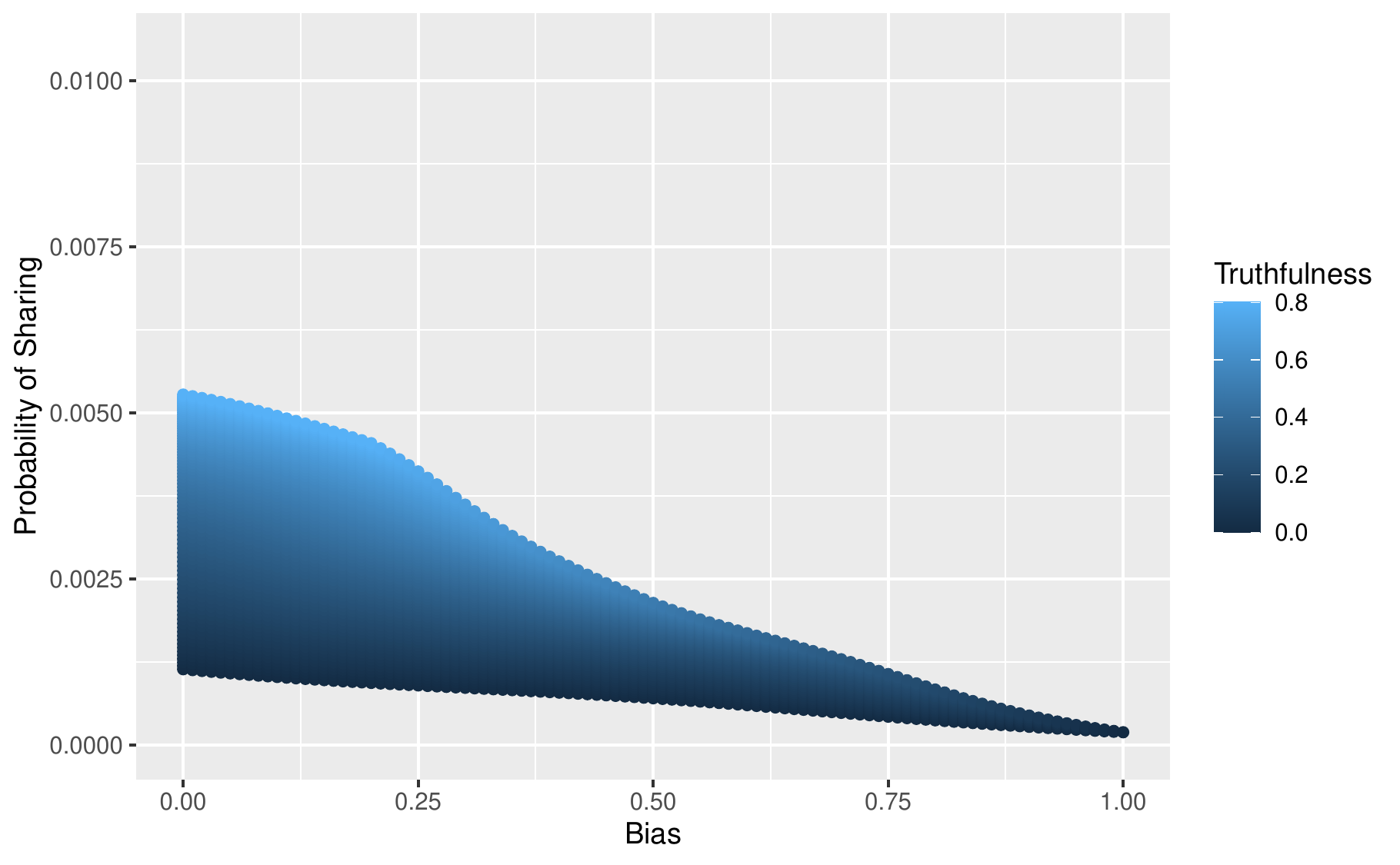}
    }
    \caption{The probability of sharing an article as a function of right political bias ($x$-axis) and truthfulness (gradient). Fitting parameters combination: $f_l$ - low, $k_l$ - high, $f_r$ - high, $k_r$ - high}
    \label{fig:biasbimrightlohihihi}

\end{figure}

\begin{figure}[H]
    \centering
    \subfloat[The empirical distribution of readers' political belief]{
    \includegraphics[width = 0.33\textwidth]{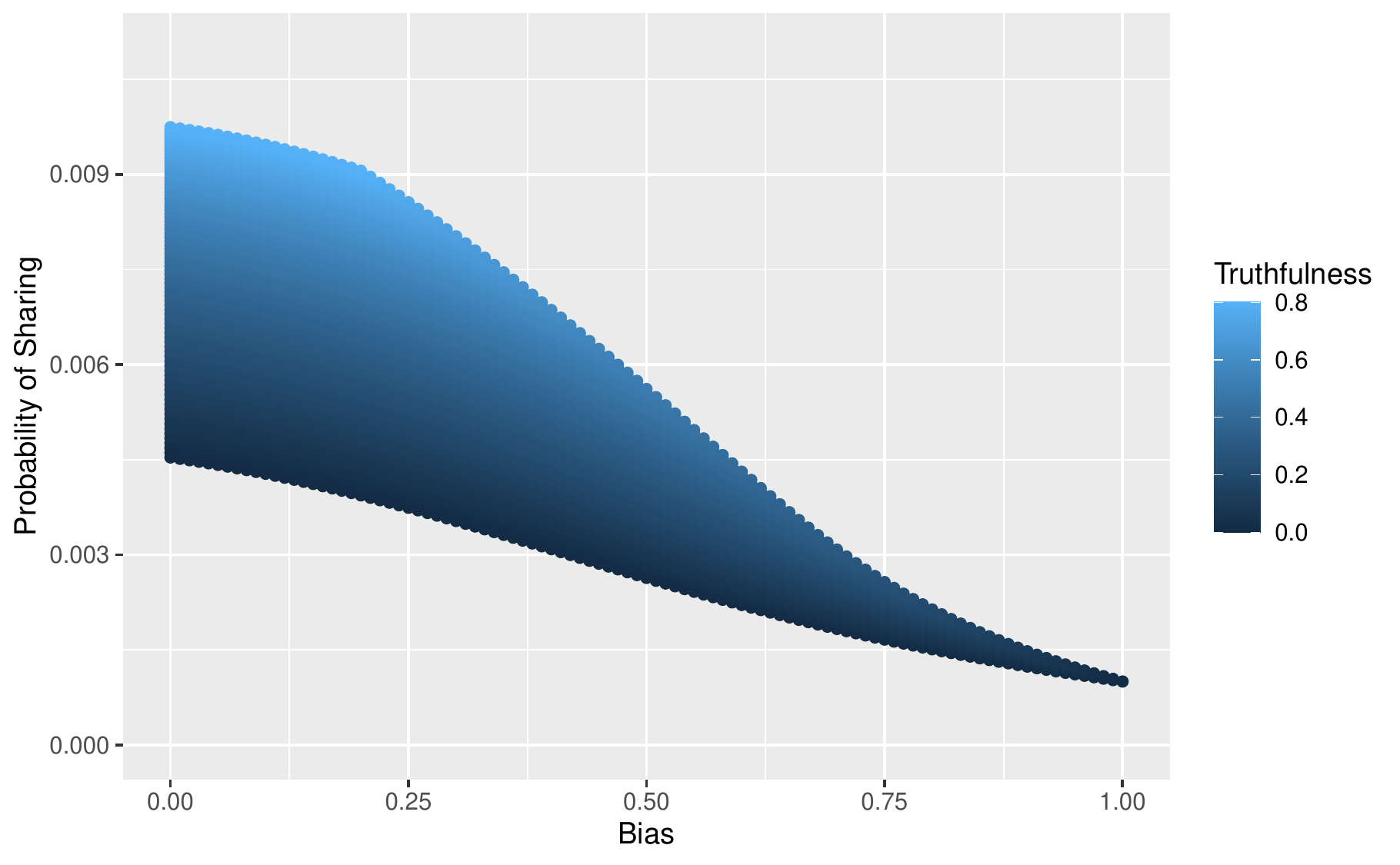}
    }
    \subfloat[A partisan bimodal distribution of readers' political belief 
    ]{
    \includegraphics[width = 0.33\textwidth]{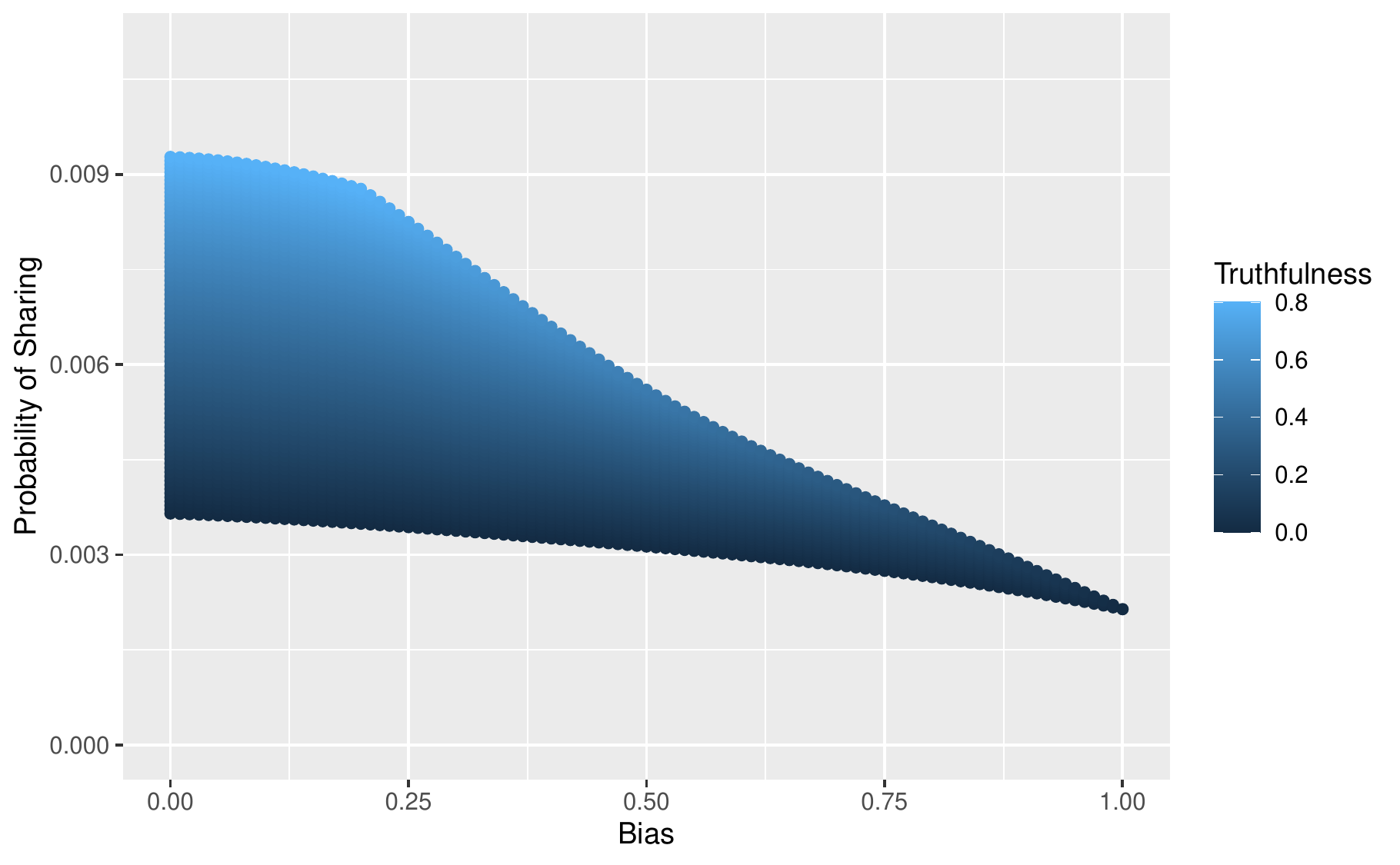}
    }
    \subfloat[A hyperpartisan bimodal distribution of readers' political belief 
    ]{
    \includegraphics[width = 0.33\textwidth]{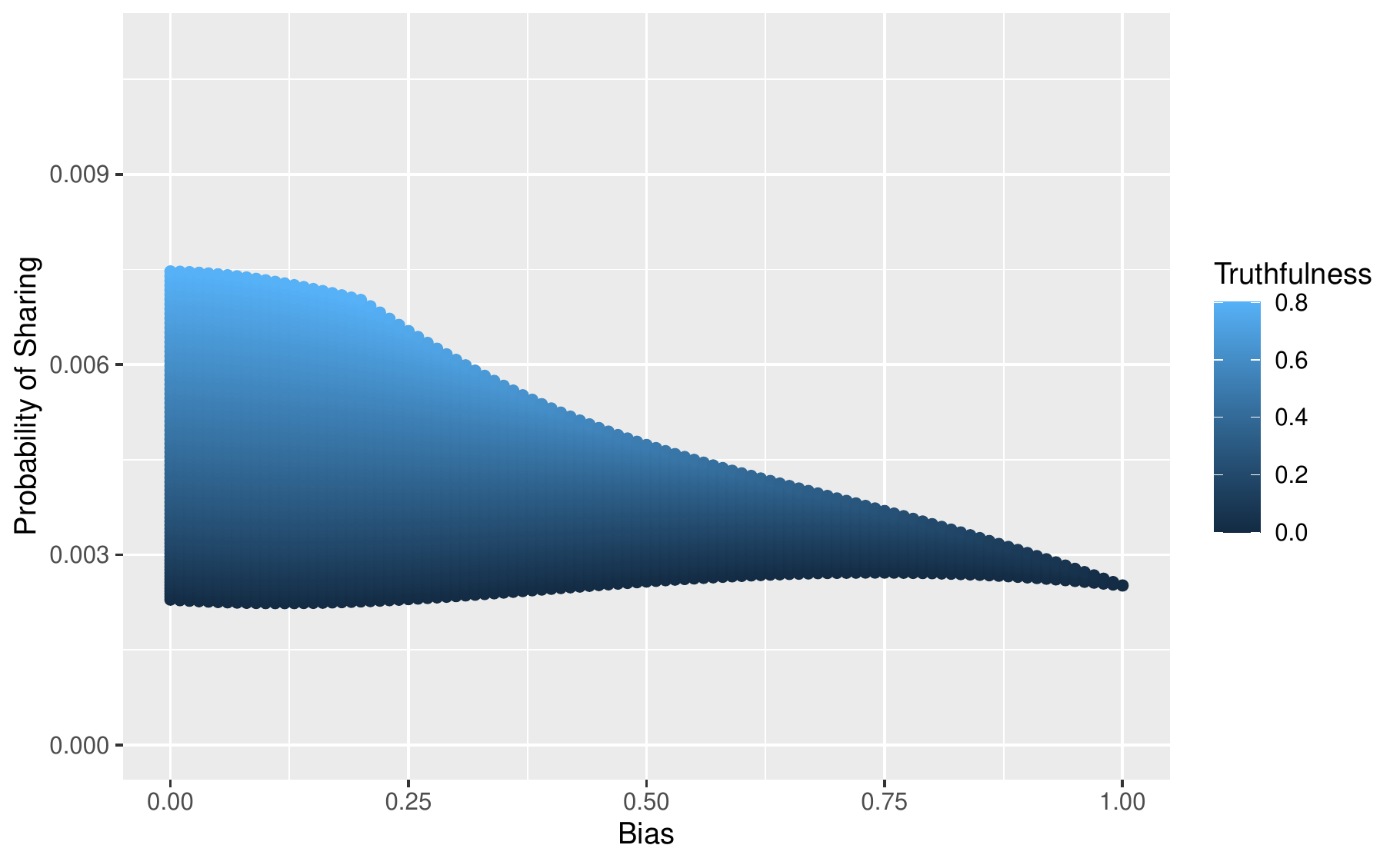}
    }
    
    \subfloat[A right-unimodal distribution of readers' political belief 
    ]{
    \includegraphics[width = 0.33\textwidth]{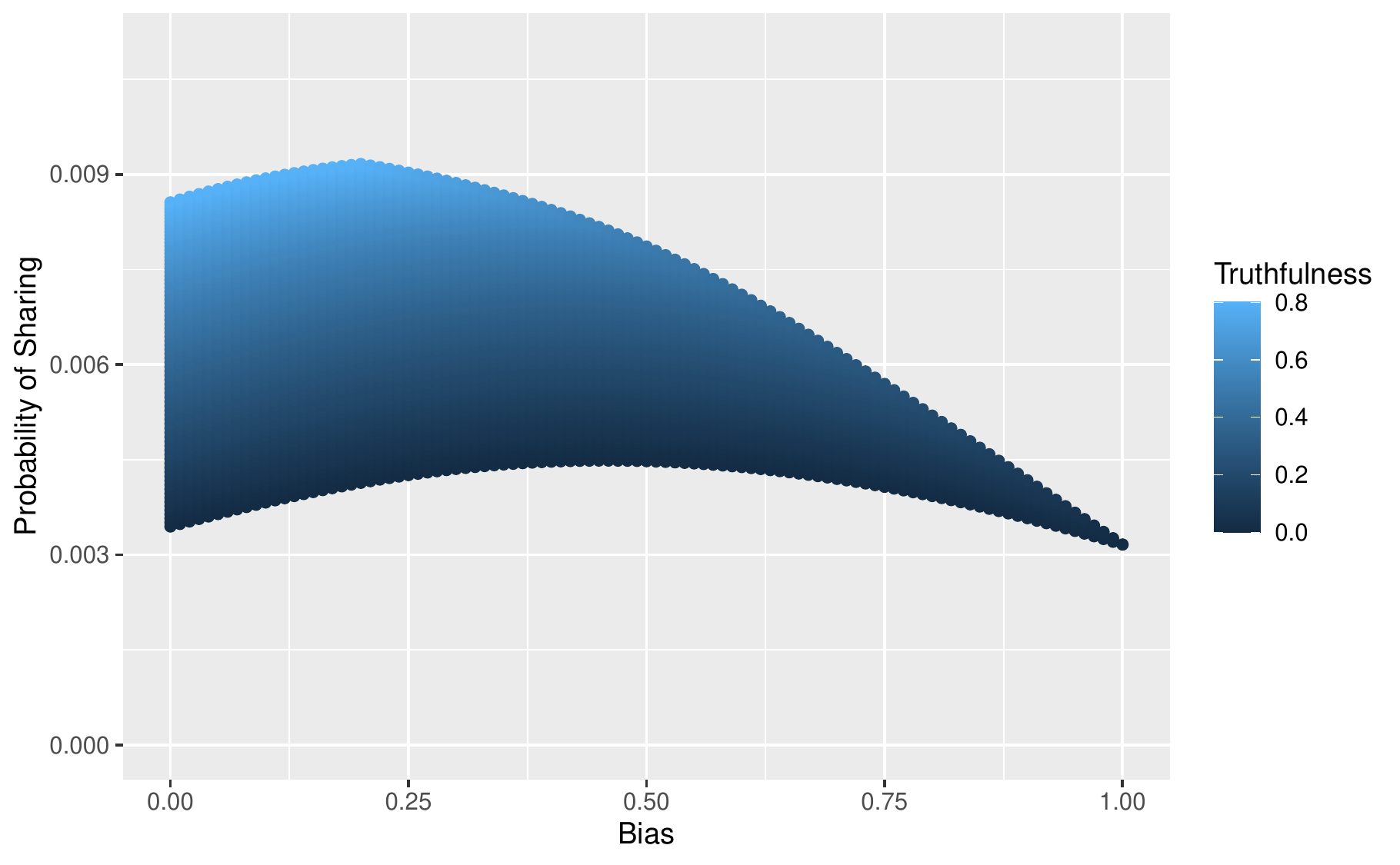}
    }
    \subfloat[A centrist-unimodal distribution of readers' political belief 
    ]{
    \includegraphics[width = 0.33\textwidth]{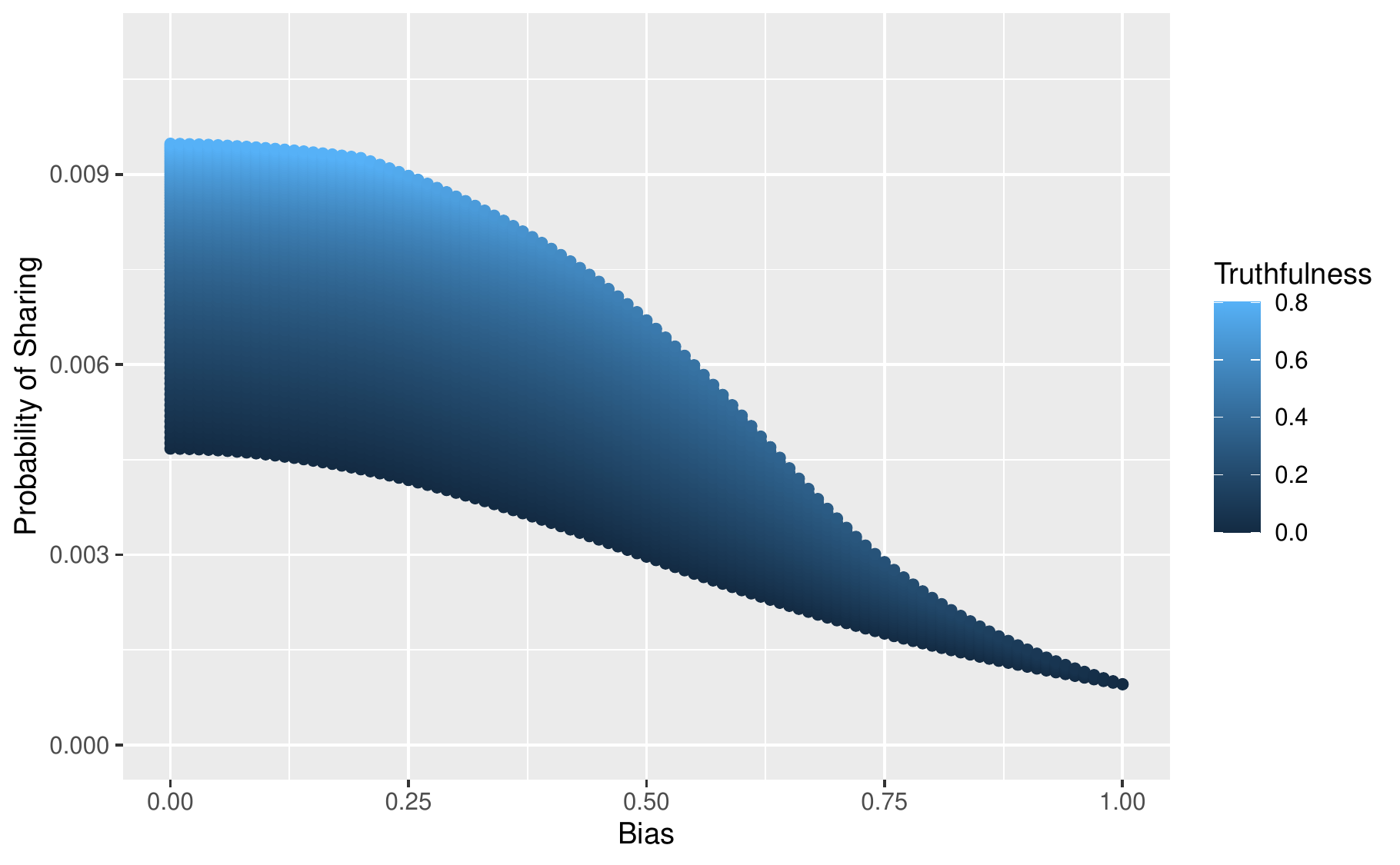}
    }
    \subfloat[A left-unimodal distribution of readers' political belief 
    ]{
    \includegraphics[width = 0.33\textwidth]{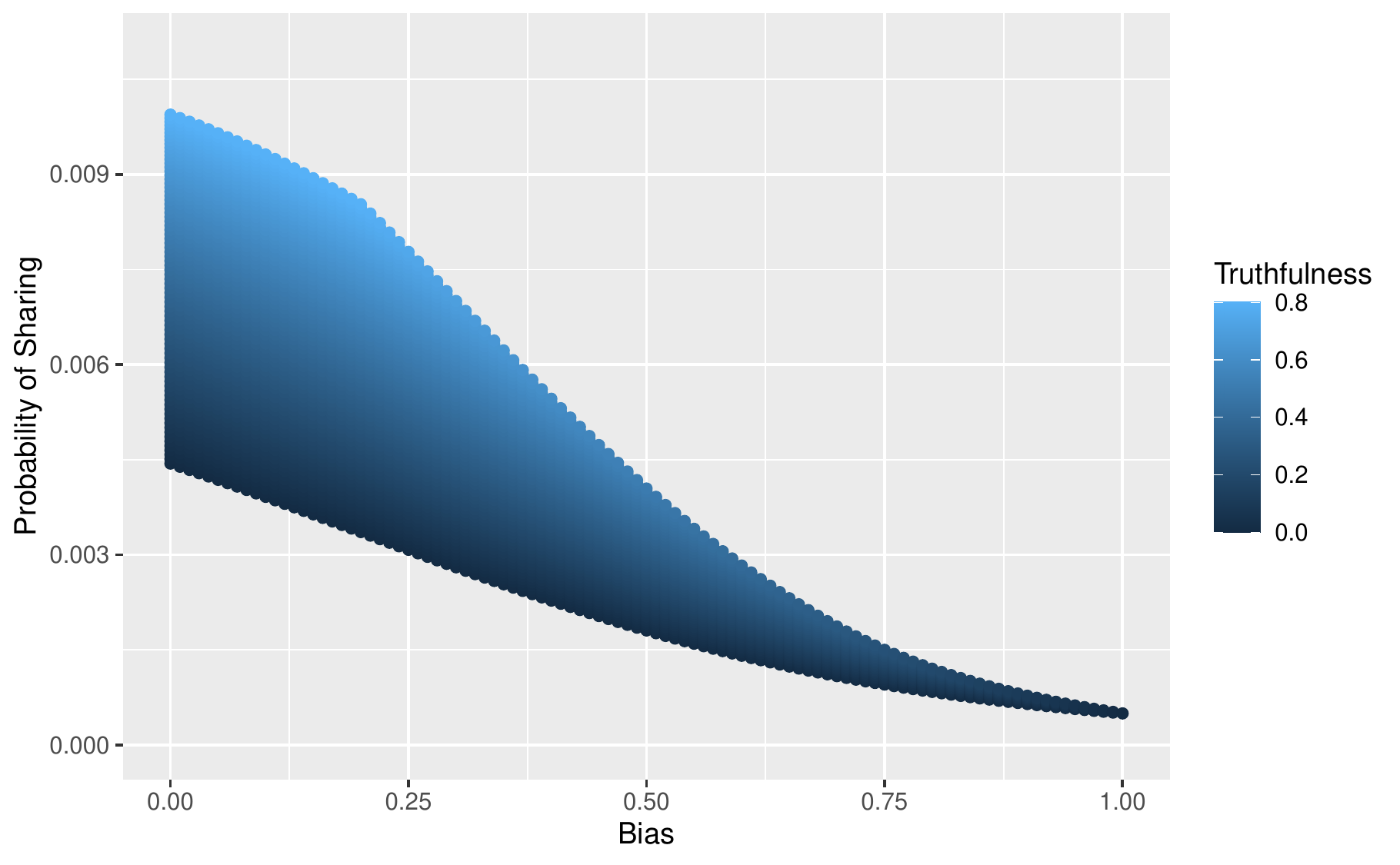}
    }
    \caption{The probability of sharing an article as a function of right political bias ($x$-axis) and truthfulness (gradient). Fitting parameters combination: $f_l$ - high, $k_l$ - low, $f_r$ - high, $k_r$ - low}
    \label{fig:biasbimrighthilohilo}

\end{figure}

\begin{figure}[H]
    \centering
    \subfloat[The empirical distribution of readers' political belief]{
    \includegraphics[width = 0.33\textwidth]{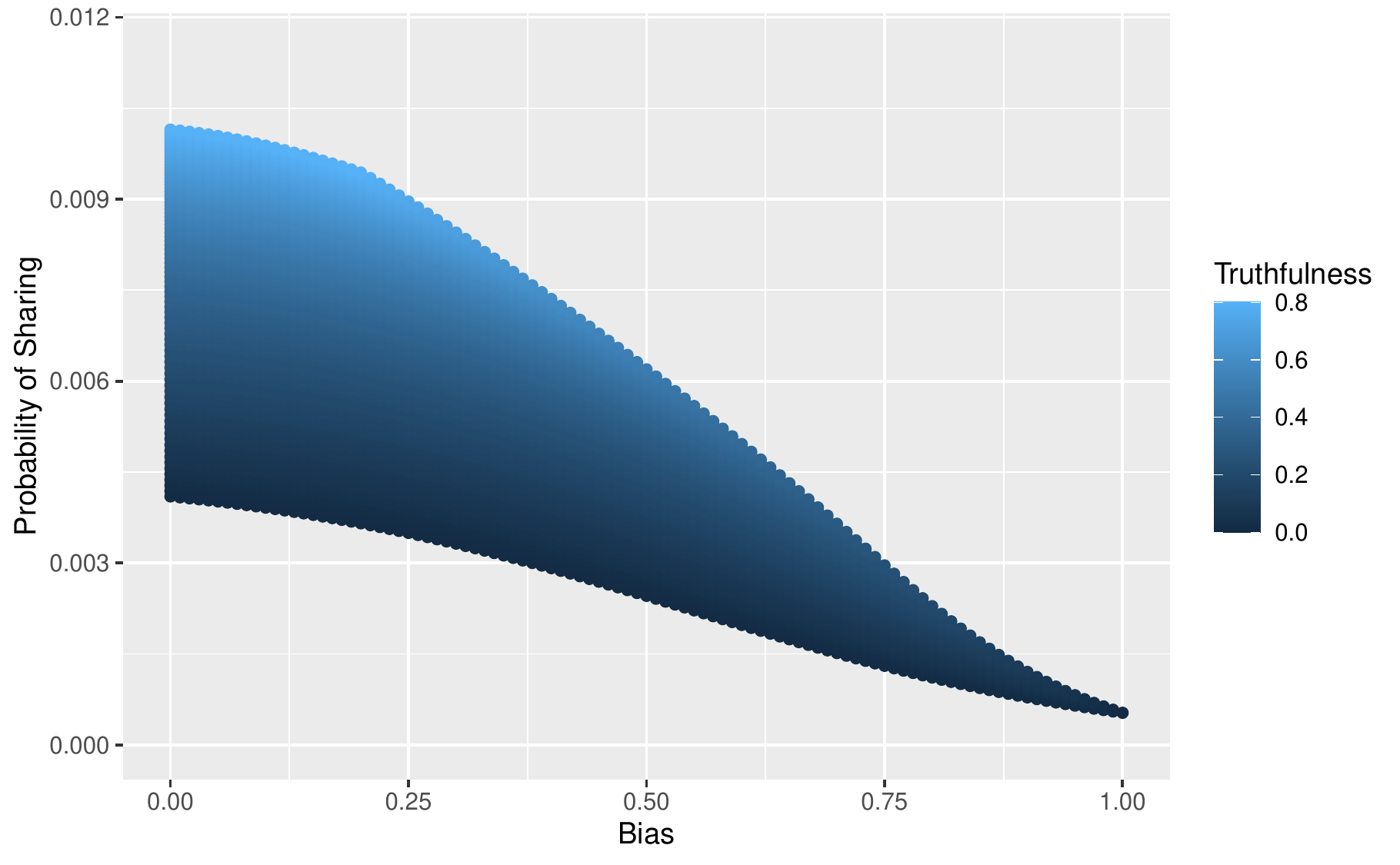}
    }
    \subfloat[A partisan bimodal distribution of readers' political belief 
    ]{
    \includegraphics[width = 0.33\textwidth]{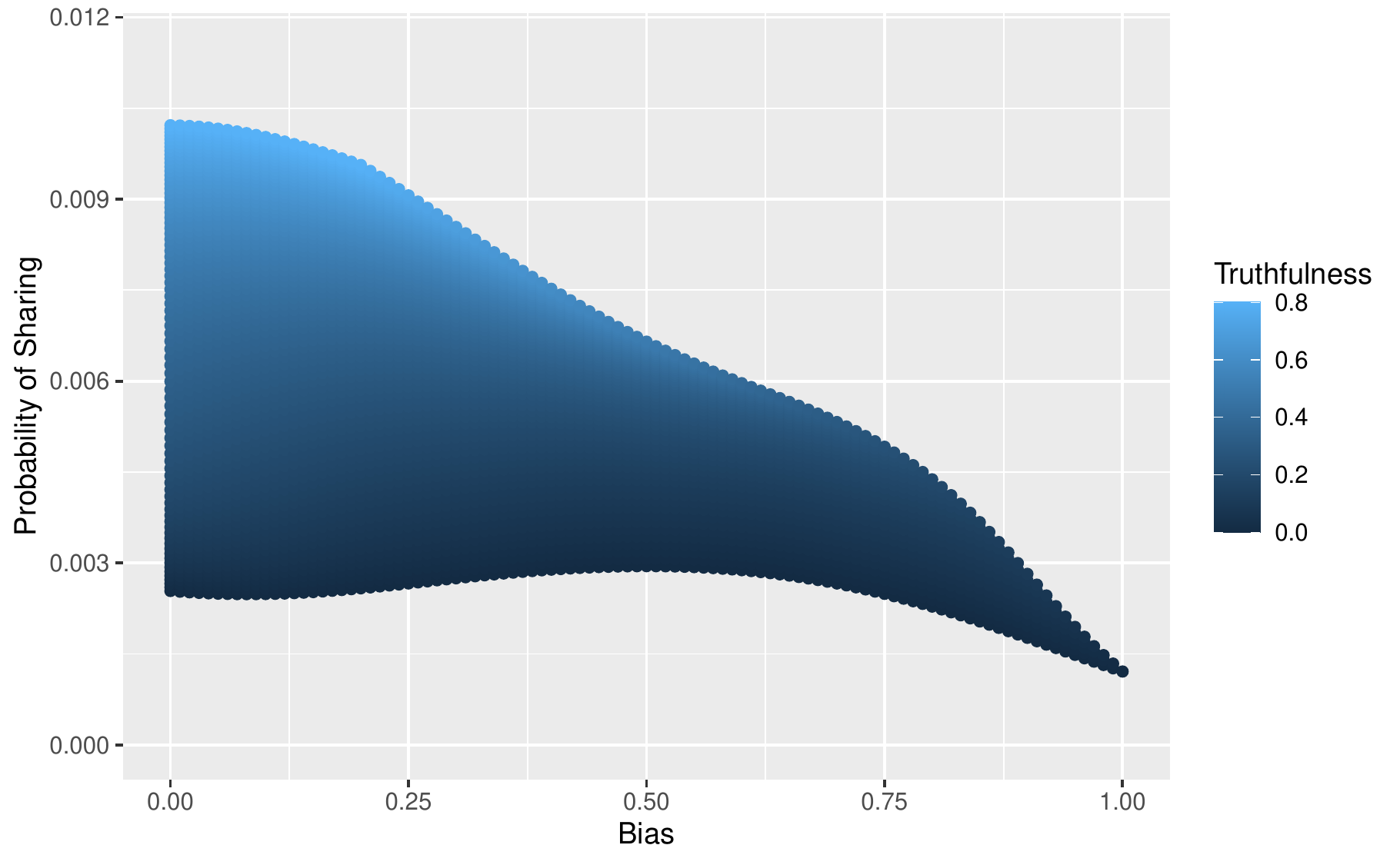}
    }
    \subfloat[A hyperpartisan bimodal distribution of readers' political belief 
    ]{
    \includegraphics[width = 0.33\textwidth]{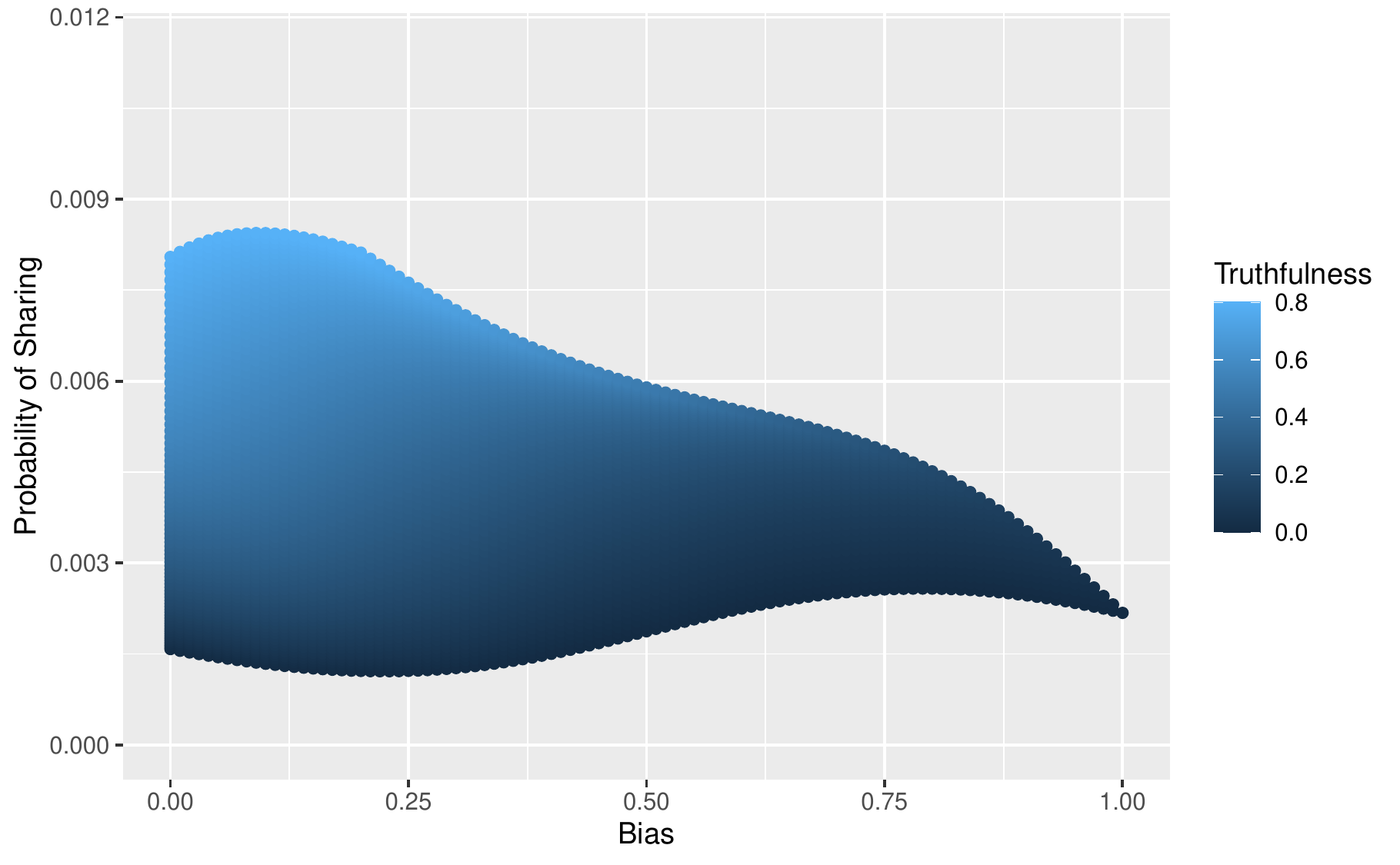}
    }
    
    \subfloat[A right-unimodal distribution of readers' political belief 
    ]{
    \includegraphics[width = 0.33\textwidth]{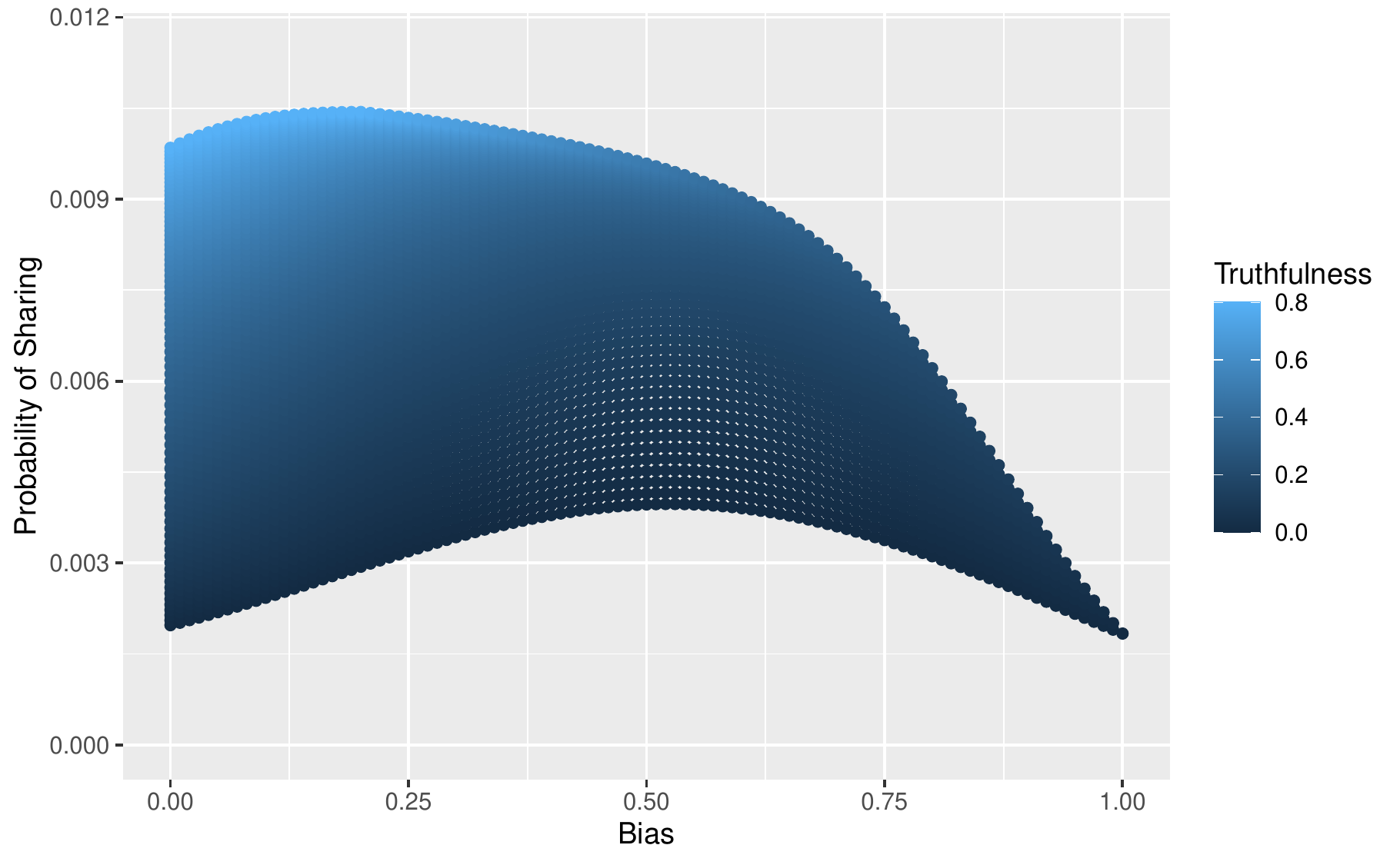}
    }
    \subfloat[A centrist-unimodal distribution of readers' political belief 
    ]{
    \includegraphics[width = 0.33\textwidth]{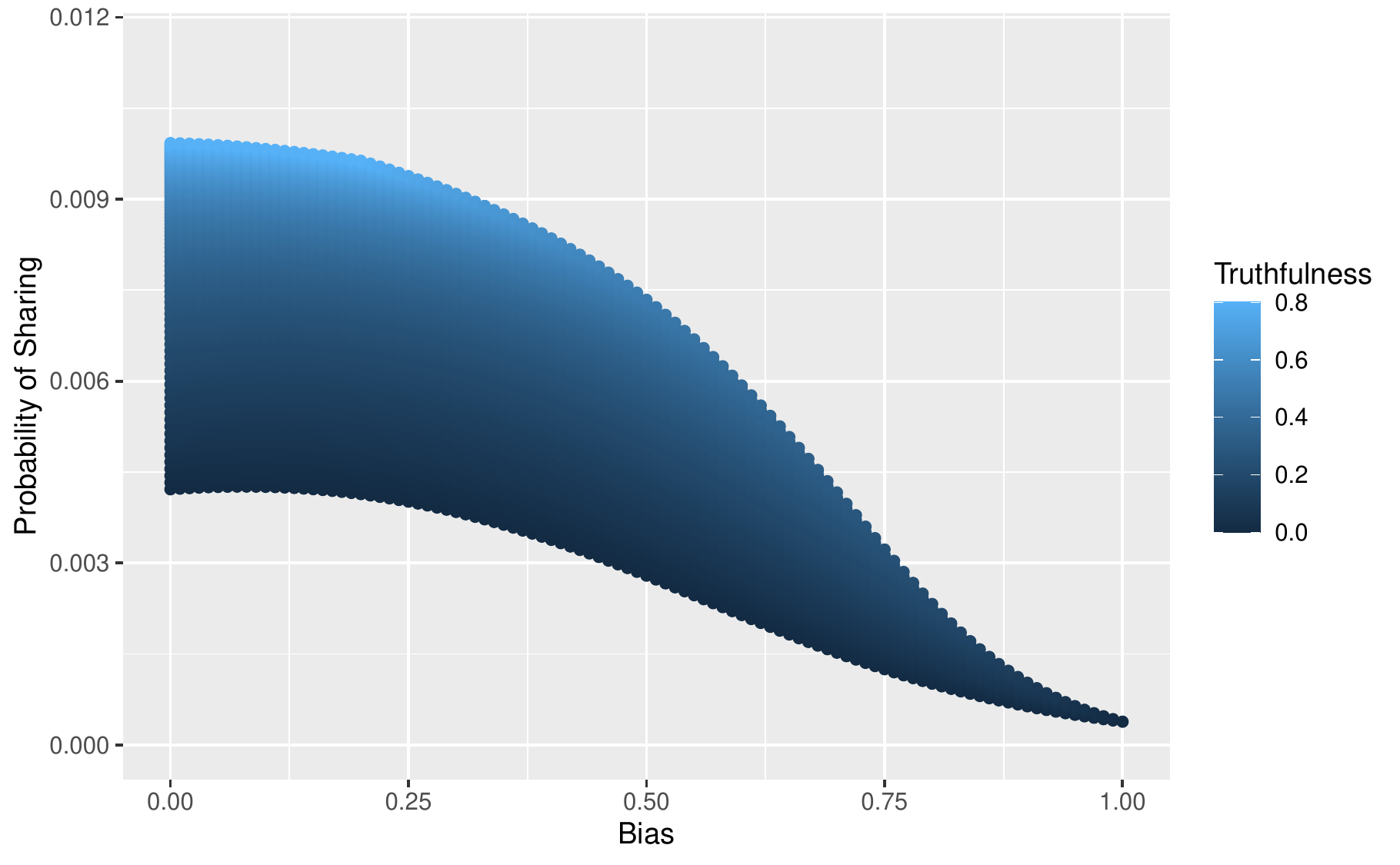}
    }
    \subfloat[A left-unimodal distribution of readers' political belief 
    ]{
    \includegraphics[width = 0.33\textwidth]{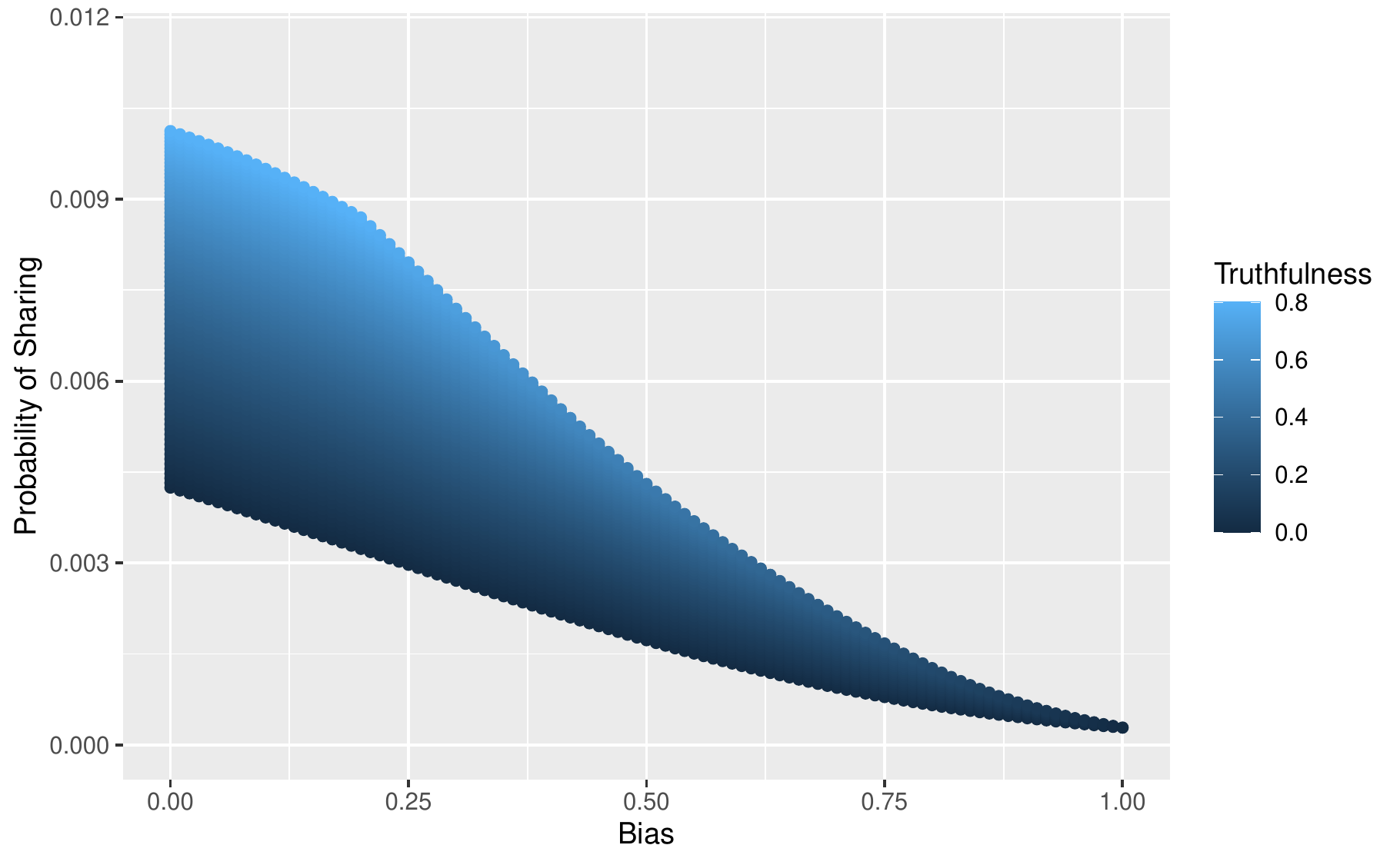}
    }
    \caption{The probability of sharing an article as a function of right political bias ($x$-axis) and truthfulness (gradient). Fitting parameters combination: $f_l$ - high, $k_l$ - low, $f_r$ - high, $k_r$ - high}
    \label{fig:biasbimrighthilohihi}

\end{figure}

\begin{figure}[H]
    \centering
    \subfloat[The empirical distribution of readers' political belief]{
    \includegraphics[width = 0.33\textwidth]{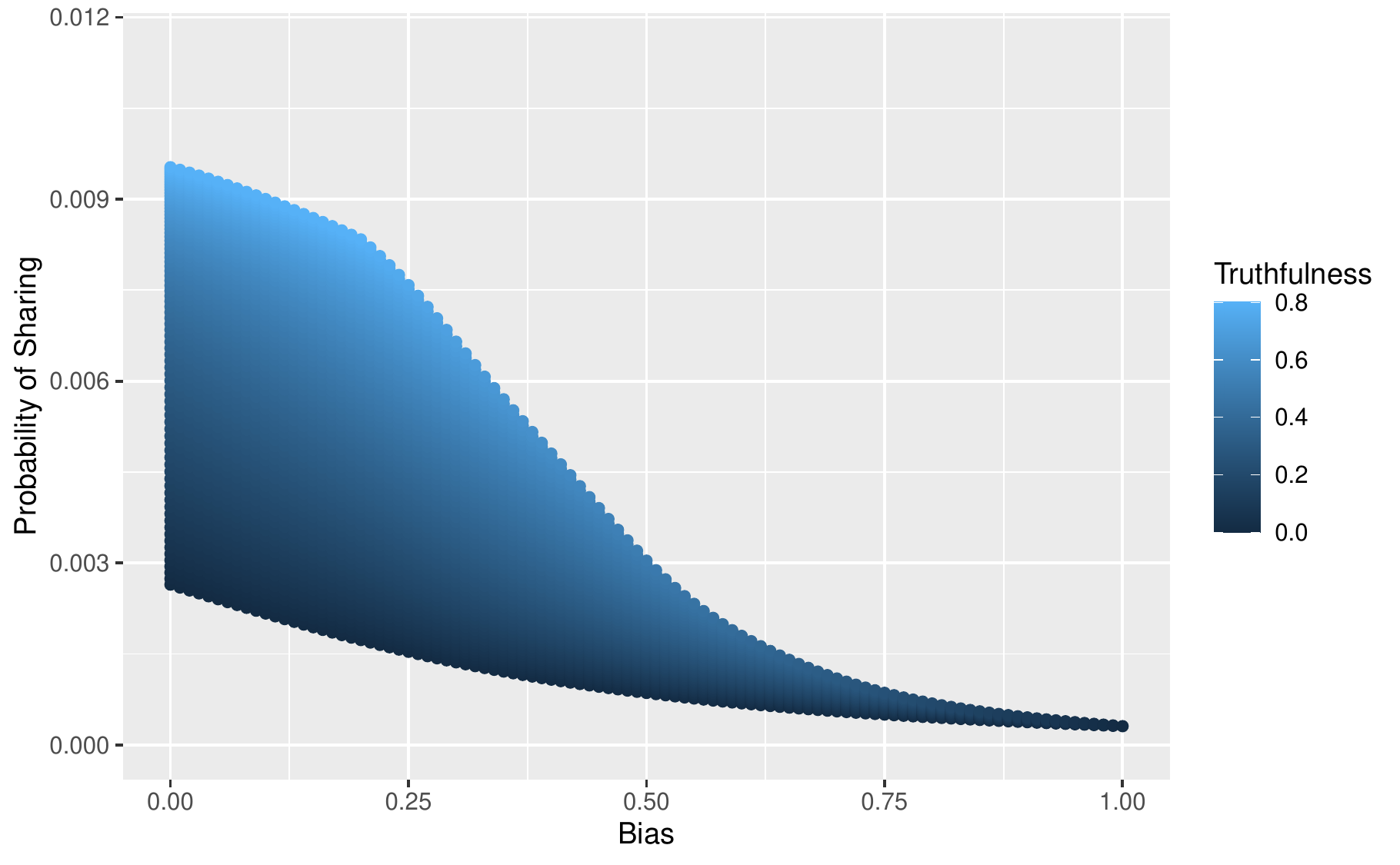}
    }
    \subfloat[A partisan bimodal distribution of readers' political belief 
    ]{
    \includegraphics[width = 0.33\textwidth]{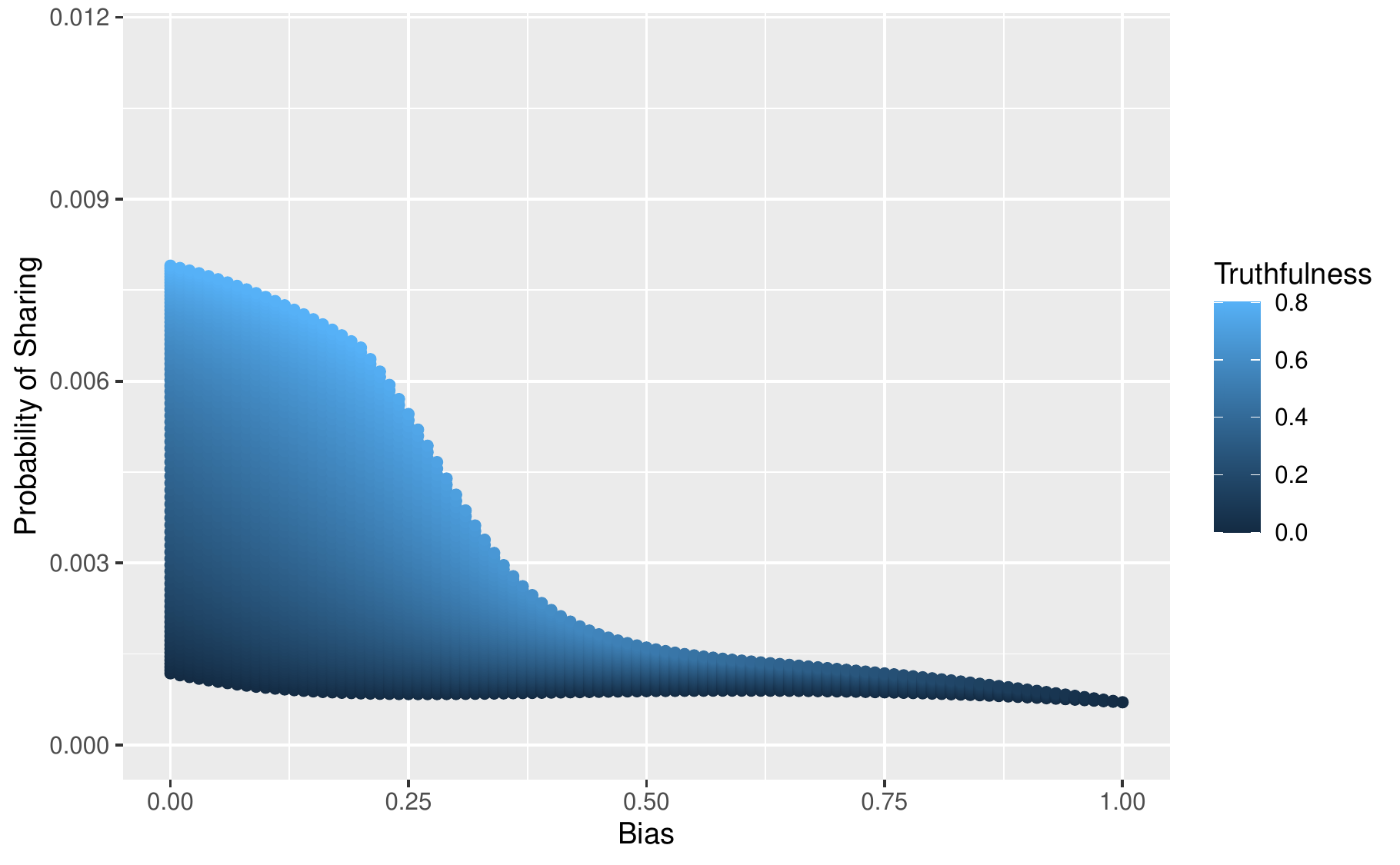}
    }
    \subfloat[A hyperpartisan bimodal distribution of readers' political belief 
    ]{
    \includegraphics[width = 0.33\textwidth]{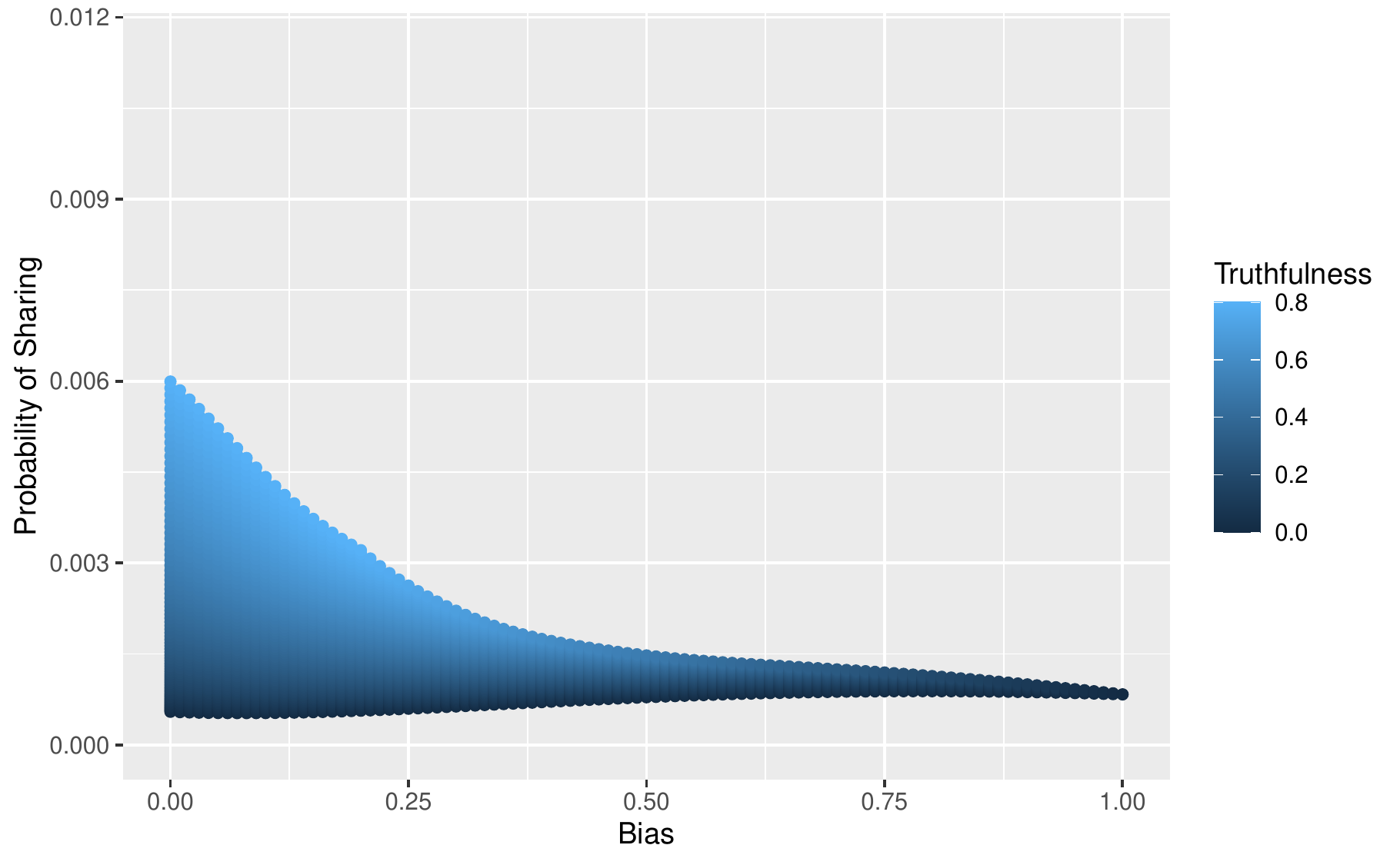}
    }
    
    \subfloat[A right-unimodal distribution of readers' political belief 
    ]{
    \includegraphics[width = 0.33\textwidth]{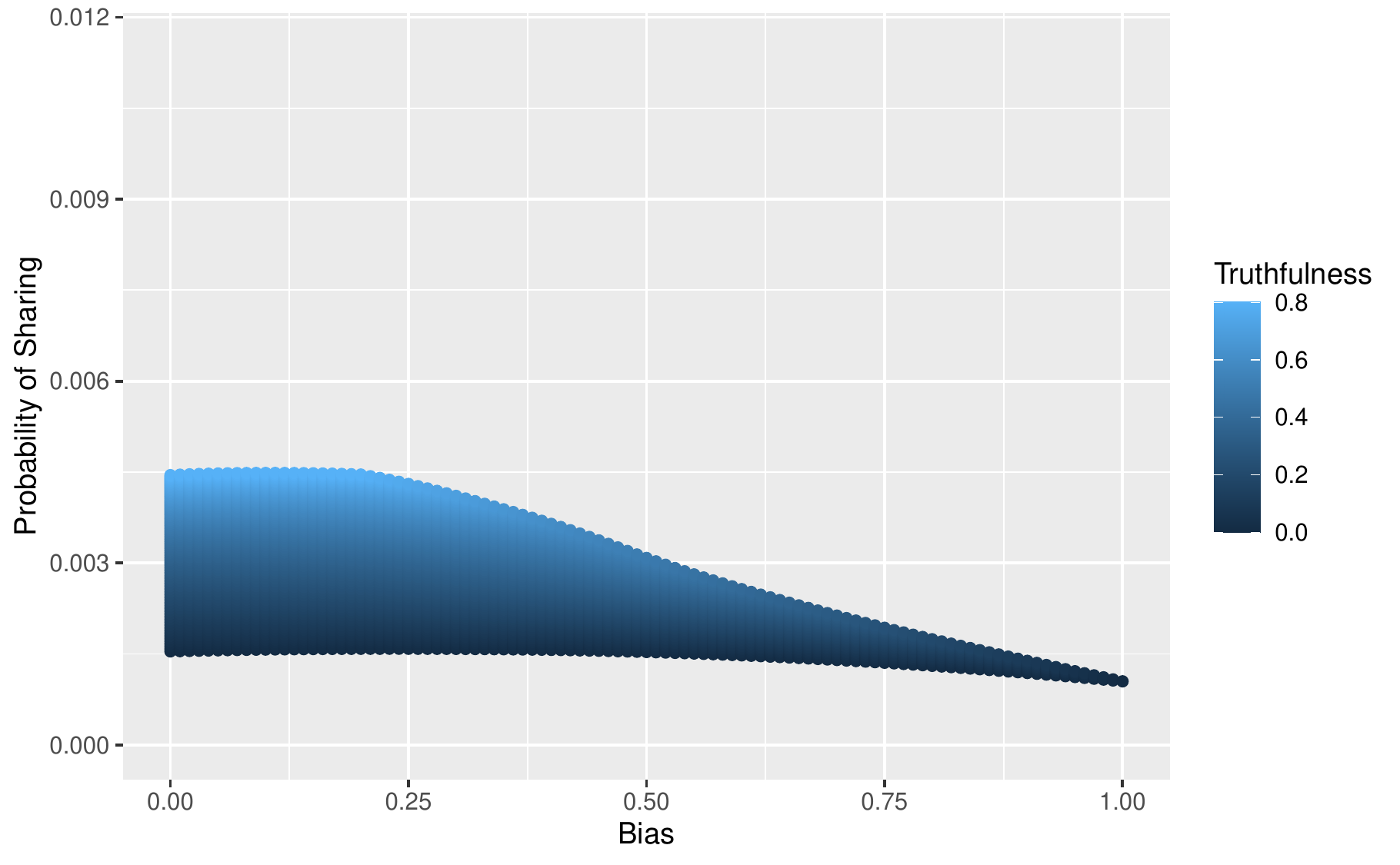}
    }
    \subfloat[A centrist-unimodal distribution of readers' political belief 
    ]{
    \includegraphics[width = 0.33\textwidth]{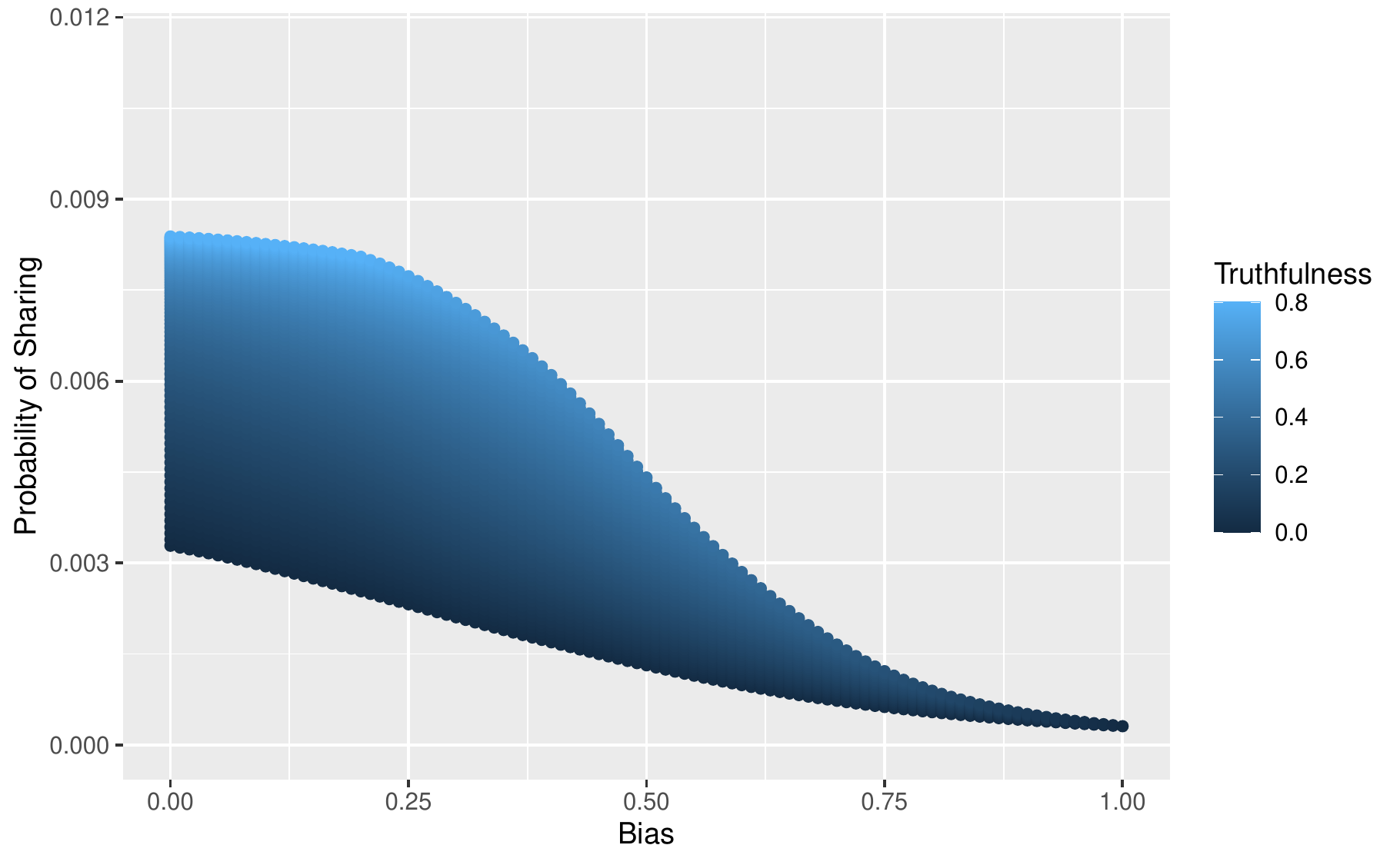}
    }
    \subfloat[A left-unimodal distribution of readers' political belief 
    ]{
    \includegraphics[width = 0.33\textwidth]{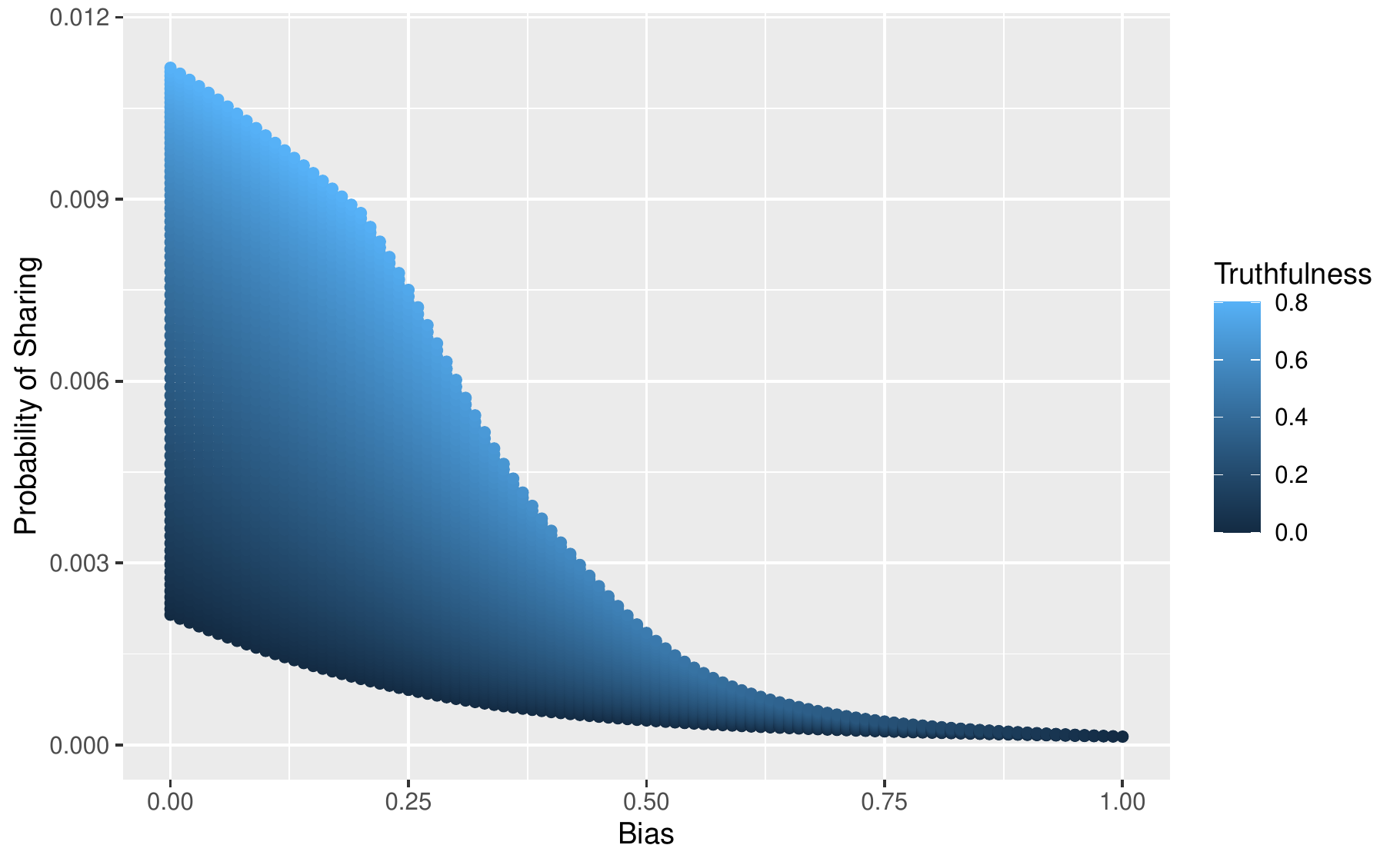}
    }
    \caption{The probability of sharing an article as a function of right political bias ($x$-axis) and truthfulness (gradient). Fitting parameters combination: $f_l$ - high, $k_l$ - high, $f_r$ - low, $k_r$ - low}
    \label{fig:biasbimrighthihilolo}

\end{figure}

\begin{figure}[H]
    \centering
    \subfloat[The empirical distribution of readers' political belief]{
    \includegraphics[width = 0.33\textwidth]{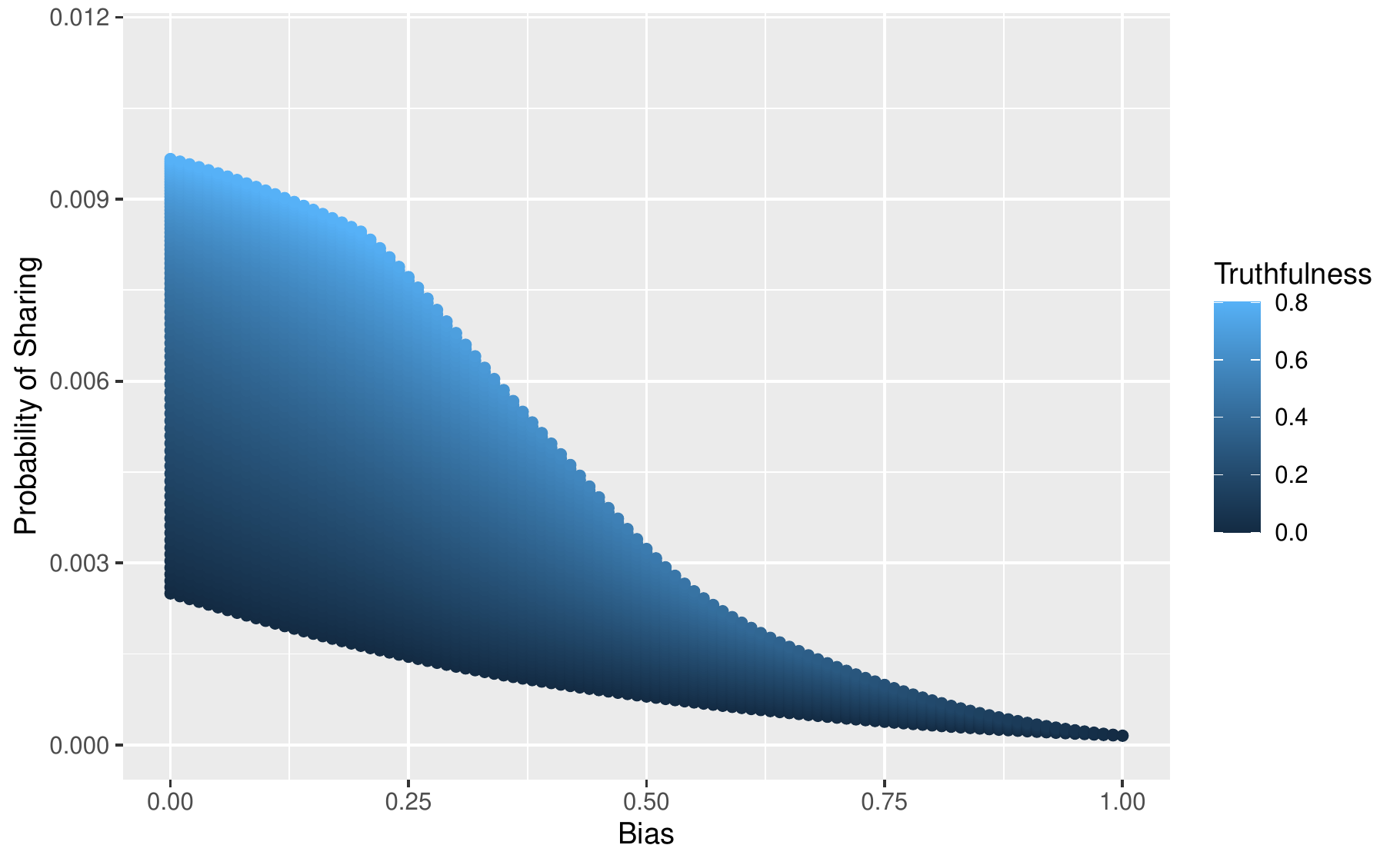}
    }
    \subfloat[A partisan bimodal distribution of readers' political belief 
    ]{
    \includegraphics[width = 0.33\textwidth]{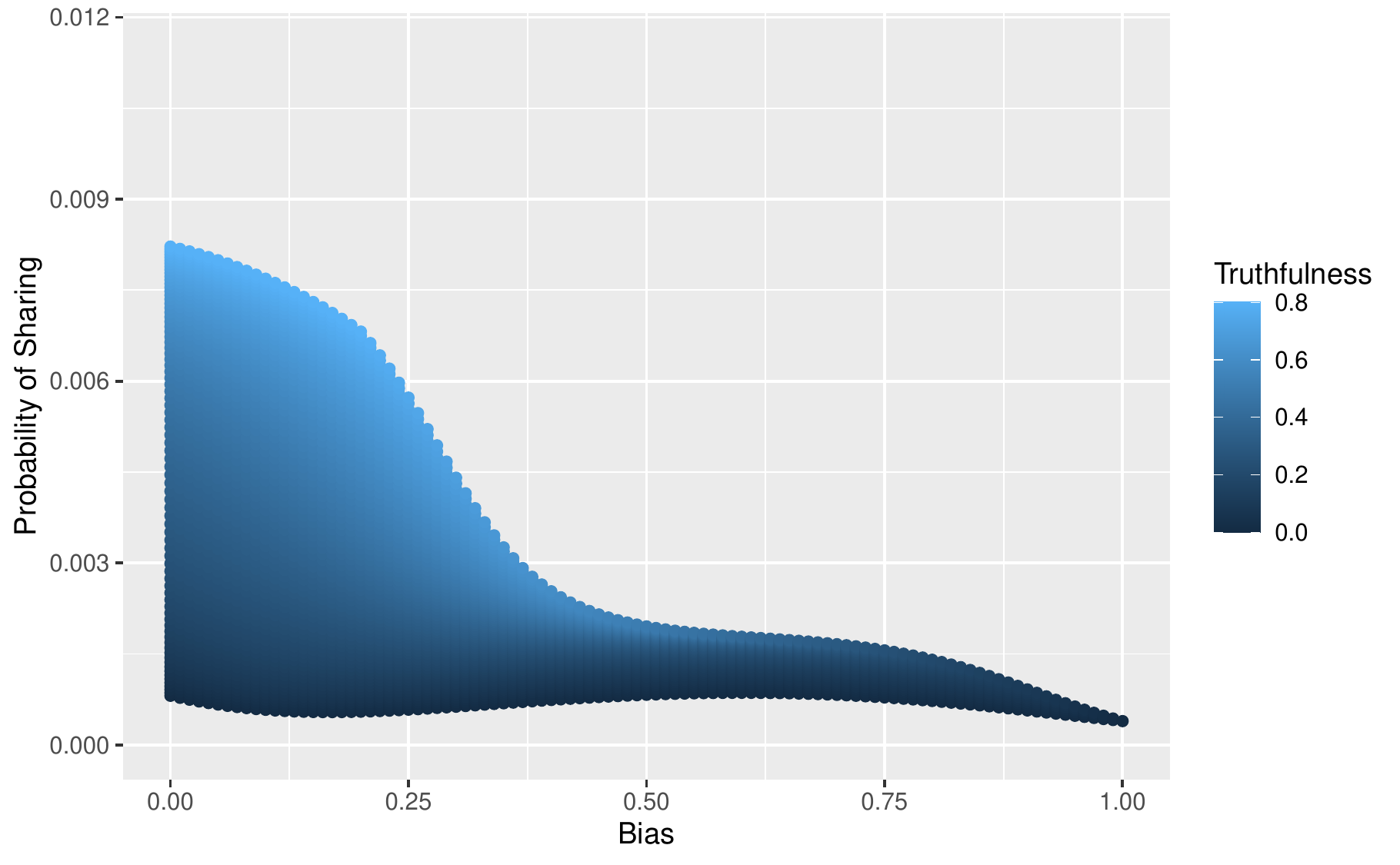}
    }
    \subfloat[A hyperpartisan bimodal distribution of readers' political belief 
    ]{
    \includegraphics[width = 0.33\textwidth]{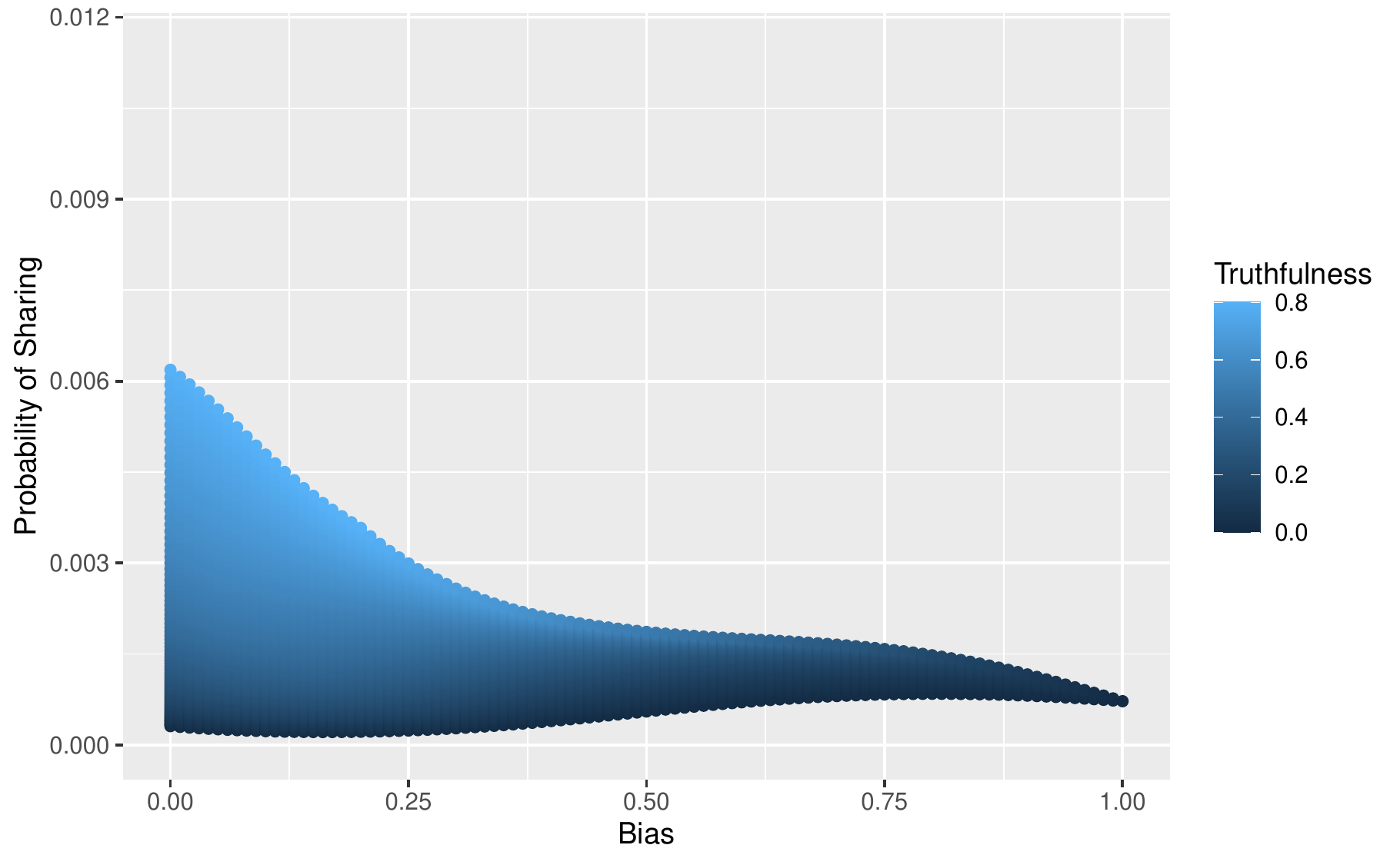}
    }
    
     \subfloat[A right-unimodal distribution of readers' political belief 
    ]{
    \includegraphics[width = 0.33\textwidth]{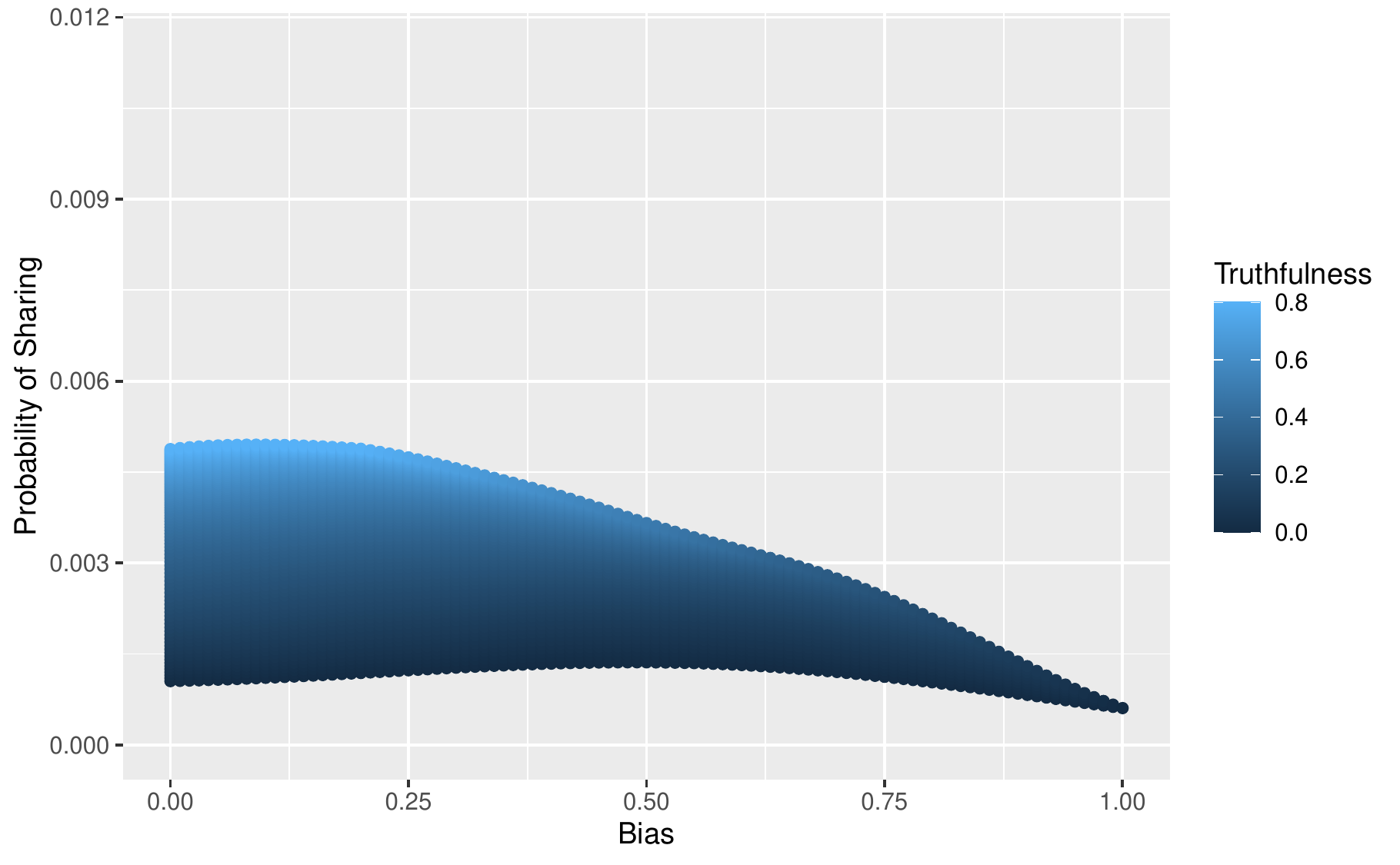}
    }
    \subfloat[A centrist-unimodal distribution of readers' political belief 
    ]{
    \includegraphics[width = 0.33\textwidth]{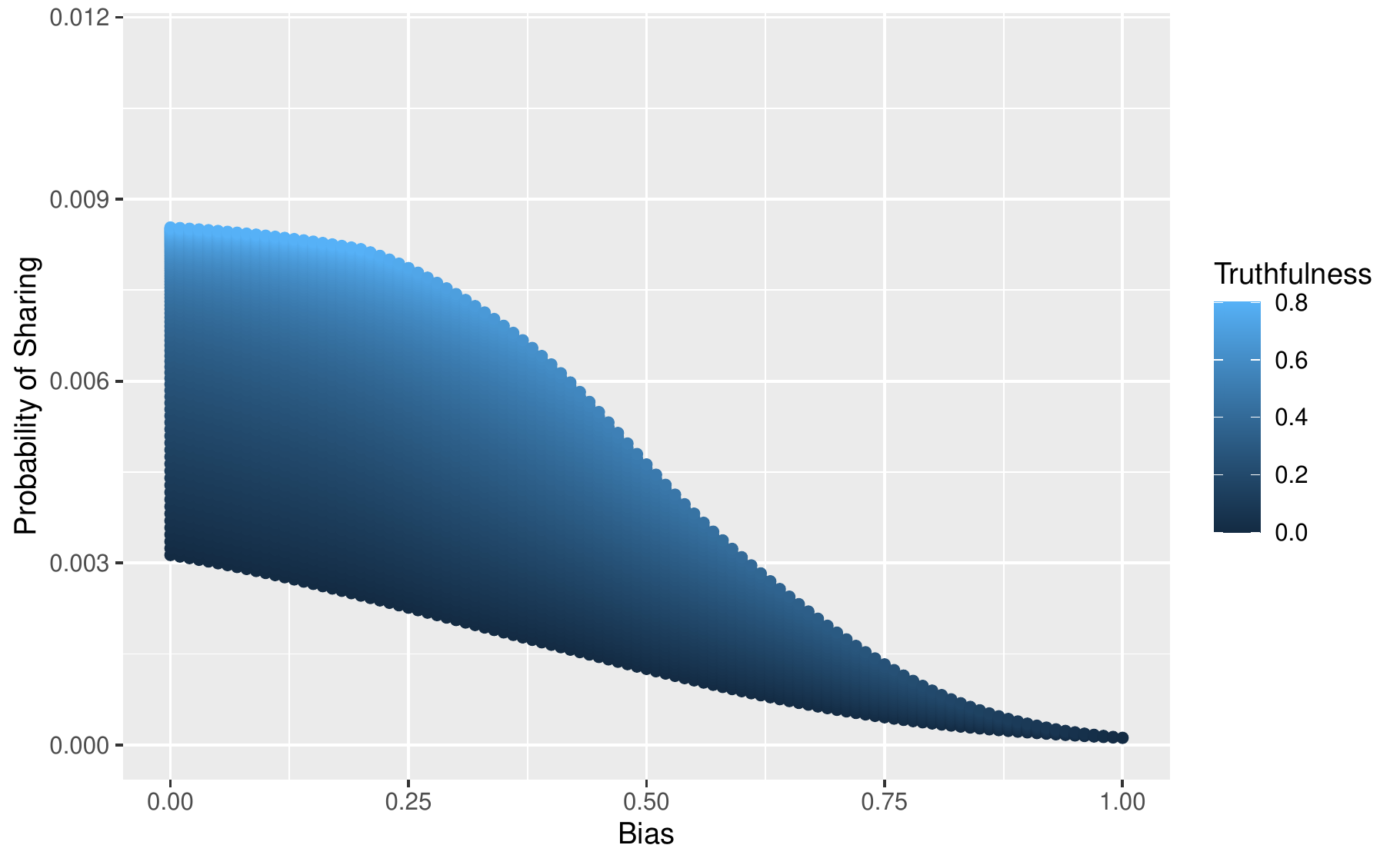}
    }
    \subfloat[A left-unimodal distribution of readers' political belief 
    ]{
    \includegraphics[width = 0.33\textwidth]{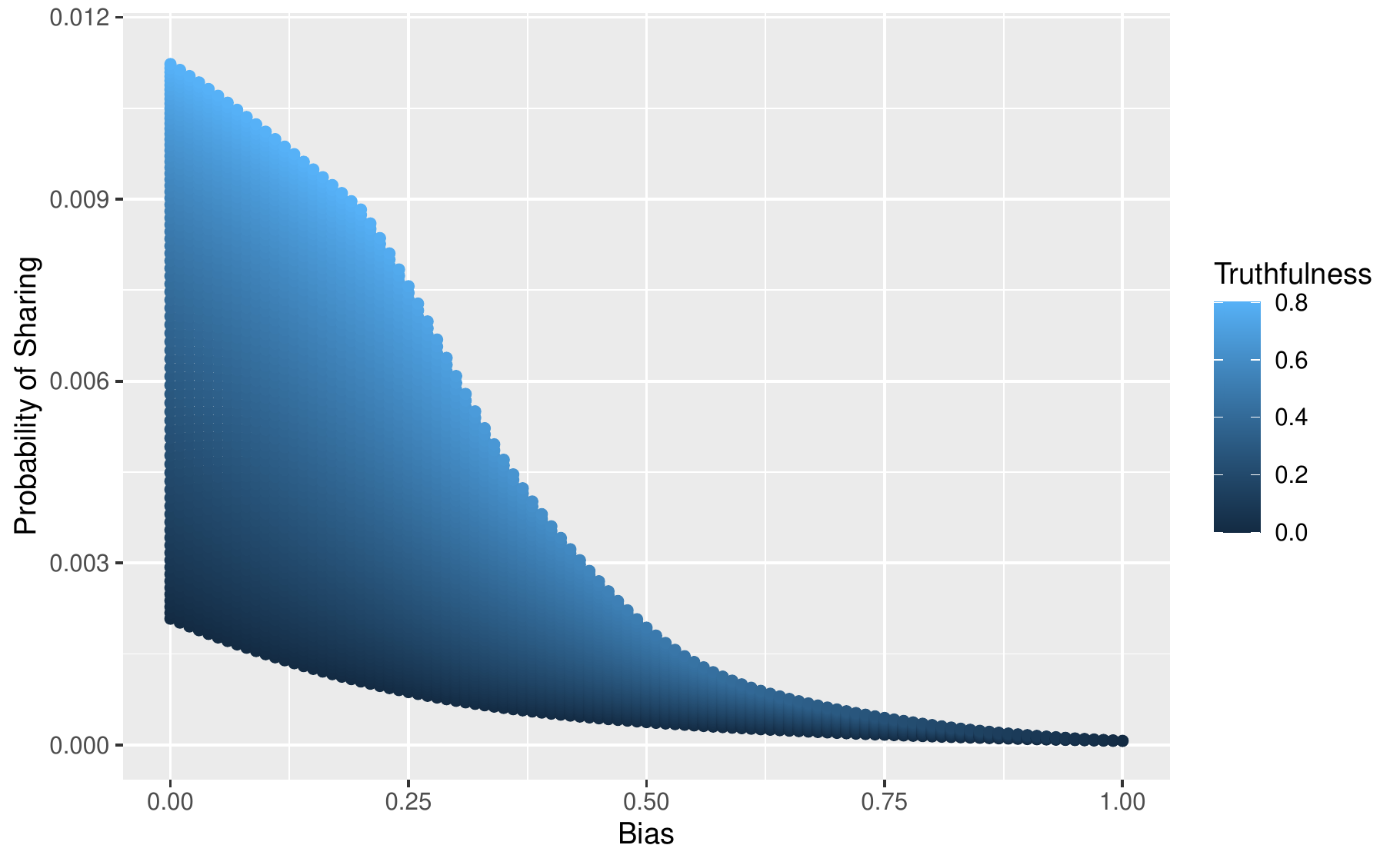}
    }
    \caption{The probability of sharing an article as a function of right political bias ($x$-axis) and truthfulness (gradient). Fitting parameters combination: $f_l$ - high, $k_l$ - high, $f_r$ - low, $k_r$ - high}
    \label{fig:biasbimrighthihilohi}

\end{figure}

\begin{figure}[H]
    \centering
    \subfloat[The empirical distribution of readers' political belief]{
    \includegraphics[width = 0.33\textwidth]{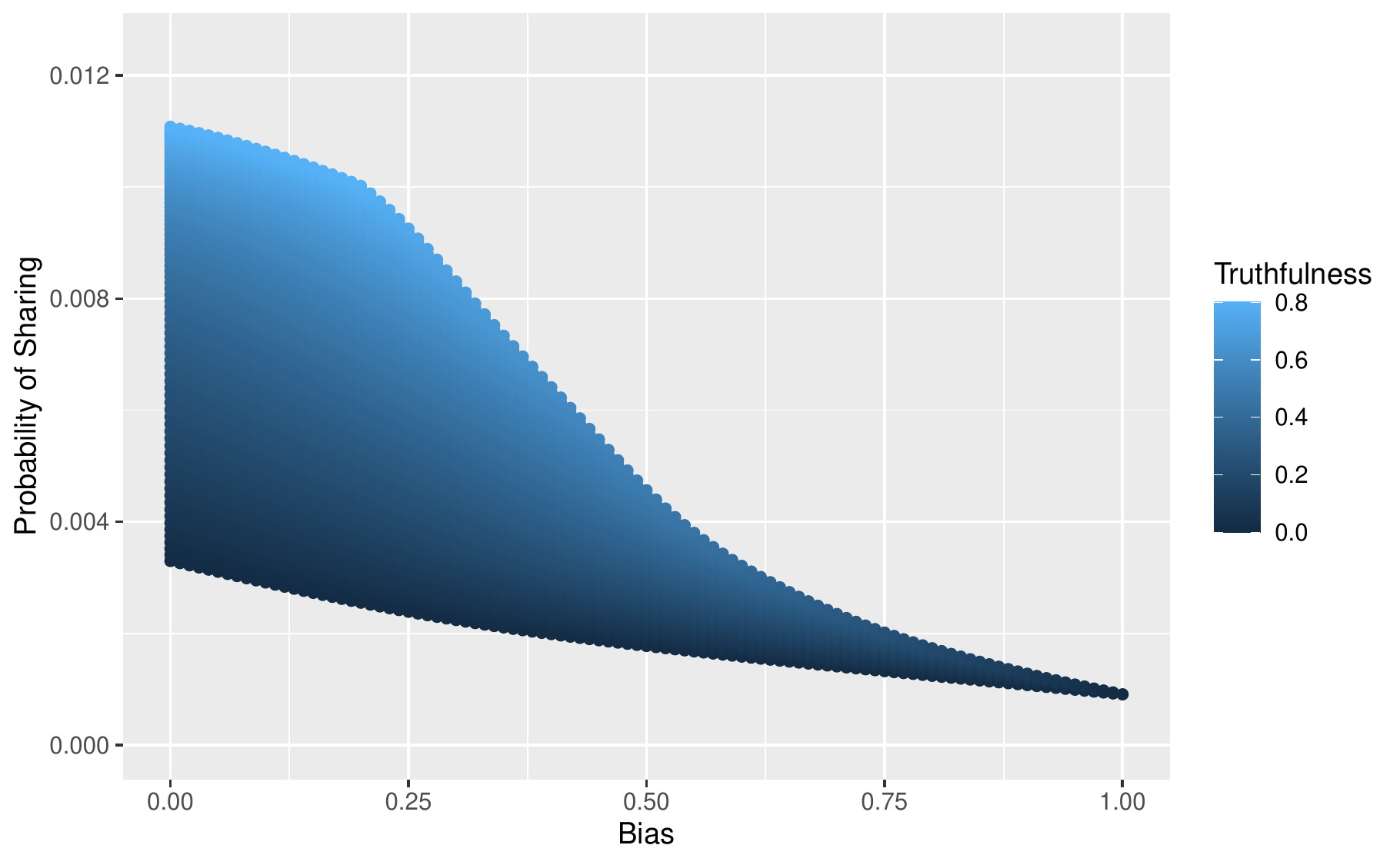}
    }
    \subfloat[A partisan bimodal distribution of readers' political belief 
    ]{
    \includegraphics[width = 0.33\textwidth]{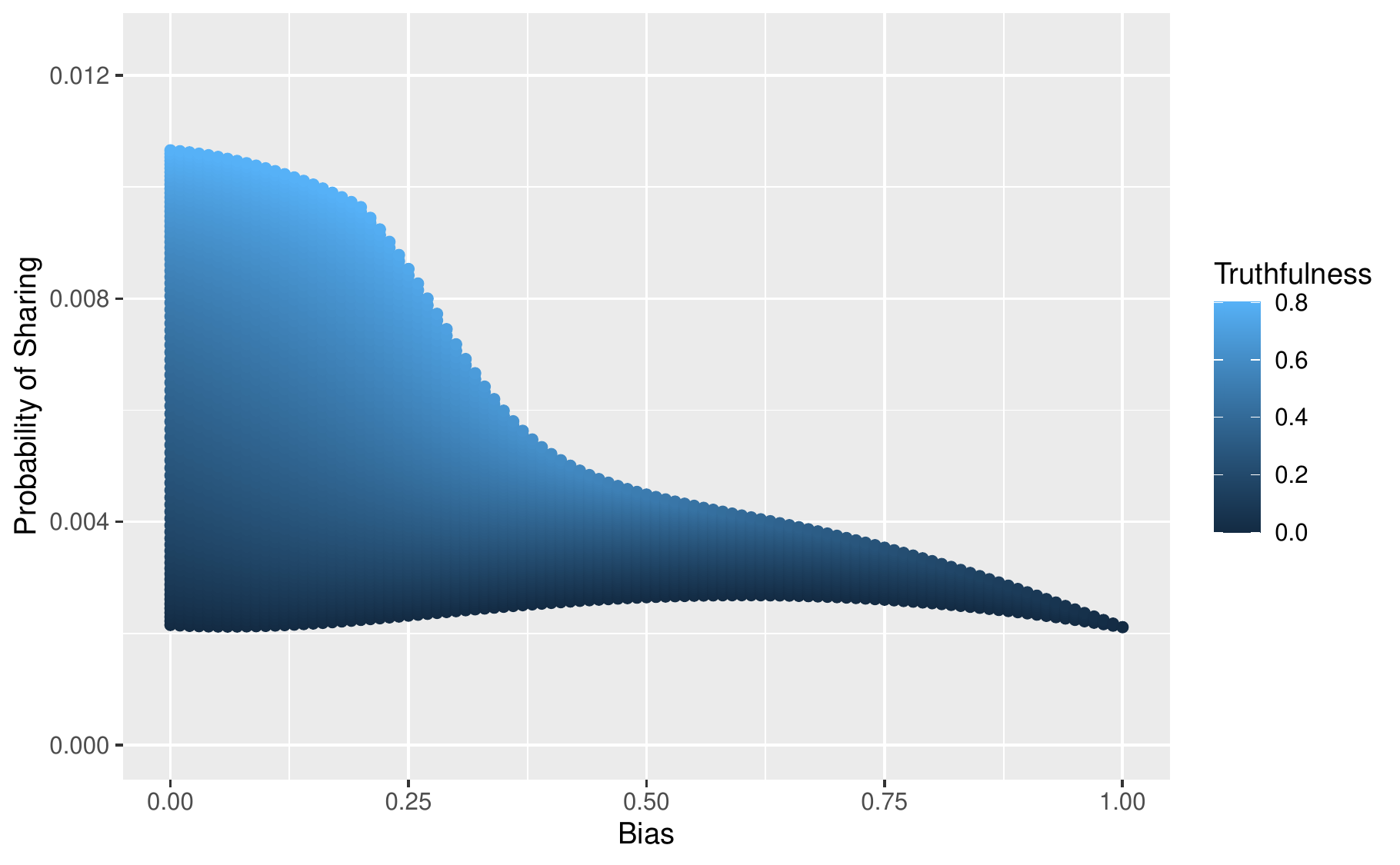}
    }
    \subfloat[A hyperpartisan bimodal distribution of readers' political belief 
    ]{
    \includegraphics[width = 0.33\textwidth]{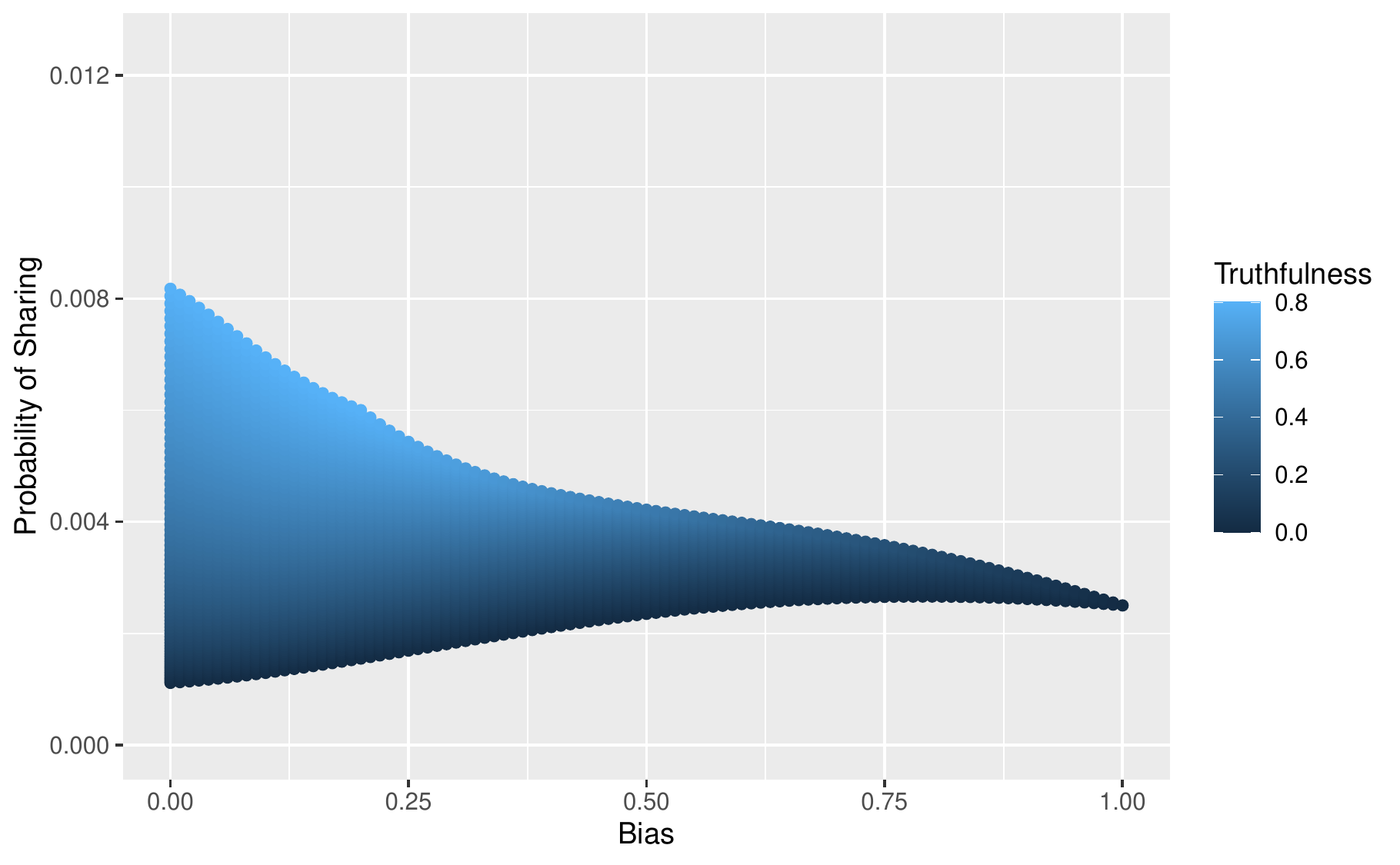}
    }
    
    \subfloat[A right-unimodal distribution of readers' political belief 
    ]{
    \includegraphics[width = 0.33\textwidth]{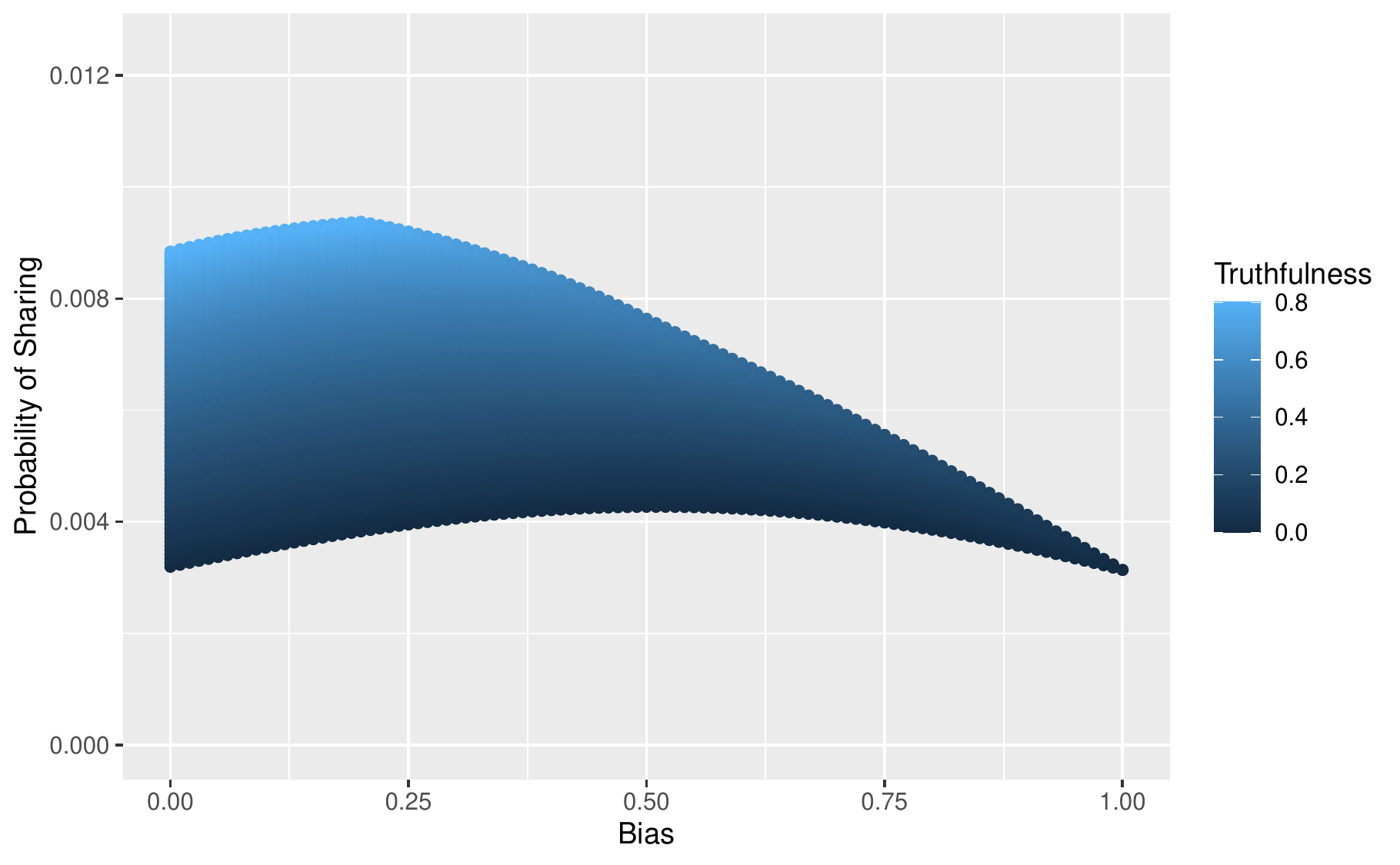}
    }
    \subfloat[A centrist-unimodal distribution of readers' political belief 
    ]{
    \includegraphics[width = 0.33\textwidth]{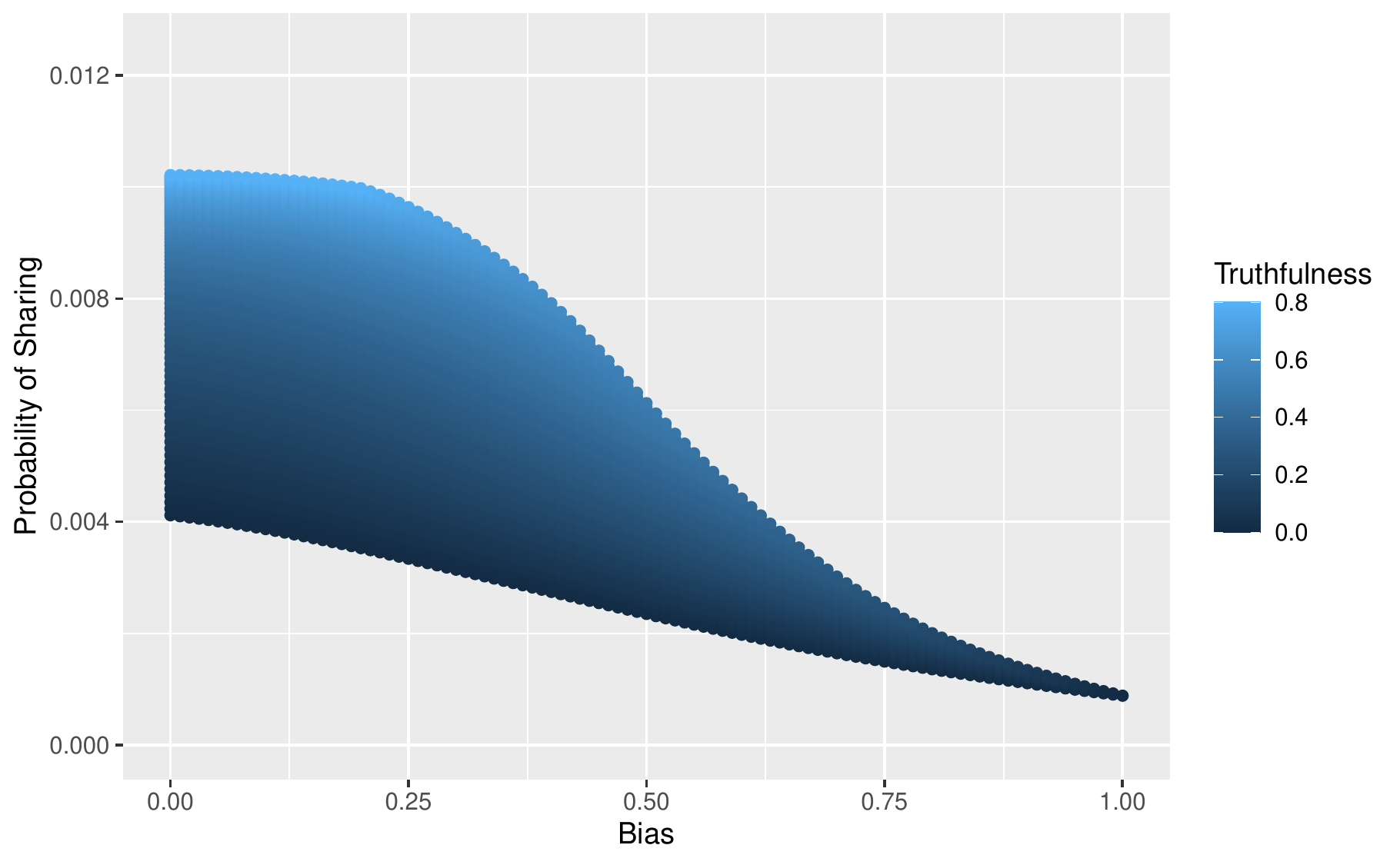}
    }
    \subfloat[A left-unimodal distribution of readers' political belief 
    ]{
    \includegraphics[width = 0.33\textwidth]{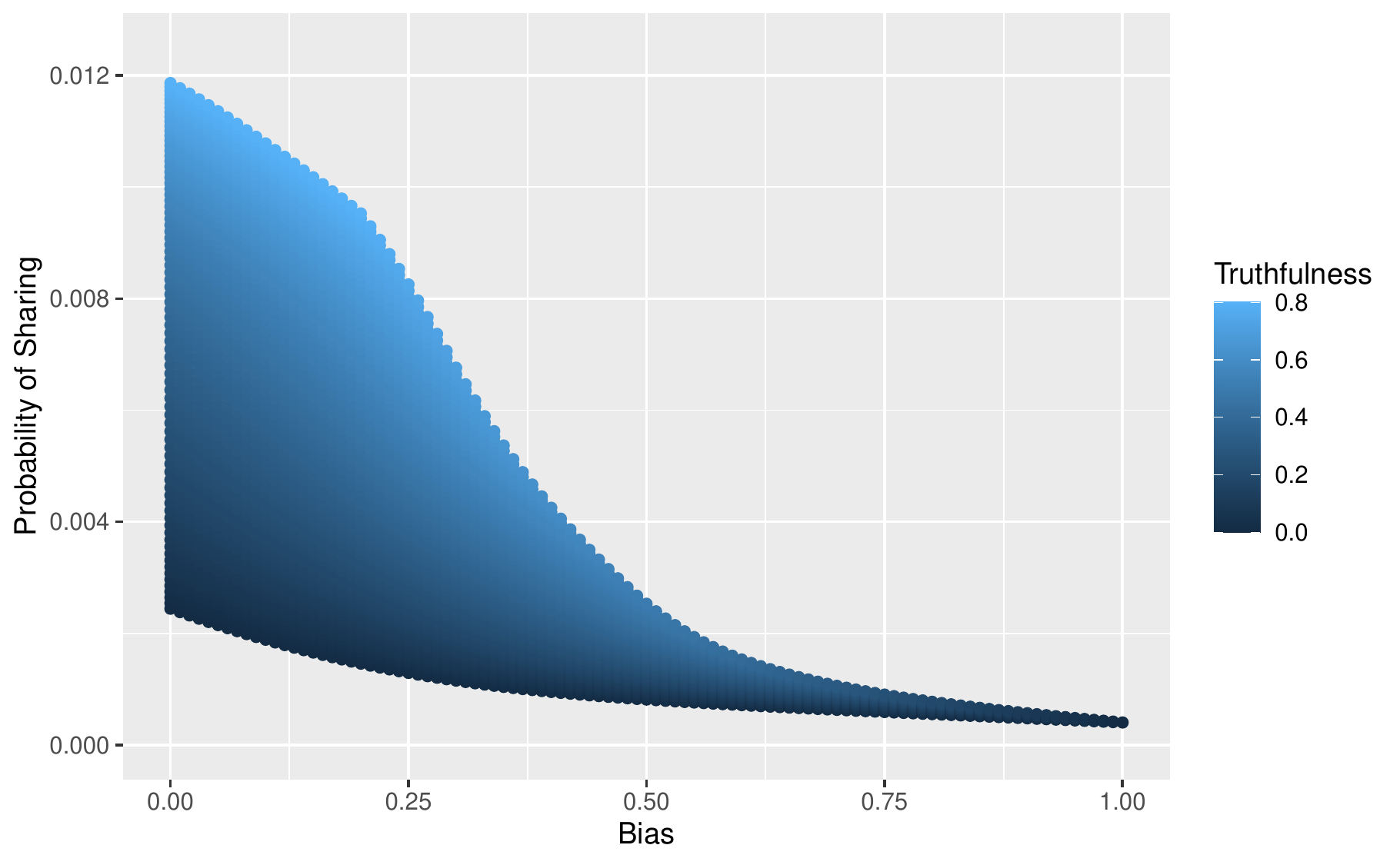}
    }
    \caption{The probability of sharing an article as a function of right political bias ($x$-axis) and truthfulness (gradient). Fitting parameters combination: $f_l$ - high, $k_l$ - high, $f_r$ - high, $k_r$ - low}
    \label{fig:biasbimrighthihihilo}

\end{figure}

\begin{figure}[H]
    \centering
    \subfloat[The empirical distribution of readers' political belief]{
    \includegraphics[width = 0.33\textwidth]{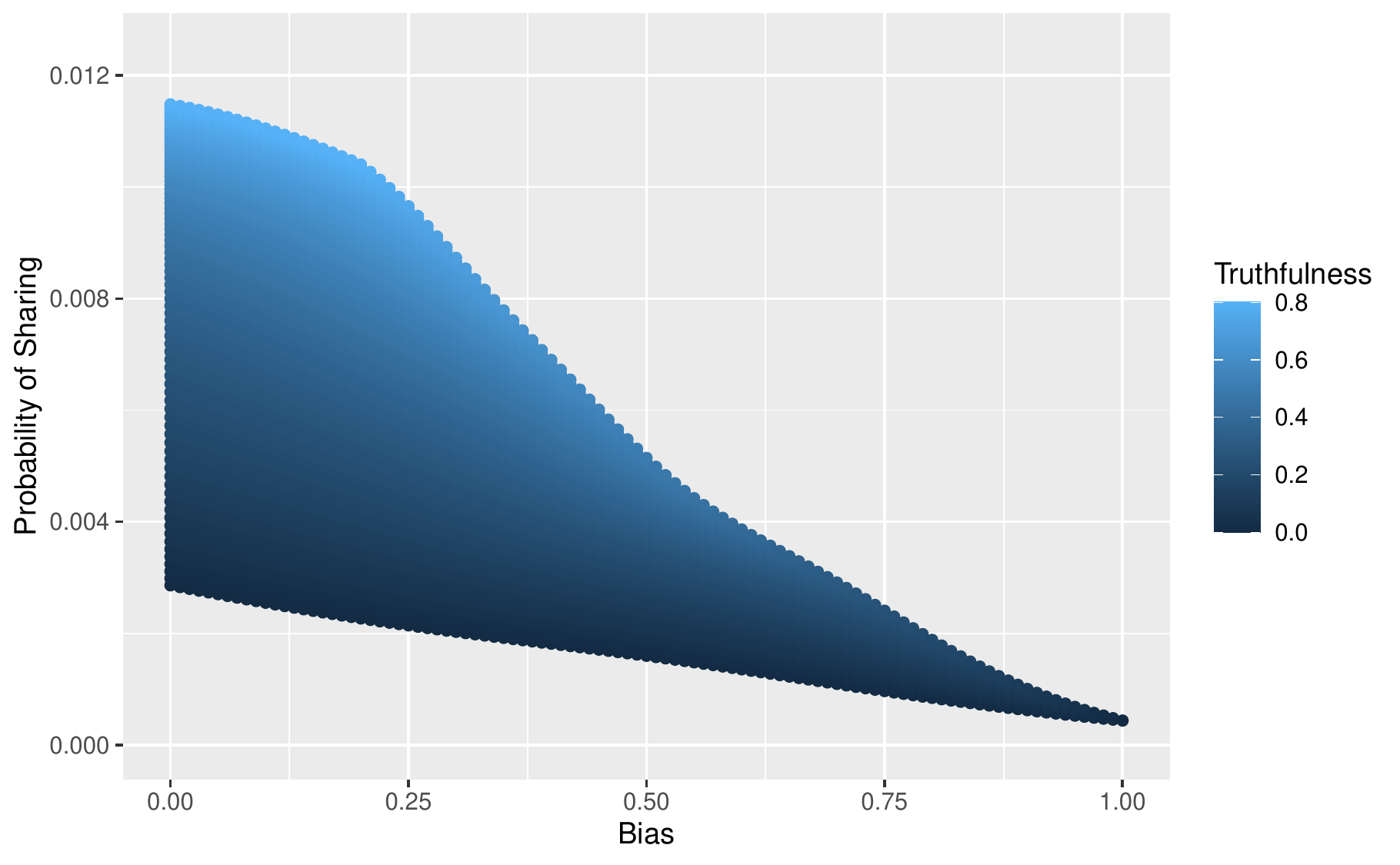}
    }
    \subfloat[A partisan bimodal distribution of readers' political belief 
    ]{
    \includegraphics[width = 0.33\textwidth]{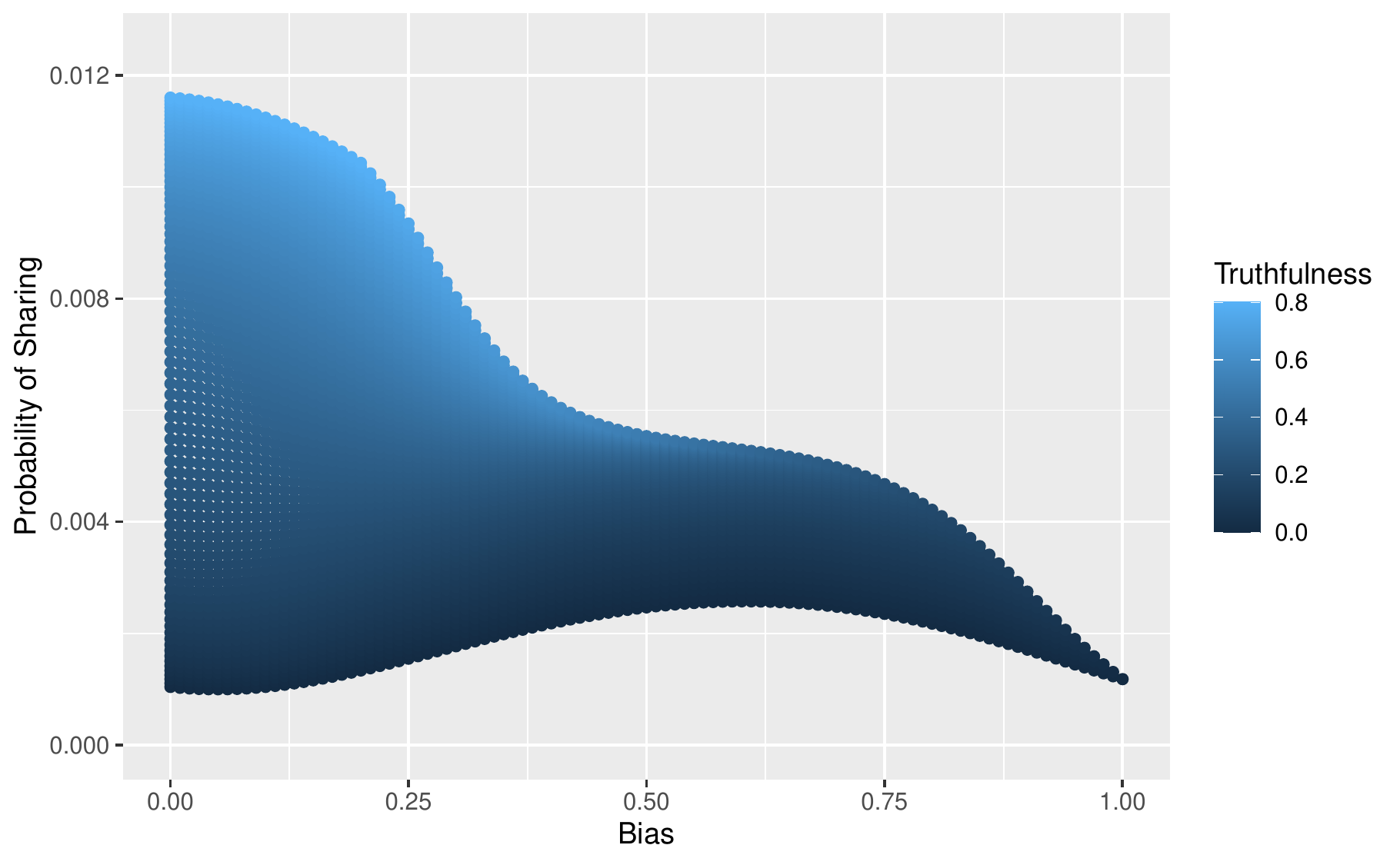}
    }
    \subfloat[A hyperpartisan bimodal distribution of readers' political belief 
    ]{
    \includegraphics[width = 0.33\textwidth]{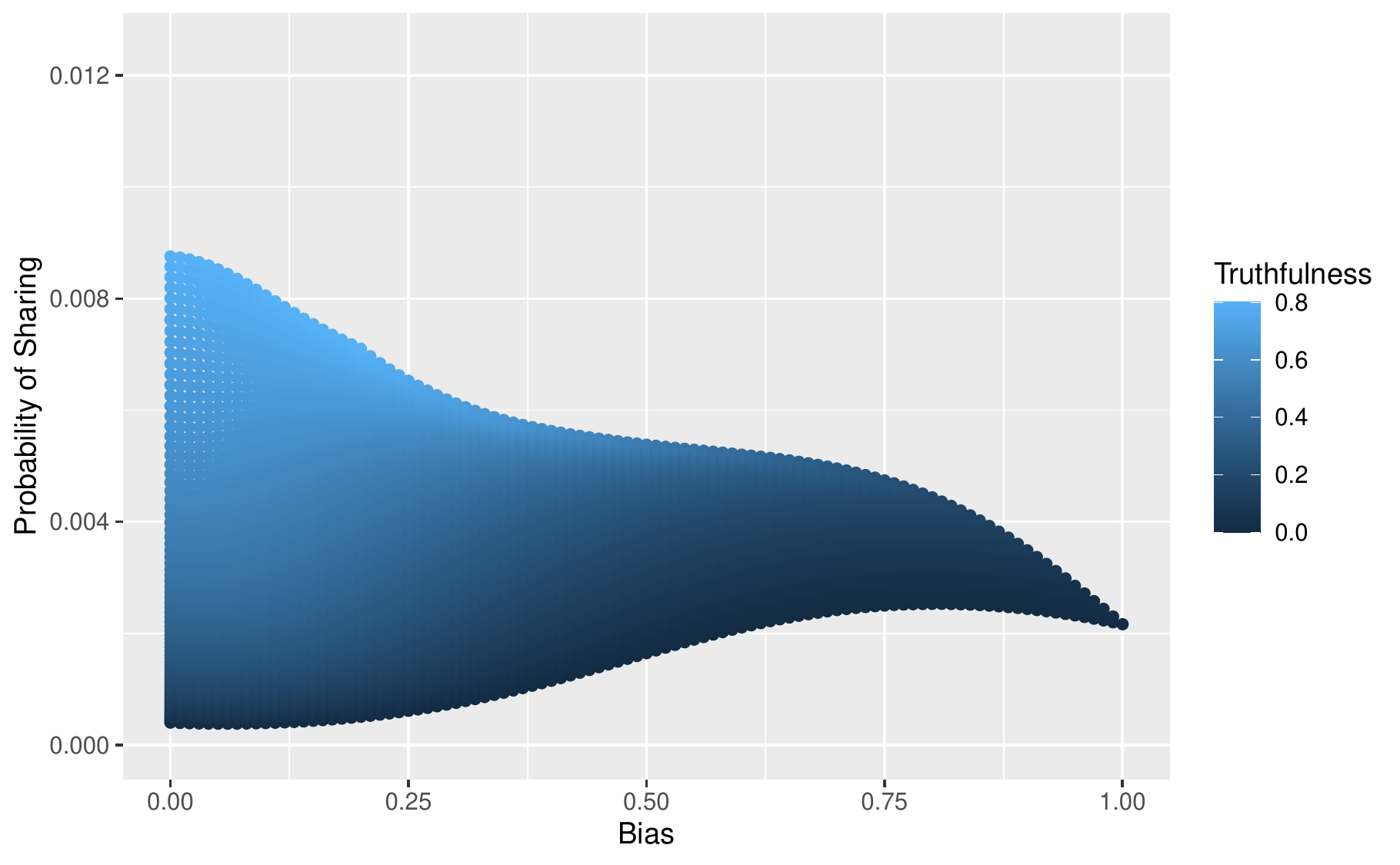}
    }
    
    \subfloat[A right-unimodal distribution of readers' political belief 
    ]{
    \includegraphics[width = 0.33\textwidth]{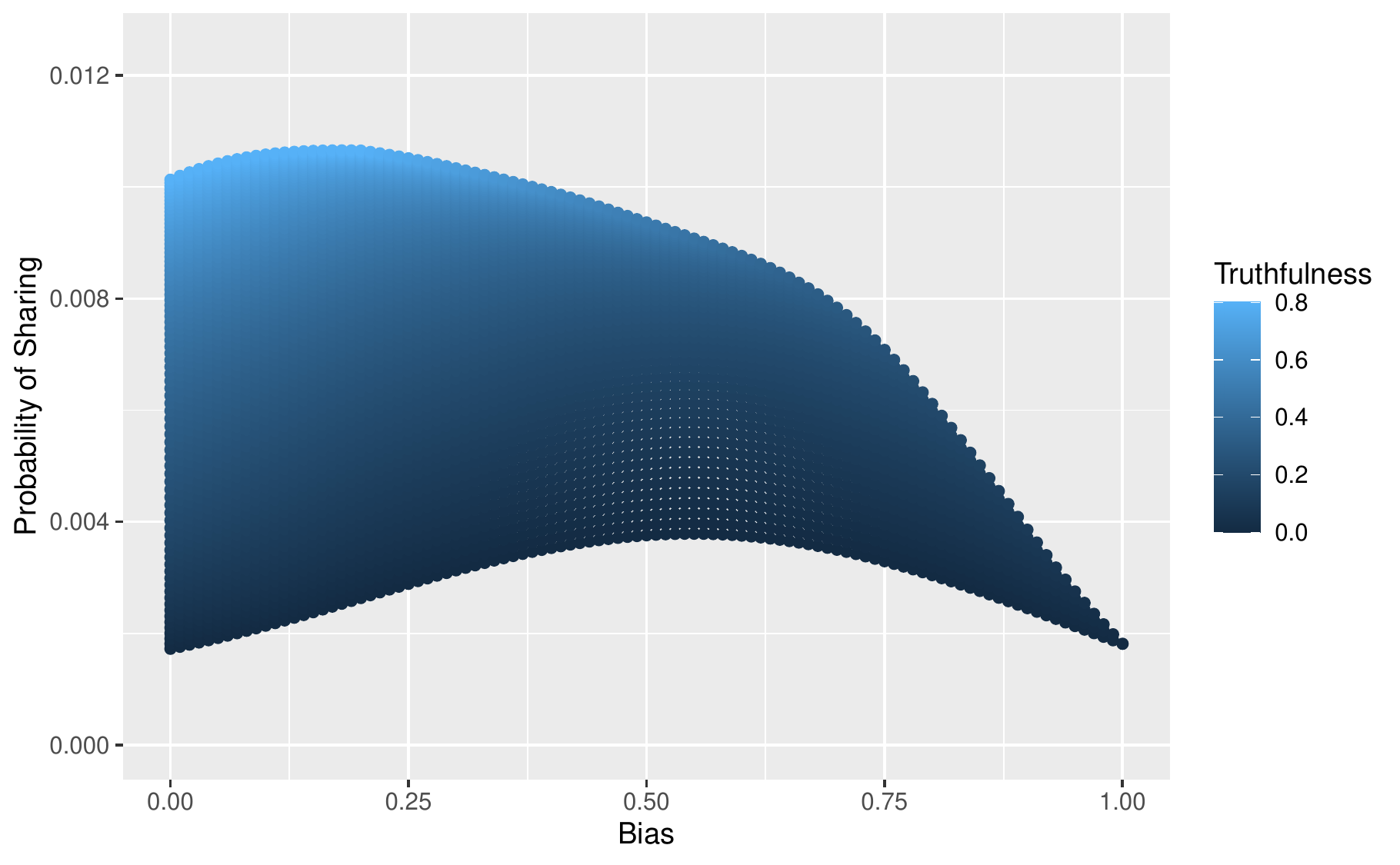}
    }
    \subfloat[A centrist-unimodal distribution of readers' political belief 
    ]{
    \includegraphics[width = 0.33\textwidth]{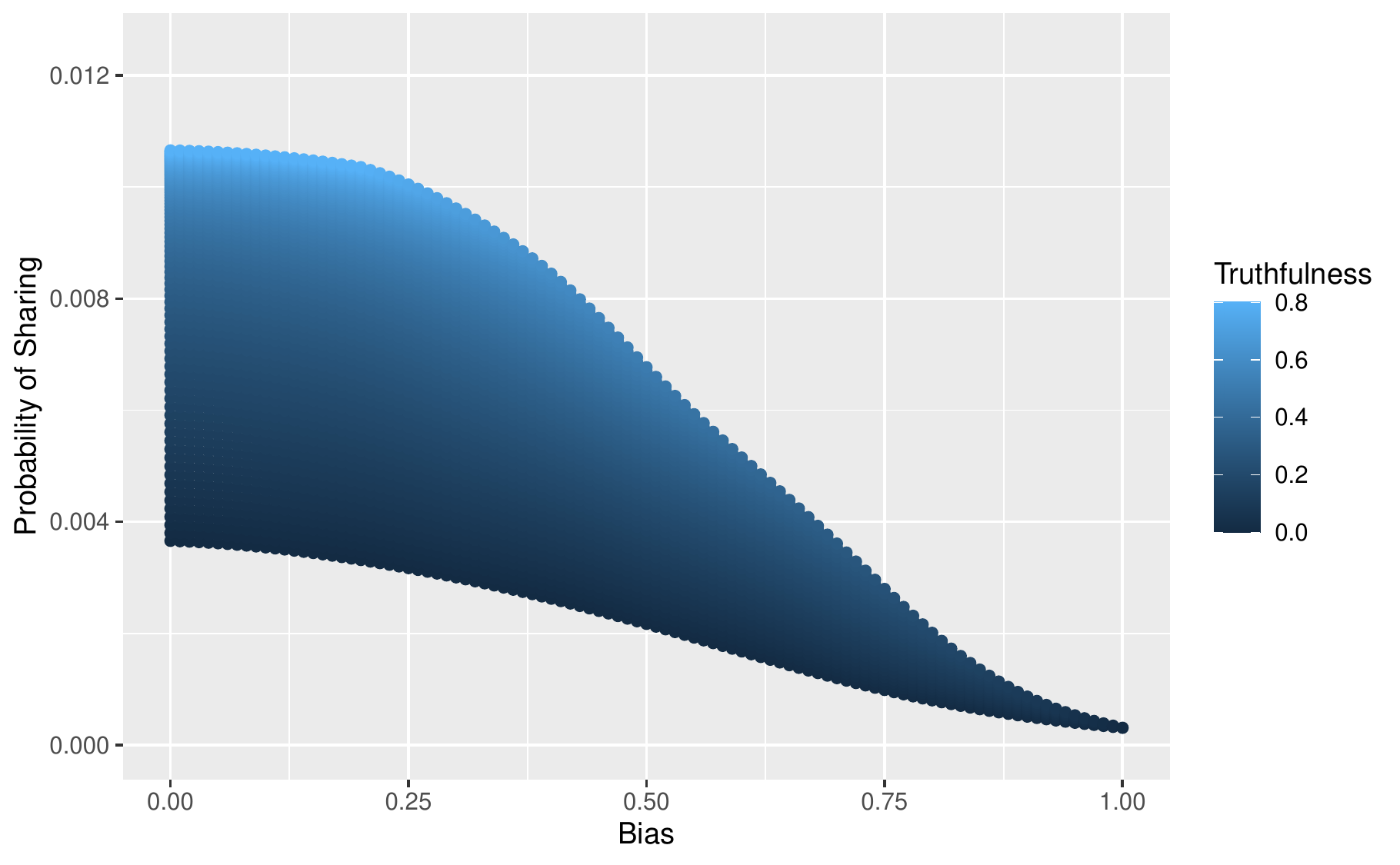}
    }
    \subfloat[A left-unimodal distribution of readers' political belief 
    ]{
    \includegraphics[width = 0.33\textwidth]{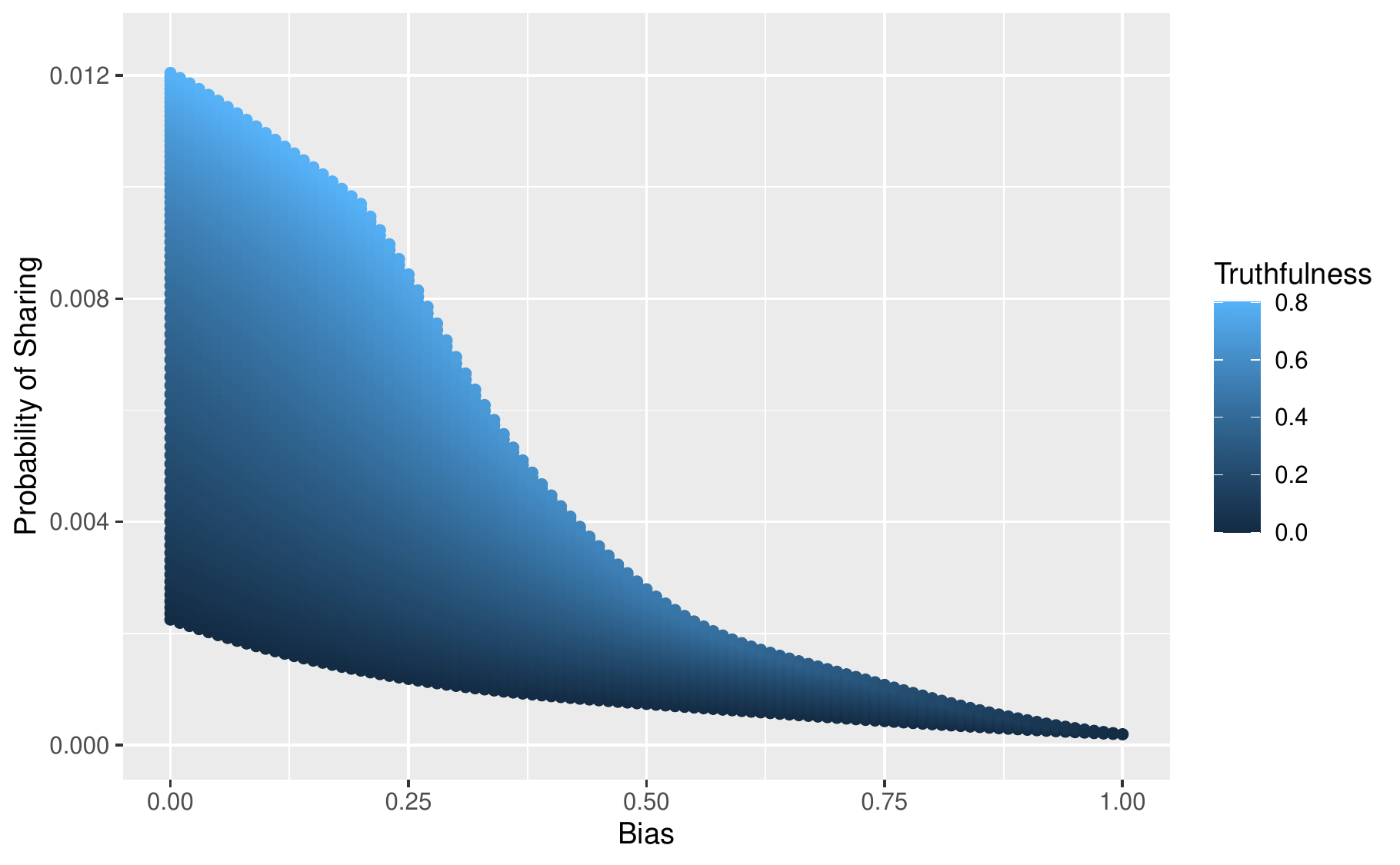}
    }
    \caption{The probability of sharing an article as a function of right political bias ($x$-axis) and truthfulness (gradient). Fitting parameters combination: $f_l$ - high, $k_l$ - high, $f_r$ - high, $k_r$ - high}
    \label{fig:biasbimrighthihihihi}

\end{figure}

\section{Proof that Population Diversity Reduces Sharing Rates}
\label{app:proof}

In Section \ref{s:discussion} we conclude from our empirical analysis that if ``users engage only with users of similar political beliefs, then the probability of sharing news, fake or true, increases. If users engage with users of opposing political beliefs, then the probability of sharing decreases, and the decrease is more pronounced for fake news.''  We can support this assertion mathematically with a toy example involving two, simple, hypothetical populations.

Consider population $U$ (for unimodal) wherein $100\%$ of the population has political belief $B_j$, and population $P$ (for partisan) wherein proportion $q$ of the population has political belief $B_j$ and the other proportion $1-q$ has political belief $-B_j$.  Consider an article having truthfulness $t_i$ and political bias $b_i$ such that $sign(b_i)=sign(B_j)$. Assume that the probability function parameters are equal on both sides of the political spectrum: $f_l = f_r$ and $k_l = k_r$.

We use equation (\ref{eq:probofbiandti}) to compare the probability of sharing this article in population $U$ to that in population $P$.
$$p^U(b_i, t_i) = p(b_i, t_i, B_j) = qp(b_i, t_i, B_j)+(1-q)p(b_i, t_i, B_j).$$  
$$p^P(b_i, t_i) = qp(b_i, t_i, B_j)+(1-q)p(b_i, t_i, -B_j).$$ Thus to compare $p^U(b_i, t_i)$ and $p^P(b_i, t_i)$, we need only compare $p(b_i, t_i, B_j)$ and $p(b_i, t_i, -B_j)$.

By the first assumption of our probability function, the probability of sharing an article decreases as the difference between the article's bias and the reader's belief increases.  $$|b_i - B_j| \leq |b_i| + |-B_j| 
= |b_i - (-B_j)|, \text{ when $b_i$ and $B_j$ have the same sign.}$$  Therefore $p(b_i, t_i, B_j) \geq p(b_i, t_i, -B_j)$, demonstrating that $p^U(b_i, t_i) \geq p^P(b_i, t_i)$.  Propagation is higher, for any truthfulness value, in the unimodal population than in the partisan population.  Moreover, because the deviation between $b_i$ and $B_j$ is attenuated multiplicatively by the term $e^{-kt_i}$ in equation (\ref{eq:probfcn}), the decrease in sharing probability between the partisan and unimodal populations is greatest for low-truth content.

\textit{Acknowledgement.}
This material is based upon work supported by the National Science Foundation under Grant No. DMS-1757952. Any opinions, findings, and conclusions or recommendations expressed in this material are those of the authors and do not necessarily reflect the views of the National Science Foundation. The authors also acknowledge financial support from Harvey Mudd College and the California State University at Long Beach. The authors would like to thank Michael Gao and Steven Witkin for their contributions to earlier versions of this work. Additionally, the authors thank David Lazer and Nir Grinberg for providing access to the data set from (\cite{grinberg_2017}).  Finally, the authors thank three anonymous reviewers for detailed feedback that greatly improved this manuscript.

\printbibliography

\end{document}